# Modern Computer Arithmetic

Richard P. Brent and Paul Zimmermann

Version 0.5.1

# Contents

# Preface

This is a book about algorithms for performing arithmetic, and their implementation on modern computers. We are concerned with software more than hardware — we do not cover computer architecture or the design of computer hardware since good books are already available on these topics. Instead we focus on algorithms for efficiently performing arithmetic operations such as addition, multiplication and division, and their connections to topics such as modular arithmetic, greatest common divisors, the Fast Fourier Transform (FFT), and the computation of special functions.

The algorithms that we present are mainly intended for arbitrary-precision arithmetic. That is, they are not limited by the computer wordsize of 32 or 64 bits, only by the memory and time available for the computation. We consider both integer and real (floating-point) computations.

The book is divided into four main chapters, plus one short chapter (essentially an appendix). Chapter 1 covers integer arithmetic. This has, of course, been considered in many other books and papers. However, there has been much recent progress, inspired in part by the application to public key cryptography, so most of the published books are now partly out of date or incomplete. Our aim is to present the latest developments in a concise manner. At the same time, we provide a self-contained introduction for the reader who is not an expert in the field.

Chapter 2 is concerned with modular arithmetic and the FFT, and their applications to computer arithmetic. We consider different number representations, fast algorithms for multiplication, division and exponentiation, and the use of the Chinese Remainder Theorem (CRT).

Chapter 3 covers floating-point arithmetic. Our concern is with high-precision floating-point arithmetic, implemented in software if the precision provided by the hardware (typically IEEE standard 53-bit significand) is inadequate. The algorithms described in this chapter focus on *correct rounding*, extending the IEEE standard to arbitrary precision.

Chapter 4 deals with the computation, to arbitrary precision, of functions such as sqrt, exp, ln, sin, cos, and more generally functions defined by power series or continued fractions. Of course, the computation of special functions is a huge topic so we have had to be selective. In particular, we have concentrated on methods that are efficient and suitable for arbitrary-precision computations.

The last chapter contains pointers to implementations, useful web sites, mailing lists, and so on. Finally, at the end there is a one-page *Summary of Complexities* which should be a useful *aide-mémoire*.

The chapters are fairly self-contained, so it is possible to read them out of order. For example, Chapter 4 could be read before Chapters 1–3, and Chapter 5 can be consulted at any time. Some topics, such as Newton's method, appear in different guises in several chapters. Cross-references are given where appropriate.

For details that are omitted we give pointers in the *Notes and References* sections of each chapter, as well as in the bibliography. We have tried, as far as possible, to keep the main text uncluttered by footnotes and references, so most references are given in the *Notes and References* sections.

The book is intended for anyone interested in the design and implementation of efficient algorithms for computer arithmetic, and more generally efficient numerical algorithms. We did our best to present algorithms that are ready to implement in your favorite language, while keeping a high-level description and not getting too involved in low-level or machine-dependent details. An alphabetical list of algorithms can be found in the index.

Although the book is not specifically intended as a textbook, it could be used in a graduate course in mathematics or computer science, and for this reason, as well as to cover topics that could not be discussed at length in the text, we have included exercises at the end of each chapter. The exercises vary considerably in difficulty, from easy to small research projects, but we have not attempted to assign them a numerical rating. For solutions to the exercises, please contact the authors.

*We welcome comments and corrections. Please send them to either of the authors.*

Richard Brent and Paul Zimmermann
MCA@rpbrent.com
Paul.Zimmermann@inria.fr
Canberra and Nancy, February 2010

# Acknowledgements

We thank the French National Institute for Research in Computer Science and Control (INRIA), the Australian National University (ANU), and the Australian Research Council (ARC), for their support. The book could not have been written without the contributions of many friends and colleagues, too numerous to mention here, but acknowledged in the text and in the *Notes and References* sections at the end of each chapter.

We also thank those who have sent us comments on and corrections to earlier versions of this book: Jörg Arndt, Marco Bodrato, Wolfgang Ehrhardt (with special thanks), Steven Galbraith, Torbjörn Granlund, Guillaume Hanrot, Marc Mezzarobba, Jean-Michel Muller, Denis Roegel, Wolfgang Schmid, Arnold Schönhage, Sidi Mohamed Sedjelmaci, Emmanuel Thomé, and Mark Wezelenburg. Two anonymous reviewers provided very helpful suggestions.

The *Mathematics Genealogy Project* (`http://www.genealogy.ams.org/`) and Don Knuth's *The Art of Computer Programming* [143] were useful resources for details of entries in the index.

We also thank the authors of the LaTeX program, which allowed us to produce this book, the authors of the `gnuplot` program, and the authors of the GNU MP library, which helped us to illustrate several algorithms with concrete figures.

Finally, we acknowledge the contribution of Erin Brent, who first suggested writing the book; and thank our wives, Judy-anne and Marie, for their patience and encouragement.

# Notation

$\mathbb{C}$          set of complex numbers
$\widehat{\mathbb{C}}$          set of extended complex numbers $\mathbb{C} \cup \{\infty\}$
$\mathbb{N}$          set of natural numbers (nonnegative integers)
$\mathbb{N}^*$          set of positive integers $\mathbb{N} \backslash \{0\}$
$\mathbb{Q}$          set of rational numbers
$\mathbb{R}$          set of real numbers
$\mathbb{Z}$          set of integers
$\mathbb{Z}/n\mathbb{Z}$          ring of residues modulo $n$
$C^n$          set of (real or complex) functions with $n$ continuous derivatives
          in the region of interest

$\Re(z)$          real part of a complex number $z$
$\Im(z)$          imaginary part of a complex number $z$
$\bar{z}$          conjugate of a complex number $z$
$|z|$          Euclidean norm of a complex number $z$,
          or absolute value of a scalar $z$

$B_n$          Bernoulli numbers, $\sum_{n \geq 0} B_n z^n / n! = z/(e^z - 1)$

$C_n$          scaled Bernoulli numbers, $C_n = B_{2n}/(2n)!$, $\sum C_n z^{2n} = (z/2)/\tanh(z/2)$

$T_n$          tangent numbers, $\sum T_n z^{2n-1}/(2n-1)! = \tan z$

$H_n$          harmonic number $\sum_{j=1}^n 1/j$   (0 if $n \leq 0$)

$\binom{n}{k}$          binomial coefficient "$n$ choose $k$" $= n!/(k!\,(n-k)!)$   (0 if $k < 0$ or $k > n$)

$\beta$              "word" base (usually $2^{32}$ or $2^{64}$) or "radix" (floating-point)

$n$              "precision": number of base $\beta$ digits in an integer or in a
                floating-point significand, or a free variable

$\varepsilon$              "machine precision" $\beta^{1-n}/2$ (in complexity bounds)
                an arbitrarily small positive constant

$\eta$              smallest positive subnormal number

$\circ(x), \circ_n(x)$    rounding of real number $x$ in precision $n$ (Definition 3.1.1)

$\text{ulp}(x)$          for a floating-point number $x$, one unit in the last place

$M(n)$            time to multiply $n$-bit integers, or polynomials of
                degree $n-1$, depending on the context

$\sim M(n)$          a function $f(n)$ such that $f(n)/M(n) \to 1$ as $n \to \infty$
                (we sometimes lazily omit the "$\sim$" if the meaning is clear)

$M(m, n)$          time to multiply an $m$-bit integer by an $n$-bit integer

$D(n)$            time to divide a $2n$-bit integer by an $n$-bit integer,
                giving quotient and remainder

$D(m, n)$          time to divide an $m$-bit integer by an $n$-bit integer,
                giving quotient and remainder

$a|b$              $a$ is a divisor of $b$, that is $b = ka$ for some $k \in \mathbb{Z}$

$a = b \bmod m$        modular equality, $m|(a-b)$

$q \leftarrow a \text{ div } b$      assignment of integer quotient to $q$ ($0 \le a - qb < b$)

$r \leftarrow a \bmod b$      assignment of integer remainder to $r$ ($0 \le r = a - qb < b$)

$(a, b)$            greatest common divisor of $a$ and $b$

$\left(\frac{a}{b}\right)$ or $(a|b)$    Jacobi symbol ($b$ odd and positive)

iff              if and only if

$i \wedge j$            bitwise *and* of integers $i$ and $j$,
                or logical *and* of two Boolean expressions

$i \vee j$            bitwise *or* of integers $i$ and $j$,
                or logical *or* of two Boolean expressions

$i \oplus j$            bitwise *exclusive-or* of integers $i$ and $j$

$i \ll k$            integer $i$ multiplied by $2^k$

$i \gg k$            quotient of division of integer $i$ by $2^k$

$a \cdot b, \quad a \times b$    product of scalars $a$, $b$

$a * b$            cyclic convolution of vectors $a$, $b$

$\nu(n)$            2-valuation: largest $k$ such that $2^k$ divides $n$ ($\nu(0) = \infty$)

$\sigma(e)$            length of the shortest addition chain to compute $e$

$\phi(n)$            Euler's totient function, $\#\{m : 0 < m \le n \wedge (m, n) = 1\}$

| | |
|---|---|
| $\deg(A)$ | for a polynomial $A$, the degree of $A$ |
| $\mathrm{ord}(A)$ | for a power series $A = \sum_j a_j z^j$, |
| | $\mathrm{ord}(A) = \min\{j : a_j \neq 0\}$ $(\mathrm{ord}(0) = +\infty)$ |
| $\exp(x)$ or $e^x$ | exponential function |
| $\ln(x)$ | natural logarithm |
| $\log_b(x)$ | base-$b$ logarithm $\ln(x)/\ln(b)$ |
| $\lg(x)$ | base-2 logarithm $\ln(x)/\ln(2) = \log_2(x)$ |
| $\log(x)$ | logarithm to any fixed base |
| $\log^k(x)$ | $(\log x)^k$ |
| $\lceil x \rceil$ | ceiling function, $\min\{n \in \mathbb{Z} : n \geq x\}$ |
| $\lfloor x \rfloor$ | floor function, $\max\{n \in \mathbb{Z} : n \leq x\}$ |
| $\lfloor x \rceil$ | nearest integer function, $\lfloor x + 1/2 \rfloor$ |
| $\mathrm{sign}(n)$ | $+1$ if $n > 0$, $-1$ if $n < 0$, and $0$ if $n = 0$ |
| $\mathrm{nbits}(n)$ | $\lfloor \lg(n) \rfloor + 1$ if $n > 0$, $0$ if $n = 0$ |
| $[a, b]$ | closed interval $\{x \in \mathbb{R} : a \leq x \leq b\}$ (empty if $a > b$) |
| $(a, b)$ | open interval $\{x \in \mathbb{R} : a < x < b\}$ (empty if $a \geq b$) |
| $[a, b), \quad (a, b]$ | half-open intervals, $a \leq x < b$, $a < x \leq b$ respectively |
| $^t[a, b]$ or $[a, b]^t$ | column vector $\begin{pmatrix} a \\ b \end{pmatrix}$ |
| $[a, b; c, d]$ | $2 \times 2$ matrix $\begin{pmatrix} a & b \\ c & d \end{pmatrix}$ |
| $\widehat{a_j}$ | element of the (forward) Fourier transform of vector $a$ |
| $\widetilde{a_j}$ | element of the backward Fourier transform of vector $a$ |
| $f(n) = O(g(n))$ | $\exists c, n_0$ such that $|f(n)| \leq cg(n)$ for all $n \geq n_0$ |
| $f(n) = \Omega(g(n))$ | $\exists c > 0, n_0$ such that $|f(n)| \geq cg(n)$ for all $n \geq n_0$ |
| $f(n) = \Theta(g(n))$ | $f(n) = O(g(n))$ and $g(n) = O(f(n))$ |
| $f(n) \sim g(n)$ | $f(n)/g(n) \to 1$ as $n \to \infty$ |
| $f(n) = o(g(n))$ | $f(n)/g(n) \to 0$ as $n \to \infty$ |
| $f(n) \ll g(n)$ | $f(n) = O(g(n))$ |
| $f(n) \gg g(n)$ | $g(n) \ll f(n)$ |
| $f(x) \sim \sum_0^n a_j/x^j$ | $f(x) - \sum_0^n a_j/x^j = o(1/x^n)$ as $x \to +\infty$ |
| $123\,456\,789$ | 123456789 (for large integers, we may use a space after every third digit) |
| $xxx.yyy_\rho$ | a number $xxx.yyy$ written in base $\rho$; for example, the decimal number 3.25 is $11.01_2$ in binary |
| $\frac{a}{b+}\,\frac{c}{d+}\,\frac{e}{f+}\,\cdots$ | continued fraction $a/(b + c/(d + e/(f + \cdots)))$ |
| $\lvert A \rvert$ | determinant of a matrix $A$, e.g. $\begin{vmatrix} a & b \\ c & d \end{vmatrix} = ad - bc$ |
| $\mathrm{PV}\int_a^b f(x)\,\mathrm{d}x$ | Cauchy principal value integral, defined by a limit if $f$ has a singularity in $(a, b)$ |
| $s \,\|\, t$ | concatenation of strings $s$ and $t$ |
| $\triangleright$ `<text>` | comment in an algorithm |
| $\square$ | end of a proof |

# Chapter 1

# Integer Arithmetic

In this chapter our main topic is integer arithmetic. However, we shall see that many algorithms for polynomial arithmetic are similar to the corresponding algorithms for integer arithmetic, but simpler due to the lack of carries in polynomial arithmetic. Consider for example addition: the sum of two polynomials of degree $n$ always has degree at most $n$, whereas the sum of two $n$-digit integers may have $n + 1$ digits. Thus we often describe algorithms for polynomials as an aid to understanding the corresponding algorithms for integers.

## 1.1   Representation and Notations

We consider in this chapter algorithms working on integers. We distinguish between the logical — or mathematical — representation of an integer, and its physical representation on a computer. Our algorithms are intended for "large" integers — they are not restricted to integers that can be represented in a single computer word.

Several physical representations are possible. We consider here only the most common one, namely a dense representation in a fixed base. Choose an integral *base* $\beta > 1$. (In case of ambiguity, $\beta$ will be called the *internal* base.) A positive integer $A$ is represented by the length $n$ and the digits $a_i$ of its base $\beta$ expansion:

$$A = a_{n-1}\beta^{n-1} + \cdots + a_1\beta + a_0,$$

where $0 \le a_i \le \beta - 1$, and $a_{n-1}$ is sometimes assumed to be non-zero.

Since the base $\beta$ is usually fixed in a given program, only the length $n$ and the integers $(a_i)_{0 \leq i < n}$ need to be stored. Some common choices for $\beta$ are $2^{32}$ on a 32-bit computer, or $2^{64}$ on a 64-bit machine; other possible choices are respectively $10^9$ and $10^{19}$ for a decimal representation, or $2^{53}$ when using double-precision floating-point registers. Most algorithms given in this chapter work in any base; the exceptions are explicitly mentioned.

We assume that the sign is stored separately from the absolute value. This is known as the "sign-magnitude" representation. Zero is an important special case; to simplify the algorithms we assume that $n = 0$ if $A = 0$, and we usually assume that this case is treated separately.

Except when explicitly mentioned, we assume that all operations are *off-line*, i.e., all inputs (resp. outputs) are completely known at the beginning (resp. end) of the algorithm. Different models include *lazy* and *relaxed* algorithms, and are discussed in the Notes and References (§1.9).

## 1.2   Addition and Subtraction

As an explanatory example, here is an algorithm for integer addition. In the algorithm, $d$ is a *carry* bit.

Our algorithms are given in a language which mixes mathematical notation and syntax similar to that found in many high-level computer languages. It should be straightforward to translate into a language such as C. Note that ":=" indicates a definition, and "←" indicates assignment. Line numbers are included if we need to refer to individual lines in the description or analysis of the algorithm.

---

**Algorithm 1.1** IntegerAddition

**Input:** $A = \sum_0^{n-1} a_i \beta^i$, $B = \sum_0^{n-1} b_i \beta^i$, carry-in $0 \leq d_{\text{in}} \leq 1$

**Output:** $C := \sum_0^{n-1} c_i \beta^i$ and $0 \leq d \leq 1$ such that $A + B + d_{\text{in}} = d\beta^n + C$

1: $d \leftarrow d_{\text{in}}$
2: **for** $i$ **from** $0$ **to** $n - 1$ **do**
3:     $s \leftarrow a_i + b_i + d$
4:     $(d, c_i) \leftarrow (s \text{ div } \beta, s \text{ mod } \beta)$
5: return $C, d$.

---

Let $T$ be the number of different values taken by the data type representing the coefficients $a_i, b_i$. (Clearly $\beta \leq T$ but equality does not necessarily hold, e.g., $\beta = 10^9$ and $T = 2^{32}$.) At step 3, the value of $s$ can be as large as $2\beta - 1$, which is not representable if $\beta = T$. Several workarounds are possible: either use a machine instruction that gives the possible carry of $a_i + b_i$; or use the fact that, if a carry occurs in $a_i + b_i$, then the computed sum — if performed modulo $T$ — equals $t := a_i + b_i - T < a_i$; thus comparing $t$ and $a_i$ will determine if a carry occurred. A third solution is to keep a bit in reserve, taking $\beta \leq T/2$.

The subtraction code is very similar. Step 3 simply becomes $s \leftarrow a_i - b_i + d$, where $d \in \{-1, 0\}$ is the *borrow* of the subtraction, and $-\beta \leq s < \beta$. The other steps are unchanged, with the invariant $A - B + d_{in} = d\beta^n + C$.

We use the *arithmetic complexity* model, where *cost* is measured by the number of machine instructions performed, or equivalently (up to a constant factor) the *time* on a single processor.

Addition and subtraction of $n$-word integers cost $O(n)$, which is negligible compared to the multiplication cost. However, it is worth trying to reduce the constant factor implicit in this $O(n)$ cost. We shall see in §1.3 that "fast" multiplication algorithms are obtained by replacing multiplications by additions (usually more additions than the multiplications that they replace). Thus, the faster the additions are, the smaller will be the thresholds for changing over to the "fast" algorithms.

## 1.3  Multiplication

A nice application of large integer multiplication is the *Kronecker-Schönhage trick*, also called *segmentation* or *substitution* by some authors. Assume we want to multiply two polynomials $A(x)$ and $B(x)$ with non-negative integer coefficients (see Exercise 1.1 for negative coefficients). Assume both polynomials have degree less than $n$, and coefficients are bounded by $\rho$. Now take a power $X = \beta^k > n\rho^2$ of the base $\beta$, and multiply the integers $a = A(X)$ and $b = B(X)$ obtained by evaluating $A$ and $B$ at $x = X$. If $C(x) = A(x)B(x) = \sum c_i x^i$, we clearly have $C(X) = \sum c_i X^i$. Now since the $c_i$ are bounded by $n\rho^2 < X$, the coefficients $c_i$ can be retrieved by simply "reading" blocks of $k$ words in $C(X)$. Assume for example that we want to compute

$$(6x^5 + 6x^4 + 4x^3 + 9x^2 + x + 3)(7x^4 + x^3 + 2x^2 + x + 7),$$

with degree less than $n = 6$, and coefficients bounded by $\rho = 9$. We can take $X = 10^3 > n\rho^2$, and perform the integer multiplication:

$$6\,006\,004\,009\,001\,003 \times 7\,001\,002\,001\,007$$
$$= 42\,048\,046\,085\,072\,086\,042\,070\,010\,021,$$

from which we can read off the product

$$42x^9 + 48x^8 + 46x^7 + 85x^6 + 72x^5 + 86x^4 + 42x^3 + 70x^2 + 10x + 21.$$

Conversely, suppose we want to multiply two integers $a = \sum_{0 \le i < n} a_i \beta^i$ and $b = \sum_{0 \le j < n} b_j \beta^j$. Multiply the polynomials $A(x) = \sum_{0 \le i < n} a_i x^i$ and $B(x) = \sum_{0 \le j < n} b_j x^j$, obtaining a polynomial $C(x)$, then evaluate $C(x)$ at $x = \beta$ to obtain $ab$. Note that the coefficients of $C(x)$ may be larger than $\beta$, in fact they may be up to about $n\beta^2$. For example, with $a = 123$, $b = 456$, and $\beta = 10$, we obtain $A(x) = x^2 + 2x + 3$, $B(x) = 4x^2 + 5x + 6$, with product $C(x) = 4x^4 + 13x^3 + 28x^2 + 27x + 18$, and $C(10) = 56088$. These examples demonstrate the analogy between operations on polynomials and integers, and also show the limits of the analogy.

A common and very useful notation is to let $M(n)$ denote the time to multiply $n$-bit integers, or polynomials of degree $n-1$, depending on the context. In the polynomial case, we assume that the cost of multiplying coefficients is constant; this is known as the *arithmetic complexity* model, whereas the *bit complexity* model also takes into account the cost of multiplying coefficients, and thus their bit-size.

## 1.3.1   Naive Multiplication

---

**Algorithm 1.2** BasecaseMultiply

---

**Input:** $A = \sum_0^{m-1} a_i \beta^i$, $B = \sum_0^{n-1} b_j \beta^j$
**Output:** $C = AB := \sum_0^{m+n-1} c_k \beta^k$

1: $C \leftarrow A \cdot b_0$
2: **for** $j$ **from** 1 **to** $n-1$ **do**
3:     $C \leftarrow C + \beta^j (A \cdot b_j)$
4: return $C$.

---

**Theorem 1.3.1** *Algorithm* **BasecaseMultiply** *computes the product AB correctly, and uses* $\Theta(mn)$ *word operations.*

The multiplication by $\beta^j$ at step 3 is trivial with the chosen dense representation: it simply requires shifting by $j$ words towards the most significant words. The main operation in Algorithm **BasecaseMultiply** is the computation of $A \cdot b_j$ and its accumulation into $C$ at step 3. Since all fast algorithms rely on multiplication, the most important operation to optimize in multiple-precision software is thus the multiplication of an array of $m$ words by one word, with accumulation of the result in another array of $m + 1$ words.

We sometimes call Algorithm **BasecaseMultiply** *schoolbook multiplication* since it is close to the "long multiplication" algorithm that used to be taught at school.

Since multiplication with accumulation usually makes extensive use of the pipeline, it is best to give it arrays that are as long as possible, which means that $A$ rather than $B$ should be the operand of larger size (i.e., $m \geq n$).

## 1.3.2 Karatsuba's Algorithm

Karatsuba's algorithm is a "divide and conquer" algorithm for multiplication of integers (or polynomials). The idea is to reduce a multiplication of length $n$ to three multiplications of length $n/2$, plus some overhead that costs $O(n)$.

In the following, $n_0 \geq 2$ denotes the threshold between naive multiplication and Karatsuba's algorithm, which is used for $n_0$-word and larger inputs. The optimal "Karatsuba threshold" $n_0$ can vary from about 10 to about 100 words, depending on the processor and on the relative cost of multiplication and addition (see Exercise 1.6).

**Theorem 1.3.2** *Algorithm* **KaratsubaMultiply** *computes the product AB correctly, using* $K(n) = O(n^\alpha)$ *word multiplications, with* $\alpha = \lg 3 \approx 1.585$.

**Proof.** Since $s_A |A_0 - A_1| = A_0 - A_1$ and $s_B |B_0 - B_1| = B_0 - B_1$, we have $s_A s_B |A_0 - A_1||B_0 - B_1| = (A_0 - A_1)(B_0 - B_1)$, and thus $C = A_0 B_0 + (A_0 B_1 + A_1 B_0)\beta^k + A_1 B_1 \beta^{2k}$.

Since $A_0$, $B_0$, $|A_0 - A_1|$ and $|B_0 - B_1|$ have (at most) $\lceil n/2 \rceil$ words, and $A_1$ and $B_1$ have (at most) $\lfloor n/2 \rfloor$ words, the number $K(n)$ of word multiplications satisfies the recurrence $K(n) = n^2$ for $n < n_0$, and $K(n) = 2K(\lceil n/2 \rceil) + K(\lfloor n/2 \rfloor)$ for $n \geq n_0$. Assume $2^{\ell-1} n_0 < n \leq 2^\ell n_0$ with $\ell \geq 1$. Then $K(n)$

---

**Algorithm 1.3** KaratsubaMultiply

---

**Input:** $A = \sum_0^{n-1} a_i \beta^i$, $B = \sum_0^{n-1} b_j \beta^j$
**Output:** $C = AB := \sum_0^{2n-1} c_k \beta^k$
    **if** $n < n_0$ **then** return **BasecaseMultiply**$(A, B)$
    $k \leftarrow \lceil n/2 \rceil$
    $(A_0, B_0) := (A, B) \bmod \beta^k$, $(A_1, B_1) := (A, B) \operatorname{div} \beta^k$
    $s_A \leftarrow \operatorname{sign}(A_0 - A_1)$, $s_B \leftarrow \operatorname{sign}(B_0 - B_1)$
    $C_0 \leftarrow$ **KaratsubaMultiply**$(A_0, B_0)$
    $C_1 \leftarrow$ **KaratsubaMultiply**$(A_1, B_1)$
    $C_2 \leftarrow$ **KaratsubaMultiply**$(|A_0 - A_1|, |B_0 - B_1|)$
    return $C := C_0 + (C_0 + C_1 - s_A s_B C_2)\beta^k + C_1\beta^{2k}$.

---

is the sum of three $K(j)$ values with $j \leq 2^{\ell-1} n_0$, so at most $3^\ell \, K(j)$ with $j \leq n_0$. Thus $K(n) \leq 3^\ell \max(K(n_0), (n_0 - 1)^2)$, which gives $K(n) \leq Cn^\alpha$ with $C = 3^{1 - \lg(n_0)} \max(K(n_0), (n_0 - 1)^2)$. □

Different variants of Karatsuba's algorithm exist; the variant presented here is known as the *subtractive* version. Another classical one is the *additive* version, which uses $A_0 + A_1$ and $B_0 + B_1$ instead of $|A_0 - A_1|$ and $|B_0 - B_1|$. However, the subtractive version is more convenient for integer arithmetic, since it avoids the possible carries in $A_0 + A_1$ and $B_0 + B_1$, which require either an extra word in these sums, or extra additions.

The efficiency of an implementation of Karatsuba's algorithm depends heavily on memory usage. It is important to avoid allocating memory for the intermediate results $|A_0 - A_1|$, $|B_0 - B_1|$, $C_0$, $C_1$, and $C_2$ at each step (although modern compilers are quite good at optimising code and removing unnecessary memory references). One possible solution is to allow a large temporary storage of $m$ words, used both for the intermediate results and for the recursive calls. It can be shown that an auxiliary space of $m = 2n$ words — or even $m = O(\log n)$ — is sufficient (see Exercises 1.7 and 1.8).

Since the product $C_2$ is used only once, it may be faster to have auxiliary routines **KaratsubaAddmul** and **KaratsubaSubmul** that accumulate their result, calling themselves recursively, together with **KaratsubaMultiply** (see Exercise 1.10).

The version presented here uses $\sim 4n$ additions (or subtractions): $2 \times (n/2)$ to compute $|A_0 - A_1|$ and $|B_0 - B_1|$, then $n$ to add $C_0$ and $C_1$, again $n$ to add or subtract $C_2$, and $n$ to add $(C_0 + C_1 - s_A s_B C_2)\beta^k$ to $C_0 + C_1\beta^{2k}$. An

improved scheme uses only $\sim 7n/2$ additions (see Exercise 1.9).

When considered as algorithms on polynomials, most fast multiplication algorithms can be viewed as evaluation/interpolation algorithms. Karatsuba's algorithm regards the inputs as polynomials $A_0 + A_1 x$ and $B_0 + B_1 x$ evaluated at $x = \beta^k$; since their product $C(x)$ is of degree 2, Lagrange's interpolation theorem says that it is sufficient to evaluate $C(x)$ at three points. The subtractive version evaluates[1] $C(x)$ at $x = 0, -1, \infty$, whereas the additive version uses $x = 0, +1, \infty$.

### 1.3.3 Toom-Cook Multiplication

Karatsuba's idea readily generalizes to what is known as Toom-Cook $r$-way multiplication. Write the inputs as $a_0 + \cdots + a_{r-1} x^{r-1}$ and $b_0 + \cdots + b_{r-1} x^{r-1}$, with $x = \beta^k$, and $k = \lceil n/r \rceil$. Since their product $C(x)$ is of degree $2r - 2$, it suffices to evaluate it at $2r - 1$ distinct points to be able to recover $C(x)$, and in particular $C(\beta^k)$. If $r$ is chosen optimally, Toom-Cook multiplication of $n$-word numbers takes time $n^{1+O(1/\sqrt{\log n})}$.

Most references, when describing subquadratic multiplication algorithms, only describe Karatsuba and FFT-based algorithms. Nevertheless, the Toom-Cook algorithm is quite interesting in practice.

Toom-Cook $r$-way reduces one $n$-word product to $2r - 1$ products of about $n/r$ words, thus costs $O(n^\nu)$ with $\nu = \log(2r-1)/\log r$. However, the constant hidden by the big-$O$ notation depends strongly on the evaluation and interpolation formulæ, which in turn depend on the chosen points. One possibility is to take $-(r-1), \ldots, -1, 0, 1, \ldots, (r-1)$ as evaluation points.

The case $r = 2$ corresponds to Karatsuba's algorithm (§1.3.2). The case $r = 3$ is known as Toom-Cook 3-way, sometimes simply called "the Toom-Cook algorithm". Algorithm **ToomCook3** uses evaluation points $0, 1, -1, 2, \infty$, and tries to optimize the evaluation and interpolation formulæ.

The divisions at step 8 are exact; if $\beta$ is a power of two, the division by 6 can be done using a division by 2 — which consists of a single shift — followed by a division by 3 (see §1.4.7).

Toom-Cook $r$-way has to invert a $(2r-1) \times (2r-1)$ Vandermonde matrix with parameters the evaluation points; if one chooses consecutive integer points, the determinant of that matrix contains all primes up to $2r - 2$. This proves that division by (a multiple of) 3 can not be avoided for Toom-Cook

---

[1] Evaluating $C(x)$ at $\infty$ means computing the product $A_1 B_1$ of the leading coefficients.

---

**Algorithm 1.4** ToomCook3

---

**Input:** two integers $0 \leq A, B < \beta^n$

**Output:** $AB := c_0 + c_1 \beta^k + c_2 \beta^{2k} + c_3 \beta^{3k} + c_4 \beta^{4k}$ with $k = \lceil n/3 \rceil$

**Require:** a threshold $n_1 \geq 3$

1: **if** $n < n_1$ **then** return **KaratsubaMultiply**$(A, B)$
2: write $A = a_0 + a_1 x + a_2 x^2$, $B = b_0 + b_1 x + b_2 x^2$ with $x = \beta^k$.
3: $v_0 \leftarrow$ **ToomCook3**$(a_0, b_0)$
4: $v_1 \leftarrow$ **ToomCook3**$(a_{02} + a_1, b_{02} + b_1)$ where $a_{02} \leftarrow a_0 + a_2, b_{02} \leftarrow b_0 + b_2$
5: $v_{-1} \leftarrow$ **ToomCook3**$(a_{02} - a_1, b_{02} - b_1)$
6: $v_2 \leftarrow$ **ToomCook3**$(a_0 + 2a_1 + 4a_2, b_0 + 2b_1 + 4b_2)$
7: $v_\infty \leftarrow$ **ToomCook3**$(a_2, b_2)$
8: $t_1 \leftarrow (3v_0 + 2v_{-1} + v_2)/6 - 2v_\infty, t_2 \leftarrow (v_1 + v_{-1})/2$
9: $c_0 \leftarrow v_0, c_1 \leftarrow v_1 - t_1, c_2 \leftarrow t_2 - v_0 - v_\infty, c_3 \leftarrow t_1 - t_2, c_4 \leftarrow v_\infty.$

---

3-way with consecutive integer points. See Exercise 1.14 for a generalization of this result.

### 1.3.4 Use of the Fast Fourier Transform (FFT)

Most subquadratic multiplication algorithms can be seen as evaluation-interpolation algorithms. They mainly differ in the number of evaluation points, and the values of those points. However, the evaluation and interpolation formulæ become intricate in Toom-Cook $r$-way for large $r$, since they involve $O(r^2)$ scalar operations. The Fast Fourier Transform (FFT) is a way to perform evaluation and interpolation efficiently for some special points (roots of unity) and special values of $r$. This explains why multiplication algorithms with the best known asymptotic complexity are based on the Fast Fourier transform.

There are different flavours of FFT multiplication, depending on the ring where the operations are performed. The Schönhage-Strassen algorithm, with a complexity of $O(n \log n \log \log n)$, works in the ring $\mathbb{Z}/(2^n + 1)\mathbb{Z}$. Since it is based on modular computations, we describe it in Chapter 2.

Other commonly used algorithms work with floating-point complex numbers. A drawback is that, due to the inexact nature of floating-point computations, a careful error analysis is required to guarantee the correctness of the implementation, assuming an underlying arithmetic with rigorous error bounds. See Theorem 3.3.2 in Chapter 3.

We say that multiplication is *in the FFT range* if $n$ is large and the multiplication algorithm satisfies $M(2n) \sim 2M(n)$. For example, this is true if the Schönhage-Strassen multiplication algorithm is used, but not if the classical algorithm or Karatsuba's algorithm is used.

### 1.3.5 Unbalanced Multiplication

The subquadratic algorithms considered so far (Karatsuba and Toom-Cook) work with equal-size operands. How do we efficiently multiply integers of different sizes with a subquadratic algorithm? This case is important in practice but is rarely considered in the literature. Assume the larger operand has size $m$, and the smaller has size $n \leq m$, and denote by $M(m, n)$ the corresponding multiplication cost.

If evaluation-interpolation algorithms are used, the cost depends mainly on the size of the result, that is $m + n$, so we have $M(m, n) \leq M((m+n)/2)$, at least approximately. We can do better than $M((m+n)/2)$ if $n$ is much smaller than $m$, for example $M(m, 1) = O(m)$.

When $m$ is an exact multiple of $n$, say $m = kn$, a trivial strategy is to cut the larger operand into $k$ pieces, giving $M(kn, n) = kM(n) + O(kn)$. However, this is not always the best strategy, see Exercise 1.16.

When $m$ is not an exact multiple of $n$, several strategies are possible:

- split the two operands into an equal number of pieces of unequal sizes;

- or split the two operands into different numbers of pieces.

Each strategy has advantages and disadvantages. We discuss each in turn.

### First Strategy: Equal Number of Pieces of Unequal Sizes

Consider for example Karatsuba multiplication, and let $K(m, n)$ be the number of word-products for an $m \times n$ product. Take for example $m = 5$, $n = 3$. A natural idea is to pad the smallest operand to the size of the largest one. However there are several ways to perform this padding, as shown in the following figure, where the "Karatsuba cut" is represented by a double column:

| $a_4$ | $a_3$ | $a_2$ | $a_1$ | $a_0$ |
|------|------|------|------|------|
|      |      | $b_2$ | $b_1$ | $b_0$ |

$A \times B$

| $a_4$ | $a_3$ | $a_2$ | $a_1$ | $a_0$ |
|------|------|------|------|------|
|      | $b_2$ | $b_1$ | $b_0$ |      |

$A \times (\beta B)$

| $a_4$ | $a_3$ | $a_2$ | $a_1$ | $a_0$ |
|------|------|------|------|------|
| $b_2$ | $b_1$ | $b_0$ |      |      |

$A \times (\beta^2 B)$

The left variant leads to two products of size 3, i.e., $2K(3,3)$, the middle one to $K(2,1)+K(3,2)+K(3,3)$, and the right one to $K(2,2)+K(3,1)+K(3,3)$, which give respectively 14, 15, 13 word products.

However, whenever $m/2 \leq n \leq m$, any such "padding variant" will require $K(\lceil m/2 \rceil, \lceil m/2 \rceil)$ for the product of the differences (or sums) of the low and high parts from the operands, due to a "wrap-around" effect when subtracting the parts from the smaller operand; this will ultimately lead to a cost similar to that of an $m \times m$ product. The "odd-even scheme" of Algorithm **OddEvenKaratsuba** (see also Exercise 1.13) avoids this wrap-around. Here is an example of this algorithm for $m = 3$ and $n = 2$. Take

---

**Algorithm 1.5** OddEvenKaratsuba

**Input:** $A = \sum_0^{m-1} a_i x^i$, $B = \sum_0^{n-1} b_j x^j$, $m \geq n \geq 1$
**Output:** $A \cdot B$
    **if** $n = 1$ **then** return $\sum_0^{m-1} a_i b_0 x^i$
    $k \leftarrow \lceil m/2 \rceil$, $\ell \leftarrow \lceil n/2 \rceil$
    write $A = A_0(x^2) + xA_1(x^2)$, $B = B_0(x^2) + xB_1(x^2)$
    $C_0 \leftarrow$ **OddEvenKaratsuba**$(A_0, B_0)$
    $C_1 \leftarrow$ **OddEvenKaratsuba**$(A_0 + A_1, B_0 + B_1)$
    $C_2 \leftarrow$ **OddEvenKaratsuba**$(A_1, B_1)$
    return $C_0(x^2) + x(C_1 - C_0 - C_2)(x^2) + x^2 C_2(x^2)$.

---

$A = a_2 x^2 + a_1 x + a_0$ and $B = b_1 x + b_0$. This yields $A_0 = a_2 x + a_0$, $A_1 = a_1$, $B_0 = b_0$, $B_1 = b_1$, thus $C_0 = (a_2 x + a_0)b_0$, $C_1 = (a_2 x + a_0 + a_1)(b_0 + b_1)$, $C_2 = a_1 b_1$. We thus get $K(3,2) = 2K(2,1) + K(1) = 5$ with the odd-even scheme. The general recurrence for the odd-even scheme is:

$$K(m, n) = 2K(\lceil m/2 \rceil, \lceil n/2 \rceil) + K(\lfloor m/2 \rfloor, \lfloor n/2 \rfloor),$$

instead of

$$K(m, n) = 2K(\lceil m/2 \rceil, \lceil m/2 \rceil) + K(\lfloor m/2 \rfloor, n - \lceil m/2 \rceil)$$

for the classical variant, assuming $n > m/2$. We see that the second parameter in $K(\cdot, \cdot)$ only depends on the smaller size $n$ for the odd-even scheme.

As for the classical variant, there are several ways of padding with the odd-even scheme. Consider $m = 5$, $n = 3$, and write $A := a_4 x^4 + a_3 x^3 + a_2 x^2 + a_1 x + a_0 = x A_1(x^2) + A_0(x^2)$, with $A_1(x) = a_3 x + a_1$, $A_0(x) = a_4 x^2 + a_2 x + a_0$; and $B := b_2 x^2 + b_1 x + b_0 = x B_1(x^2) + B_0(x^2)$, with $B_1(x) = b_1$, $B_0(x) = b_2 x + b_0$. Without padding, we write $AB = x^2(A_1 B_1)(x^2) + x((A_0 + A_1)(B_0 + B_1) - A_1 B_1 - A_0 B_0)(x^2) + (A_0 B_0)(x^2)$, which gives $K(5, 3) = K(2, 1) + 2K(3, 2) = 12$. With padding, we consider $xB = x B_1'(x^2) + B_0'(x^2)$, with $B_1'(x) = b_2 x + b_0$, $B_0' = b_1 x$. This gives $K(2, 2) = 3$ for $A_1 B_1'$, $K(3, 2) = 5$ for $(A_0 + A_1)(B_0' + B_1')$, and $K(3, 1) = 3$ for $A_0 B_0'$ — taking into account the fact that $B_0'$ has only one non-zero coefficient — thus a total of 11 only.

Note that when the variable $x$ corresponds to say $\beta = 2^{64}$, Algorithm **OddEvenKaratsuba** as presented above is not very practical in the integer case, because of a problem with carries. For example, in the sum $A_0 + A_1$ we have $\lfloor m/2 \rfloor$ carries to store. A workaround is to consider $x$ to be say $\beta^{10}$, in which case we have to store only one carry bit for 10 words, instead of one carry bit per word.

The first strategy, which consists in cutting the operands into an equal number of pieces of unequal sizes, does not scale up nicely. Assume for example that we want to multiply a number of 999 words by another number of 699 words, using Toom-Cook 3-way. With the classical variant — without padding — and a "large" base of $\beta^{333}$, we cut the larger operand into three pieces of 333 words and the smaller one into two pieces of 333 words and one small piece of 33 words. This gives four full $333 \times 333$ products — ignoring carries — and one unbalanced $333 \times 33$ product (for the evaluation at $x = \infty$). The "odd-even" variant cuts the larger operand into three pieces of 333 words, and the smaller operand into three pieces of 233 words, giving rise to five equally unbalanced $333 \times 233$ products, again ignoring carries.

## Second Strategy: Different Number of Pieces of Equal Sizes

Instead of splitting unbalanced operands into an equal number of pieces — which are then necessarily of different sizes — an alternative strategy is to split the operands into a different number of pieces, and use a multiplication algorithm which is naturally unbalanced. Consider again the example of multiplying two numbers of 999 and 699 words. Assume we have a multi-plication algorithm, say Toom-$(3, 2)$, which multiplies a number of $3n$ words by another number of $2n$ words; this requires four products of numbers of about $n$ words. Using $n = 350$, we can split the larger number into two pieces of 350 words, and one piece of 299 words, and the smaller number into one piece of 350 words and one piece of 349 words.

Similarly, for two inputs of 1000 and 500 words, we can use a Toom-$(4, 2)$ algorithm which multiplies two numbers of $4n$ and $2n$ words, with $n = 250$. Such an algorithm requires five evaluation points; if we choose the same points as for Toom 3-way, then the interpolation phase can be shared between both implementations.

It seems that this second strategy is not compatible with the "odd-even" variant, which requires that both operands are cut into the same number of pieces. Consider for example the "odd-even" variant modulo 3. It writes the numbers to be multiplied as $A = a(\beta)$ and $B = b(\beta)$ with $a(t) = a_0(t^3) + ta_1(t^3) + t^2a_2(t^3)$, and similarly $b(t) = b_0(t^3) + tb_1(t^3) + t^2b_2(t^3)$. We see that the number of pieces of each operand is the chosen modulus, here 3 (see Exercise 1.11).

### Asymptotic complexity of unbalanced multiplication

Suppose $m \geq n$ and $n$ is large. To use an evaluation-interpolation scheme we need to evaluate the product at $m + n$ points, whereas balanced $k$ by $k$ multiplication needs $2k$ points. Taking $k \approx (m+n)/2$, we see that $M(m, n) \leq M((m + n)/2)(1 + o(1))$ as $n \to \infty$. On the other hand, from the discussion above, we have $M(m, n) \leq \lceil m/n \rceil M(n)$. This explains the upper bound on $M(m, n)$ given in the *Summary of Complexities* at the end of the book.

## 1.3.6  Squaring

In many applications, a significant proportion of the multiplications have equal operands, i.e., are squarings. Hence it is worth tuning a special squaring implementation as much as the implementation of multiplication itself, bearing in mind that the best possible speedup is two (see Exercise 1.17).

For naive multiplication, Algorithm **BasecaseMultiply** (§1.3.1) can be modified to obtain a theoretical speedup of two, since only about half of the products $a_ib_j$ need to be computed.

Subquadratic algorithms like Karatsuba and Toom-Cook $r$-way can be specialized for squaring too. In general, the threshold obtained is larger than the corresponding multiplication threshold. For example, on a modern 64-bit computer, one can expect a threshold between the naive quadratic squaring and Karatsuba's algorithm in the 30-word range, between Karatsuba's and Toom-Cook 3-way in the 100-word range, between Toom-Cook 3-way and Toom-Cook 4-way in the 150-word range, and between Toom-Cook 4-way and the FFT in the 2500-word range.

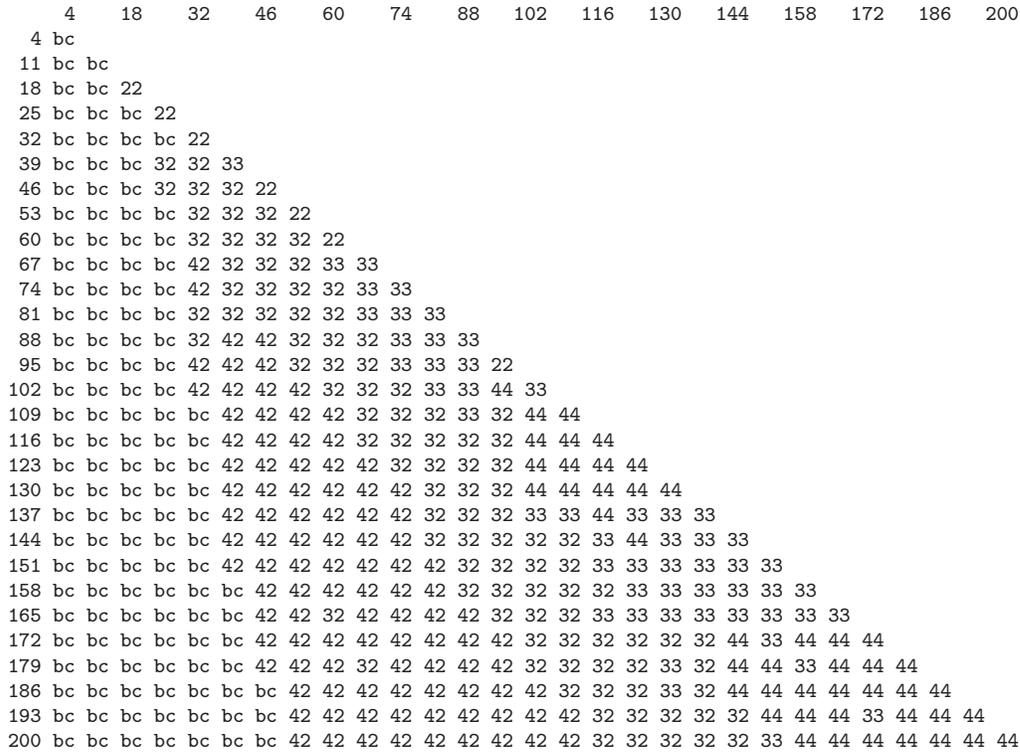

Figure 1.1: The best algorithm to multiply two numbers of $x$ and $y$ words for $4 \leq x \leq y \leq 200$: `bc` is schoolbook multiplication, `22` is Karatsuba's algorithm, `33` is Toom-3, `32` is Toom-$(3,2)$, `44` is Toom-4, and `42` is Toom-$(4,2)$. This graph was obtained on a Core 2, with GMP 5.0.0, and GCC 4.4.2. Note that for $x \leq (y+3)/4$, only the schoolbook multiplication is available; since we did not consider the algorithm that cuts the larger operand into several pieces, this explains why `bc` is best for say $x = 46$ and $y = 200$.

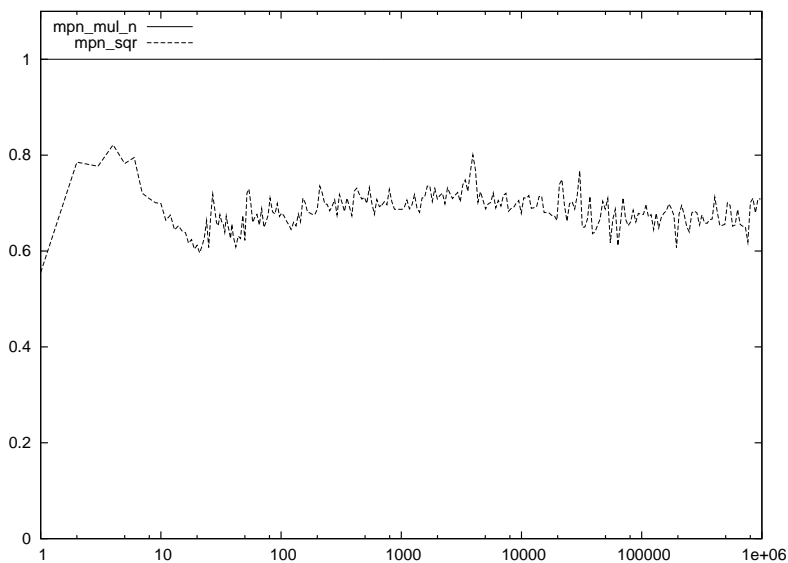

Figure 1.2: Ratio of the squaring and multiplication time for the GNU MP library, version 5.0.0, on a Core 2 processor, up to one million words.

Figure 1.2 compares the multiplication and squaring time with the GNU MP library. It shows that whatever the word range, a good rule of thumb is to count 2/3 of the cost of a product for a squaring.

The classical approach for fast squaring is to take a fast multiplication algorithm, say Toom-Cook $r$-way, and to replace the $2r - 1$ recursive products by $2r - 1$ recursive squarings. For example, starting from Algorithm **ToomCook3**, we obtain five recursive squarings $a_0^2$, $(a_0 + a_1 + a_2)^2$, $(a_0 - a_1 + a_2)^2$, $(a_0 + 2a_1 + 4a_2)^2$, and $a_2^2$. A different approach, called *asymmetric squaring*, is to allow products which are not squares in the recursive calls. For example, the square of $a_2\beta^2 + a_1\beta + a_0$ is $c_4\beta^4 + c_3\beta^3 + c_2\beta^2 + c_1\beta + c_0$, where $c_4 = a_2^2$, $c_3 = 2a_1a_2$, $c_2 = c_0 + c_4 - s$, $c_1 = 2a_1a_0$, and $c_0 = a_0^2$, where $s = (a_0 - a_2 + a_1)(a_0 - a_2 - a_1)$. This formula performs two squarings, and three normal products. Such asymmetric squaring formulæ are not asymptotically optimal, but might be faster in some medium range, due to simpler evaluation or interpolation phases.

### 1.3.7 Multiplication by a Constant

It often happens that the same multiplier is used in several consecutive operations, or even for a complete calculation. If this constant multiplier is small, i.e., less than the base $\beta$, not much speedup can be obtained compared to the usual product. We thus consider here a "large" constant multiplier.

When using evaluation-interpolation algorithms, like Karatsuba or Toom-Cook (see §1.3.2–1.3.3), one may store the evaluations for that fixed multiplier at the different points chosen.

Special-purpose algorithms also exist. These algorithms differ from classical multiplication algorithms because they take into account the *value* of the given constant multiplier, and not only its size in bits or digits. They also differ in the model of complexity used. For example, R. Bernstein's algorithm [27], which is used by several compilers to compute addresses in data structure records, considers as basic operation $x, y \mapsto 2^i x \pm y$, with a cost assumed to be independent of the integer $i$.

For example, Bernstein's algorithm computes $20061x$ in five steps:

$$
\begin{aligned}
x_1 := 31x &= 2^5 x - x \\
x_2 := 93x &= 2^1 x_1 + x_1 \\
x_3 := 743x &= 2^3 x_2 - x \\
x_4 := 6687x &= 2^3 x_3 + x_3 \\
20061x &= 2^1 x_4 + x_4.
\end{aligned}
$$

## 1.4 Division

Division is the next operation to consider after multiplication. Optimizing division is almost as important as optimizing multiplication, since division is usually more expensive, thus the speedup obtained on division will be more significant. On the other hand, one usually performs more multiplications than divisions.

One strategy is to avoid divisions when possible, or replace them by multiplications. An example is when the same divisor is used for several consecutive operations; one can then precompute its inverse (see §2.4.1).

We distinguish several kinds of division: *full division* computes both quotient and remainder, while in other cases only the quotient (for example, when dividing two floating-point significands) or remainder (when multiplying two residues modulo $n$) is needed. We also discuss *exact division* — when the remainder is known to be zero — and the problem of dividing by a single word.

### 1.4.1  Naive Division

In all division algorithms, we assume that divisors are normalized. We say that $B := \sum_0^{n-1} b_j \beta^j$ is *normalized* when its most significant word $b_{n-1}$ satisfies $b_{n-1} \geq \beta/2$. This is a stricter condition (for $\beta > 2$) than simply requiring that $b_{n-1}$ be nonzero.

---

**Algorithm 1.6** BasecaseDivRem

---

**Input:** $A = \sum_0^{n+m-1} a_i \beta^i$, $B = \sum_0^{n-1} b_j \beta^j$, $B$ normalized, $m \geq 0$
**Output:** quotient $Q$ and remainder $R$ of $A$ divided by $B$
  1: **if** $A \geq \beta^m B$ **then** $q_m \leftarrow 1$, $A \leftarrow A - \beta^m B$ **else** $q_m \leftarrow 0$
  2: **for** $j$ **from** $m - 1$ **downto** 0 **do**
  3:      $q_j^* \leftarrow \lfloor (a_{n+j}\beta + a_{n+j-1})/b_{n-1} \rfloor$           ▷ quotient selection step
  4:      $q_j \leftarrow \min(q_j^*, \beta - 1)$
  5:      $A \leftarrow A - q_j \beta^j B$
  6:      **while** $A < 0$ **do**
  7:          $q_j \leftarrow q_j - 1$
  8:          $A \leftarrow A + \beta^j B$
  9: **return** $Q = \sum_0^m q_j \beta^j$, $R = A$.
(Note: in step 3, $a_i$ denotes the *current* value of the $i$-th word of $A$, which may be modified at steps 5 and 8.)

---

If $B$ is not normalized, we can compute $A' = 2^k A$ and $B' = 2^k B$ so that $B'$ is normalized, then divide $A'$ by $B'$ giving $A' = Q'B' + R'$; the quotient and remainder of the division of $A$ by $B$ are respectively $Q := Q'$ and $R := R'/2^k$, the latter division being exact.

**Theorem 1.4.1** *Algorithm* **BasecaseDivRem** *correctly computes the quotient and remainder of the division of A by a normalized B, in $O(n(m+1))$ word operations.*

**Proof.** We prove that the invariant $A < \beta^{j+1}B$ holds at step 2. This holds trivially for $j = m-1$: $B$ being normalized, $A < 2\beta^m B$ initially.

First consider the case $q_j = q_j^*$: then $q_j b_{n-1} \geq a_{n+j}\beta + a_{n+j-1} - b_{n-1} + 1$, thus

$$A - q_j \beta^j B \leq (b_{n-1} - 1)\beta^{n+j-1} + (A \bmod \beta^{n+j-1}),$$

which ensures that the new $a_{n+j}$ vanishes, and $a_{n+j-1} < b_{n-1}$, thus $A < \beta^j B$ after step 5. Now $A$ may become negative after step 5, but since $q_j b_{n-1} \leq a_{n+j}\beta + a_{n+j-1}$, we have:

$$\begin{aligned} A - q_j \beta^j B &> (a_{n+j}\beta + a_{n+j-1})\beta^{n+j-1} - q_j(b_{n-1}\beta^{n-1} + \beta^{n-1})\beta^j \\ &\geq -q_j \beta^{n+j-1}. \end{aligned}$$

Therefore $A - q_j \beta^j B + 2\beta^j B \geq (2b_{n-1} - q_j)\beta^{n+j-1} > 0$, which proves that the while-loop at steps 6-8 is performed at most twice [143, Theorem 4.3.1.B]. When the while-loop is entered, $A$ may increase only by $\beta^j B$ at a time, hence $A < \beta^j B$ at exit.

In the case $q_j \neq q_j^*$, i.e., $q_j^* \geq \beta$, we have before the while-loop: $A < \beta^{j+1}B - (\beta-1)\beta^j B = \beta^j B$, thus the invariant holds. If the while-loop is entered, the same reasoning as above holds.

We conclude that when the for-loop ends, $0 \leq A < B$ holds, and since $(\sum_j^m q_j \beta^j)B + A$ is invariant throughout the algorithm, the quotient $Q$ and remainder $R$ are correct.

The most expensive part is step 5, which costs $O(n)$ operations for $q_j B$ (the multiplication by $\beta^j$ is simply a word-shift); the total cost is $O(n(m+1))$. (For $m = 0$ we need $O(n)$ work if $A \geq B$, and even if $A < B$ to compare the inputs in the case $A = B - 1$.) □

Here is an example of algorithm **BasecaseDivRem** for the inputs $A = 766\,970\,544\,842\,443\,844$ and $B = 862\,664\,913$, with $\beta = 1000$, which gives quotient $Q = 889\,071\,217$ and remainder $R = 778\,334\,723$.

| $j$ | $A$ | $q_j$ | $A - q_j B\beta^j$ | after correction |
|---|---|---|---|---|
| 2 | $766\,970\,544\,842\,443\,844$ | 889 | $61\,437\,185\,443\,844$ | no change |
| 1 | $61\,437\,185\,443\,844$ | 071 | $187\,976\,620\,844$ | no change |
| 0 | $187\,976\,620\,844$ | 218 | $-84\,330\,190$ | $778\,334\,723$ |

Algorithm **BasecaseDivRem** simplifies when $A < \beta^m B$: remove step 1, and change $m$ into $m - 1$ in the return value $Q$. However, the more general form we give is more convenient for a computer implementation, and will be used below.

A possible variant when $q_j^* \geq \beta$ is to let $q_j = \beta$; then $A - q_j \beta^j B$ at step 5 reduces to a single subtraction of $B$ shifted by $j + 1$ words. However in this case the while-loop will be performed at least once, which corresponds to the identity $A - (\beta - 1)\beta^j B = A - \beta^{j+1} B + \beta^j B$.

If instead of having $B$ normalized, i.e., $b_n \geq \beta/2$, one has $b_n \geq \beta/k$, there can be up to $k$ iterations of the while-loop (and step 1 has to be modified).

A drawback of Algorithm **BasecaseDivRem** is that the test $A < 0$ at line 6 is true with non-negligible probability, therefore branch prediction algorithms available on modern processors will fail, resulting in wasted cycles. A workaround is to compute a more accurate partial quotient, in order to decrease the proportion of corrections to almost zero (see Exercise 1.20).

### 1.4.2   Divisor Preconditioning

Sometimes the quotient selection — step 3 of Algorithm **BasecaseDivRem** — is quite expensive compared to the total cost, especially for small sizes. Indeed, some processors do not have a machine instruction for the division of two words by one word; one way to compute $q_j^*$ is then to precompute a one-word approximation of the inverse of $b_{n-1}$, and to multiply it by $a_{n+j}\beta + a_{n+j-1}$.

Svoboda's algorithm makes the quotient selection trivial, after preconditioning the divisor. The main idea is that if $b_{n-1}$ equals the base $\beta$ in Algorithm **BasecaseDivRem**, then the quotient selection is easy, since it suffices to take $q_j^* = a_{n+j}$. (In addition, $q_j^* \leq \beta - 1$ is then always fulfilled, thus step 4 of **BasecaseDivRem** can be avoided, and $q_j^*$ replaced by $q_j$.)

With the example of §1.4.1, Svoboda's algorithm would give $k = 1160$, $B' = 1\,000\,691\,299\,080$:

| $j$ | $A$ | $q_j$ | $A - q_j B' \beta^j$ | after correction |
|---|---|---|---|---|
| 2 | $766\,970\,544\,842\,443\,844$ | 766 | $441\,009\,747\,163\,844$ | no change |
| 1 | $441\,009\,747\,163\,844$ | 441 | $-295\,115\,730\,436$ | $705\,575\,568\,644$ |

---

**Algorithm 1.7** SvobodaDivision

---

**Input:** $A = \sum_0^{n+m-1} a_i \beta^i$, $B = \sum_0^{n-1} b_j \beta^j$ normalized, $A < \beta^m B$, $m \geq 1$

**Output:** quotient $Q$ and remainder $R$ of $A$ divided by $B$

1: $k \leftarrow \lceil \beta^{n+1}/B \rceil$
2: $B' \leftarrow kB = \beta^{n+1} + \sum_0^{n-1} b_j' \beta^j$
3: **for** $j$ **from** $m-1$ **downto** $1$ **do**
4:     $q_j \leftarrow a_{n+j}$                            ▷ current value of $a_{n+j}$
5:     $A \leftarrow A - q_j \beta^{j-1} B'$
6:     **if** $A < 0$ **then**
7:         $q_j \leftarrow q_j - 1$
8:         $A \leftarrow A + \beta^{j-1} B'$
9: $Q' = \sum_1^{m-1} q_j \beta^j$, $R' = A$
10: $(q_0, R) \leftarrow (R' \text{ div } B, R' \text{ mod } B)$             ▷ using **BasecaseDivRem**
11: **return** $Q = kQ' + q_0$, $R$.

---

We thus get $Q' = 766\,440$ and $R' = 705\,575\,568\,644$. The final division of step 10 gives $R' = 817B + 778\,334\,723$, thus we get $Q = 1\,160 \cdot 766\,440 + 817 = 889\,071\,217$, and $R = 778\,334\,723$, as in §1.4.1.

Svoboda's algorithm is especially interesting when only the remainder is needed, since then one can avoid the "deconditioning" $Q = kQ' + q_0$. Note that when only the quotient is needed, dividing $A' = kA$ by $B' = kB$ is another way to compute it.

## 1.4.3   Divide and Conquer Division

The base-case division of §1.4.1 determines the quotient word by word. A natural idea is to try getting several words at a time, for example replacing the quotient selection step in Algorithm **BasecaseDivRem** by:

$$q_j^* \leftarrow \left\lfloor \frac{a_{n+j}\beta^3 + a_{n+j-1}\beta^2 + a_{n+j-2}\beta + a_{n+j-3}}{b_{n-1}\beta + b_{n-2}} \right\rfloor.$$

Since $q_j^*$ has then two words, fast multiplication algorithms (§1.3) might speed up the computation of $q_j B$ at step 5 of Algorithm **BasecaseDivRem**.

More generally, the most significant half of the quotient — say $Q_1$, of $\ell = m-k$ words — mainly depends on the $\ell$ most significant words of the dividend and divisor. Once a good approximation to $Q_1$ is known, fast multiplication algorithms can be used to compute the partial remainder $A - Q_1 B \beta^k$. The second idea of the divide and conquer algorithm **RecursiveDivRem** is to compute the corresponding remainder together with the partial quotient $Q_1$; in such a way, one only has to subtract the product of $Q_1$ by the low part of the divisor, before computing the low part of the quotient.

---

**Algorithm 1.8** RecursiveDivRem

---

**Input:** $A = \sum_0^{n+m-1} a_i \beta^i$, $B = \sum_0^{n-1} b_j \beta^j$, $B$ normalized, $n \geq m$
**Output:** quotient $Q$ and remainder $R$ of $A$ divided by $B$
 1: **if** $m < 2$ **then** return **BasecaseDivRem**$(A, B)$
 2: $k \leftarrow \lfloor m/2 \rfloor$, $B_1 \leftarrow B$ div $\beta^k$, $B_0 \leftarrow B$ mod $\beta^k$
 3: $(Q_1, R_1) \leftarrow$ **RecursiveDivRem**$(A$ div $\beta^{2k}, B_1)$
 4: $A' \leftarrow R_1 \beta^{2k} + (A$ mod $\beta^{2k}) - Q_1 B_0 \beta^k$
 5: **while** $A' < 0$ **do** $Q_1 \leftarrow Q_1 - 1$, $A' \leftarrow A' + \beta^k B$
 6: $(Q_0, R_0) \leftarrow$ **RecursiveDivRem**$(A'$ div $\beta^k, B_1)$
 7: $A'' \leftarrow R_0 \beta^k + (A'$ mod $\beta^k) - Q_0 B_0$
 8: **while** $A'' < 0$ **do** $Q_0 \leftarrow Q_0 - 1$, $A'' \leftarrow A'' + B$
 9: return $Q := Q_1 \beta^k + Q_0$, $R := A''$.

---

In Algorithm **RecursiveDivRem**, one may replace the condition $m < 2$ at step 1 by $m < T$ for any integer $T \geq 2$. In practice, $T$ is usually in the range 50 to 200.

One can not require $A < \beta^m B$ at input, since this condition may not be satisfied in the recursive calls. Consider for example $A = 5517$, $B = 56$ with $\beta = 10$: the first recursive call will divide 55 by 5, which yields a two-digit quotient 11. Even $A \leq \beta^m B$ is not recursively fulfilled, as this example shows. The weakest possible input condition is that the $n$ most significant words of $A$ do not exceed those of $B$, i.e., $A < \beta^m(B + 1)$. In that case, the quotient is bounded by $\beta^m + \lfloor(\beta^m - 1)/B\rfloor$, which yields $\beta^m + 1$ in the case $n = m$ (compare Exercise 1.19). See also Exercise 1.22.

**Theorem 1.4.2** *Algorithm* **RecursiveDivRem** *is correct, and uses $D(n+m, n)$ operations, where $D(n+m, n) = 2D(n, n-m/2) + 2M(m/2) + O(n)$. In particular $D(n) := D(2n, n)$ satisfies $D(n) = 2D(n/2) + 2M(n/2) + O(n)$, which gives $D(n) \sim M(n)/(2^{\alpha-1} - 1)$ for $M(n) \sim n^\alpha$, $\alpha > 1$.*

**Proof.** We first check the assumption for the recursive calls: $B_1$ is normalized since it has the same most significant word than $B$.

After step 3, we have $A = (Q_1 B_1 + R_1)\beta^{2k} + (A \bmod \beta_{2k})$, thus after step 4: $A' = A - Q_1 \beta^k B$, which still holds after step 5. After step 6, we have $A' = (Q_0 B_1 + R_0)\beta^k + (A' \bmod \beta^k)$, thus after step 7: $A'' = A' - Q_0 B$, which still holds after step 8. At step 9 we thus have $A = QB + R$.

$A$ div $\beta^{2k}$ has $m+n-2k$ words, while $B_1$ has $n-k$ words, thus $0 \le Q_1 < 2\beta^{m-k}$ and $0 \le R_1 < B_1 < \beta^{n-k}$. Thus at step 4, $-2\beta^{m+k} < A' < \beta^k B$. Since $B$ is normalized, the while-loop at step 5 is performed at most four times (this can happen only when $n = m$). At step 6 we have $0 \le A' < \beta^k B$, thus $A'$ div $\beta^k$ has at most $n$ words.

It follows $0 \le Q_0 < 2\beta^k$ and $0 \le R_0 < B_1 < \beta^{n-k}$. Hence at step 7, $-2\beta^{2k} < A'' < B$, and after at most four iterations at step 8, we have $0 \le A'' < B$. □

Theorem 1.4.2 gives $D(n) \sim 2M(n)$ for Karatsuba multiplication, and $D(n) \sim 2.63M(n)$ for Toom-Cook 3-way; in the FFT range, see Exercise 1.23.

The same idea as in Exercise 1.20 applies: to decrease the probability that the estimated quotients $Q_1$ and $Q_0$ are too large, use one extra word of the truncated dividend and divisors in the recursive calls to **RecursiveDivRem**.

A graphical view of Algorithm **RecursiveDivRem** in the case $m = n$ is given in Figure 1.3, which represents the multiplication $Q \cdot B$: one first computes the lower left corner in $D(n/2)$ (step 3), second the lower right corner in $M(n/2)$ (step 4), third the upper left corner in $D(n/2)$ (step 6), and finally the upper right corner in $M(n/2)$ (step 7).

**Unbalanced Division**

The condition $n \ge m$ in Algorithm **RecursiveDivRem** means that the dividend $A$ is at most twice as large as the divisor $B$.

When $A$ is more than twice as large as $B$ ($m > n$ with the notation above), a possible strategy (see Exercise 1.24) computes $n$ words of the quotient at a time. This reduces to the base-case algorithm, replacing $\beta$ by $\beta^n$.

Figure 1.4 compares unbalanced multiplication and division in GNU MP. As expected, multiplying $x$ words by $n - x$ words takes the same time as multiplying $n - x$ words by $n$ words. However, there is no symmetry for the division, since dividing $n$ words by $x$ words for $x < n/2$ is more expensive, at least for the version of GMP that we used, than dividing $n$ words by $n-x$ words.

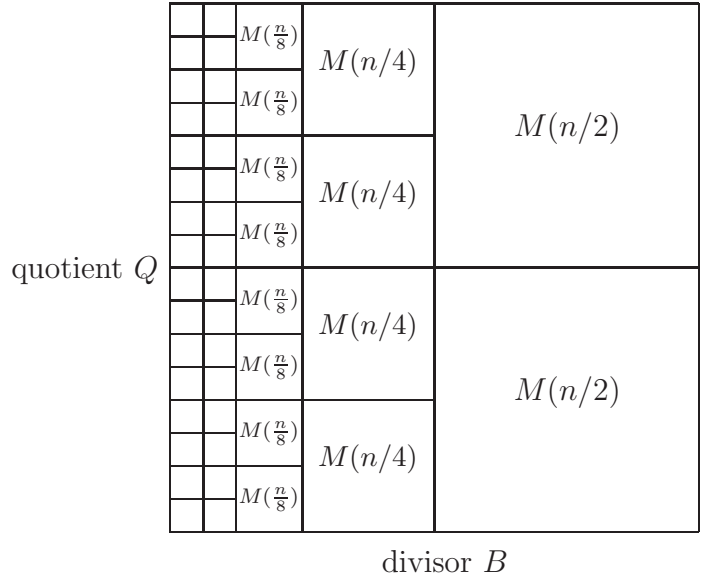

Figure 1.3: Divide and conquer division: a graphical view (most significant parts at the lower left corner).

---

**Algorithm 1.9** UnbalancedDivision

---

**Input:** $A = \sum_0^{n+m-1} a_i \beta^i$, $B = \sum_0^{n-1} b_j \beta^j$, $B$ normalized, $m > n$

**Output:** quotient $Q$ and remainder $R$ of $A$ divided by $B$

$\quad Q \leftarrow 0$

$\quad$**while** $m > n$ **do**

$\quad\quad (q, r) \leftarrow$ **RecursiveDivRem**$(A \text{ div } \beta^{m-n}, B)$ $\qquad \triangleright 2n$ by $n$ division

$\quad\quad Q \leftarrow Q\beta^n + q$

$\quad\quad A \leftarrow r\beta^{m-n} + A \bmod \beta^{m-n}$

$\quad\quad m \leftarrow m - n$

$\quad (q, r) \leftarrow$ **RecursiveDivRem**$(A, B)$

$\quad$return $Q := Q\beta^m + q$, $R := r$.

---

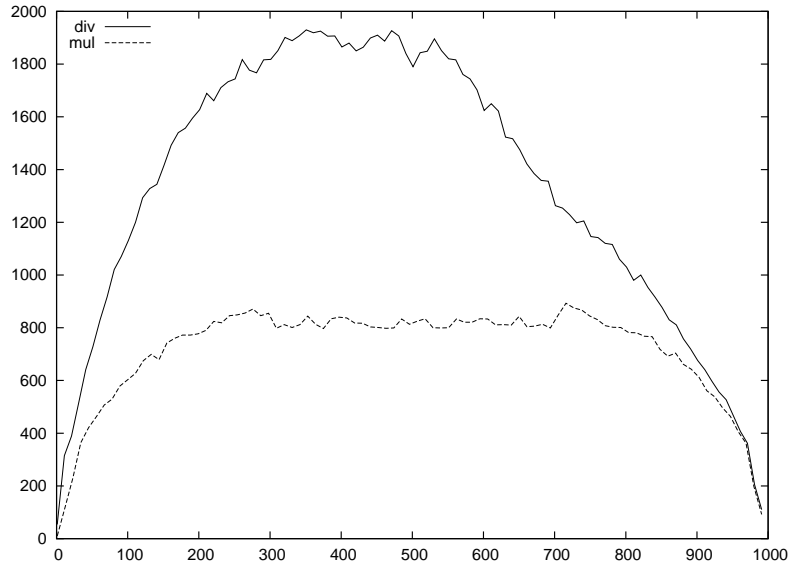

Figure 1.4: Time in $10^{-5}$ seconds for the multiplication (lower curve) of $x$ words by $1000 - x$ words and for the division (upper curve) of 1000 words by $x$ words, with GMP 5.0.0 on a Core 2 running at 2.83GHz.

### 1.4.4   Newton's Method

Newton's iteration gives the division algorithm with best asymptotic complexity. One basic component of Newton's iteration is the computation of an approximate inverse. We refer here to Chapter 4. The *p*-adic version of Newton's method, also called Hensel lifting, is used in §1.4.5 for exact division.

### 1.4.5   Exact Division

A division is *exact* when the remainder is zero. This happens, for example, when normalizing a fraction $a/b$: one divides both $a$ and $b$ by their greatest common divisor, and both divisions are exact. If the remainder is known *a priori* to be zero, this information is useful to speed up the computation of the quotient. Two strategies are possible:

- use MSB (most significant bits first) division algorithms, without computing the lower part of the remainder. Here, one has to take care

of rounding errors, in order to guarantee the correctness of the final result; or

- use LSB (least significant bits first) algorithms. If the quotient is known to be less than $\beta^n$, computing $a/b \bmod \beta^n$ will reveal it.

Subquadratic algorithms can use both strategies. We describe a least significant bit algorithm using Hensel lifting, which can be viewed as a $p$-adic version of Newton's method:

---

**Algorithm 1.10** ExactDivision

---

**Input:** $A = \sum_0^{n-1} a_i \beta^i$, $B = \sum_0^{n-1} b_j \beta^j$
**Output:** quotient $Q = A/B \bmod \beta^n$
**Require:** $\gcd(b_0, \beta) = 1$
1: $C \leftarrow 1/b_0 \bmod \beta$
2: **for** $i$ **from** $\lceil \lg n \rceil - 1$ **downto** 1 **do**
3:     $k \leftarrow \lceil n/2^i \rceil$
4:     $C \leftarrow C + C(1 - BC) \bmod \beta^k$
5: $Q \leftarrow AC \bmod \beta^k$
6: $Q \leftarrow Q + C(A - BQ) \bmod \beta^n$.

---

Algorithm **ExactDivision** uses the Karp-Markstein trick: lines 1-4 compute $1/B \bmod \beta^{\lceil n/2 \rceil}$, while the two last lines incorporate the dividend to obtain $A/B \bmod \beta^n$. Note that the *middle product* (§3.3.2) can be used in lines 4 and 6, to speed up the computation of $1 - BC$ and $A - BQ$ respectively.

A further gain can be obtained by using both strategies simultaneously: compute the most significant $n/2$ bits of the quotient using the MSB strategy, and the least significant $n/2$ bits using the LSB strategy. Since a division of size $n$ is replaced by two divisions of size $n/2$, this gives a speedup of up to two for quadratic algorithms (see Exercise 1.27).

## 1.4.6  Only Quotient or Remainder Wanted

When both the quotient and remainder of a division are needed, it is best to compute them simultaneously. This may seem to be a trivial statement, nevertheless some high-level languages provide both **div** and **mod**, but no single instruction to compute both quotient and remainder.

Once the quotient is known, the remainder can be recovered by a single multiplication as $A - QB$; on the other hand, when the remainder is known, the quotient can be recovered by an exact division as $(A - R)/B$ (§1.4.5).

However, it often happens that only one of the quotient or remainder is needed. For example, the division of two floating-point numbers reduces to the quotient of their significands (see Chapter 3). Conversely, the multiplication of two numbers modulo $N$ reduces to the remainder of their product after division by $N$ (see Chapter 2). In such cases, one may wonder if faster algorithms exist.

For a dividend of $2n$ words and a divisor of $n$ words, a significant speedup — up to a factor of two for quadratic algorithms — can be obtained when only the quotient is needed, since one does not need to update the low $n$ words of the current remainder (step 5 of Algorithm **BasecaseDivRem**).

It seems difficult to get a similar speedup when only the remainder is required. One possibility is to use Svoboda's algorithm, but this requires some precomputation, so is only useful when several divisions are performed with the same divisor. The idea is the following: precompute a multiple $B_1$ of $B$, having $3n/2$ words, the $n/2$ most significant words being $\beta^{n/2}$. Then reducing $A \bmod B_1$ requires a single $n/2 \times n$ multiplication. Once $A$ is reduced to $A_1$ of $3n/2$ words by Svoboda's algorithm with cost $2M(n/2)$, use **RecursiveDivRem** on $A_1$ and $B$, which costs $D(n/2) + M(n/2)$. The total cost is thus $3M(n/2) + D(n/2)$, instead of $2M(n/2) + 2D(n/2)$ for a full division with **RecursiveDivRem**. This gives $5M(n)/3$ for Karatsuba and $2.04M(n)$ for Toom-Cook 3-way, instead of $2M(n)$ and $2.63M(n)$ respectively. A similar algorithm is described in §2.4.2 (Subquadratic Montgomery Reduction) with further optimizations.

## 1.4.7 Division by a Single Word

We assume here that we want to divide a multiple precision number by a one-word integer $c$. As for multiplication by a one-word integer, this is an important special case. It arises for example in Toom-Cook multiplication, where one has to perform an exact division by 3 (§1.3.3). One could of course use a classical division algorithm (§1.4.1). When $\gcd(c, \beta) = 1$, Algorithm **DivideByWord** might be used to compute a modular division:

$$A + b\beta^n = cQ,$$

where the "carry" $b$ will be zero when the division is exact.

---

**Algorithm 1.11** DivideByWord

---

**Input:** $A = \sum_0^{n-1} a_i \beta^i$, $0 \le c < \beta$, $\gcd(c, \beta) = 1$
**Output:** $Q = \sum_0^{n-1} q_i \beta^i$ and $0 \le b < c$ such that $A + b\beta^n = cQ$

1: $d \leftarrow 1/c \bmod \beta$                              ▷ might be precomputed
2: $b \leftarrow 0$
3: **for** $i$ **from** $0$ **to** $n - 1$ **do**
4:     **if** $b \le a_i$ **then** $(x, b') \leftarrow (a_i - b, 0)$
5:         **else** $(x, b') \leftarrow (a_i - b + \beta, 1)$
6:     $q_i \leftarrow dx \bmod \beta$
7:     $b'' \leftarrow (q_i c - x)/\beta$
8:     $b \leftarrow b' + b''$
9: return $\sum_0^{n-1} q_i \beta^i$, $b$.

---

**Theorem 1.4.3** *The output of Alg.* ***DivideByWord*** *satisfies* $A + b\beta^n = cQ$.

**Proof.** We show that after step $i$, $0 \le i < n$, we have $A_i + b\beta^{i+1} = cQ_i$, where $A_i := \sum_{j=0}^{i} a_i \beta^i$ and $Q_i := \sum_{j=0}^{i} q_i \beta^i$. For $i = 0$, this is $a_0 + b\beta = cq_0$, which is just line 7: since $q_0 = a_0/c \bmod \beta$, $q_0 c - a_0$ is divisible by $\beta$. Assume now that $A_{i-1} + b\beta^i = cQ_{i-1}$ holds for $1 \le i < n$. We have $a_i - b + b'\beta = x$, so $x + b''\beta = cq_i$, thus $A_i + (b' + b'')\beta^{i+1} = A_{i-1} + \beta^i(a_i + b'\beta + b''\beta) = cQ_{i-1} - b\beta^i + \beta^i(x + b - b'\beta + b'\beta + b''\beta) = cQ_{i-1} + \beta^i(x + b''\beta) = cQ_i$. □

REMARK: at step 7, since $0 \le x < \beta$, $b''$ can also be obtained as $\lfloor q_i c/\beta \rfloor$.

Algorithm **DivideByWord** is just a special case of Hensel's division, which is the topic of the next section; it can easily be extended to divide by integers of a few words.

## 1.4.8   Hensel's Division

Classical division involves cancelling the most significant part of the dividend by a multiple of the divisor, while Hensel's division cancels the least significant part (Figure 1.5). Given a dividend $A$ of $2n$ words and a divisor $B$ of $n$ words, the classical or MSB (most significant bit) division computes a quotient $Q$ and a remainder $R$ such that $A = QB + R$, while Hensel's or LSB (least significant bit) division computes a LSB-quotient $Q'$ and a LSB-remainder $R'$ such that $A = Q'B + R'\beta^n$. While MSB division requires the most significant bit of $B$ to be set, LSB division requires $B$ to be relatively prime to the word base $\beta$, i.e., $B$ to be odd for $\beta$ a power of two.

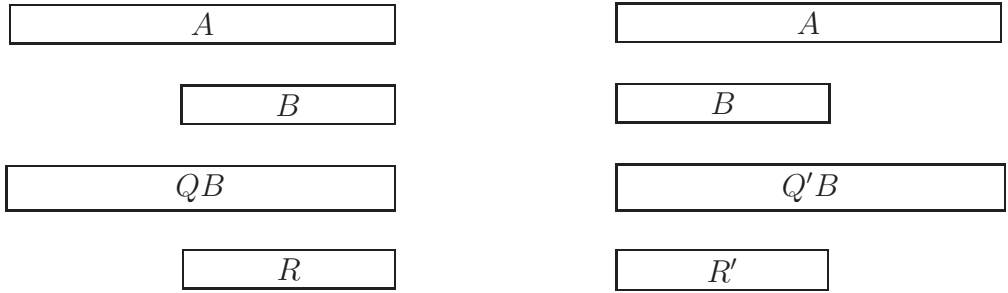

Figure 1.5: Classical/MSB division (left) vs Hensel/LSB division (right).

The LSB-quotient is uniquely defined by $Q' = A/B \bmod \beta^n$, with $0 \le Q' < \beta^n$. This in turn uniquely defines the LSB-remainder $R' = (A - Q'B)\beta^{-n}$, with $-B < R' < \beta^n$.

Most MSB-division variants (naive, with preconditioning, divide and conquer, Newton's iteration) have their LSB-counterpart. For example, LSB preconditioning involves using a multiple $kB$ of the divisor such that $kB = 1 \bmod \beta$, and Newton's iteration is called Hensel lifting in the LSB case. The exact division algorithm described at the end of §1.4.5 uses both MSB- and LSB-division simultaneously. One important difference is that LSB-division does not need any correction step, since the carries go in the direction opposite to the cancelled bits.

When only the remainder is wanted, Hensel's division is usually known as Montgomery reduction (see §2.4.2).

## 1.5 Roots

### 1.5.1 Square Root

The "paper and pencil" method once taught at school to extract square roots is very similar to "paper and pencil" division. It decomposes an integer $m$ of the form $s^2 + r$, taking two digits of $m$ at a time, and finding one digit of $s$ for each two digits of $m$. It is based on the following idea. If $m = s^2 + r$ is the current decomposition, then taking two more digits of the argument, we have a decomposition of the form $100m + r' = 100s^2 + 100r + r'$ with $0 \le r' < 100$. Since $(10s + t)^2 = 100s^2 + 20st + t^2$, a good approximation to the next digit $t$ can be found by dividing $10r$ by $2s$.

Algorithm **SqrtRem** generalizes this idea to a power $\beta^\ell$ of the internal base close to $m^{1/4}$: one obtains a divide and conquer algorithm, which is in fact an error-free variant of Newton's method (*cf* Chapter 4):

---

**Algorithm 1.12** SqrtRem

---

**Input:** $m = a_{n-1}\beta^{n-1} + \cdots + a_1\beta + a_0$ with $a_{n-1} \ne 0$

**Output:** $(s, r)$ such that $s^2 \le m = s^2 + r < (s + 1)^2$

**Require:** a base-case routine **BasecaseSqrtRem**

  $\ell \leftarrow \lfloor (n-1)/4 \rfloor$

  **if** $\ell = 0$ **then** return **BasecaseSqrtRem**$(m)$

  write $m = a_3\beta^{3\ell} + a_2\beta^{2\ell} + a_1\beta^\ell + a_0$ with $0 \le a_2, a_1, a_0 < \beta^\ell$

  $(s', r') \leftarrow$ **SqrtRem**$(a_3\beta^\ell + a_2)$

  $(q, u) \leftarrow$ **DivRem**$(r'\beta^\ell + a_1, 2s')$

  $s \leftarrow s'\beta^\ell + q$

  $r \leftarrow u\beta^\ell + a_0 - q^2$

  **if** $r < 0$ **then**

    $r \leftarrow r + 2s - 1, \quad s \leftarrow s - 1$

  return $(s, r)$.

---

**Theorem 1.5.1** *Algorithm **SqrtRem** correctly returns the integer square root $s$ and remainder $r$ of the input $m$, and has complexity $R(2n) \sim R(n) + D(n) + S(n)$ where $D(n)$ and $S(n)$ are the complexities of the division with remainder and squaring respectively. This gives $R(n) \sim n^2/2$ with naive multiplication, $R(n) \sim 4K(n)/3$ with Karatsuba's multiplication, assuming $S(n) \sim 2M(n)/3$.*

As an example, assume Algorithm **SqrtRem** is called on $m = 123\,456\,789$ with $\beta = 10$. One has $n = 9$, $\ell = 2$, $a_3 = 123$, $a_2 = 45$, $a_1 = 67$, and $a_0 = 89$. The recursive call for $a_3\beta^\ell + a_2 = 12\,345$ yields $s' = 111$ and $r' = 24$. The **DivRem** call yields $q = 11$ and $u = 25$, which gives $s = 11\,111$ and $r = 2\,468$.

Another nice way to compute the integer square root of an integer $m$, i.e., $\lfloor m^{1/2} \rfloor$, is Algorithm **SqrtInt**, which is an all-integer version of Newton's method (§4.2).

Still with input $123\,456\,789$, we successively get $s = 61\,728\,395$, $30\,864\,198$, $15\,432\,100$, $7\,716\,053$, $3\,858\,034$, $1\,929\,032$, $964\,547$, $482\,337$, $241\,296$, $120\,903$,

---

**Algorithm 1.13** SqrtInt

---

**Input:** an integer $m \geq 1$
**Output:** $s = \lfloor m^{1/2} \rfloor$
  1: $u \leftarrow m$                                  ▷ any value $u \geq \lfloor m^{1/2} \rfloor$ works
  2: **repeat**
  3:     $s \leftarrow u$
  4:     $t \leftarrow s + \lfloor m/s \rfloor$
  5:     $u \leftarrow \lfloor t/2 \rfloor$
  6: **until** $u \geq s$
  7: return $s$.

---

$60\,962$, $31\,493$, $17\,706$, $12\,339$, $11\,172$, $11\,111$, $11\,111$. Convergence is slow because the initial value of $u$ assigned at line 1 is much too large. However, any initial value greater than or equal to $\lfloor m^{1/2} \rfloor$ works (see the proof of Algorithm **RootInt** below): starting from $s = 12\,000$, one gets $s = 11\,144$ then $s = 11\,111$. See Exercise 1.28.

## 1.5.2    $k$-th Root

The idea of Algorithm **SqrtRem** for the integer square root can be generalized to any power: if the current decomposition is $m = m'\beta^k + m''\beta^{k-1} + m'''$, first compute a $k$-th root of $m'$, say $m' = s^k + r$, then divide $r\beta + m''$ by $ks^{k-1}$ to get an approximation of the next root digit $t$, and correct it if needed. Unfortunately the computation of the remainder, which is easy for the square root, involves $O(k)$ terms for the $k$-th root, and this method may be slower than Newton's method with floating-point arithmetic (§4.2.3).

Similarly, Algorithm **SqrtInt** can be generalized to the $k$-th root (see Algorithm **RootInt**).

**Theorem 1.5.2** *Algorithm* **RootInt** *terminates and returns* $\lfloor m^{1/k} \rfloor$.

**Proof.** As long as $u < s$ in step 6, the sequence of $s$-values is decreasing, thus it suffices to consider what happens when $u \geq s$. First it is easy so see that $u \geq s$ implies $m \geq s^k$, because $t \geq ks$ thus $(k-1)s + m/s^{k-1} \geq ks$. Consider now the function $f(t) := [(k-1)t + m/t^{k-1}]/k$ for $t > 0$; its derivative is negative for $t < m^{1/k}$, and positive for $t > m^{1/k}$, thus $f(t) \geq f(m^{1/k}) = m^{1/k}$. This proves that $s \geq \lfloor m^{1/k} \rfloor$. Together with $s \leq m^{1/k}$, this proves that $s = \lfloor m^{1/k} \rfloor$ at the end of the algorithm. $\qquad\square$

---

**Algorithm 1.14** RootInt

---

**Input:** integers $m \geq 1$, and $k \geq 2$

**Output:** $s = \lfloor m^{1/k} \rfloor$

  1: $u \leftarrow m$                                             ▷ any value $u \geq \lfloor m^{1/k} \rfloor$ works

  2: **repeat**

  3:     $s \leftarrow u$

  4:     $t \leftarrow (k-1)s + \lfloor m/s^{k-1} \rfloor$

  5:     $u \leftarrow \lfloor t/k \rfloor$

  6: **until** $u \geq s$

  7: return $s$.

---

Note that any initial value greater than or equal to $\lfloor m^{1/k} \rfloor$ works at step 1. Incidentally, we have proved the correctness of Algorithm **SqrtInt**, which is just the special case $k = 2$ of Algorithm **RootInt**.

### 1.5.3 Exact Root

When a $k$-th root is known to be exact, there is of course no need to compute exactly the final remainder in "exact root" algorithms, which saves some computation time. However, one has to check that the remainder is sufficiently small that the computed root is correct.

When a root is known to be exact, one may also try to compute it starting from the least significant bits, as for exact division. Indeed, if $s^k = m$, then $s^k = m \bmod \beta^\ell$ for any integer $\ell$. However, in the case of exact division, the equation $a = qb \bmod \beta^\ell$ has only one solution $q$ as soon as $b$ is relatively prime to $\beta$. Here, the equation $s^k = m \bmod \beta^\ell$ may have several solutions, so the lifting process is not unique. For example, $x^2 = 1 \bmod 2^3$ has four solutions $1, 3, 5, 7$.

Suppose we have $s^k = m \bmod \beta^\ell$, and we want to lift to $\beta^{\ell+1}$. This implies $(s + t\beta^\ell)^k = m + m'\beta^\ell \bmod \beta^{\ell+1}$ where $0 \leq t, m' < \beta$. Thus

$$kt = m' + \frac{m - s^k}{\beta^\ell} \bmod \beta.$$

This equation has a unique solution $t$ when $k$ is relatively prime to $\beta$. For example, we can extract cube roots in this way for $\beta$ a power of two. When $k$ is relatively prime to $\beta$, we can also compute the root simultaneously from the most significant and least significant ends, as for exact division.

**Unknown Exponent**

Assume now that one wants to check if a given integer $m$ is an exact power, without knowing the corresponding exponent. For example, some primality testing or factorization algorithms fail when given an exact power, so this has to be checked first. Algorithm **IsPower** detects exact powers, and returns the largest corresponding exponent (or 1 if the input is not an exact power).

---

**Algorithm 1.15** IsPower

---

**Input:** a positive integer $m$
**Output:** $k \geq 2$ when $m$ is an exact $k$-th power, 1 otherwise
 1: **for** $k$ **from** $\lfloor \lg m \rfloor$ **downto** 2 **do**
 2:    **if** $m$ is a $k$-th power **then** return $k$
 3: return 1.

---

To quickly detect non-$k$-th powers at step 2, one may use modular algorithms when $k$ is relatively prime to the base $\beta$ (see above).

REMARK: in Algorithm **IsPower**, one can limit the search to prime exponents $k$, but then the algorithm does not necessarily return the largest exponent, and we might have to call it again. For example, taking $m = 117649$, the modified algorithm first returns 3 because $117649 = 49^3$, and when called again with $m = 49$ it returns 2.

## 1.6 Greatest Common Divisor

Many algorithms for computing gcds may be found in the literature. We can distinguish between the following (non-exclusive) types:

- left-to-right (MSB) versus right-to-left (LSB) algorithms: in the former the actions depend on the most significant bits, while in the latter the actions depend on the least significant bits;

- naive algorithms: these $O(n^2)$ algorithms consider one word of each operand at a time, trying to guess from them the first quotients; we count in this class algorithms considering double-size words, namely Lehmer's algorithm and Sorenson's $k$-ary reduction in the left-to-right and right-to-left cases respectively; algorithms not in this class consider a number of words that depends on the input size $n$, and are often subquadratic;

- subtraction-only algorithms: these algorithms trade divisions for subtractions, at the cost of more iterations;

- plain versus extended algorithms: the former just compute the gcd of the inputs, while the latter express the gcd as a linear combination of the inputs.

## 1.6.1   Naive GCD

For completeness we mention Euclid's algorithm for finding the gcd of two non-negative integers $u, v$.

---

**Algorithm 1.16** EuclidGcd

**Input:** $u, v$ nonnegative integers (not both zero)
**Output:** $\gcd(u, v)$
   **while** $v \neq 0$ **do**
      $(u, v) \leftarrow (v, u \bmod v)$
  **return** $u$.

---

Euclid's algorithm is discussed in many textbooks, and we do not recommend it in its simplest form, except for testing purposes. Indeed, it is usually a slow way to compute a gcd. However, Euclid's algorithm does show the connection between gcds and continued fractions. If $u/v$ has a regular continued fraction of the form

$$u/v = q_0 + \frac{1}{q_1+} \frac{1}{q_2+} \frac{1}{q_3+} \cdots,$$

then the quotients $q_0, q_1, \ldots$ are precisely the quotients $u$ div $v$ of the divisions performed in Euclid's algorithm. For more on continued fractions, see §4.6.

**Double-Digit Gcd.**   A first improvement comes from Lehmer's observation: the first few quotients in Euclid's algorithm usually can be determined from the most significant words of the inputs. This avoids expensive divisions that give small quotients most of the time (see [143, §4.5.3]). Consider for example $a = 427\,419\,669\,081$ and $b = 321\,110\,693\,270$ with 3-digit words.

The first quotients are $1, 3, 48, \ldots$ Now if we consider the most significant words, namely 427 and 321, we get the quotients $1, 3, 35, \ldots$. If we stop after the first two quotients, we see that we can replace the initial inputs by $a - b$ and $-3a + 4b$, which gives $106\,308\,975\,811$ and $2\,183\,765\,837$.

Lehmer's algorithm determines cofactors from the most significant words of the input integers. Those cofactors usually have size only half a word. The **DoubleDigitGcd** algorithm — which should be called "double-word" — uses the *two* most significant words instead, which gives cofactors $t, u, v, w$ of one full-word each, such that $\gcd(a, b) = \gcd(ta + ub, va + wb)$. This is optimal for the computation of the four products $ta$, $ub$, $va$, $wb$. With the above example, if we consider $427\,419$ and $321\,110$, we find that the first five quotients agree, so we can replace $a, b$ by $-148a + 197b$ and $441a - 587b$, i.e., $695\,550\,202$ and $97\,115\,231$.

---

**Algorithm 1.17** DoubleDigitGcd

---

**Input:** $a := a_{n-1}\beta^{n-1} + \cdots + a_0$, $b := b_{m-1}\beta^{m-1} + \cdots + b_0$
**Output:** $\gcd(a, b)$
    **if** $b = 0$ **then** return $a$
    **if** $m < 2$ **then** return **BasecaseGcd**$(a, b)$
    **if** $a < b$ or $n > m$ **then** return **DoubleDigitGcd**$(b, a \bmod b)$
    $(t, u, v, w) \leftarrow$ **HalfBezout**$(a_{n-1}\beta + a_{n-2}, b_{n-1}\beta + b_{n-2})$
    return **DoubleDigitGcd**$(|ta + ub|, |va + wb|)$.

---

The subroutine **HalfBezout** takes as input two 2-word integers, performs Euclid's algorithm until the smallest remainder fits in one word, and returns the corresponding matrix $[t, u; v, w]$.

**Binary Gcd.** A better algorithm than Euclid's, though also of $O(n^2)$ complexity, is the *binary* algorithm. It differs from Euclid's algorithm in two ways: it consider least significant bits first, and it avoids divisions, except for divisions by two (which can be implemented as shifts on a binary computer). See Algorithm **BinaryGcd**. Note that the first three "while" loops can be omitted if the inputs $a$ and $b$ are odd.

**Sorenson's $k$-ary reduction**

The binary algorithm is based on the fact that if $a$ and $b$ are both odd, then $a - b$ is even, and we can remove a factor of two since $\gcd(a, b)$ is odd.

---

**Algorithm 1.18** BinaryGcd

---

**Input:** $a, b > 0$
**Output:** $\gcd(a, b)$

   $t \leftarrow 1$
   **while** $a \bmod 2 = b \bmod 2 = 0$ **do**
      $(t, a, b) \leftarrow (2t, a/2, b/2)$
   **while** $a \bmod 2 = 0$ **do**
      $a \leftarrow a/2$
   **while** $b \bmod 2 = 0$ **do**
      $b \leftarrow b/2$                        $\triangleright$ now $a$ and $b$ are both odd
   **while** $a \neq b$ **do**
      $(a, b) \leftarrow (|a - b|, \min(a, b))$
      $a \leftarrow a/2^{\nu(a)}$              $\triangleright$ $\nu(a)$ is the 2-valuation of $a$
  return $ta$.

---

Sorenson's $k$-ary reduction is a generalization of that idea: given $a$ and $b$ odd, we try to find small integers $u, v$ such that $ua - vb$ is divisible by a large power of two.

**Theorem 1.6.1** [227] *If $a, b > 0$, $m > 1$ with $\gcd(a, m) = \gcd(b, m) = 1$, there exist $u, v$, $0 < |u|, v < \sqrt{m}$ such that $ua = vb \bmod m$.*

Algorithm **ReducedRatMod** finds such a pair $(u, v)$; it is a simple variation of the extended Euclidean algorithm; indeed, the $u_i$ are quotients in the continued fraction expansion of $c/m$.

When $m$ is a prime power, the inversion $1/b \bmod m$ at step 1 of Algorithm **ReducedRatMod** can be performed efficiently using Hensel lifting (§2.5).

Given two integers $a, b$ of say $n$ words, Algorithm **ReducedRatMod** with $m = \beta^2$ returns two integers $u, v$ such that $vb - ua$ is a multiple of $\beta^2$. Since $u, v$ have at most one word each, $a' = (vb - ua)/\beta^2$ has at most $n - 1$ words — plus possibly one bit — therefore with $b' = b \bmod a'$ we obtain $\gcd(a, b) = \gcd(a', b')$, where both $a'$ and $b'$ have about one word less than $\max(a, b)$. This gives an LSB variant of the double-digit (MSB) algorithm.

---

**Algorithm 1.19** ReducedRatMod

---

**Input:** $a, b > 0$, $m > 1$ with $\gcd(a, m) = \gcd(b, m) = 1$

**Output:** $(u, v)$ such that $0 < |u|, v < \sqrt{m}$ and $ua = vb \bmod m$

1: $c \leftarrow a/b \bmod m$
2: $(u_1, v_1) \leftarrow (0, m)$
3: $(u_2, v_2) \leftarrow (1, c)$
4: **while** $v_2 \geq \sqrt{m}$ **do**
5: $\quad q \leftarrow \lfloor v_1/v_2 \rfloor$
6: $\quad (u_1, u_2) \leftarrow (u_2, u_1 - qu_2)$
7: $\quad (v_1, v_2) \leftarrow (v_2, v_1 - qv_2)$
8: return $(u_2, v_2)$.

---

## 1.6.2 Extended GCD

Algorithm **ExtendedGcd** solves the *extended* greatest common divisor problem: given two integers $a$ and $b$, it computes their gcd $g$, and also two integers $u$ and $v$ (called *Bézout coefficients* or sometimes *cofactors* or *multipliers*) such that $g = ua + vb$.

---

**Algorithm 1.20** ExtendedGcd

---

**Input:** positive integers $a$ and $b$

**Output:** integers $(g, u, v)$ such that $g = \gcd(a, b) = ua + vb$

1: $(u, w) \leftarrow (1, 0)$
2: $(v, x) \leftarrow (0, 1)$
3: **while** $b \neq 0$ **do**
4: $\quad (q, r) \leftarrow \textbf{DivRem}(a, b)$
5: $\quad (a, b) \leftarrow (b, r)$
6: $\quad (u, w) \leftarrow (w, u - qw)$
7: $\quad (v, x) \leftarrow (x, v - qx)$
8: return $(a, u, v)$.

---

If $a_0$ and $b_0$ are the input numbers, and $a, b$ the current values, the following invariants hold at the start of each iteration of the while loop and after the while loop: $a = ua_0 + vb_0$, and $b = wa_0 + xb_0$. (See Exercise 1.30 for a bound on the cofactor $u$.)

An important special case is modular inversion (see Chapter 2): given an integer $n$, one wants to compute $1/a \bmod n$ for $a$ relatively prime to $n$. One then simply runs Algorithm **ExtendedGcd** with input $a$ and $b = n$: this yields $u$ and $v$ with $ua + vn = 1$, thus $1/a = u \bmod n$. Since $v$ is not needed here, we can simply avoid computing $v$ and $x$, by removing steps 2 and 7.

It may also be worthwhile to compute only $u$ in the general case, as the cofactor $v$ can be recovered from $v = (g - ua)/b$, this division being exact (see §1.4.5).

All known algorithms for subquadratic gcd rely on an extended gcd subroutine which is called recursively, so we discuss the subquadratic extended gcd in the next section.

### 1.6.3   Half Binary GCD, Divide and Conquer GCD

Designing a subquadratic integer gcd algorithm that is both mathematically correct and efficient in practice is a challenging problem.

A first remark is that, starting from $n$-bit inputs, there are $O(n)$ terms in the remainder sequence $r_0 = a$, $r_1 = b$, ..., $r_{i+1} = r_{i-1} \bmod r_i$, ..., and the size of $r_i$ decreases linearly with $i$. Thus, computing all the partial remainders $r_i$ leads to a quadratic cost, and a fast algorithm should avoid this.

However, the partial quotients $q_i = r_{i-1} \operatorname{div} r_i$ are usually small: the main idea is thus to compute them without computing the partial remainders. This can be seen as a generalization of the **DoubleDigitGcd** algorithm: instead of considering a fixed base $\beta$, adjust it so that the inputs have four "big words". The cofactor-matrix returned by the **HalfBezout** subroutine will then reduce the input size to about $3n/4$. A second call with the remaining two most significant "big words" of the new remainders will reduce their size to half the input size. See Exercise 1.31.

The same method applies in the LSB case, and is in fact simpler to turn into a correct algorithm. In this case, the terms $r_i$ form a *binary remainder sequence*, which corresponds to the iteration of the **BinaryDivide** algorithm, with starting values $a, b$.

The integer $q$ is the *binary quotient* of $a$ and $b$, and $r$ is the *binary remainder*.

This right-to-left division defines a right-to-left remainder sequence $a_0 = a$, $a_1 = b$, ..., where $a_{i+1} = $ **BinaryRemainder** $(a_{i-1}, a_i)$, and $\nu(a_{i+1}) < \nu(a_i)$. It can be shown that this sequence eventually reaches $a_{i+1} = 0$ for some index $i$. Assuming $\nu(a) = 0$, then $\gcd(a, b)$ is the odd part of $a_i$. Indeed, in Algorithm **BinaryDivide**, if some odd prime divides both $a$ and $b$, it certainly divides $2^{-j}b$ which is an integer, and thus it divides $a + q2^{-j}b$.

---

**Algorithm 1.21** BinaryDivide

---

**Input:** $a, b \in \mathbb{Z}$ with $\nu(b) - \nu(a) = j > 0$
**Output:** $|q| < 2^j$ and $r = a + q2^{-j}b$ such that $\nu(b) < \nu(r)$
    $b' \leftarrow 2^{-j}b$
    $q \leftarrow -a/b' \bmod 2^{j+1}$
    **if** $q \geq 2^j$ **then** $q \leftarrow q - 2^{j+1}$
    return $q, r = a + q2^{-j}b$.

---

Conversely, if some odd prime divides both $b$ and $r$, it divides also $2^{-j}b$, thus it divides $a = r - q2^{-j}b$; this shows that no spurious factor appears, unlike in some other gcd algorithms.

EXAMPLE: let $a = a_0 = 935$ and $b = a_1 = 714$, so $\nu(b) = \nu(a)+1$. Algorithm **BinaryDivide** computes $b' = 357$, $q = 1$, and $a_2 = a + q2^{-j}b = 1292$. The next step gives $a_3 = 1360$, then $a_4 = 1632$, $a_5 = 2176$, $a_6 = 0$. Since $2176 = 2^7 \cdot 17$, we conclude that the gcd of 935 and 714 is 17. Note that the binary remainder sequence might contain negative terms and terms larger than $a, b$. For example, starting from $a = 19$ and $b = 2$, we get $19, 2, 20, -8, 16, 0$.

An asymptotically fast GCD algorithm with complexity $O(M(n) \log n)$ can be constructed with Algorithm **HalfBinaryGcd**.

**Theorem 1.6.2** *Given $a, b \in \mathbb{Z}$ with $\nu(a) = 0$ and $\nu(b) > 0$, and an integer $k \geq 0$, Algorithm **HalfBinaryGcd** returns an integer $0 \leq j \leq k$ and a matrix $R$ such that, if $c = 2^{-2j}(R_{1,1}a + R_{1,2}b)$ and $d = 2^{-2j}(R_{2,1}a + R_{2,2}b)$:*

1. *$c$ and $d$ are integers with $\nu(c) = 0$ and $\nu(d) > 0$;*

2. *$c^* = 2^j c$ and $d^* = 2^j d$ are two consecutive terms from the binary remainder sequence of $a, b$ with $\nu(c^*) \leq k < \nu(d^*)$.*

**Proof.** We prove the theorem by induction on $k$. If $k = 0$, the algorithm returns $j = 0$ and the identity matrix, thus we have $c = a$ and $d = b$, and the statement is true. Now suppose $k > 0$, and assume that the theorem is true up to $k - 1$.

The first recursive call uses $k_1 < k$, since $k_1 = \lfloor k/2 \rfloor < k$. After step 5, by induction $a'_1 = 2^{-2j_1}(R_{1,1}a_1 + R_{1,2}b_1)$ and $b'_1 = 2^{-2j_1}(R_{2,1}a_1 + R_{2,2}b_1)$ are integers with $\nu(a'_1) = 0 < \nu(b'_1)$, and $2^{j_1}a'_1, 2^{j_1}b'_1$ are two consecutive terms

---

**Algorithm 1.22** HalfBinaryGcd

---

**Input:** $a, b \in \mathbb{Z}$ with $0 = \nu(a) < \nu(b)$, a non-negative integer $k$
**Output:** an integer $j$ and a $2 \times 2$ matrix $R$ satisfying Theorem 1.6.2

1: **if** $\nu(b) > k$ **then**

2:       return $0, \begin{pmatrix} 1 & 0 \\ 0 & 1 \end{pmatrix}$

3: $k_1 \leftarrow \lfloor k/2 \rfloor$

4: $a_1 \leftarrow a \bmod 2^{2k_1+1}, \quad b_1 \leftarrow b \bmod 2^{2k_1+1}$

5: $j_1, R \leftarrow \textbf{HalfBinaryGcd}(a_1, b_1, k_1)$

6: $a' \leftarrow 2^{-2j_1}(R_{1,1}a + R_{1,2}b), \quad b' \leftarrow 2^{-2j_1}(R_{2,1}a + R_{2,2}b)$

7: $j_0 \leftarrow \nu(b')$

8: **if** $j_0 + j_1 > k$ **then**

9:       return $j_1, R$

10: $q, r \leftarrow \textbf{BinaryDivide}(a', b')$

11: $k_2 \leftarrow k - (j_0 + j_1)$

12: $a_2 \leftarrow b'/2^{j_0} \bmod 2^{2k_2+1}, \quad b_2 \leftarrow r/2^{j_0} \bmod 2^{2k_2+1}$

13: $j_2, S \leftarrow \textbf{HalfBinaryGcd}(a_2, b_2, k_2)$

14: return $j_1 + j_0 + j_2, S \times \begin{pmatrix} 0 & 2^{j_0} \\ 2^{j_0} & q \end{pmatrix} \times R.$

---

from the binary remainder sequence of $a_1, b_1$. Lemma 7 of [209] says that the quotients of the remainder sequence of $a, b$ coincide with those of $a_1, b_1$ up to $2^{j_1}a'$ and $2^{j_1}b'$. This proves that $2^{j_1}a', 2^{j_1}b'$ are two consecutive terms of the remainder sequence of $a, b$. Since $a$ and $a_1$ differ by a multiple of $2^{2k_1+1}$, $a'$ and $a_1'$ differ by a multiple of $2^{2k_1+1-2j_1} \geq 2$ since $j_1 \leq k_1$ by induction. It follows that $\nu(a') = 0$. Similarly, $b'$ and $b_1'$ differ by a multiple of 2, thus $j_0 = \nu(b') > 0$.

The second recursive call uses $k_2 < k$, since by induction $j_1 \geq 0$ and we just showed $j_0 > 0$. It easily follows that $j_1 + j_0 + j_2 > 0$, and thus $j \geq 0$. If we exit at step 9, we have $j = j_1 \leq k_1 < k$. Otherwise $j = j_1 + j_0 + j_2 = k - k_2 + j_2 \leq k$ by induction.

If $j_0 + j_1 > k$, we have $\nu(2^{j_1}b') = j_0 + j_1 > k$, we exit the algorithm and the statement holds. Now assume $j_0 + j_1 \leq k$. We compute an extra term $r$ of the remainder sequence from $a', b'$, which up to multiplication by $2^{j_1}$, is an extra term of the remainder sequence of $a, b$. Since $r = a' + q2^{-j_0}b'$, we have

$$\begin{pmatrix} b' \\ r \end{pmatrix} = 2^{-j_0} \begin{pmatrix} 0 & 2^{j_0} \\ 2^{j_0} & q \end{pmatrix} \begin{pmatrix} a' \\ b' \end{pmatrix}.$$

The new terms of the remainder sequence are $b'/2^{j_0}$ and $r/2^{j_0}$, adjusted so that $\nu(b'/2^{j_0}) = 0$. The same argument as above holds for the second recursive call, which stops when the 2-valuation of the sequence starting from $a_2, b_2$ exceeds $k_2$; this corresponds to a 2-valuation larger than $j_0 + j_1 + k_2 = k$ for the $a, b$ remainder sequence. □

Given two $n$-bit integers $a$ and $b$, and $k = n/2$, **HalfBinaryGcd** yields two consecutive elements $c^*, d^*$ of their binary remainder sequence with bit-size about $n/2$ (for their odd part).

EXAMPLE: let $a = 1\,889\,826\,700\,059$ and $b = 421\,872\,857\,844$, with $k = 20$. The first recursive call with $a_1 = 1\,243\,931$, $b_1 = 1\,372\,916$, $k_1 = 10$ gives $j_1 = 8$ and $R = \begin{pmatrix} 352 & 280 \\ 260 & 393 \end{pmatrix}$, which corresponds to $a' = 11\,952\,871\,683$ and $b' = 10\,027\,328\,112$, with $j_0 = 4$. The binary division yields the new term $r = 8\,819\,331\,648$, and we have $k_2 = 8$, $a_2 = 52\,775$, $b_2 = 50\,468$. The second recursive call gives $j_2 = 8$ and $S = \begin{pmatrix} 64 & 272 \\ 212 & -123 \end{pmatrix}$, which finally gives $j = 20$ and the matrix $\begin{pmatrix} 1\,444\,544 & 1\,086\,512 \\ 349\,084 & 1\,023\,711 \end{pmatrix}$, which corresponds to the remainder terms $r_8 = 2\,899\,749 \cdot 2^j$, $r_9 = 992\,790 \cdot 2^j$. With the same $a, b$ values, but with $k = 41$, which corresponds to the bit-size of $a$, we get as final values of the algorithm $r_{15} = 3 \cdot 2^{41}$ and $r_{16} = 0$, which proves that $\gcd(a, b) = 3$.

Let $H(n)$ be the complexity of **HalfBinaryGcd** for inputs of $n$ bits and $k = n/2$; $a_1$ and $b_1$ have $\sim n/2$ bits, the coefficients of $R$ have $\sim n/4$ bits, and $a', b'$ have $\sim 3n/4$ bits. The remainders $a_2, b_2$ have $\sim n/2$ bits, the coefficients of $S$ have $\sim n/4$ bits, and the final values $c, d$ have $\sim n/2$ bits. The main costs are the matrix-vector product at step 6, and the final matrix-matrix product. We obtain $H(n) \sim 2H(n/2) + 4M(n/4, n) + 7M(n/4)$, assuming we use Strassen's algorithm to multiply two $2 \times 2$ matrices with 7 scalar products, i.e., $H(n) \sim 2H(n/2) + 17M(n/4)$, assuming that we compute each $M(n/4, n)$ product with a single FFT transform of width $5n/4$, which gives cost about $M(5n/8) \sim 0.625M(n)$ in the FFT range. Thus $H(n) = O(M(n) \log n)$.

For the plain gcd, we call **HalfBinaryGcd** with $k = n$, and instead of computing the final matrix product, we multiply $2^{-2j_2}S$ by $(b', r)$ — the components have $\sim n/2$ bits — to obtain the final $c, d$ values. The first recursive call has $a_1, b_1$ of size $n$ with $k_1 \approx n/2$, and corresponds to $H(n)$; the matrix $R$ and $a', b'$ have $n/2$ bits, and $k_2 \approx n/2$, thus the second recursive call corresponds to a plain gcd of size $n/2$. The cost $G(n)$ satisfies $G(n) = H(n) + G(n/2) + 4M(n/2, n) + 4M(n/2) \sim H(n) + G(n/2) + 10M(n/2)$. Thus $G(n) = O(M(n) \log n)$.

An application of the half gcd *per se* in the MSB case is the *rational reconstruction* problem. Assume one wants to compute a rational $p/q$ where $p$ and $q$ are known to be bounded by some constant $c$. Instead of computing with rationals, one may perform all computations modulo some integer $n > c^2$. Hence one will end up with $p/q = m \bmod n$, and the problem is now to find the unknown $p$ and $q$ from the known integer $m$. To do this, one starts an extended gcd from $m$ and $n$, and one stops as soon as the current $a$ and $u$ values — as in **ExtendedGcd** — are smaller than $c$: since we have $a = um + vn$, this gives $m = a/u \bmod n$. This is exactly what is called a half-gcd; a subquadratic version in the LSB case is given above.

## 1.7  Base Conversion

Since computers usually work with binary numbers, and human prefer decimal representations, input/output base conversions are needed. In a typical computation, there are only a few conversions, compared to the total number of operations, so optimizing conversions is less important than optimizing other aspects of the computation. However, when working with huge numbers, naive conversion algorithms may slow down the whole computation.

In this section we consider that numbers are represented internally in base $\beta$ — usually a power of 2 — and externally in base $B$ — say a power of 10. When both bases are *commensurable*, i.e., both are powers of a common integer, like $\beta = 8$ and $B = 16$, conversions of $n$-digit numbers can be performed in $O(n)$ operations. We assume here that $\beta$ and $B$ are not commensurable.

One might think that only one algorithm is needed, since input and output are symmetric by exchanging bases $\beta$ and $B$. Unfortunately, this is not true, since computations are done only in base $\beta$ (see Exercise 1.37).

## 1.7.1  Quadratic Algorithms

Algorithms **IntegerInput** and **IntegerOutput** respectively read and write $n$-word integers, both with a complexity of $O(n^2)$.

---

**Algorithm 1.23** IntegerInput

**Input:** a string $S = s_{m-1} \ldots s_1 s_0$ of digits in base $B$
**Output:** the value $A$ in base $\beta$ of the integer represented by $S$

   $A \leftarrow 0$
   **for** $i$ **from** $m - 1$ **downto** $0$ **do**
      $A \leftarrow BA + \mathrm{val}(s_i)$            $\triangleright$ $\mathrm{val}(s_i)$ is the value of $s_i$ in base $\beta$
   return $A$.

---

**Algorithm 1.24** IntegerOutput

**Input:** $A = \sum_0^{n-1} a_i \beta^i > 0$
**Output:** a string $S$ of characters, representing $A$ in base $B$

   $m \leftarrow 0$
   **while** $A \neq 0$ **do**
      $s_m \leftarrow \mathrm{char}(A \bmod B)$    $\triangleright$ $s_m$: character corresponding to $A \bmod B$
      $A \leftarrow A \ \mathrm{div} \ B$
      $m \leftarrow m + 1$
   return $S = s_{m-1} \ldots s_1 s_0$.

---

## 1.7.2  Subquadratic Algorithms

Fast conversions routines are obtained using a "divide and conquer" strategy. Given two strings $s$ and $t$, we let $s \,\|\, t$ denote the concatenation of $s$ and $t$.

For integer input, if the given string decomposes as $S = S_{\text{hi}} \,||\, S_{\text{lo}}$ where $S_{\text{lo}}$ has $k$ digits in base $B$, then

$$\text{Input}(S, B) = \text{Input}(S_{\text{hi}}, B)B^k + \text{Input}(S_{\text{lo}}, B),$$

where $\text{Input}(S, B)$ is the value obtained when reading the string $S$ in the external base $B$. Algorithm **FastIntegerInput** shows one way to implement this: if the output $A$ has $n$ words, Algorithm **FastIntegerInput** has com-

---

**Algorithm 1.25** FastIntegerInput

---

**Input:** a string $S = s_{m-1} \ldots s_1 s_0$ of digits in base $B$
**Output:** the value $A$ of the integer represented by $S$

   $\ell \leftarrow [\text{val}(s_0), \text{val}(s_1), \ldots, \text{val}(s_{m-1})]$
   $(b, k) \leftarrow (B, m)$            ▷ Invariant: $\ell$ has $k$ elements $\ell_0, \ldots, \ell_{k-1}$
   **while** $k > 1$ **do**
      **if** $k$ even **then** $\ell \leftarrow [\ell_0 + b\ell_1, \ell_2 + b\ell_3, \ldots, \ell_{k-2} + b\ell_{k-1}]$
         **else** $\ell \leftarrow [\ell_0 + b\ell_1, \ell_2 + b\ell_3, \ldots, \ell_{k-1}]$
      $(b, k) \leftarrow (b^2, \lceil k/2 \rceil)$
   return $\ell_0$.

---

plexity $O(M(n) \log n)$, more precisely $\sim M(n/4) \lg n$ for $n$ a power of two in the FFT range (see Exercise 1.34).

   For integer output, a similar algorithm can be designed, replacing multiplications by divisions. Namely, if $A = A_{\text{hi}}B^k + A_{\text{lo}}$, then

$$\text{Output}(A, B) = \text{Output}(A_{\text{hi}}, B) \,||\, \text{Output}(A_{\text{lo}}, B),$$

where $\text{Output}(A, B)$ is the string resulting from writing the integer $A$ in the external base $B$, and it is assumed that $\text{Output}(A_{\text{lo}}, B)$ has exactly $k$ digits, after possibly padding with leading zeros.

   If the input $A$ has $n$ words, Algorithm **FastIntegerOutput** has complexity $O(M(n) \log n)$, more precisely $\sim D(n/4) \lg n$ for $n$ a power of two in the FFT range, where $D(n)$ is the cost of dividing a $2n$-word integer by an $n$-word integer. Depending on the cost ratio between multiplication and division, integer output may thus be from 2 to 5 times slower than integer input; see however Exercise 1.35.

---

**Algorithm 1.26** FastIntegerOutput

---

**Input:** $A = \sum_0^{n-1} a_i \beta^i$
**Output:** a string $S$ of characters, representing $A$ in base $B$
  **if** $A < B$ **then**
    return char$(A)$
  **else**
      find $k$ such that $B^{2k-2} \le A < B^{2k}$
      $(Q, R) \leftarrow$ **DivRem**$(A, B^k)$
      $r \leftarrow$ **FastIntegerOutput**$(R)$
      return **FastIntegerOutput**$(Q) \,\|\, 0^{k-\mathrm{len}(r)} \,\|\, r$.

---

# 1.8 Exercises

**Exercise 1.1** Extend the Kronecker-Schönhage trick mentioned at the beginning of §1.3 to negative coefficients, assuming the coefficients are in the range $[-\rho, \rho]$.

**Exercise 1.2 (Harvey** [114]**)** For multiplying two polynomials of degree less than $n$, with non-negative integer coefficients bounded above by $\rho$, the Kronecker-Schönhage trick performs one integer multiplication of size about $2n \lg \rho$, assuming $n$ is small compared to $\rho$. Show that it is possible to perform two integer multiplications of size $n \lg \rho$ instead, and even four integer multiplications of size $(n/2) \lg \rho$.

**Exercise 1.3** Assume your processor provides an instruction `fmaa`$(a, b, c, d)$ returning $h, \ell$ such that $ab + c + d = h\beta + \ell$ where $0 \le a, b, c, d, \ell, h < \beta$. Rewrite Algorithm **BasecaseMultiply** using `fmaa`.

**Exercise 1.4 (Harvey, Khachatrian *et al.*** [139]**)** For $A = \sum_{i=0}^{n-1} a_i \beta^i$ and $B = \sum_{j=0}^{n-1} b_j \beta^j$, prove the formula:

$$AB = \sum_{i=1}^{n-1} \sum_{j=0}^{i-1} (a_i + a_j)(b_i + b_j)\beta^{i+j} + 2\sum_{i=0}^{n-1} a_i b_i \beta^{2i} - \sum_{i=0}^{n-1} \beta^i \sum_{j=0}^{n-1} a_j b_j \beta^j.$$

Deduce a new algorithm for schoolbook multiplication.

**Exercise 1.5 (Hanrot)** Prove that the number $K(n)$ of word products (as defined in the proof of Thm. 1.3.2) in Karatsuba's algorithm is non-decreasing, provided $n_0 = 2$. Plot the graph of $K(n)/n^{\lg 3}$ with a logarithmic scale for $n$, for $2^7 \le n \le 2^{10}$, and find experimentally where the maximum appears.

**Exercise 1.6 (Ryde)** Assume the basecase multiply costs $M(n) = an^2 + bn$, and that Karatsuba's algorithm costs $K(n) = 3K(n/2) + cn$. Show that dividing $a$ by two increases the Karatsuba threshold $n_0$ by a factor of two, and on the contrary decreasing $b$ and $c$ decreases $n_0$.

**Exercise 1.7 (Maeder [158], Thomé [216])** Show that an auxiliary memory of $2n + o(n)$ words is enough to implement Karatsuba's algorithm in-place, for an $n$-word$\times n$-word product. In the polynomial case, prove that an auxiliary space of $n$ coefficients is enough, in addition to the $n+n$ coefficients of the input polynomials, and the $2n - 1$ coefficients of the product. [You can use the $2n$ result words, but must not destroy the $n + n$ input words.]

**Exercise 1.8 (Roche [191])** If Exercise 1.7 was too easy for you, design a Karatsuba-like algorithm using only $O(\log n)$ extra space (you are allowed to read and write in the $2n$ output words, but the $n + n$ input words are read-only).

**Exercise 1.9 (Quercia, McLaughlin)** Modify Algorithm **KaratsubaMultiply** to use only $\sim 7n/2$ additions/subtractions. [Hint: decompose each of $C_0$, $C_1$ and $C_2$ into two parts.]

**Exercise 1.10** Design an in-place version of **KaratsubaMultiply** (see Exercise 1.7) that accumulates the result in $c_0, \ldots, c_{n-1}$, and returns a carry bit.

**Exercise 1.11 (Vuillemin)** Design an algorithm to multiply $a_2 x^2 + a_1 x + a_0$ by $b_1 x + b_0$ using 4 multiplications. Can you extend it to a $6 \times 6$ product using 16 multiplications?

**Exercise 1.12 (Weimerskirch, Paar)** Extend the Karatsuba trick to compute an $n \times n$ product in $n(n + 1)/2$ multiplications. For which $n$ does this win over the classical Karatsuba algorithm?

**Exercise 1.13 (Hanrot)** In Algorithm **OddEvenKaratsuba**, if both $m$ and $n$ are odd, one combines the larger parts $A_0$ and $B_0$ together, and the smaller parts $A_1$ and $B_1$ together. Find a way to get instead:

$$K(m, n) = K(\lceil m/2 \rceil, \lfloor n/2 \rfloor) + K(\lfloor m/2 \rfloor, \lceil n/2 \rceil) + K(\lceil m/2 \rceil, \lceil n/2 \rceil).$$

**Exercise 1.14** Prove that if 5 integer evaluation points are used for Toom-Cook 3-way (§1.3.3), the division by (a multiple of) 3 can not be avoided. Does this remain true if only 4 integer points are used together with $\infty$?

**Exercise 1.15 (Quercia, Harvey)** In Toom–Cook 3-way (§1.3.3), take as evaluation point $2^w$ instead of 2, where $w$ is the number of bits per word (usually $w = 32$ or $64$). Which division is then needed? Similarly for the evaluation point $2^{w/2}$.

**Exercise 1.16** For an integer $k \geq 2$ and multiplication of two numbers of size $kn$ and $n$, show that the trivial strategy which performs $k$ multiplications, each $n \times n$, is not the best possible in the FFT range.

**Exercise 1.17 (Karatsuba, Zuras [236])** Assuming the multiplication has superlinear cost, show that the speedup of squaring with respect to multiplication can not significantly exceed 2.

**Exercise 1.18 (Thomé, Quercia)** Consider two sets $A = \{a, b, c, \ldots\}$ and $U = \{u, v, w, \ldots\}$, and a set $X = \{x, y, z, \ldots\}$ of sums of products of elements of $A$ and $U$ (assumed to be in some field $F$). We can ask "what is the least number of multiplies required to compute all elements of $X$?". In general, this is a difficult problem, related to the problem of computing tensor rank, which is NP-complete (see for example Håstad [119] and the book by Bürgisser *et al.* [59]). Special cases include integer/polynomial multiplication, the middle product, and matrix multiplication (for matrices of fixed size). As a specific example, can we compute $x = au + cw$, $y = av + bw$, $z = bu + cv$ in fewer than 6 multiplies? Similarly for $x = au - cw$, $y = av - bw$, $z = bu - cv$.

**Exercise 1.19** In Algorithm **BasecaseDivRem** (§1.4.1), prove that $q_j^* \leq \beta + 1$. Can this bound be reached? In the case $q_j^* \geq \beta$, prove that the while-loop at steps 6-8 is executed at most once. Prove that the same holds for Svoboda's algorithm, i.e., that $A \geq 0$ after step 8 of Algorithm **SvobodaDivision** (§1.4.2).

**Exercise 1.20 (Granlund, Möller)** In Algorithm **BasecaseDivRem**, estimate the probability that $A < 0$ is true at step 6, assuming the remainder $r_j$ from the division of $a_{n+j}\beta + a_{n+j-1}$ by $b_{n-1}$ is uniformly distributed in $[0, b_{n-1} - 1]$, $A \bmod \beta^{n+j-1}$ is uniformly distributed in $[0, \beta^{n+j-1} - 1]$, and $B \bmod \beta^{n-1}$ is uniformly distributed in $[0, \beta^{n-1} - 1]$. Then replace the computation of $q_j^*$ by a division of the three most significant words of $A$ by the two most significant words of $B$. Prove the algorithm is still correct. What is the maximal number of corrections, and the probability that $A < 0$?

**Exercise 1.21 (Montgomery [172])** Let $0 < b < \beta$, and $0 \leq a_4, \ldots, a_0 < \beta$. Prove that $a_4(\beta^4 \bmod b) + \cdots + a_1(\beta \bmod b) + a_0 < \beta^2$, provided $b < \beta/3$. Use this fact to design an efficient algorithm dividing $A = a_{n-1}\beta^{n-1} + \cdots + a_0$ by $b$. Does the algorithm extend to division by the least significant digits?

**Exercise 1.22** In Algorithm **RecursiveDivRem**, find inputs that require 1, 2, 3 or 4 corrections in step 8. [Hint: consider $\beta = 2$.] Prove that when $n = m$ and $A < \beta^m(B + 1)$, at most two corrections occur.

**Exercise 1.23** Find the complexity of Algorithm **RecursiveDivRem** in the FFT range.

**Exercise 1.24** Consider the division of $A$ of $kn$ words by $B$ of $n$ words, with integer $k \geq 3$, and the alternate strategy that consists of extending the divisor with zeros so that it has half the size of the dividend. Show that this is always slower than Algorithm **UnbalancedDivision** [assuming that division has superlinear cost].

**Exercise 1.25** An important special base of division is when the divisor is of the form $b^k$. For example, this is useful for an integer output routine (§1.7). Can one design a fast algorithm for this case?

**Exercise 1.26 (Sedoglavic)** Does the Kronecker-Schönhage trick to reduce polynomial multiplication to integer multiplication (§1.3) also work — in an efficient way — for division? Assume that you want to divide a degree-$2n$ polynomial $A(x)$ by a monic degree-$n$ polynomial $B(x)$, both polynomials having integer coefficients bounded by $\rho$.

**Exercise 1.27** Design an algorithm that performs an exact division of a $4n$-bit integer by a $2n$-bit integer, with a quotient of $2n$ bits, using the idea mentioned in the last paragraph of §1.4.5. Prove that your algorithm is correct.

**Exercise 1.28** Improve the initial speed of convergence of Algorithm **SqrtInt** (§1.5.1) by using a better starting approximation at step 1. Your approximation should be in the interval $[\lfloor \sqrt{m} \rfloor, \lceil 2\sqrt{m} \rceil]$.

**Exercise 1.29 (Luschny)** Devise a fast algorithm for computing the binomial coefficient

$$C(n, k) = \binom{n}{k} = \frac{n!}{k!(n-k)!}$$

for integers $n$, $k$, $0 \leq k \leq n$. The algorithm should use exact integer arithmetic and compute the exact answer.

**Exercise 1.30 (Shoup)** Show that in Algorithm **ExtendedGcd**, if $a \geq b > 0$, and $g = \gcd(a, b)$, then the cofactor $u$ satisfies $-b/(2g) < u \leq b/(2g)$.

**Exercise 1.31** (a) Devise a subquadratic GCD algorithm **HalfGcd** along the lines outlined in the first three paragraphs of §1.6.3 (most-significant bits first). The input is two integers $a \geq b > 0$. The output is a $2 \times 2$ matrix $R$ and integers $a'$, $b'$ such that $[a'\ b']^t = R[a\ b]^t$. If the inputs have size $n$ bits, then the elements of $R$ should have at most $n/2 + O(1)$ bits, and the outputs $a'$, $b'$ should have at most $3n/4 + O(1)$ bits. (b) Construct a plain GCD algorithm which calls **HalfGcd** until the arguments are small enough to call a naive algorithm. (c) Compare this approach with the use of **HalfBinaryGcd** in §1.6.3.

**Exercise 1.32 (Galbraith, Schönhage, Stehlé)** The Jacobi symbol $(a|b)$ of an integer $a$ and a positive odd integer $b$ satisfies $(a|b) = (a \bmod b|b)$, the law of quadratic reciprocity $(a|b)(b|a) = (-1)^{(a-1)(b-1)/4}$ for $a$ odd and positive, together with $(-1|b) = (-1)^{(b-1)/2}$, and $(2|b) = (-1)^{(b^2-1)/8}$. This looks very much like the gcd recurrence: $\gcd(a, b) = \gcd(a \bmod b, b)$ and $\gcd(a, b) = \gcd(b, a)$. Can you design an $O(M(n)\log n)$ algorithm to compute the Jacobi symbol of two $n$-bit integers?

**Exercise 1.33** Show that $B$ and $\beta$ are commensurable, in the sense defined in §1.7, iff $\ln(B)/\ln(\beta) \in \mathbb{Q}$.

**Exercise 1.34** Find a formula $T(n)$ for the asymptotic complexity of Algorithm **FastIntegerInput** when $n = 2^k$ (§1.7.2). Show that, for general $n$, your formula is within a factor of two of $T(n)$. [Hint: consider the binary expansion of $n$.]

**Exercise 1.35** Show that the integer output routine can be made as fast (asymptotically) as the integer input routine **FastIntegerInput**. Do timing experiments with your favorite multiple-precision software. [Hint: use D. Bernstein's scaled remainder tree [21] and the middle product.]

**Exercise 1.36** If the internal base $\beta$ and the external base $B$ share a nontrivial common divisor — as in the case $\beta = 2^\ell$ and $B = 10$ — show how one can exploit this to speed up the subquadratic input and output routines.

**Exercise 1.37** Assume you are given two $n$-digit integers in base ten, but you have fast arithmetic only in base two. Can you multiply the integers in time $O(M(n))$?

# 1.9    Notes and References

"On-line" (as opposed to "off-line") algorithms are considered in many books and papers, see for example the book by Borodin and El-Yaniv [33]. "Relaxed" algorithms were introduced by van der Hoeven. For references and a discussion of the differences between "lazy", "zealous" and "relaxed" algorithms, see [124].

An example of an implementation with "guard bits" to avoid overflow problems in integer addition (§1.2) is the block-wise modular arithmetic of Lenstra and Dixon on the MasPar [87]. They used $\beta = 2^{30}$ with 32-bit words.

The observation that polynomial multiplication reduces to integer multiplication is due to both Kronecker and Schönhage, which explains the name "Kronecker-Schönhage trick". More precisely, Kronecker [147, pp. 941–942] (also [148, §4]) reduced the irreducibility test for factorization of multivariate polynomials to the univariate case, and Schönhage [197] reduced the univariate case to the integer case. The Kronecker-Schönhage trick is improved in Harvey [114] (see Exercise 1.2), and some nice applications of it are given in Steel [207].

Karatsuba's algorithm was first published in [136]. Very little is known about its *average* complexity. What is clear is that no simple asymptotic equivalent can be obtained, since the ratio $K(n)/n^\alpha$ does not converge (see Exercise 1.5).

Andrei Toom[218] discovered the class of Toom-Cook algorithms, and they were discussed by Stephen Cook in his thesis [76, pp. 51–77]. A very good description of these algorithms can be found in the book by Crandall and Pomerance [81, §9.5.1]. In particular it describes how to generate the evaluation and interpolation formulæ symbolically. Zuras [236] considers the 4-way and 5-way variants, together with squaring. Bodrato and Zanoni [31] show that the Toom-Cook 3-way interpolation scheme of §1.3.3 is close to optimal for the points $0, 1, -1, 2, \infty$; they also exhibit efficient 4-way and 5-way schemes. Bodrato and Zanoni also introduced the Toom-2.5 and Toom-3.5 notations for what we call Toom-$(3, 2)$ and Toom-$(4, 3)$, these algorithms being useful for unbalanced multiplication using a different number of pieces. They noticed that Toom-$(4, 2)$ only differs from Toom 3-way in the evaluation phase, thus most of the implementation can be shared.

The Schönhage-Strassen algorithm first appeared in [200], and is described in §2.3.3. Algorithms using floating-point complex numbers are discussed in Knuth's classic [143, §4.3.3.C]. See also §3.3.1.

The odd-even scheme is described in Hanrot and Zimmermann [112], and was independently discovered by Andreas Enge. The asymmetric squaring formula given in §1.3.6 was invented by Chung and Hasan (see their paper [66] for other asymmetric formulæ). Exercise 1.4 was suggested by David Harvey, who independently discovered the algorithm of Khachatrian *et al.* [139].

See Lefèvre [153] for a comparison of different algorithms for the problem of multiplication by an integer constant.

Svoboda's algorithm was introduced in [212]. The exact division algorithm starting from least significant bits is due to Jebelean [130]. Jebelean and Krandick invented the "bidirectional" algorithm [145]. The Karp-Markstein trick to speed up Newton's iteration (or Hensel lifting over $p$-adic numbers) is described in [138]. The "recursive division" of §1.4.3 is from Burnikel and Ziegler [60], although earlier but not-so-detailed ideas can be found in Jebelean [132], and even earlier in Moenck and Borodin [167]. The definition of Hensel's division used here is due to Shand and Vuillemin [202], who also point out the duality with Euclidean division.

Algorithm **SqrtRem** (§1.5.1) was first described in Zimmermann [235], and proved correct in Bertot *et al.* [29]. Algorithm **SqrtInt** is described in [73]; its generalization to $k$-th roots (Algorithm **RootInt**) is due to Keith Briggs. The detection of exact powers is discussed in Bernstein, Lenstra and Pila [23] and earlier in Bernstein [17] and Cohen [73]. It is necessary, for example, in the AKS primality test [2].

The classical (quadratic) Euclidean algorithm has been considered by many authors — a good reference is Knuth [143]. The Gauss-Kuz'min theorem[2] gives the distribution of quotients in the regular continued fraction of almost all real numbers, and hence is a good guide to the distribution of quotients in the Euclidean algorithm for large, random inputs. Lehmer's original algorithm is described in [155]. The binary gcd is almost as old as the classical Euclidean algorithm — Knuth [143] has traced it back to a first-century AD Chinese text *Chiu Chang Suan Shu* (see also Mikami [166]). It was rediscovered several times in the 20th century, and it is usually attributed to Stein [210]. The binary gcd has been analysed by Brent [44, 50], Knuth [143], Maze [160] and Vallée [222]. A parallel (systolic) version that runs in $O(n)$ time using $O(n)$ processors was given by Brent and Kung [53].

The double-digit gcd is due to Jebelean [131]. The $k$-ary gcd reduction is due to Sorenson [206], and was improved and implemented in GNU MP by Weber. Weber also invented Algorithm **ReducedRatMod** [227], inspired by previous work of Wang.

The first subquadratic gcd algorithm was published by Knuth [142], but his complexity analysis was suboptimal — he gave $O(n \log^5 n \log \log n)$. The correct complexity $O(n \log^2 n \log \log n)$ was given by Schönhage [196]; for this reason the algorithm is sometimes called the Knuth-Schönhage algorithm. A description for the polynomial case can be found in Aho, Hopcroft and Ullman [3], and a detailed (but incorrect) description for the integer case in Yap [233]. The subquadratic binary gcd given in §1.6.3 is due to Stehlé and Zimmermann [209]. Möller [169]

---

[2]According to the Gauss-Kuz'min theorem [140], the probability of a quotient $q \in \mathbb{N}^*$ is $\lg(1 + 1/q) - \lg(1 + 1/(q+1))$.

compares various subquadratic algorithms, and gives a nice algorithm without "repair steps".

Several authors mention an $O(n \log^2 n \log \log n)$ algorithm for the computation of the Jacobi symbol [89, 201]. The earliest reference that we know is a paper by Bach [8], which gives the basic idea (due to Gauss [101, p. 509]). Details are given in the book by Bach and Shallit [9, Solution of Exercise 5.52], where the algorithm is said to be "folklore", with the ideas going back to Bachmann [10] and Gauss. The existence of such an algorithm is mentioned in Schönhage's book [199, §7.2.3], but without details. See also Exercise 1.32.

# Chapter 2

# Modular Arithmetic and the FFT

In this chapter our main topic is modular arithmetic, i.e., how to compute efficiently modulo a given integer $N$. In most applications, the modulus $N$ is fixed, and special-purpose algorithms benefit from some precomputations, depending only on $N$, to speed up arithmetic modulo $N$.

There is an overlap between Chapter 1 and this chapter. For example, integer division and modular multiplication are closely related. In Chapter 1 we present algorithms where no (or only a few) precomputations with respect to the modulus $N$ are performed. In this chapter we consider algorithms which benefit from such precomputations.

Unless explicitly stated, we consider that the modulus $N$ occupies $n$ words in the word-base $\beta$, i.e., $\beta^{n-1} \leq N < \beta^n$.

## 2.1 Representation

We consider in this section the different possible representations of residues modulo $N$. As in Chapter 1, we consider mainly dense representations.

### 2.1.1 Classical Representation

The classical representation stores a residue (class) $a$ as an integer $0 \leq a < N$. Residues are thus always fully reduced, i.e., in *canonical* form.

Another non-redundant form consists in choosing a symmetric representation, say $-N/2 \leq a < N/2$. This form might save some reductions in additions or subtractions (see §2.2). Negative numbers might be stored either with a separate sign (sign-magnitude representation) or with a two's-complement representation.

Since $N$ takes $n$ words in base $\beta$, an alternative *redundant* representation chooses $0 \leq a < \beta^n$ to represent a residue class. If the underlying arithmetic is word-based, this will yield no slowdown compared to the canonical form. An advantage of this representation is that, when adding two residues, it suffices to compare their sum to $\beta^n$ in order to decide whether the sum has to be reduced, and the result of this comparison is simply given by the carry bit of the addition (see Algorithm **IntegerAddition** in §1.2), instead of by comparing the sum with $N$. However, in the case that the sum has to be reduced, one or more further comparisons are needed.

## 2.1.2   Montgomery's Form

Montgomery's form is another representation widely used when several modular operations have to be performed modulo the same integer $N$ (additions, subtractions, modular multiplications). It implies a small overhead to convert — if needed — from the classical representation to Montgomery's and vice-versa, but this overhead is often more than compensated by the speedup obtained in the modular multiplication.

The main idea is to represent a residue $a$ by $a' = aR \bmod N$, where $R = \beta^n$, and $N$ takes $n$ words in base $\beta$. Thus Montgomery is not concerned with the *physical* representation of a residue class, but with the *meaning* associated to a given physical representation. (As a consequence, the different choices mentioned above for the physical representation are all possible.) Addition and subtraction are unchanged, but (modular) multiplication translates to a different, much simpler, algorithm (§2.4.2).

In most applications using Montgomery's form, all inputs are first converted to Montgomery's form, using $a' = aR \bmod N$, then all computations are performed in Montgomery's form, and finally all outputs are converted back — if needed — to the classical form, using $a = a'/R \bmod N$. We need to assume that $(R, N) = 1$, or equivalently that $(\beta, N) = 1$, to ensure the existence of $1/R \bmod N$. This is not usually a problem because $\beta$ is a power of two and $N$ can be assumed to be odd.

| classical (MSB) | *p*-adic (LSB) |
|---|---|
| Euclidean division | Hensel division, Montgomery reduction |
| Svoboda's algorithm | Montgomery-Svoboda |
| Euclidean gcd | binary gcd |
| Newton's method | Hensel lifting |

Figure 2.1: Equivalence between LSB and MSB algorithms.

### 2.1.3 Residue Number Systems

In a *Residue Number System*, a residue $a$ is represented by a list of residues $a_i$ modulo $N_i$, where the moduli $N_i$ are coprime and their product is $N$. The integers $a_i$ can be efficiently computed from $a$ using a remainder tree, and the unique integer $0 \le a < N = N_1 N_2 \cdots$ is computed from the $a_i$ by an Explicit Chinese Remainder Theorem (§2.7). The residue number system is interesting since addition and multiplication can be performed in parallel on each small residue $a_i$. This representation requires that $N$ factors into convenient moduli $N_1, N_2, \ldots$, which is not always the case (see however §2.9). Conversion to/from the RNS representation costs $O(M(n) \log n)$, see §2.7.

### 2.1.4 MSB vs LSB Algorithms

Many classical (most significant bits first or MSB) algorithms have a *p*-adic (least significant bits first or LSB) equivalent form. Thus several algorithms in this chapter are just LSB-variants of algorithms discussed in Chapter 1 (see Figure 2.1).

### 2.1.5 Link with Polynomials

As in Chapter 1, a strong link exists between modular arithmetic and arithmetic on polynomials. One way of implementing finite fields $\mathbb{F}_q$ with $q = p^n$ elements is to work with polynomials in $\mathbb{F}_p[x]$, which are reduced modulo a monic irreducible polynomial $f(x) \in \mathbb{F}_p[x]$ of degree $n$. In this case modular reduction happens both at the coefficient level (in $\mathbb{F}_p$) and at the polynomial level (modulo $f(x)$).

Some algorithms work in the ring $(\mathbb{Z}/N\mathbb{Z})[x]$, where $N$ is a composite integer. An important case is the Schönhage-Strassen multiplication algorithm, where $N$ has the form $2^{\ell} + 1$.

In both domains $\mathbb{F}_p[x]$ and $(\mathbb{Z}/N\mathbb{Z})[x]$, the Kronecker-Schönhage trick (§1.3) can be applied efficiently. Since the coefficients are known to be bounded, by $p$ and $N$ respectively, and thus have a fixed size, the segmentation is quite efficient. If polynomials have degree $d$ and coefficients are bounded by $N$, the product coefficients are bounded by $dN^2$, and one obtains $O(M(d\log(Nd)))$ operations, instead of $O(M(d)M(\log N))$ with the classical approach. Also, the implementation is simpler, because we only have to implement fast arithmetic for large integers instead of fast arithmetic at both the polynomial level and the coefficient level (see also Exercises 1.2 and 2.4).

## 2.2   Modular Addition and Subtraction

The addition of two residues in classical representation can be done as in Algorithm **ModularAdd**.

---

**Algorithm 2.1** ModularAdd

---

**Input:** residues $a, b$ with $0 \le a, b < N$
**Output:** $c = a + b \bmod N$
$\quad c \leftarrow a + b$
$\quad$**if** $c \ge N$ **then**
$\quad\quad c \leftarrow c - N.$

---

Assuming that $a$ and $b$ are uniformly distributed in $\mathbb{Z} \cap [0, N-1]$, the subtraction $c \leftarrow c - N$ is performed with probability $(1 - 1/N)/2$. If we use instead a symmetric representation in $[-N/2, N/2)$, the probability that we need to add or subtract $N$ drops to $1/4 + O(1/N^2)$ at the cost of an additional test. This extra test might be expensive for small $N$ — say one or two words — but should be relatively cheap if $N$ is large enough, say at least ten words.

## 2.3   The Fourier Transform

In this section we introduce the discrete Fourier transform (DFT). An important application of the DFT is in computing convolutions via the *Convo-* *lution Theorem*. In general, the convolution of two vectors can be computed using three DFTs (for details see §2.9). Here we show how to compute the DFT efficiently (via the *Fast Fourier Transform* or FFT), and show how it can be used to multiply two $n$-bit integers in time $O(n \log n \log \log n)$ (the Schönhage-Strassen algorithm, see §2.3.3).

## 2.3.1 Theoretical Setting

Let $R$ be a ring, $K \geq 2$ an integer, and $\omega$ a $K$-th principal root of unity in $R$, i.e., such that $\omega^K = 1$ and $\sum_{j=0}^{K-1} \omega^{ij} = 0$ for $1 \leq i < K$. The *Fourier transform* (or *forward (Fourier) transform*) of a vector $\mathbf{a} = [a_0, a_1, \ldots, a_{K-1}]$ of $K$ elements from $R$ is the vector $\widehat{\mathbf{a}} = [\widehat{a}_0, \widehat{a}_1, \ldots, \widehat{a}_{K-1}]$ such that

$$\widehat{a}_i = \sum_{j=0}^{K-1} \omega^{ij} a_j. \tag{2.1}$$

If we transform the vector $\mathbf{a}$ twice, we get back to the initial vector, apart from a multiplicative factor $K$ and a permutation of the elements of the vector. Indeed, for $0 \leq i < K$,

$$\widehat{\widehat{a}}_i = \sum_{j=0}^{K-1} \omega^{ij} \widehat{a}_j = \sum_{j=0}^{K-1} \omega^{ij} \sum_{\ell=0}^{K-1} \omega^{j\ell} a_\ell = \sum_{\ell=0}^{K-1} a_\ell \left( \sum_{j=0}^{K-1} \omega^{(i+\ell)j} \right).$$

Let $\tau = \omega^{i+\ell}$. If $i + \ell \neq 0 \bmod K$, i.e., if $i + \ell$ is not 0 or $K$, the sum $\sum_{j=0}^{K-1} \tau^j$ vanishes since $\omega$ is principal. For $i + \ell \in \{0, K\}$ we have $\tau = 1$ and the sum equals $K$. It follows that

$$\widehat{\widehat{a}}_i = K \sum_{\substack{\ell=0 \\ i+\ell \in \{0,K\}}}^{K-1} a_\ell = K a_{(-i) \bmod K}.$$

Thus we have $\widehat{\widehat{\mathbf{a}}} = K[a_0, a_{K-1}, a_{K-2}, \ldots, a_2, a_1]$.

If we transform the vector $\mathbf{a}$ twice, but use $\omega^{-1}$ instead of $\omega$ for the second transform (which is then called a *backward transform*), we get:

$$\widetilde{\widehat{a}}_i = \sum_{j=0}^{K-1} \omega^{-ij} \widehat{a}_j = \sum_{j=0}^{K-1} \omega^{-ij} \sum_{\ell=0}^{K-1} \omega^{j\ell} a_\ell = \sum_{\ell=0}^{K-1} a_\ell \left( \sum_{j=0}^{K-1} \omega^{(\ell-i)j} \right).$$

The sum $\sum_{j=0}^{K-1} \omega^{(\ell-i)j}$ vanishes unless $\ell = i$, in which case it equals $K$. Thus we have $\widehat{\widehat{a}}_i = K a_i$. Apart from the multiplicative factor $K$, the backward transform is the inverse of the forward transform, as might be expected from the names.

### 2.3.2   The Fast Fourier Transform

If evaluated naively, Eqn. (2.1) requires $\Omega(K^2)$ operations to compute the Fourier transform of a vector of $K$ elements. The *Fast Fourier Transform* or FFT is an efficient way to evaluate Eqn. (2.1), using only $O(K \log K)$ operations. From now on we assume that $K$ is a power of two, since this is the most common case and simplifies the description of the FFT (see §2.9 for the general case).

Let us illustrate the FFT for $K = 8$. Since $\omega^8 = 1$, we have reduced the exponents modulo 8 in the following. We want to compute:

$$
\begin{aligned}
\widehat{a}_0 &= a_0 + a_1 + a_2 + a_3 + a_4 + a_5 + a_6 + a_7, \\
\widehat{a}_1 &= a_0 + \omega a_1 + \omega^2 a_2 + \omega^3 a_3 + \omega^4 a_4 + \omega^5 a_5 + \omega^6 a_6 + \omega^7 a_7, \\
\widehat{a}_2 &= a_0 + \omega^2 a_1 + \omega^4 a_2 + \omega^6 a_3 + a_4 + \omega^2 a_5 + \omega^4 a_6 + \omega^6 a_7, \\
\widehat{a}_3 &= a_0 + \omega^3 a_1 + \omega^6 a_2 + \omega a_3 + \omega^4 a_4 + \omega^7 a_5 + \omega^2 a_6 + \omega^5 a_7, \\
\widehat{a}_4 &= a_0 + \omega^4 a_1 + a_2 + \omega^4 a_3 + a_4 + \omega^4 a_5 + a_6 + \omega^4 a_7, \\
\widehat{a}_5 &= a_0 + \omega^5 a_1 + \omega^2 a_2 + \omega^7 a_3 + \omega^4 a_4 + \omega a_5 + \omega^6 a_6 + \omega^3 a_7, \\
\widehat{a}_6 &= a_0 + \omega^6 a_1 + \omega^4 a_2 + \omega^2 a_3 + a_4 + \omega^6 a_5 + \omega^4 a_6 + \omega^2 a_7, \\
\widehat{a}_7 &= a_0 + \omega^7 a_1 + \omega^6 a_2 + \omega^5 a_3 + \omega^4 a_4 + \omega^3 a_5 + \omega^2 a_6 + \omega a_7.
\end{aligned}
$$

We see that we can share some computations. For example, the sum $a_0 + a_4$ appears in four places: in $\widehat{a}_0$, $\widehat{a}_2$, $\widehat{a}_4$ and $\widehat{a}_6$. Let us define $a_{0,4} = a_0 + a_4$, $a_{1,5} = a_1 + a_5$, $a_{2,6} = a_2 + a_6$, $a_{3,7} = a_3 + a_7$, $a_{4,0} = a_0 + \omega^4 a_4$, $a_{5,1} = a_1 + \omega^4 a_5$, $a_{6,2} = a_2 + \omega^4 a_6$, $a_{7,3} = a_3 + \omega^4 a_7$. Then we have, using the fact that $\omega^8 = 1$:

$$
\begin{aligned}
\widehat{a}_0 &= a_{0,4} + a_{1,5} + a_{2,6} + a_{3,7}, &\quad \widehat{a}_1 &= a_{4,0} + \omega a_{5,1} + \omega^2 a_{6,2} + \omega^3 a_{7,3}, \\
\widehat{a}_2 &= a_{0,4} + \omega^2 a_{1,5} + \omega^4 a_{2,6} + \omega^6 a_{3,7}, &\quad \widehat{a}_3 &= a_{4,0} + \omega^3 a_{5,1} + \omega^6 a_{6,2} + \omega a_{7,3}, \\
\widehat{a}_4 &= a_{0,4} + \omega^4 a_{1,5} + a_{2,6} + \omega^4 a_{3,7}, &\quad \widehat{a}_5 &= a_{4,0} + \omega^5 a_{5,1} + \omega^2 a_{6,2} + \omega^7 a_{7,3}, \\
\widehat{a}_6 &= a_{0,4} + \omega^6 a_{1,5} + \omega^4 a_{2,6} + \omega^2 a_{3,7}, &\quad \widehat{a}_7 &= a_{4,0} + \omega^7 a_{5,1} + \omega^6 a_{6,2} + \omega^5 a_{7,3}.
\end{aligned}
$$

Now the sum $a_{0,4}+a_{2,6}$ appears at two different places. Let $a_{0,4,2,6} = a_{0,4}+a_{2,6}$, $a_{1,5,3,7} = a_{1,5} + a_{3,7}$, $a_{2,6,0,4} = a_{0,4} + \omega^4 a_{2,6}$, $a_{3,7,1,5} = a_{1,5} + \omega^4 a_{3,7}$, $a_{4,0,6,2} =$

$a_{4,0}+\omega^2 a_{6,2}$, $a_{5,1,7,3} = a_{5,1}+\omega^2 a_{7,3}$, $a_{6,2,4,0} = a_{4,0}+\omega^6 a_{6,2}$, $a_{7,3,5,1} = a_{5,1}+\omega^6 a_{7,3}$.
Then we have

$$\begin{array}{llll}
\widehat{a}_0 & = & a_{0,4,2,6} + a_{1,5,3,7}, & \widehat{a}_1 & = & a_{4,0,6,2} + \omega a_{5,1,7,3}, \\
\widehat{a}_2 & = & a_{2,6,0,4} + \omega^2 a_{3,7,1,5}, & \widehat{a}_3 & = & a_{6,2,4,0} + \omega^3 a_{7,3,5,1}, \\
\widehat{a}_4 & = & a_{0,4,2,6} + \omega^4 a_{1,5,3,7}, & \widehat{a}_5 & = & a_{4,0,6,2} + \omega^5 a_{5,1,7,3}, \\
\widehat{a}_6 & = & a_{2,6,0,4} + \omega^6 a_{3,7,1,5}, & \widehat{a}_7 & = & a_{6,2,4,0} + \omega^7 a_{7,3,5,1}.
\end{array}$$

In summary, after a first stage where we have computed 8 intermediary variables $a_{0,4}$ to $a_{7,3}$, and a second stage with 8 extra intermediary variables $a_{0,4,2,6}$ to $a_{7,3,5,1}$, we are able to compute the transformed vector in 8 extra steps. The total number of steps is thus $24 = 8 \lg 8$, where each step has the form $a \leftarrow b + \omega^j c$.

If we take a closer look, we can group operations in pairs $(a, a')$ which have the form $a = b + \omega^j c$ and $a' = b + \omega^{j+4} c$. For example, in the first stage we have $a_{1,5} = a_1 + a_5$ and $a_{5,1} = a_1 + \omega^4 a_5$; in the second stage we have $a_{4,0,6,2} = a_{4,0} + \omega^2 a_{6,2}$ and $a_{6,2,4,0} = a_{4,0} + \omega^6 a_{6,2}$. Since $\omega^4 = -1$, this can also be written $(a, a') = (b + \omega^j c, b - \omega^j c)$, where $\omega^j c$ needs to be computed only once. A pair of two such operations is called a *butterfly* operation.

The FFT can be performed *in place*. Indeed, the result of the butterfly between $a_0$ and $a_4$, that is $(a_{0,4}, a_{4,0}) = (a_0 + a_4, a_0 - a_4)$, can overwrite $(a_0, a_4)$, since the values of $a_0$ and $a_4$ are no longer needed.

Algorithm **ForwardFFT** is a recursive and in-place implementation of the forward FFT. It uses an auxiliary function bitrev$(j, K)$ which returns the *bit-reversal* of the integer $j$, considered as an integer of $\lg K$ bits. For example, bitrev$(j, 8)$ gives $0, 4, 2, 6, 1, 5, 3, 7$ for $j = 0, \ldots, 7$.

---

**Algorithm 2.2** ForwardFFT

**Input:** vector $\mathbf{a} = [a_0, a_1, \ldots, a_{K-1}]$, $\omega$ principal $K$-th root of unity, $K = 2^k$
**Output:** in-place transformed vector $\widehat{\mathbf{a}}$, bit-reversed

1: **if** $K = 2$ **then**
2:     $[a_0, a_1] \leftarrow [a_0 + a_1, a_0 - a_1]$
3: **else**
4:     $[a_0, a_2, ..., a_{K-2}] \leftarrow$ **ForwardFFT**$([a_0, a_2, ..., a_{K-2}], \omega^2, K/2)$
5:     $[a_1, a_3, ..., a_{K-1}] \leftarrow$ **ForwardFFT**$([a_1, a_3, ..., a_{K-1}], \omega^2, K/2)$
6:     **for** $j$ from $0$ to $K/2 - 1$ **do**
7:         $[a_{2j}, a_{2j+1}] \leftarrow [a_{2j} + \omega^{\text{bitrev}(j,K/2)} a_{2j+1}, a_{2j} - \omega^{\text{bitrev}(j,K/2)} a_{2j+1}]$.

---

**Theorem 2.3.1** *Given an input vector* $\mathbf{a} = [a_0, a_1, \ldots, a_{K-1}]$, *Algorithm* **ForwardFFT** *replaces it by its Fourier transform, in bit-reverse order, in* $O(K \log K)$ *operations in the ring* $R$.

**Proof.** We prove the statement by induction on $K = 2^k$. For $K = 2$, the Fourier transform of $[a_0, a_1]$ is $[a_0 + a_1, a_0 + \omega a_1]$, and the bit-reverse order coincides with the normal order; since $\omega = -1$, the statement follows. Now assume the statement is true for $K/2$. Let $0 \leq j < K/2$, and write $j' := \text{bitrev}(j, K/2)$. Let $\mathbf{b} = [b_0, ..., b_{K/2-1}]$ be the vector obtained at step 4, and $\mathbf{c} = [c_0, ..., c_{K/2-1}]$ be the vector obtained at step 5. By induction:

$$b_j = \sum_{\ell=0}^{K/2-1} \omega^{2j'\ell} a_{2\ell}, \quad c_j = \sum_{\ell=0}^{K/2-1} \omega^{2j'\ell} a_{2\ell+1}.$$

Since $b_j$ is stored at $a_{2j}$ and $c_j$ at $a_{2j+1}$, we compute at step 7:

$$a_{2j} = b_j + \omega^{j'} c_j = \sum_{\ell=0}^{K/2-1} \omega^{2j'\ell} a_{2\ell} + \omega^{j'} \sum_{\ell=0}^{K/2-1} \omega^{2j'\ell} a_{2\ell+1} = \sum_{\ell=0}^{K-1} \omega^{j'\ell} a_\ell = \widehat{a}_{j'}.$$

Similarly, since $-\omega^{j'} = \omega^{K/2+j'}$:

$$\begin{aligned} a_{2j+1} &= \sum_{\ell=0}^{K/2-1} \omega^{2j'\ell} a_{2\ell} + \omega^{K/2+j'} \sum_{\ell=0}^{K/2-1} \omega^{2j'\ell} a_{2\ell+1} \\ &= \sum_{\ell=0}^{K-1} \omega^{(K/2+j')\ell} a_\ell = \widehat{a}_{K/2+j'}, \end{aligned}$$

where we used the fact that $\omega^{2j'} = \omega^{2(j'+K/2)}$. Since $\text{bitrev}(2j, K) = \text{bitrev}(j, K/2)$ and $\text{bitrev}(2j+1, K) = K/2 + \text{bitrev}(j, K/2)$, the first part of the theorem follows. The complexity bound follows from the fact that the cost $T(K)$ satisfies the recurrence $T(K) \leq 2T(K/2) + O(K)$. ☐

**Theorem 2.3.2** *Given an input vector* $\mathbf{a} = [a_0, a_{K/2}, \ldots, a_{K-1}]$ *in bit-reverse order, Algorithm* **BackwardFFT** *replaces it by its backward Fourier transform, in normal order, in* $O(K \log K)$ *operations in* $R$.

---

**Algorithm 2.3** BackwardFFT

---

**Input:** vector $\mathbf{a}$ bit-reversed, $\omega$ principal $K$-th root of unity, $K = 2^k$
**Output:** in-place transformed vector $\widetilde{\mathbf{a}}$, normal order

1: **if** $K = 2$ **then**
2:      $[a_0, a_1] \leftarrow [a_0 + a_1, a_0 - a_1]$
3: **else**
4:      $[a_0, ..., a_{K/2-1}] \leftarrow \textbf{BackwardFFT}([a_0, ..., a_{K/2-1}], \omega^2, K/2)$
5:      $[a_{K/2}, ..., a_{K-1}] \leftarrow \textbf{BackwardFFT}([a_{K/2}, ..., a_{K-1}], \omega^2, K/2)$
6:      **for** $j$ **from** 0 **to** $K/2 - 1$ **do**      $\triangleright \omega^{-j} = \omega^{K-j}$
7:          $[a_j, a_{K/2+j}] \leftarrow [a_j + \omega^{-j} a_{K/2+j}, a_j - \omega^{-j} a_{K/2+j}]$.

---

**Proof.** The complexity bound follows as in the proof of Theorem 2.3.1. For the correctness result, we again use induction on $K = 2^k$. For $K = 2$ the backward Fourier transform $\widetilde{\mathbf{a}} = [a_0 + a_1, a_0 + \omega^{-1}a_1]$ is exactly what the algorithm returns, since $\omega = \omega^{-1} = -1$ in that case. Assume now $K \geq 4$, a power of two. The first half, say $\mathbf{b}$, of the vector $\mathbf{a}$ corresponds to the bit-reversed vector of the even indices, since bitrev$(2j, K) = $ bitrev$(j, K/2)$. Similarly, the second half, say $\mathbf{c}$, corresponds to the bit-reversed vector of the odd indices, since bitrev$(2j+1, K) = K/2 + $bitrev$(j, K/2)$. Thus we can apply the theorem by induction to $\mathbf{b}$ and $\mathbf{c}$. It follows that $\mathbf{b}$ is the backward transform of length $K/2$ with $\omega^2$ for the even indices (in normal order), and similarly $\mathbf{c}$ is the backward transform of length $K/2$ for the odd indices:

$$b_j = \sum_{\ell=0}^{K/2-1} \omega^{-2j\ell} a_{2\ell}, \quad c_j = \sum_{\ell=0}^{K/2-1} \omega^{-2j\ell} a_{2\ell+1}.$$

Since $b_j$ is stored in $a_j$ and $c_j$ in $a_{K/2+j}$, we have:

$$
\begin{aligned}
a_j = b_j + \omega^{-j} c_j &= \sum_{\ell=0}^{K/2-1} \omega^{-2j\ell} a_{2\ell} + \omega^{-j} \sum_{\ell=0}^{K/2-1} \omega^{-2j\ell} a_{2\ell+1} \\
&= \sum_{\ell=0}^{K-1} \omega^{-j\ell} a_\ell = \widetilde{a}_j,
\end{aligned}
$$

and similarly, using $-\omega^{-j} = \omega^{-K/2-j}$ and $\omega^{-2j} = \omega^{-2(K/2+j)}$:

$$
\begin{aligned}
a_{K/2+j} &= \sum_{\ell=0}^{K/2-1} \omega^{-2j\ell} a_{2\ell} + \omega^{-K/2-j} \sum_{\ell=0}^{K/2-1} \omega^{-2j\ell} a_{2\ell+1} \\
&= \sum_{\ell=0}^{K-1} \omega^{-(K/2+j)\ell} a_\ell = \widetilde{a}_{K/2+j}.
\end{aligned}
$$

□

### 2.3.3   The Schönhage-Strassen Algorithm

We now describe the Schönhage-Strassen $O(n \log n \log \log n)$ algorithm to multiply two integers of $n$ bits. The heart of the algorithm is a routine to multiply two integers modulo $2^n + 1$.

---

**Algorithm 2.4** FFTMulMod

---

**Input:** $0 \le A, B < 2^n + 1$, an integer $K = 2^k$ such that $n = MK$

**Output:** $C = A \cdot B \bmod (2^n + 1)$

1: decompose $A = \sum_{j=0}^{K-1} a_j 2^{jM}$ with $0 \le a_j < 2^M$, except $0 \le a_{K-1} \le 2^M$
2: decompose $B$ similarly
3: choose $n' \ge 2n/K + k$, $n'$ multiple of $K$; let $\theta = 2^{n'/K}, \omega = \theta^2$
4: **for** $j$ **from** 0 **to** $K - 1$ **do**
5:     $(a_j, b_j) \leftarrow (\theta^j a_j, \theta^j b_j) \bmod (2^{n'} + 1)$
6: $\mathbf{a} \leftarrow \textbf{ForwardFFT}(\mathbf{a}, \omega, K),$    $\mathbf{b} \leftarrow \textbf{ForwardFFT}(\mathbf{b}, \omega, K)$
7: **for** $j$ **from** 0 **to** $K - 1$ **do**                ▷ call FFTMulMod
8:     $c_j \leftarrow a_j b_j \bmod (2^{n'} + 1)$           ▷ recursively if $n'$ is large
9: $\mathbf{c} \leftarrow \textbf{BackwardFFT}(\mathbf{c}, \omega, K)$
10: **for** $j$ **from** 0 **to** $K - 1$ **do**
11:     $c_j \leftarrow c_j/(K\theta^j) \bmod (2^{n'} + 1)$
12:     **if** $c_j \ge (j+1)2^{2M}$ **then**
13:         $c_j \leftarrow c_j - (2^{n'} + 1)$
14: $C = \sum_{j=0}^{K-1} c_j 2^{jM}$.

---

**Theorem 2.3.3** *Given* $0 \le A, B < 2^n + 1$, *Algorithm* **FFTMulMod** *correctly returns* $A \cdot B \bmod (2^n + 1)$, *and it costs* $O(n \log n \log \log n)$ *bit-operations if* $K = \Theta(\sqrt{n})$.

**Proof.** The proof is by induction on $n$, because at step 8 we call FFT-MulMod recursively unless $n'$ is sufficiently small that a simpler algorithm (classical, Karatsuba or Toom-Cook) can be used. There is no difficulty in starting the induction.

With $a_j, b_j$ the values at steps 1 and 2, we have $A = \sum_{j=0}^{K-1} a_j 2^{jM}$ and $B = \sum_{j=0}^{K-1} b_j 2^{jM}$, thus $A \cdot B = \sum_{j=0}^{K-1} c_j 2^{jM} \bmod (2^n + 1)$ with

$$c_j = \sum_{\substack{\ell,m=0 \\ \ell+m=j}}^{K-1} a_\ell b_m - \sum_{\substack{\ell,m=0 \\ \ell+m=K+j}}^{K-1} a_\ell b_m. \tag{2.2}$$

We have $(j+1-K)2^{2M} \leq c_j < (j+1)2^{2M}$, since the first sum contains $j+1$ terms, the second sum $K - (j+1)$ terms, and at least one of $a_\ell$ and $b_m$ is less than $2^M$ in the first sum.

Let $a_j'$ be the value of $a_j$ after step 5: $a_j' = \theta^j a_j \bmod (2^{n'} + 1)$, and similarly for $b_j'$. Using Theorem 2.3.1, after step 6 we have $a_{\mathrm{bitrev}(j,K)} = \sum_{\ell=0}^{K-1} \omega^{\ell j} a_\ell' \bmod (2^{n'} + 1)$, and similarly for $b$. Thus at step 8:

$$c_{\mathrm{bitrev}(j,K)} = \left( \sum_{\ell=0}^{K-1} \omega^{\ell j} a_\ell' \right) \left( \sum_{m=0}^{K-1} \omega^{mj} b_m' \right).$$

After step 9, using Theorem 2.3.2:

$$\begin{aligned}
c_i' &= \sum_{j=0}^{K-1} \omega^{-ij} \left( \sum_{\ell=0}^{K-1} \omega^{\ell j} a_\ell' \right) \left( \sum_{m=0}^{K-1} \omega^{mj} b_m' \right) \\
&= K \sum_{\substack{\ell,m=0 \\ \ell+m=i}}^{K-1} a_\ell' b_m' + K \sum_{\substack{\ell,m=0 \\ \ell+m=K+i}}^{K-1} a_\ell' b_m'.
\end{aligned}$$

The first sum equals $\theta^i \sum_{\ell+m=i} a_\ell b_m$; the second is $\theta^{K+i} \sum_{\ell+m=K+i} a_\ell b_m$. Since $\theta^K = -1 \bmod (2^{n'} + 1)$, after step 11 we have:

$$c_i = \sum_{\substack{\ell,m=0 \\ \ell+m=i}}^{K-1} a_\ell b_m - \sum_{\substack{\ell,m=0 \\ \ell+m=K+i}}^{K-1} a_\ell b_m \bmod (2^{n'} + 1).$$

The correction at step 13 ensures that $c_i$ lies in the correct interval, as given by Eqn. (2.2).

For the complexity analysis, assume that $K = \Theta(\sqrt{n})$. Thus we have $n' = \Theta(\sqrt{n})$. Steps 1 and 2 cost $O(n)$; step 5 also costs $O(n)$ (counting the cumulated cost for all values of $j$). Step 6 costs $O(K \log K)$ times the cost of one butterfly operation mod $(2^{n'} + 1)$, which is $O(n')$, thus a total of $O(Kn' \log K) = O(n \log n)$. Step 8, using the same algorithm recursively, costs $O(n' \log n' \log \log n')$ per value of $j$ by the induction hypothesis, giving a total of $O(n \log n \log \log n)$. The backward FFT costs $O(n \log n)$ too, and the final steps cost $O(n)$, giving a total cost of $O(n \log n \log \log n)$. The $\log \log n$ term is the depth of the recursion, each level reducing $n$ to $n' = O(\sqrt{n})$. □

EXAMPLE: to multiply two integers modulo $(2^{1\,048\,576} + 1)$, we can take $K = 2^{10} = 1024$, and $n' = 3072$. We recursively compute 1024 products modulo $(2^{3072} + 1)$. Alternatively, we can take the smaller value $K = 512$, with 512 recursive products modulo $(2^{4608} + 1)$.

REMARK 1: the "small" products at step 8 (mod $(2^{3072}+1)$ or mod $(2^{4608}+1)$ in our example) can be performed by the same algorithm applied recursively, but at some point (determined by details of the implementation) it will be more efficient to use a simpler algorithm, such as the classical or Karatsuba algorithm (see §1.3). In practice the depth of recursion is a small constant, typically 1 or 2. Thus, for practical purposes, the $\log \log n$ term can be regarded as a constant. For a theoretical way of avoiding the $\log \log n$ term, see the comments on Fürer's algorithm in §2.9.

REMARK 2: if we replace $\theta$ by 1 in Algorithm **FFTMulMod**, i.e., remove step 5, replace step 11 by $c_j \leftarrow c_j/K \mod (2^{n'}+1)$, and replace the condition at step 12 by $c_j \geq K \cdot 2^{2M}$, then we compute $C = A \cdot B \mod (2^n - 1)$ instead of mod $(2^n + 1)$. This is useful, for example, in McLaughlin's algorithm (§2.4.3).

Algorithm **FFTMulMod** enables us to multiply two integers modulo $(2^n + 1)$ in $O(n \log n \log \log n)$ operations, for a suitable $n$ and a corresponding FFT length $K = 2^k$. Since we should have $K \approx \sqrt{n}$ and $K$ must divide $n$, suitable values of $n$ are the integers with the low-order half of their bits zero; there is no shortage of such integers. To multiply two integers of at most $n$ bits, we first choose a suitable bit size $m \geq 2n$. We consider the integers as residues modulo $(2^m + 1)$, then Algorithm **FFTMulMod** gives their integer product. The resulting complexity is $O(n \log n \log \log n)$, since $m = O(n)$. In practice the $\log \log n$ term can be regarded as a constant; theoretically it can be replaced by an extremely slowly-growing function (see Remark 1 above).

In this book, we sometimes implicitly assume that $n$-bit integer multiplication costs the same as three FFTs of length $2n$, since this is true if an FFT-based algorithm is used for multiplication. The constant "three" can be reduced if some of the FFTs can be precomputed and reused many times, for example if some of the operands in the multiplications are fixed.

## 2.4   Modular Multiplication

Modular multiplication means computing $A \cdot B \bmod N$, where $A$ and $B$ are residues modulo $N$. Of course, once the product $C = A \cdot B$ has been computed, it suffices to perform a *modular reduction* $C \bmod N$, which itself reduces to an integer division. The reader may ask why we did not cover this topic in §1.4. There are two reasons. First, the algorithms presented below benefit from some precomputations involving $N$, and are thus specific to the case where several reductions are performed with the same modulus. Second, some algorithms avoid performing the full product $C = A \cdot B$; one such example is McLaughlin's algorithm (§2.4.3).

Algorithms with precomputations include Barrett's algorithm (§2.4.1), which computes an approximation to the inverse of the modulus, thus trading division for multiplication; Montgomery's algorithm, which corresponds to Hensel's division with remainder only (§1.4.8), and its subquadratic variant, which is the LSB-variant of Barrett's algorithm; and finally McLaughlin's algorithm (§2.4.3). The cost of the precomputations is not taken into account: it is assumed to be negligible if many modular reductions are performed. However, we assume that the amount of precomputed data uses only linear, that is $O(\log N)$, space.

As usual, we assume that the modulus $N$ has $n$ words in base $\beta$, that $A$ and $B$ have at most $n$ words, and in some cases that they are fully reduced, i.e., $0 \le A, B < N$.

### 2.4.1   Barrett's Algorithm

Barrett's algorithm is attractive when many divisions have to be made with the same divisor; this is the case when one performs computations modulo a fixed integer. The idea is to precompute an approximation to the inverse of the divisor. Thus, an approximation to the quotient is obtained with just one multiplication, and the corresponding remainder after a second multiplication. A small number of corrections suffice to convert the approximations into exact values. For the sake of simplicity, we describe Barrett's algorithm in base $\beta$, where $\beta$ might be replaced by any integer, in particular $2^n$ or $\beta^n$.

---

**Algorithm 2.5** BarrettDivRem

---

**Input:** integers $A$, $B$ with $0 \leq A < \beta^2$, $\beta/2 < B < \beta$
**Output:** quotient $Q$ and remainder $R$ of $A$ divided by $B$

  1: $I \leftarrow \lfloor \beta^2/B \rfloor$                                      ▷ precomputation
  2: $Q \leftarrow \lfloor A_1 I/\beta \rfloor$ where $A = A_1\beta + A_0$ with $0 \leq A_0 < \beta$
  3: $R \leftarrow A - QB$
  4: **while** $R \geq B$ **do**
  5:      $(Q, R) \leftarrow (Q + 1, R - B)$
  6: return $(Q, R)$.

---

**Theorem 2.4.1** *Algorithm* **BarrettDivRem** *is correct and step 5 is performed at most 3 times.*

**Proof.** Since $A = QB + R$ is invariant in the algorithm, we just need to prove that $0 \leq R < B$ at the end. We first consider the value of $Q, R$ before the while-loop. Since $\beta/2 < B < \beta$, we have $\beta < \beta^2/B < 2\beta$, thus $\beta \leq I < 2\beta$. We have $Q \leq A_1 I/\beta \leq A_1\beta/B \leq A/B$. This ensures that $R$ is nonnegative. Now $I > \beta^2/B - 1$, which gives

$$IB > \beta^2 - B.$$

Similarly, $Q > A_1 I/\beta - 1$ gives

$$\beta Q > A_1 I - \beta.$$

This yields $\beta QB > A_1 IB - \beta B > A_1(\beta^2 - B) - \beta B = \beta(A - A_0) - B(\beta + A_1) > \beta A - 4\beta B$ since $A_0 < \beta < 2B$ and $A_1 < \beta$. We conclude that $A < B(Q + 4)$, thus at most 3 corrections are needed. □

The bound of 3 corrections is tight: it is attained for $A = 1980$, $B = 36$, $\beta = 64$. In this example $I = 113$, $A_1 = 30$, $Q = 52$, $R = 108 = 3B$.

    The multiplications at steps 2 and 3 may be replaced by short products, more precisely the multiplication at step 2 by a high short product, and that at step 3 by a low short product (see §3.3).

    Barrett's algorithm can also be used for an unbalanced division, when dividing $(k + 1)n$ words by $n$ words for $k \geq 2$, which amounts to $k$ divisions of $2n$ words by the same $n$-word divisor. In this case, we say that the divisor is *implicitly invariant*.

**Complexity of Barrett's Algorithm**

If the multiplications at steps 2 and 3 are performed using full products, Barrett's algorithm costs $2M(n)$ for a divisor of size $n$. In the FFT range, this cost might be lowered to $1.5M(n)$ using the "wrap-around trick" (§3.4.1); moreover, if the forward transforms of $I$ and $B$ are stored, the cost decreases to $M(n)$, assuming $M(n)$ is the cost of three FFTs.

## 2.4.2 Montgomery's Multiplication

Montgomery's algorithm is very efficient for modular arithmetic modulo a fixed modulus $N$. The main idea is to replace a residue $A \bmod N$ by $A' = \lambda A \bmod N$, where $A'$ is the "Montgomery form" corresponding to the residue $A$, with $\lambda$ an integer constant such that $\gcd(N, \lambda) = 1$. Addition and subtraction are unchanged, since $\lambda A + \lambda B = \lambda(A + B) \bmod N$. The multiplication of two residues in Montgomery form does not give exactly what we want: $(\lambda A)(\lambda B) \neq \lambda(AB) \bmod N$. The trick is to replace the classical modular multiplication by "Montgomery's multiplication":

$$\textbf{MontgomeryMul}(A', B') = \frac{A'B'}{\lambda} \bmod N.$$

For some values of $\lambda$, **MontgomeryMul**$(A', B')$ can easily be computed, in particular for $\lambda = \beta^n$, where $N$ uses $n$ words in base $\beta$. Algorithm 2.6 is a quadratic algorithm (**REDC**) to compute **MontgomeryMul**(A', B') in this case, and a subquadratic reduction (**FastREDC**) is given in Algorithm 2.7.

Another view of Montgomery's algorithm for $\lambda = \beta^n$ is to consider that it computes the remainder of Hensel's division (§1.4.8).

---

**Algorithm 2.6** REDC (quadratic non-interleaved version). The $c_i$ form the current base-$\beta$ decomposition of $C$, i.e., they are defined by $C = \sum_0^{2n-1} c_i \beta^i$.

**Input:** $0 \leq C < \beta^{2n}$, $N < \beta^n$, $\mu \leftarrow -N^{-1} \bmod \beta$, $(\beta, N) = 1$
**Output:** $0 \leq R < \beta^n$ such that $R = C\beta^{-n} \bmod N$
  1: **for** $i$ **from** 0 **to** $n-1$ **do**
  2:     $q_i \leftarrow \mu c_i \bmod \beta$                    $\triangleright$ quotient selection
  3:     $C \leftarrow C + q_i N \beta^i$
  4: $R \leftarrow C\beta^{-n}$                                  $\triangleright$ trivial exact division
  5: **if** $R \geq \beta^n$ **then** return $R - N$ **else** return $R$.

---

**Theorem 2.4.2** *Algorithm* **REDC** *is correct.*

**Proof.** We first prove that $R = C\beta^{-n} \bmod N$: $C$ is only modified in step 3, which does not change $C \bmod N$, thus at step 4 we have $R = C\beta^{-n} \bmod N$, and this remains true in the last step.

Assume that, for a given $i$, we have $C = 0 \bmod \beta^i$ when entering step 2. Since $q_i = -c_i/N \bmod \beta$, we have $C + q_i N \beta^i = 0 \bmod \beta^{i+1}$ at the next step, so the next value of $c_i$ is 0. Thus, on exiting the for-loop, $C$ is a multiple of $\beta^n$, and $R$ is an integer at step 4.

Still at step 4, we have $C < \beta^{2n} + (\beta - 1)N(1 + \beta + \cdots + \beta^{n-1}) = \beta^{2n} + N(\beta^n - 1)$, thus $R < \beta^n + N$ and $R - N < \beta^n$. $\qquad\square$

Compared to classical division (Algorithm **BasecaseDivRem**, §1.4.1), Montgomery's algorithm has two significant advantages: the quotient selection is performed by a multiplication modulo the word base $\beta$, which is more efficient than a division by the most significant word $b_{n-1}$ of the divisor as in **BasecaseDivRem**; and there is no repair step *inside* the for-loop — the repair step is at the very end.

For example, with inputs $C = 766\,970\,544\,842\,443\,844$, $N = 862\,664\,913$, and $\beta = 1000$, Algorithm **REDC** precomputes $\mu = 23$; then we have $q_0 = 412$, which yields $C \leftarrow C + 412N = 766\,970\,900\,260\,388\,000$; then $q_1 = 924$, which yields $C \leftarrow C + 924N\beta = 767\,768\,002\,640\,000\,000$; then $q_2 = 720$, which yields $C \leftarrow C + 720N\beta^2 = 1\,388\,886\,740\,000\,000\,000$. At step 4, $R = 1\,388\,886\,740$, and since $R \geq \beta^3$, **REDC** returns $R - N = 526\,221\,827$.

Since Montgomery's algorithm — i.e., Hensel's division with remainder only — can be viewed as an LSB variant of classical division, Svoboda's divisor preconditioning (§1.4.2) also translates to the LSB context. More precisely, in Algorithm **REDC**, one wants to modify the divisor $N$ so that the quotient selection $q \leftarrow \mu c_i \bmod \beta$ at step 2 becomes trivial. The multiplier $k$ used in Svoboda division is simply the parameter $\mu$ in **REDC**. A natural choice is $\mu = 1$, which corresponds to $N = -1 \bmod \beta$. This motivates the Montgomery-Svoboda algorithm, which is as follows:

1. first compute $N' = \mu N$, with $N' < \beta^{n+1}$, where $\mu = -1/N \bmod \beta$;

2. perform the $n - 1$ first loops of **REDC**, replacing $\mu$ by 1, and $N$ by $N'$;

3. perform a final classical loop with $\mu$ and $N$, and the last steps (4–5) from **REDC**.

Quotient selection in the Montgomery-Svoboda algorithm simply involves "reading" the word of weight $\beta^i$ in the divisor $C$.

For the example above, we get $N' = 19\,841\,292\,999$; $q_0$ is the least significant word of $C$, i.e., $q_0 = 844$, so $C \leftarrow C + 844N' = 766\,987\,290\,893\,735\,000$; then $q_1 = 735$ and $C \leftarrow C + 735N'\beta = 781\,570\,641\,248\,000\,000$. The last step gives $q_2 = 704$ and $C \leftarrow C + 704N\beta^2 = 1\,388\,886\,740\,000\,000\,000$, which is what we found previously.

**Subquadratic Montgomery Reduction**

A subquadratic version **FastREDC** of Algorithm **REDC** is obtained by taking $n = 1$, and considering $\beta$ as a "giant base" (alternatively, replace $\beta$ by $\beta^n$ below):

---

**Algorithm 2.7** FastREDC (subquadratic Montgomery reduction)

**Input:** $0 \leq C < \beta^2$, $N < \beta$, $\mu \leftarrow -1/N \bmod \beta$
**Output:** $0 \leq R < \beta$ such that $R = C/\beta \bmod N$

1: $Q \leftarrow \mu C \bmod \beta$
2: $R \leftarrow (C + QN)/\beta$
3: **if** $R \geq \beta$ **then** return $R - N$ **else** return $R$.

---

This is exactly the 2-adic counterpart of Barrett's subquadratic algorithm; steps 1–2 might be performed by a low short product and a high short product respectively.

When combined with Karatsuba's multiplication, assuming the products of steps 1–2 are full products, the reduction requires 2 multiplications of size $n$, i.e., 6 multiplications of size $n/2$ ($n$ denotes the size of $N$, $\beta$ being a giant base). With some additional precomputation, the reduction might be performed with 5 multiplications of size $n/2$, assuming $n$ is even. This is simply the Montgomery-Svoboda algorithm with $N$ having two big words in base $\beta^{n/2}$: The cost of the algorithm is $M(n, n/2)$ to compute $q_0N'$ (even if $N'$ has in principle $3n/2$ words, we know $N' = H\beta^{n/2} - 1$ with $H < \beta^n$, thus it suffices to multiply $q_0$ by $H$), $M(n/2)$ to compute $\mu C \bmod \beta^{n/2}$, and again $M(n, n/2)$ to compute $q_1N$, thus a total of $5M(n/2)$ if each $n \times (n/2)$ product is realized by two $(n/2) \times (n/2)$ products.

---

**Algorithm 2.8** MontgomerySvoboda2

---

**Input:** $0 \leq C < \beta^{2n}$, $N < \beta^n$, $\mu \leftarrow -1/N \bmod \beta^{n/2}$, $N' = \mu N$

**Output:** $0 \leq R < \beta^n$ such that $R = C/\beta^n \bmod N$

1: $q_0 \leftarrow C \bmod \beta^{n/2}$
2: $C \leftarrow (C + q_0 N')/\beta^{n/2}$
3: $q_1 \leftarrow \mu C \bmod \beta^{n/2}$
4: $R \leftarrow (C + q_1 N)/\beta^{n/2}$
5: **if** $R \geq \beta^n$ **then** return $R - N$ **else** return $R$.

---

The algorithm is quite similar to the one described at the end of §1.4.6, where the cost was $3M(n/2) + D(n/2)$ for a division of $2n$ by $n$ with remainder only. The main difference here is that, thanks to Montgomery's form, the last classical division $D(n/2)$ in Svoboda's algorithm is replaced by multiplications of total cost $2M(n/2)$, which is usually faster.

Algorithm **MontgomerySvoboda2** can be extended as follows. The value $C$ obtained after step 2 has $3n/2$ words, i.e., an excess of $n/2$ words. Instead of reducing that excess with **REDC**, one could reduce it using Svoboda's technique with $\mu' = -1/N \bmod \beta^{n/4}$, and $N'' = \mu' N$. This would reduce the low $n/4$ words from $C$ at the cost of $M(n, n/4)$, and a last **REDC** step would reduce the final excess of $n/4$, which would give $D(2n, n) = M(n, n/2) + M(n, n/4) + M(n/4) + M(n, n/4)$. This "folding" process can be generalized to $D(2n, n) = M(n, n/2) + \cdots + M(n, n/2^k) + M(n/2^k) + M(n, n/2^k)$. If $M(n, n/2^k)$ reduces to $2^k M(n/2^k)$, this gives:

$$D(n) = 2M(n/2) + 4M(n/4) + \cdots + 2^{k-1}M(n/2^{k-1}) + (2^{k+1} + 1)M(n/2^k).$$

Unfortunately, the resulting multiplications become more and more unbalanced, and we need to store $k$ precomputed multiples $N', N'', \ldots$ of $N$, each requiring at least $n$ words. Figure 2.2 shows that the single-folding algorithm is the best one.

Exercise 2.6 discusses further possible improvements in the Montgomery-Svoboda algorithm, achieving $D(n) \approx 1.58M(n)$ in the case of Karatsuba multiplication.

## 2.4.3   McLaughlin's Algorithm

McLaughlin's algorithm assumes one can perform fast multiplication modulo both $2^n - 1$ and $2^n + 1$, for sufficiently many values of $n$. This assumption is true for example with the Schönhage-Strassen algorithm: the original version multiplies two numbers modulo $2^n + 1$, but discarding the "twist" operations before and after the Fourier transforms computes their product modulo $2^n - 1$. (This has to be done at the top level only: the recursive operations compute modulo $2^{n'} + 1$ in both cases. See Remark 2 on page 62.)

| Algorithm | Karatsuba | Toom-Cook 3-way | Toom-Cook 4-way |
|:---:|:---:|:---:|:---:|
| $D(n)$ | $2.00M(n)$ | $2.63M(n)$ | $3.10M(n)$ |
| 1-folding | $1.67M(n)$ | $1.81M(n)$ | $1.89M(n)$ |
| 2-folding | $1.67M(n)$ | $1.91M(n)$ | $2.04M(n)$ |
| 3-folding | $1.74M(n)$ | $2.06M(n)$ | $2.25M(n)$ |

Figure 2.2: Theoretical complexity of subquadratic REDC with 1-, 2- and 3-folding, for different multiplication algorithms.

The key idea in McLaughlin's algorithm is to avoid the classical "multiply and divide" method for modular multiplication. Instead, assuming that $N$ is relatively prime to $2^n - 1$, it determines $AB/(2^n - 1) \bmod N$ with convolutions modulo $2^n \pm 1$, which can be performed in an efficient way using the FFT.

---

**Algorithm 2.9** MultMcLaughlin

---

**Input:** $A, B$ with $0 \le A, B < N < 2^n$, $\mu = -N^{-1} \bmod (2^n - 1)$

**Output:** $AB/(2^n - 1) \bmod N$

1: $m \leftarrow AB\mu \bmod (2^n - 1)$
2: $S \leftarrow (AB + mN) \bmod (2^n + 1)$
3: $w \leftarrow -S \bmod (2^n + 1)$
4: **if** $2|w$ **then** $s \leftarrow w/2$ **else** $s \leftarrow (w + 2^n + 1)/2$
5: **if** $AB + mN = s \bmod 2$ **then** $t \leftarrow s$ **else** $t \leftarrow s + 2^n + 1$
6: **if** $t < N$ **then** return $t$ **else** return $t - N$.

---

**Theorem 2.4.3** *Algorithm* **MultMcLaughlin** *computes* $AB/(2^n - 1) \bmod N$ *correctly, in* $\sim 1.5M(n)$ *operations, assuming multiplication modulo* $2^n \pm 1$ *costs* $\sim M(n/2)$, *or the same as 3 Fourier transforms of size* $n$.

**Proof.** Step 1 is similar to step 1 of Algorithm **FastREDC**, with $\beta$ replaced by $2^n - 1$. It follows that $AB + mN = 0 \bmod (2^n - 1)$, therefore we have $AB + mN = k(2^n - 1)$ with $0 \le k < 2N$. Step 2 computes $S = -2k \bmod (2^n + 1)$, then step 3 gives $w = 2k \bmod (2^n+1)$, and $s = k \bmod (2^n+1)$ in step 4. Now, since $0 \le k < 2^{n+1}$, the value $s$ does not uniquely determine $k$, whose missing bit is determined from the least significant bit from $AB + mN$ (step 5). Finally, the last step reduces $t = k$ modulo $N$.

The cost of the algorithm is mainly that of the four multiplications $AB \bmod (2^n \pm 1)$, $(AB)\mu \bmod (2^n - 1)$ and $mN \bmod (2^n + 1)$, which cost $4M(n/2)$ altogether. However, in $(AB)\mu \bmod (2^n-1)$ and $mN \bmod (2^n+1)$, the operands $\mu$ and $N$ are invariant, therefore their Fourier transforms can be precomputed, which saves $2M(n/2)/3$ altogether. A further saving of $M(n/2)/3$ is obtained since we perform only one backward Fourier transform in step 2. Accounting for the savings gives $(4 - 2/3 - 1/3)M(n/2) = 3M(n/2) \sim 1.5M(n)$.  □

The $\sim 1.5M(n)$ cost of McLaughlin's algorithm is quite surprising, since it means that a modular multiplication can be performed faster than two multiplications. In other words, since a modular multiplication is basically a multiplication followed by a division, this means that (at least in this case) the "division" can be performed for half the cost of a multiplication!

### 2.4.4   Special Moduli

For special moduli $N$ faster algorithms may exist. The ideal case is $N = \beta^n \pm 1$. This is precisely the kind of modulus used in the Schönhage-Strassen algorithm based on the Fast Fourier Transform (FFT). In the FFT range, a multiplication modulo $\beta^n \pm 1$ is used to perform the product of two integers of at most $n/2$ words, and a multiplication modulo $\beta^n \pm 1$ costs $\sim M(n/2) \sim M(n)/2$.

For example, in elliptic curve cryptography (ECC), one almost always uses a special modulus, for example a pseudo-Mersenne prime like $2^{192}-2^{64}-1$ or $2^{256} - 2^{224} + 2^{192} + 2^{96} - 1$. However, in most applications the modulus can not be chosen, and there is no reason for it to have a special form.

We refer to §2.9 for further information about special moduli.

## 2.5   Modular Division and Inversion

We have seen above that modular multiplication reduces to integer division, since to compute $ab \bmod N$, the classical method consists of dividing $ab$ by $N$ to obtain $ab = qN + r$, then $ab = r \bmod N$. In the same vein, modular division reduces to an (extended) integer gcd. More precisely, the division $a/b \bmod N$ is usually computed as $a \cdot (1/b) \bmod N$, thus a modular inverse is followed by a modular multiplication. We concentrate on modular inversion in this section.

We have seen in Chapter 1 that computing an extended gcd is expensive, both for small sizes, where it usually costs the same as several multiplications, and for large sizes, where it costs $O(M(n) \log n)$. Therefore modular inversions should be avoided if possible; we explain at the end of this section how this can be done.

Algorithm 2.10 (**ModularInverse**) is just Algorithm **ExtendedGcd** (§1.6.2), with $(a, b) \to (b, N)$ and the lines computing the cofactors of $N$ omitted.

---

**Algorithm 2.10** ModularInverse

---

**Input:** integers $b$ and $N$, $b$ prime to $N$
**Output:** integer $u = 1/b \bmod N$
    $(u, w) \leftarrow (1, 0)$, $c \leftarrow N$
    **while** $c \neq 0$ **do**
        $(q, r) \leftarrow \textbf{DivRem}(b, c)$
        $(b, c) \leftarrow (c, r)$
        $(u, w) \leftarrow (w, u - qw)$
    return $u$.

---

Algorithm **ModularInverse** is the naive version of modular inversion, with complexity $O(n^2)$ if $N$ takes $n$ words in base $\beta$. The subquadratic $O(M(n) \log n)$ algorithm is based on the **HalfBinaryGcd** algorithm (§1.6.3).

When the modulus $N$ has a special form, faster algorithms may exist. In particular for $N = p^k$, $O(M(n))$ algorithms exist, based on Hensel lifting, which can be seen as the $p$-adic variant of Newton's method (§4.2). To compute $1/b \bmod N$, we use a $p$-adic version of the iteration (4.5):

$$x_{j+1} = x_j + x_j(1 - bx_j) \bmod p^k. \tag{2.3}$$

Assume $x_j$ approximates $1/b$ to "$p$-adic precision" $\ell$, i.e., $bx_j = 1 + \varepsilon p^\ell$, and $k = 2\ell$. Then, modulo $p^k$: $bx_{j+1} = bx_j(2 - bx_j) = (1 + \varepsilon p^\ell)(1 - \varepsilon p^\ell) = 1 - \varepsilon^2 p^{2\ell}$. Therefore $x_{j+1}$ approximates $1/b$ to double precision (in the $p$-adic sense).

As an example, assume one wants to compute the inverse of an odd integer $b$ modulo $2^{32}$. The initial approximation $x_0 = 1$ satisfies $x_0 = 1/b \bmod 2$, thus five iterations are enough. The first iteration is $x_1 \leftarrow x_0 + x_0(1 - bx_0) \bmod 2^2$, which simplifies to $x_1 \leftarrow 2 - b \bmod 4$ since $x_0 = 1$. Now whether $b = 1 \bmod 4$ or $b = 3 \bmod 4$, we have $2 - b = b \bmod 4$, thus one can immediately start the second iteration with $x_1 = b$ implicit:

$$x_2 \leftarrow b(2 - b^2) \bmod 2^4, \qquad x_3 \leftarrow x_2(2 - bx_2) \bmod 2^8,$$
$$x_4 \leftarrow x_3(2 - bx_3) \bmod 2^{16}, \quad x_5 \leftarrow x_4(2 - bx_4) \bmod 2^{32}.$$

Consider for example $b = 17$. The above algorithm yields $x_2 = 1$, $x_3 = 241$, $x_4 = 61\,681$ and $x_5 = 4\,042\,322\,161$. Of course, any computation mod $p^\ell$ might be computed modulo $p^k$ for $k \geq \ell$. In particular, all the above computations might be performed modulo $2^{32}$. On a 32-bit computer, arithmetic on basic integer types is usually performed modulo $2^{32}$, thus the reduction comes for free, and one can write in the C language (using **unsigned** variables and the same variable x for $x_2, \ldots, x_5$):

```
x = b*(2 - b*b); x *= 2 - b*x; x *= 2 - b*x; x *= 2 - b*x;
```

Another way to perform modular division when the modulus has a special form is Hensel's division (§1.4.8). For a modulus $N = \beta^n$, given two integers $A, B$, we compute $Q$ and $R$ such that

$$A = QB + R\beta^n.$$

Therefore we have $A/B = Q \bmod \beta^n$. While Montgomery's modular multiplication only computes the remainder $R$ of Hensel's division, modular division computes the quotient $Q$, thus Hensel's division plays a central role in modular arithmetic modulo $\beta^n$.

## 2.5.1   Several Inversions at Once

A modular inversion, which reduces to an extended gcd (§1.6.2), is usually much more expensive than a multiplication. This is true not only in the FFT range, where a gcd takes time $\Theta(M(n)\log n)$, but also for smaller numbers. When several inversions are to be performed modulo the same number, Algorithm **MultipleInversion** is usually faster.

**Theorem 2.5.1** *Algorithm* **MultipleInversion** *is correct.*

**Proof.** We have $z_i = x_1 x_2 \ldots x_i \bmod N$, thus at the beginning of step 6 for a given $i$, $q = (x_1 \ldots x_i)^{-1} \bmod N$, which indeed gives $y_i = 1/x_i \bmod N$.  □

---

**Algorithm 2.11** MultipleInversion

---

**Input:** $0 < x_1, \ldots, x_k < N$
**Output:** $y_1 = 1/x_1 \bmod N, \ldots, y_k = 1/x_k \bmod N$

1: $z_1 \leftarrow x_1$
2: **for** $i$ **from** 2 **to** $k$ **do**
3:      $z_i \leftarrow z_{i-1}x_i \bmod N$
4: $q \leftarrow 1/z_k \bmod N$
5: **for** $i$ **from** $k$ **downto** 2 **do**
6:      $y_i \leftarrow qz_{i-1} \bmod N$
7:      $q \leftarrow qx_i \bmod N$
8: $y_1 \leftarrow q.$

---

This algorithm uses only one modular inversion (step 4), and $3(k-1)$ modular multiplications. Thus it is faster than $k$ inversions when a modular inversion is more than three times as expensive as a product. Figure 2.3 shows a recursive variant of the algorithm, with the same number of modular multiplications: one for each internal node when going up the (product) tree, and two for each internal node when going down the (remainder) tree. The recursive variant might be performed in parallel in $O(\log k)$ operations using $O(k/\log k)$ processors.

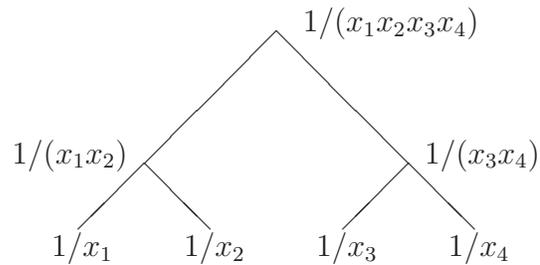

Figure 2.3: A recursive variant of Algorithm **MultipleInversion**. First go up the tree, building $x_1x_2 \bmod N$ from $x_1$ and $x_2$ in the left branch, $x_3x_4 \bmod N$ in the right branch, and $x_1x_2x_3x_4 \bmod N$ at the root of the tree. Then invert the root of the tree. Finally go down the tree, multiplying $1/(x_1x_2x_3x_4)$ by the stored value $x_3x_4$ to get $1/(x_1x_2)$, and so on.

A dual case is when there are several moduli but the number to invert is fixed. Say we want to compute $1/x \bmod N_1, \ldots, 1/x \bmod N_k$. We illustrate a possible algorithm in the case $k = 4$. First compute $N = N_1 \ldots N_k$ using a product tree like that in Figure 2.3, for example first compute $N_1 N_2$ and $N_3 N_4$, then multiply both to get $N = (N_1 N_2)(N_3 N_4)$. Then compute $y = 1/x \bmod N$, and go down the tree, while reducing the residue at each node. In our example we compute $z = y \bmod (N_1 N_2)$ in the left branch, then $z \bmod N_1$ yields $1/x \bmod N_1$. An important difference between this algorithm and the algorithm illustrated in Figure 2.3 is that here, the numbers grow while going up the tree. Thus, depending on the sizes of $x$ and the $N_j$, this algorithm might be of theoretical interest only.

## 2.6   Modular Exponentiation

Modular exponentiation is the most time-consuming mathematical operation in several cryptographic algorithms. The well-known RSA public-key cryptosystem is based on the fact that computing

$$c = a^e \bmod N \tag{2.4}$$

is relatively easy, but recovering $a$ from $c$, $e$ and $N$ is difficult when $N$ has at least two (unknown) large prime factors. The discrete logarithm problem is similar: here $c$, $a$ and $N$ are given, and one looks for $e$ satisfying Eqn. (2.4). In this case the problem is difficult when $N$ has at least one large prime factor (for example, $N$ could be prime). The discrete logarithm problem is the basis of the El Gamal cryptosystem, and a closely related problem is the basis of the Diffie-Hellman key exchange protocol.

When the exponent $e$ is fixed (or known to be small), an optimal sequence of squarings and multiplications might be computed in advance. This is related to the classical *addition chain* problem: What is the smallest chain of additions to reach the integer $e$, starting from 1? For example, if $e = 15$, a possible chain is:

$$1, 1 + 1 = 2, 1 + 2 = 3, 1 + 3 = 4, 3 + 4 = 7, 7 + 7 = 14, 1 + 14 = 15.$$

The length of a chain is defined to be the number of additions needed to compute it (the above chain has length 6). An addition chain readily translates to a *multiplication chain*:

$$a, a \cdot a = a^2, a \cdot a^2 = a^3, a \cdot a^3 = a^4, a^3 \cdot a^4 = a^7, a^7 \cdot a^7 = a^{14}, a \cdot a^{14} = a^{15}.$$

A shorter chain for $e = 15$ is:

$$1, 1 + 1 = 2, 1 + 2 = 3, 2 + 3 = 5, 5 + 5 = 10, 5 + 10 = 15.$$

This chain is the shortest possible for $e = 15$, so we write $\sigma(15) = 5$, where in general $\sigma(e)$ denotes the length of the shortest addition chain for $e$. In the case where $e$ is small, and an addition chain of shortest length $\sigma(e)$ is known for $e$, computing $a^e \bmod N$ may be performed in $\sigma(e)$ modular multiplications.

When $e$ is large and $(a, N) = 1$, then $e$ might be reduced modulo $\phi(N)$, where $\phi(N)$ is Euler's totient function, i.e., the number of integers in $[1, N]$ which are relatively prime to $N$. This is because $a^{\phi(N)} = 1 \bmod N$ whenever $(a, N) = 1$ (Fermat's little theorem).

Since $\phi(N)$ is a multiplicative function, it is easy to compute $\phi(N)$ if we know the prime factorisation of $N$. For example,

$$\phi(1001) = \phi(7 \cdot 11 \cdot 13) = (7 - 1)(11 - 1)(13 - 1) = 720,$$

and $2009 = 569 \bmod 720$, so $17^{2009} = 17^{569} \bmod 1001$.

Assume now that $e$ is smaller than $\phi(N)$. Since a lower bound on the length $\sigma(e)$ of the addition chain for $e$ is $\lg e$, this yields a lower bound $(\lg e)M(n)$ for modular exponentiation, where $n$ is the size of $N$. When $e$ is of size $k$, a modular exponentiation costs $O(kM(n))$. For $k = n$, the cost $O(nM(n))$ of modular exponentiation is much more than the cost of operations considered in Chapter 1, with $O(M(n) \log n)$ for the more expensive ones there. The different algorithms presented in this section save only a constant factor compared to binary exponentiation (§2.6.1).

REMARK: when $a$ fits in one word but $N$ does not, the shortest addition chain for $e$ might not be the best way to compute $a^e \bmod N$, since in this case computing $a \cdot a^j \bmod N$ is cheaper than computing $a^i \cdot a^j \bmod N$ for $i \geq 2$.

## 2.6.1 Binary Exponentiation

A simple (and not far from optimal) algorithm for modular exponentiation is *binary (modular) exponentiation*. Two variants exist: left-to-right and right-to-left. We give the former in Algorithm **LeftToRightBinaryExp** and leave the latter as an exercise for the reader.

Left-to-right binary exponentiation has two advantages over right-to-left exponentiation:

- it requires only one auxiliary variable, instead of two for the right-to-left exponentiation: one to store successive values of $a^{2^i}$, and one to store the result;

---

**Algorithm 2.12** LeftToRightBinaryExp

---

**Input:** $a, e, N$ positive integers

**Output:** $x = a^e \bmod N$

1: let $(e_\ell e_{\ell-1} \ldots e_1 e_0)$ be the binary representation of $e$, with $e_\ell = 1$
2: $x \leftarrow a$
3: **for** $i$ **from** $\ell - 1$ **downto** $0$ **do**
4:     $x \leftarrow x^2 \bmod N$
5:     **if** $e_i = 1$ **then** $x \leftarrow ax \bmod N$.

---

- in the case where $a$ is small, the multiplications $ax$ at step 5 always involve a small operand.

If $e$ is a random integer of $\ell + 1$ bits, step 5 will be performed on average $\ell/2$ times, giving average cost $3\ell M(n)/2$.

EXAMPLE: for the exponent $e = 3\,499\,211\,612$, which is

$$(11\,010\,000\,100\,100\,011\,011\,101\,101\,011\,100)_2$$

in binary, Algorithm **LeftToRightBinaryExp** performs 31 squarings and 15 multiplications (one for each 1-bit, except the most significant one).

## 2.6.2   Exponentiation With a Larger Base

Compared to binary exponentiation, base $2^k$ exponentiation reduces the number of multiplications $ax \bmod N$ (Algorithm **LeftToRightBinaryExp**, step 5). The idea is to precompute small powers of $a \bmod N$:

---

**Algorithm 2.13** BaseKExp

---

**Input:** $a, e, N$ positive integers

**Output:** $x = a^e \bmod N$

1: precompute $t[i] := a^i \bmod N$ for $1 \leq i < 2^k$
2: let $(e_\ell e_{\ell-1} \ldots e_1 e_0)$ be the base $2^k$ representation of $e$, with $e_\ell \neq 0$
3: $x \leftarrow t[e_\ell]$
4: **for** $i$ **from** $\ell - 1$ **downto** $0$ **do**
5:     $x \leftarrow x^{2^k} \bmod N$
6:     **if** $e_i \neq 0$ **then** $x \leftarrow t[e_i]x \bmod N$.

---

The precomputation cost is $(2^k - 2)M(n)$, and if the digits $e_i$ are random and uniformly distributed in $\mathbb{Z} \cap [0, 2^k)$, then the modular multiplication at step 6 of **BaseKExp** is performed with probability $1 - 2^{-k}$. If $e$ has $n$ bits, the number of loops is about $n/k$. Ignoring the squares at step 5 whose total cost depends on $k\ell \approx n$ (independent of $k$), the total expected cost in terms of multiplications modulo $N$ is:

$$2^k - 2 + n(1 - 2^{-k})/k.$$

For $k = 1$ this formula gives $n/2$; for $k = 2$ it gives $3n/8 + 2$, which is faster for $n > 16$; for $k = 3$ it gives $7n/24 + 6$, which is faster than the $k = 2$ formula for $n > 48$. When $n$ is large, the optimal value of $k$ satisfies $k^2 2^k \approx n/\ln 2$. A minor disadvantage of this algorithm is its memory usage, since $\Theta(2^k)$ precomputed entries have to be stored. This is not a serious problem if we choose the optimal value of $k$ (or a smaller value), because then the number of precomputed entries to be stored is $o(n)$.

EXAMPLE: consider the exponent $e = 3\,499\,211\,612$. Algorithm **BaseKExp** performs 31 squarings independently of $k$, thus we count multiplications only. For $k = 2$, we have $e = (3\,100\,210\,123\,231\,130)_4$: Algorithm **BaseKExp** performs two multiplications to precompute $a^2$ and $a^3$, and 11 multiplications for the non-zero digits of $e$ in base 4 (except for the leading digit), thus a total of 13. For $k = 3$, we have $e = (32\,044\,335\,534)_8$, and the algorithm performs 6 multiplications to precompute $a^2, a^3, \ldots, a^7$, and 9 multiplications in step 6, thus a total of 15.

The last example illustrates two facts. First, if some digits (here 6 and 7) do not appear in the base-$2^k$ representation of $e$, then we do not need to precompute the corresponding powers of $a$. Second, when a digit is even, say $e_i = 2$, instead of doing three squarings and multiplying by $a^2$, we could do two squarings, multiply by $a$, and perform a last squaring. These considerations lead to Algorithm **BaseKExpOdd**. The correctness of steps 7–9 follows from:

$$x^{2^k} a^{2^m d} = (x^{2^{k-m}} a^d)^{2^m}.$$

On the previous example, with $k = 3$, this algorithm performs only four multiplications in step 1 (to precompute $a^2$ then $a^3, a^5, a^7$), then nine multiplications in step 8.

### 2.6.3 Sliding Window and Redundant Representation

The "sliding window" algorithm is a straightforward generalization of Algorithm **BaseKExpOdd**. Instead of cutting the exponent into fixed parts of $k$

---

**Algorithm 2.14** BaseKExpOdd

---

**Input:** $a, e, N$ positive integers
**Output:** $x = a^e \bmod N$
1: precompute $a^2$ then $t[i] := a^i \bmod N$ for $i$ odd, $1 \leq i < 2^k$
2: let $(e_\ell e_{\ell-1} \ldots e_1 e_0)$ be the base $2^k$ representation of $e$, with $e_\ell \neq 0$
3: write $e_\ell = 2^m d$ with $d$ odd
4: $x \leftarrow t[d], \quad x \leftarrow x^{2^m} \bmod N$
5: **for** $i$ **from** $\ell - 1$ **downto** 0 **do**
6: $\quad$ write $e_i = 2^m d$ with $d$ odd (if $e_i = 0$ then $m = d = 0$)
7: $\quad x \leftarrow x^{2^{k-m}} \bmod N$
8: $\quad$ **if** $e_i \neq 0$ **then** $x \leftarrow t[d]x \bmod N$
9: $\quad x \leftarrow x^{2^m} \bmod N.$

---

bits each, the idea is to divide it into windows, where two adjacent windows might be separated by a block of zero or more 0-bits. The decomposition starts from the least significant bits. For example, with $e = 3\,499\,211\,612$, or in binary:

$$\underbrace{1}_{e_8}\ \underbrace{101}_{e_7}\,00\ \underbrace{001}_{e_6}\ \underbrace{001}_{e_5}\,00\ \underbrace{011}_{e_4}\ \underbrace{011}_{e_3}\ \underbrace{101}_{e_2}\ \underbrace{101}_{e_1}\,0\ \underbrace{111}_{e_0}\,00.$$

Here there are 9 windows (indicated by $e_8, ..., e_0$ above) and we perform only 8 multiplications, an improvement of one multiplication over Algorithm **BaseKExpOdd**. On average, the sliding window base $2^k$ algorithm leads to about $n/(k+1)$ windows instead of $n/k$ with fixed windows.

Another improvement may be feasible when division is feasible (and cheap) in the underlying group. For example, if we encounter three consecutive ones, say 111, in the binary representation of $e$, we may replace some bits by $-1$, denoted by $\bar{1}$, as in $100\bar{1}$. We have thus replaced three multiplications by one multiplication and one division, in other words $x^7 = x^8 \cdot x^{-1}$. For our running example, this gives:

$$e = 11\,010\,000\,100\,100\,100\,\bar{1}00\,0\bar{1}0\,0\bar{1}0\,\bar{1}00\,\bar{1}00,$$

which has only 10 non-zero digits, apart from the leading one, instead of 15 with bits 0 and 1 only. The redundant representation with bits $\{0, 1, \bar{1}\}$ is called the *Booth representation*. It is a special case of the *Avizienis signed-digit redundant representation*. Signed-digit representations exist in any base.

For simplicity we have not distinguished between the cost of multiplication and the cost of squaring (when the two operands in the multiplication are known to be equal), but this distinction is significant in some applications (e.g., elliptic curve cryptography). Note that, when the underlying group operation is denoted by addition rather than multiplication, as is usually the case for abelian groups (such as groups defined over elliptic curves), then the discussion above applies with "multiplication" replaced by "addition", "division" by "subtraction", and "squaring" by "doubling".

## 2.7  Chinese Remainder Theorem

In applications where integer or rational results are expected, it is often worthwhile to use a "residue number system" (as in §2.1.3) and perform all computations modulo several small primes (or pairwise coprime integers). The final result can then be recovered via the Chinese Remainder Theorem (CRT). For such applications, it is important to have fast conversion routines from integer to modular representation, and vice versa.

The integer to modular conversion problem is the following: given an integer $x$, and several pairwise coprime moduli $m_i$, $1 \le i \le k$, how to efficiently compute $x_i = x \bmod m_i$, for $1 \le i \le k$? This is the remainder tree problem of Algorithm **IntegerToRNS**, which is also discussed in §2.5.1 and Exercise 1.35.

---

**Algorithm 2.15** IntegerToRNS

**Input:** integer $x$, moduli $m_1, m_2, \ldots, m_k$ pairwise coprime, $k \ge 1$
**Output:** $x_i = x \bmod m_i$ for $1 \le i \le k$

1: **if** $k \le 2$ **then**
2:     return $x_1 = x \bmod m_1, \ldots, x_k = x \bmod m_k$
3: $\ell \leftarrow \lfloor k/2 \rfloor$
4: $M_1 \leftarrow m_1 m_2 \cdots m_\ell, \quad M_2 \leftarrow m_{\ell+1} \cdots m_k$ ▷ might be precomputed
5: $x_1, \ldots, x_\ell \leftarrow$ **IntegerToRNS**$(x \bmod M_1, m_1, \ldots, m_\ell)$
6: $x_{\ell+1}, \ldots, x_k \leftarrow$ **IntegerToRNS**$(x \bmod M_2, m_{\ell+1}, \ldots, m_k)$.

---

If all moduli $m_i$ have the same size, and if the size $n$ of $x$ is comparable to that of the product $m_1 m_2 \cdots m_k$, the cost $T(k)$ of Algorithm **IntegerToRNS** satisfies the recurrence $T(n) = 2D(n/2) + 2T(n/2)$, which yields $T(n) = O(M(n) \log n)$. Such a conversion is therefore more expensive than a multiplication or division, and is comparable in complexity terms to a base conversion or a gcd.

The converse *CRT reconstruction* problem is the following: given the $x_i$, how to efficiently reconstruct the unique integer $x$, $0 \leq x < m_1 m_2 \cdots m_k$, such that $x = x_i \bmod m_i$, for $1 \leq i \leq k$? Algorithm **RNSToInteger** performs that conversion, where the values $u, v$ at step 7 might be precomputed if several conversions are made with the same moduli, and step 11 ensures that the final result $x$ lies in the interval $[0, M_1 M_2)$.

---

**Algorithm 2.16** RNSToInteger

**Input:** residues $x_i$, $0 \leq x_i < m_i$ for $1 \leq i \leq k$, $m_i$ pairwise coprime
**Output:** $0 \leq x < m_1 m_2 \cdots m_k$ with $x = x_i \bmod m_i$

1: **if** $k = 1$ **then**
2:    **return** $x_1$
3: $\ell \leftarrow \lfloor k/2 \rfloor$
4: $M_1 \leftarrow m_1 m_2 \cdots m_\ell, \quad M_2 \leftarrow m_{\ell+1} \cdots m_k$    ▷ might be precomputed
5: $X_1 \leftarrow$ **RNSToInteger**$([x_1, \ldots, x_\ell], [m_1, \ldots, m_\ell])$
6: $X_2 \leftarrow$ **RNSToInteger**$([x_{\ell+1}, \ldots, x_k], [m_{\ell+1}, \ldots, m_k])$
7: compute $u, v$ such that $uM_1 + vM_2 = 1$    ▷ might be precomputed
8: $\lambda_1 \leftarrow uX_2 \bmod M_2, \quad \lambda_2 \leftarrow vX_1 \bmod M_1$
9: $x \leftarrow \lambda_1 M_1 + \lambda_2 M_2$
10: **if** $x \geq M_1 M_2$ **then**
11:    $x \leftarrow x - M_1 M_2$.

---

To see that Algorithm **RNSToInteger** is correct, consider an integer $i$, $1 \leq i \leq k$, and show that $x = x_i \bmod m_i$. If $k = 1$, it is trivial. Assume $k \geq 2$, and without loss of generality $1 \leq i \leq \ell$. Since $M_1$ is a multiple of $m_i$, we have $x \bmod m_i = (x \bmod M_1) \bmod m_i$, where

$$x \bmod M_1 = \lambda_2 M_2 \bmod M_1 = vX_1 M_2 \bmod M_1 = X_1 \bmod M_1,$$

and the result follows from the induction hypothesis that $X_1 = x_i \bmod m_i$.

Like **IntegerToRNS**, Algorithm **RNSToInteger** costs $O(M(n) \log n)$ for $M = m_1 m_2 \cdots m_k$ of size $n$, assuming that the $m_i$ are of equal sizes.

The CRT reconstruction problem is analogous to the Lagrange polynomial interpolation problem: find a polynomial of minimal degree interpolating given values $x_i$ at $k$ points $m_i$.

A "flat" variant of the explicit Chinese remainder reconstruction is the following, taking for example $k = 3$:

$$x = \lambda_1 x_1 + \lambda_2 x_2 + \lambda_3 x_3,$$

where $\lambda_i = 1 \bmod m_i$, and $\lambda_i = 0 \bmod m_j$ for $j \neq i$. In other words, $\lambda_i$ is the reconstruction of $x_1 = 0, \ldots, x_{i-1} = 0, x_i = 1, x_{i+1} = 0, \ldots, x_k = 0$. For example, with $m_1 = 11$, $m_2 = 13$ and $m_3 = 17$ we get:

$$x = 221 x_1 + 1496 x_2 + 715 x_3.$$

To reconstruct the integer corresponding to $x_1 = 2$, $x_2 = 3$, $x_3 = 4$, we get $x = 221 \cdot 2 + 1496 \cdot 3 + 715 \cdot 4 = 7790$, which after reduction modulo $11 \cdot 13 \cdot 17 = 2431$ gives 497.

## 2.8 Exercises

**Exercise 2.1** In §2.1.3 we considered the representation of nonnegative integers using a residue number system. Show that a residue number system can also be used to represent signed integers, provided their absolute values are not too large. (Specifically, if relatively prime moduli $m_1, m_2, \ldots, m_k$ are used, and $B = m_1 m_2 \cdots m_k$, the integers $x$ should satisfy $|x| < B/2$.)

**Exercise 2.2** Suppose two nonnegative integers $x$ and $y$ are represented by their residues modulo a set of relatively prime moduli $m_1, m_2, \ldots, m_k$ as in §2.1.3. Consider the *comparison problem*: is $x < y$? Is it necessary to convert $x$ and $y$ back to a standard (non-CRT) representation in order to answer this question? Similarly, if a signed integer $x$ is represented as in Exercise 2.1, consider the *sign detection problem*: is $x < 0$?

**Exercise 2.3** Consider the use of redundant moduli in the Chinese remainder representation. In other words, using the notation of Exercise 2.2, consider the case that $x$ could be reconstructed without using all the residues. Show that this could be useful for error detection (and possibly error correction) if arithmetic operations are performed on unreliable hardware.

**Exercise 2.4** Consider the two complexity bounds $O(M(d \log(Nd)))$ and $O(M(d)M(\log N))$ given at the end of §2.1.5. Compare the bounds in three cases: (a) $d \ll N$; (b) $d \sim N$; (c) $d \gg N$. Assume two subcases for the multiplication algorithm: (i) $M(n) = O(n^2)$; (ii) $M(n) = O(n \log n)$. (For the sake of simplicity, ignore any log log factors.)

**Exercise 2.5** Show that, if a symmetric representation in $[-N/2, N/2)$ is used in Algorithm **ModularAdd** (§2.2), then the probability that we need to add or subtract $N$ is $1/4$ if $N$ is even, and $(1 - 1/N^2)/4$ if $N$ is odd (assuming in both cases that $a$ and $b$ are uniformly distributed).

**Exercise 2.6** Write down the complexity of the Montgomery-Svoboda algorithm (§2.4.2, page 67) for $k$ steps. For $k = 3$, use van der Hoeven's relaxed Karatsuba multiplication [124] to save one $M(n/3)$ product.

**Exercise 2.7** Assume you have an FFT algorithm computing products modulo $2^n + 1$. Prove that, with some preconditioning, you can perform a division with remainder of a $2n$-bit integer by an $n$-bit integer as fast as 1.5 multiplications of $n$ bits by $n$ bits.

**Exercise 2.8** Assume you know $p(x) \bmod (x^{n_1}-1)$ and $p(x) \bmod (x^{n_2}-1)$, where $p(x) \in F[x]$ has degree $n - 1$, and $n_1 > n_2$, and $F$ is a field. Up to which value of $n$ can you uniquely reconstruct $p$? Design a corresponding algorithm.

**Exercise 2.9** Consider the problem of computing the Fourier transform of a vector $\mathbf{a} = [a_0, a_1, \ldots, a_{K-1}]$, defined in Eqn. (2.1), when the size $K$ is not a power of two. For example, $K$ might be an odd prime or an odd prime power. Can you find an algorithm to do this in $O(K \log K)$ operations?

**Exercise 2.10** Consider the problem of computing the cyclic convolution of two $K$-vectors, where $K$ is not a power of two. (For the definition, with $K$ replaced by $N$, see §3.3.1.) Show that the cyclic convolution can be computed using FFTs on $2^\lambda$ points for some suitable $\lambda$, or by using DFTs on $K$ points (see Exercise 2.9). Which method is better?

**Exercise 2.11** Devise a parallel version of Algorithm **MultipleInversion** as outlined in §2.5.1. Analyse its time and space complexity. Try to minimise the number of parallel processors required while achieving a parallel time complexity of $O(\log k)$.

**Exercise 2.12** Analyse the complexity of the algorithm outlined at the end of §2.5.1 to compute $1/x \bmod N_1, \ldots, 1/x \bmod N_k$, when all the $N_i$ have size $n$, and $x$ has size $\ell$. For which values of $n, \ell$ is it faster than the naive algorithm which computes all modular inverses separately? [Assume $M(n)$ is quasi-linear, and neglect multiplicative constants.]

**Exercise 2.13** Write a **RightToLeftBinaryExp** algorithm and compare it with Algorithm **LeftToRightBinaryExp** of §2.6.1.

**Exercise 2.14** Investigate heuristic algorithms for obtaining close-to-optimal addition (or multiplication) chains when the cost of a general addition $a + b$ (or multiplication $a \cdot b$) is $\lambda$ times the cost of duplication $a + a$ (or squaring $a \cdot a$), and $\lambda$ is some fixed positive constant. (This is a reasonable model for modular exponentiation, because multiplication mod $N$ is generally more expensive than squaring mod $N$. It is also a reasonable model for operations in groups defined by elliptic curves, since in this case the formulæ for addition and duplication are usually different and have different costs.)

## 2.9 Notes and References

Several number-theoretic algorithms make heavy use of modular arithmetic, in particular integer factorization algorithms (for example: Pollard's $\rho$ algorithm and the elliptic curve method).

Another important application of modular arithmetic in computer algebra is computing the roots of a univariate polynomial over a finite field, which requires efficient arithmetic over $\mathbb{F}_p[x]$. See for example the excellent book "MCA" by von zur Gathen and Gerhard [100].

We say in §2.1.3 that residue number systems can only be used when $N$ factors into $N_1 N_2 \ldots$; this is not quite true, since Bernstein and Sorenson show in [24] how to perform modular arithmetic using a residue number system.

For notes on the Kronecker-Schönhage trick, see §1.9.

Barrett's algorithm is described in [14], which also mentions the idea of using two short products. The original description of Montgomery's REDC algorithm is [170]. It is now widely used in several applications. However, only a few authors considered using a reduction factor which is not of the form $\beta^n$, among them McLaughlin [161] and Mihailescu [165]. The Montgomery-Svoboda algorithm (§2.4.2) is also called "Montgomery tail tayloring" by Hars [113], who attributes Svoboda's algorithm — more precisely its variant with the most significant word being $\beta - 1$ instead of $\beta$ — to Quisquater. The folding optimization of REDC described in §2.4.2 (Subquadratic Montgomery Reduction) is an LSB-extension of the algorithm described in the context of Barrett's algorithm by Hasenplaugh, Gaubatz and Gopal [118]. Amongst the algorithms not covered in this book, we mention the "bipartite modular multiplication" of Kaihara and Takagi [134], which involves performing both MSB- and LSB-division in parallel.

The description of McLaughlin's algorithm in §2.4.3 follows [161, Variation 2]; McLaughlin's algorithm was reformulated in a polynomial context by Mihailescu [165].

Many authors have proposed FFT algorithms, or improvements of such algo-rithms, and applications such as fast computation of convolutions. Some references are Aho, Hopcroft and Ullman [3]; Nussbaumer [177]; Borodin and Munro [35], who describe the polynomial approach; Van Loan [223] for the linear algebra approach; and Pollard [186] for the FFT over finite fields. Rader [188] considered the case where the number of data points is a prime, and Winograd [231] generalised Rader's algorithm to prime powers. Bluestein's algorithm [30] is also applicable in these cases. In Bernstein [22, §23] the reader will find some historical remarks and several nice applications of the FFT.

The Schönhage-Strassen algorithm first appeared in [200]. Recently Fürer [98] has proposed an integer multiplication algorithm that is asymptotically faster than the Schönhage-Strassen algorithm. Fürer's algorithm *almost* achieves the conjectured best possible $\Theta(n \log n)$ running time.

Concerning special moduli, Percival considers in [184] the case $N = a \pm b$ where both $a$ and $b$ are highly composite; this is a generalization of the case $N = \beta^n \pm 1$. The pseudo-Mersenne primes of §2.4.4 are recommended by the National Institute of Standards and Technology (NIST) [75]. See also the book by Hankerson, Menezes and Vanstone [110].

Algorithm **MultipleInversion** — also known as "batch inversion" — is due to Montgomery [171]. The application of Barrett's algorithm for an implicitly invariant divisor was suggested by Granlund.

Modular exponentiation and cryptographic algorithms are described in much detail in the book by Menezes, van Oorschot and Vanstone [162, Chapter 14]. A detailed description of the best theoretical algorithms, with references, can be found in Bernstein [18]. When both the modulus and base are invariant, modular exponentiation with $k$-bit exponent and $n$-bit modulus can be performed in time $O((k/\log k)M(n))$, after a precomputation of $O(k/\log k)$ powers in time $O(kM(n))$. Take for example $b = 2^{k/t}$ in Note 14.112 and Algorithm 14.109 of [162], with $t \log t \approx k$, where the powers $a^{b^i} \bmod N$ for $0 \le i < t$ are precomputed. An algorithm of same complexity using a DBNS (Double-Base Number System) was proposed by Dimitrov, Jullien and Miller [86], however with a larger table of $\Theta(k^2)$ precomputed powers.

Original papers on Booth recoding, SRT division, *etc.*, are reprinted in the book by Swartzlander [213].

A quadratic algorithm for CRT reconstruction is discussed in [73]; Möller gives some improvements in the case of a small number of small moduli known in advance [168]. Algorithm **IntegerToRNS** can be found in Borodin and Moenck [34]. The explicit Chinese Remainder Theorem and its applications to modular exponentiation are discussed by Bernstein and Sorenson in [24].

# Chapter 3

# Floating-Point Arithmetic

This chapter discusses the basic operations — addition, subtraction, multiplication, division, square root, conversion — on arbitrary precision floating-point numbers, as Chapter 1 does for arbitrary precision integers. More advanced functions like elementary and special functions are covered in Chapter 4. This chapter largely follows the IEEE 754 standard, and extends it in a natural way to arbitrary precision; deviations from IEEE 754 are explicitly mentioned. By default IEEE 754 refers to the 2008 revision, known as IEEE 754-2008; we write IEEE 754-1985 when we explicitly refer to the 1985 initial standard. Topics not discussed here include: hardware implementations, fixed-precision implementations, special representations.

## 3.1  Representation

The classical non-redundant representation of a floating-point number $x$ in radix $\beta > 1$ is the following (other representations are discussed in §3.8):

$$x = (-1)^s \cdot m \cdot \beta^e, \qquad (3.1)$$

where $(-1)^s$, $s \in \{0, 1\}$, is the *sign*, $m \geq 0$ is the *significand*, and the integer $e$ is the *exponent* of $x$. In addition, a positive integer $n$ defines the *precision* of $x$, which means that the significand $m$ contains at most $n$ significant digits in radix $\beta$.

An important special case is $m = 0$ representing zero. In this case the sign

$s$ and exponent $e$ are irrelevant and may be used to encode other information (see for example §3.1.3).

For $m \neq 0$, several semantics are possible; the most common ones are:

- $\beta^{-1} \leq m < 1$, then $\beta^{e-1} \leq |x| < \beta^e$. In this case $m$ is an integer multiple of $\beta^{-n}$. We say that the *unit in the last place* of $x$ is $\beta^{e-n}$, and we write $\mathrm{ulp}(x) = \beta^{e-n}$. For example, $x = 3.1416$ with radix $\beta = 10$ is encoded by $m = 0.31416$ and $e = 1$. This is the convention that we will use in this chapter;

- $1 \leq m < \beta$, then $\beta^e \leq |x| < \beta^{e+1}$, and $\mathrm{ulp}(x) = \beta^{e+1-n}$. With radix ten the number $x = 3.1416$ is encoded by $m = 3.1416$ and $e = 0$. This is the convention adopted in the IEEE 754 standard;

- we can also use an integer significand $\beta^{n-1} \leq m < \beta^n$, then $\beta^{e+n-1} \leq |x| < \beta^{e+n}$, and $\mathrm{ulp}(x) = \beta^e$. With radix ten the number $x = 3.1416$ is encoded by $m = 31416$ and $e = -4$.

Note that in the above three cases, there is only one possible representation of a non-zero floating-point number: we have a *canonical* representation. In some applications, it is useful to relax the lower bound on nonzero $m$, which in the three cases above gives respectively $0 < m < 1$, $0 < m < \beta$, and $0 < m < \beta^n$, with $m$ an integer multiple of $\beta^{-n}$, $\beta^{e+1-n}$, and 1 respectively. In this case, there is no longer a canonical representation. For example, with an integer significand and a precision of 5 digits, the number 3.1400 might be encoded by $(m = 31400, e = -4)$, $(m = 03140, e = -3)$, or $(m = 00314, e = -2)$. This non-canonical representation has the drawback that the most significant non-zero digit of the significand is not known in advance. The unique encoding with a non-zero most significant digit, i.e., $(m = 31400, e = -4)$ here, is called the *normalised* — or simply *normal* — encoding.

The significand is also sometimes called the *mantissa* or *fraction*. The above examples demonstrate that the different significand semantics correspond to different positions of the decimal (or radix $\beta$) point, or equivalently to different *biases* of the exponent. We assume in this chapter that both the radix $\beta$ and the significand semantics are implicit for a given implementation, thus are not physically encoded.

The words "base" and "radix" have similar meanings. For clarity we reserve "radix" for the constant $\beta$ in a floating-point representation such as (3.1). The significand $m$ and exponent $e$ might be stored in a different base, as discussed below.

## 3.1.1 Radix Choice

Most floating-point implementations use radix $\beta = 2$ or a power of two, because this is convenient and efficient on binary computers. For a radix $\beta$ which is not a power of 2, two choices are possible:

- store the significand in base $\beta$, or more generally in base $\beta^k$ for an integer $k \geq 1$. Each digit in base $\beta^k$ requires $\lceil k \lg \beta \rceil$ bits. With such a choice, individual digits can be accessed easily. With $\beta = 10$ and $k = 1$, this is the "Binary Coded Decimal" or BCD encoding: each decimal digit is represented by 4 bits, with a memory loss of about 17% (since $\lg(10)/4 \approx 0.83$). A more compact choice is radix $10^3$, where 3 decimal digits are stored in 10 bits, instead of in 12 bits with the BCD format. This yields a memory loss of only 0.34% (since $\lg(1000)/10 \approx 0.9966$);

- store the significand in binary. This idea is used in Intel's Binary-Integer Decimal (BID) encoding, and in one of the two decimal encodings in IEEE 754-2008. Individual digits can not be accessed directly, but one can use efficient binary hardware or software to perform operations on the significand.

A drawback of the binary encoding is that, during the addition of two arbitrary-precision numbers, it is not easy to detect if the significand exceeds the maximum value $\beta^n - 1$ (when considered as an integer) and thus if rounding is required. Either $\beta^n$ is precomputed, which is only realistic if all computations involve the same precision $n$, or it is computed on the fly, which might result in increased complexity (see Chapter 1 and §2.6.1).

## 3.1.2 Exponent Range

In principle, one might consider an unbounded exponent. In other words, the exponent $e$ might be encoded by an arbitrary-precision integer (see Chapter 1). This would have the great advantage that no underflow or overflow could occur (see below). However, in most applications, an exponent encoded in 32 bits is more than enough: this enables us to represent values up to about $10^{646\,456\,993}$ for $\beta = 2$. A result exceeding this value most probably corresponds to an error in the algorithm or the implementation. Using arbitrary-precision integers for the exponent induces an extra overhead that slows down the implementation in the average case, and it usually requires more memory to store each number.

Thus, in practice the exponent nearly always has a limited range $e_{\min} \leq e \leq e_{\max}$. We say that a floating-point number is *representable* if it can be represented in the form $(-1)^s \cdot m \cdot \beta^e$ with $e_{\min} \leq e \leq e_{\max}$. The set of representable numbers clearly depends on the significand semantics. For the convention we use here, i.e., $\beta^{-1} \leq m < 1$, the smallest positive representable floating-point number is $\beta^{e_{\min}-1}$, and the largest one is $\beta^{e_{\max}}(1 - \beta^{-n})$.

Other conventions for the significand yield different exponent ranges. For example the double-precision format — called `binary64` in IEEE 754-2008 — has $e_{\min} = -1022$, $e_{\max} = 1023$ for a significand in $[1, 2)$; this corresponds to $e_{\min} = -1021$, $e_{\max} = 1024$ for a significand in $[1/2, 1)$, and $e_{\min} = -1074$, $e_{\max} = 971$ for an integer significand in $[2^{52}, 2^{53})$.

### 3.1.3   Special Values

With a bounded exponent range, if we want a complete arithmetic, we need some special values to represent very large and very small values. Very small values are naturally flushed to zero, which is a special number in the sense that its significand is $m = 0$, which is not normalised. For very large values, it is natural to introduce two special values $-\infty$ and $+\infty$, which encode large non-representable values. Since we have two infinities, it is natural to have two zeros $-0$ and $+0$, for example $1/(-\infty) = -0$ and $1/(+\infty) = +0$. This is the IEEE 754 choice. Another possibility would be to have only one infinity $\infty$ and one zero $0$, forgetting the sign in both cases.

An additional special value is *Not a Number* (NaN), which either represents an uninitialised value, or is the result of an *invalid* operation like $\sqrt{-1}$ or $(+\infty) - (+\infty)$. Some implementations distinguish between different kinds of NaN, in particular IEEE 754 defines *signalling* and *quiet* NaNs.

### 3.1.4   Subnormal Numbers

*Subnormal numbers* are required by the IEEE 754 standard, to allow what is called *gradual underflow* between the smallest (in absolute value) non-zero normalised numbers and zero. We first explain what subnormal numbers are; then we will see why they are not necessary in arbitrary precision.

Assume we have an integer significand in $[\beta^{n-1}, \beta^n)$ where $n$ is the precision, and an exponent in $[e_{\min}, e_{\max}]$. Write $\eta = \beta^{e_{\min}}$. The two smallest positive normalised numbers are $x = \beta^{n-1}\eta$ and $y = (\beta^{n-1} + 1)\eta$. The difference $y - x$ equals $\eta$, which is tiny compared to $x$. In particular, $y - x$ can not be represented exactly as a normalised number (assuming $\beta^{n-1} > 1$) and will be rounded to zero in "rounding to nearest" mode (§3.1.9). This has the unfortunate consequence that instructions like:

```
if (y != x) then
    z = 1.0/(y - x);
```

will produce a "division by zero" error when executing `1.0/(y - x)`.

Subnormal numbers solve this problem. The idea is to relax the condition $\beta^{n-1} \leq m$ for the exponent $e_{\min}$. In other words, we include all numbers of the form $m \cdot \beta^{e_{\min}}$ for $1 \leq m < \beta^{n-1}$ in the set of valid floating-point numbers. One could also permit $m = 0$, and then zero would be a subnormal number, but we continue to regard zero as a special case.

Subnormal numbers are all positive integer multiples of $\pm\eta$, with a multiplier $m$, $1 \leq m < \beta^{n-1}$. The difference between $x = \beta^{n-1}\eta$ and $y = (\beta^{n-1} + 1)\eta$ is now representable, since it equals $\eta$, the smallest positive subnormal number. More generally, all floating-point numbers are multiples of $\eta$, likewise for their sum or difference (in other words, operations in the subnormal domain correspond to fixed-point arithmetic). If the sum or difference is non-zero, it has magnitude at least $\eta$, thus can not be rounded to zero. Thus the "division by zero" problem mentioned above does not occur with subnormal numbers.

In the IEEE 754 double-precision format — called `binary64` in IEEE 754-2008 — the smallest positive normal number is $2^{-1022}$, and the smallest positive subnormal number is $2^{-1074}$. In arbitrary precision, subnormal numbers seldom occur, since usually the exponent range is huge compared to the expected exponents in a given application. Thus the only reason for implementing subnormal numbers in arbitrary precision is to provide an extension of IEEE 754 arithmetic. Of course, if the exponent range is unbounded, then there is absolutely no need for subnormal numbers, because any nonzero floating-point number can be normalised.

### 3.1.5   Encoding

The *encoding* of a floating-point number $x = (-1)^s \cdot m \cdot \beta^e$ is the way the values $s$, $m$ and $e$ are stored in the computer. Remember that $\beta$ is implicit, i.e., is considered fixed for a given implementation; as a consequence, we do not consider here *mixed radix* operations involving numbers with different radices $\beta$ and $\beta'$.

We have already seen that there are several ways to encode the significand $m$ when $\beta$ is not a power of two: in base-$\beta^k$ or in binary. For normal numbers in radix 2, i.e., $2^{n-1} \leq m < 2^n$, the leading bit of the significand is necessarily 1, thus one might choose not the encode it in memory, to gain an extra bit of precision. This is called the *implicit leading bit*, and it is the choice made in the IEEE 754 formats. For example the double-precision format has a sign bit, an exponent field of 11 bits, and a significand of 53 bits, with only 52 bits stored, which gives a total of 64 stored bits:

| sign | (biased) exponent | significand |
|------|-------------------|-------------|
| (1 bit) | (11 bits) | (52 bits, plus implicit leading bit) |

A nice consequence of this particular encoding is the following. Let $x$ be a double-precision number, neither subnormal, $\pm\infty$, NaN, nor the largest normal number in absolute value. Consider the 64-bit encoding of $x$ as a 64-bit integer, with the sign bit in the most significant bit, the exponent bits in the next most significant bits, and the explicit part of the significand in the low significant bits. Adding 1 to this 64-bit integer yields the next double-precision number to $x$, away from zero. Indeed, if the significand $m$ is smaller than $2^{53} - 1$, $m$ becomes $m + 1$ which is smaller than $2^{53}$. If $m = 2^{53} - 1$, then the lowest 52 bits are all set, and a carry occurs between the significand field and the exponent field. Since the significand field becomes zero, the new significand is $2^{52}$, taking into account the implicit leading bit. This corresponds to a change from $(2^{53} - 1) \cdot 2^e$ to $2^{52} \cdot 2^{e+1}$, which is exactly the next number away from zero. Thanks to this consequence of the encoding, an integer comparison of two words (ignoring the actual type of the operands) should give the same result as a floating-point comparison, so it is possible to sort normal positive floating-point numbers as if they were integers of the same length (64-bit for double precision).

In arbitrary precision, saving one bit is not as crucial as in fixed (small) precision, where one is constrained by the word size (usually 32 or 64 bits). Thus, in arbitrary precision, it is easier and preferable to encode the whole significand. Also, note that having an "implicit bit" is not possible in radix $\beta > 2$, since for a normal number the most significant digit might take several values, from 1 to $\beta - 1$.

When the significand occupies several words, it can be stored in a linked list, or in an array (with a separate size field). Lists are easier to extend, but accessing arrays is usually more efficient because fewer memory references are required in the inner loops and memory locality is better.

The sign $s$ is most easily encoded as a separate bit field, with a non-negative significand. This is the *sign-magnitude* encoding. Other possibilities are to have a signed significand, using either 1's complement or 2's complement, but in the latter case a special encoding is required for zero, if it is desired to distinguish $+0$ from $-0$. Finally, the exponent might be encoded as a signed word (for example, type `long` in the C language).

### 3.1.6 Precision: Local, Global, Operation, Operand

The different operands of a given operation might have different precisions, and the result of that operation might be desired with yet another precision. There are several ways to address this issue.

- The precision, say $n$ is attached to a given operation. In this case, operands with a smaller precision are automatically converted to precision $n$. Operands with a larger precision might either be left unchanged, or rounded to precision $n$. In the former case, the code implementing the operation must be able to handle operands with different precisions. In the latter case, the rounding mode to shorten the operands must be specified. Note that this rounding mode might differ from that of the operation itself, and that operand rounding might yield large errors. Consider for example $a = 1.345$ and $b = 1.234567$ with a precision of 4 digits. If $b$ is taken as exact, the exact value of $a - b$ equals $0.110433$, which when rounded to nearest becomes $0.1104$. If $b$ is first rounded to nearest to 4 digits, we get $b' = 1.235$, and $a - b' = 0.1100$ is rounded to itself.

- The precision $n$ is attached to each variable. Here again two cases may occur. If the operation destination is part of the operation inputs, as in `sub(c, a, b)`, which means $c \leftarrow \text{round}(a-b)$, then the precision of the result operand $c$ is known, thus the rounding precision is known in advance. Alternatively, if no precision is given for the result, one might choose the maximal (or minimal) precision from the input operands, or use a global variable, or request an extra precision parameter for the operation, as in `c = sub(a, b, n)`.

Of course, these different semantics are inequivalent, and may yield different results. In the following, we consider the case where each variable, including the destination variable, has its own precision, and no pre-rounding or post-rounding occurs. In other words, the operands are considered exact to their full precision.

Rounding is considered in detail in §3.1.9. Here we define what we mean by the *correct rounding* of a function.

**Definition 3.1.1** *Let $a, b, \ldots$ be floating-point numbers, $f$ a mathematical function, $n \geq 1$ an integer, and $\circ$ a rounding mode. We say that $c$ is the correct rounding of $f(a, b, \ldots)$, and we write $c = \circ_n(f(a, b, \ldots))$, if $c$ is the floating-point number closest to $f(a, b, \ldots)$ in precision $n$ and according to the given rounding mode. In case several numbers are at the same distance from $f(a, b, \ldots)$, the rounding mode must define in a deterministic way which one is "the closest". When there is no ambiguity, we omit $n$ and write simply $c = \circ(f(a, b, \ldots))$.*

## 3.1.7   Link to Integers

Most floating-point operations reduce to arithmetic on the significands, which can be considered as integers as seen at the beginning of this section. Therefore efficient arbitrary precision floating-point arithmetic requires efficient underlying integer arithmetic (see Chapter 1).

Conversely, floating-point numbers might be useful for the implementation of arbitrary precision integer arithmetic. For example, one might use hardware floating-point numbers to represent an arbitrary precision integer. Indeed, since a double-precision floating-point number has 53 bits of precision, it can represent an integer up to $2^{53} - 1$, and an integer $A$ can be represented as: $A = a_{n-1}\beta^{n-1} + \cdots + a_i\beta^i + \cdots + a_1\beta + a_0$, where $\beta = 2^{53}$, and the $a_i$ are stored in double-precision data types. Such an encoding was popular when most processors were 32-bit, and some had relatively slow integer operations in hardware. Now that most computers are 64-bit, this encoding is obsolete.

Floating-point *expansions* are a variant of the above. Instead of storing $a_i$ and having $\beta^i$ implicit, the idea is to directly store $a_i \beta^i$. Of course, this only works for relatively small $i$, i.e., whenever $a_i \beta^i$ does not exceed the format range. For example, for IEEE 754 double precision, the maximal integer precision is 1024 bits. (Alternatively, one might represent an integer as a multiple of the smallest positive number $2^{-1074}$, with a corresponding maximal precision of 2098 bits.)

Hardware floating-point numbers might also be used to implement the Fast Fourier Transform (FFT), using complex numbers with floating-point real and imaginary part (see §3.3.1).

### 3.1.8 Ziv's Algorithm and Error Analysis

A *rounding boundary* is a point at which the rounding function $\circ(x)$ is discontinuous.

In fixed precision, for basic arithmetic operations, it is sometimes possible to design one-pass algorithms that directly compute a correct rounding. However, in arbitrary precision, or for elementary or special functions, the classical method is to use Ziv's algorithm:

1. we are given an input $x$, a target precision $n$, and a rounding mode;

2. compute an approximation $y$ with precision $m > n$, and a corresponding error bound $\varepsilon$ such that $|y - f(x)| \leq \varepsilon$;

3. if $[y - \varepsilon, y + \varepsilon]$ contains a rounding boundary, increase $m$ and go to step 2;

4. output the rounding of $y$, according to the given rounding mode.

The error bound $\varepsilon$ at step 2 might be computed either *a priori*, i.e., from $x$ and $n$ only, or *dynamically*, i.e., from the different intermediate values computed by the algorithm. A dynamic bound will usually be tighter, but will require extra computations (however, those computations might be done in low precision).

Depending on the mathematical function to be implemented, one might prefer an absolute or a relative error analysis. When computing a relative error bound, at least two techniques are available: one might express the errors in terms of units in the last place (ulps), or one might express them in terms of true relative error. It is of course possible in a given analysis to mix both kinds of errors, but in general one loses a constant factor — the radix $\beta$ — when converting from one kind of relative error to the other kind.

Another important distinction is *forward* versus *backward* error analysis. Assume we want to compute $y = f(x)$. Because the input is rounded, and/or because of rounding errors during the computation, we might actually compute $y' \approx f(x')$. Forward error analysis will bound $|y' - y|$ if we have a bound on $|x' - x|$ and on the rounding errors that occur during the computation.

Backward error analysis works in the other direction. If the computed value is $y'$, then backward error analysis will give us a number $\delta$ such that, for *some* $x'$ in the ball $|x' - x| \le \delta$, we have $y' = f(x')$. This means that the error is *no worse* than might have been caused by an error of $\delta$ in the input value. Note that, if the problem is ill-conditioned, $\delta$ might be small even if $|y' - y|$ is large.

In our error analyses, we assume that no overflow or underflow occurs, or equivalently that the exponent range is unbounded, unless the contrary is explicitly stated.

### 3.1.9  Rounding

There are several possible definitions of rounding. For example *probabilistic rounding* — also called *stochastic rounding* — chooses at random a rounding towards $+\infty$ or $-\infty$ for each operation. The IEEE 754 standard defines four rounding modes: towards zero, $+\infty$, $-\infty$ and to nearest (with ties broken to even). Another useful mode is "rounding away from zero", which rounds in the opposite direction from zero: a positive number is rounded towards $+\infty$, and a negative number towards $-\infty$. If the sign of the result is known, all IEEE 754 rounding modes might be converted to either rounding to nearest, rounding towards zero, or rounding away from zero.

**Theorem 3.1.1** *Consider a floating-point system with radix $\beta$ and precision $n$. Let $u$ be the rounding to nearest of some real $x$, then the following*

*inequalities hold:*

$$|u - x| \leq \frac{1}{2}\operatorname{ulp}(u)$$

$$|u - x| \leq \frac{1}{2}\beta^{1-n}|u|$$

$$|u - x| \leq \frac{1}{2}\beta^{1-n}|x|.$$

**Proof.** For $x = 0$, necessarily $u = 0$, and the statement holds. Without loss of generality, we can assume $u$ and $x$ positive. The first inequality is the definition of rounding to nearest, and the second one follows from $\operatorname{ulp}(u) \leq \beta^{1-n}u$. (In the case $\beta = 2$, it gives $|u - x| \leq 2^{-n}|u|$.) For the last inequality, we distinguish two cases: if $u \leq x$, it follows from the second inequality. If $x < u$, then if $x$ and $u$ have the same exponent, i.e., $\beta^{e-1} \leq x < u < \beta^e$, then $\operatorname{ulp}(u) = \beta^{e-n} \leq \beta^{1-n}x$. The only remaining case is $\beta^{e-1} \leq x < u = \beta^e$. Since the floating-point number preceding $\beta^e$ is $\beta^e(1 - \beta^{-n})$, and $x$ was rounded to nearest, we have $|u - x| \leq \beta^{e-n}/2$ here too. □

In order to round according to a given rounding mode, one proceeds as follows:

1. first round as if the exponent range was unbounded, with the given rounding mode;

2. if the rounded result is within the exponent range, return this result;

3. otherwise raise the "underflow" or "overflow" exception, and return $\pm 0$ or $\pm\infty$ accordingly.

For example, assume radix 10 with precision 4, $e_{\max} = 3$, with $x = 0.9234 \cdot 10^3$, $y = 0.7656 \cdot 10^2$. The exact sum $x + y$ equals $0.99996 \cdot 10^3$. With rounding towards zero, we obtain $0.9999 \cdot 10^3$, which is representable, so there is no overflow. With rounding to nearest, $x + y$ rounds to $0.1000 \cdot 10^4$, where the exponent 4 exceeds $e_{\max} = 3$, so we get $+\infty$ as the result, with an overflow. In this model, overflow depends not only on the operands, but also on the rounding mode.

The "round to nearest" mode of IEEE 754 rounds the result of an operation to the nearest representable number. In case the result of an operation is exactly halfway between two consecutive numbers, the one with least significant bit zero is chosen (for radix 2). For example $1.1011_2$ is rounded with a precision of 4 bits to $1.110_2$, as is $1.1101_2$. However this rule does not readily extend to an arbitrary radix. Consider for example radix $\beta = 3$, a precision of 4 digits, and the number $1212.111\ldots_3$. Both $1212_3$ and $1220_3$ end in an even digit. The natural extension is to require the whole significand to be even, when interpreted as an integer in $[\beta^{n-1}, \beta^n - 1]$. In this setting, $(1212.111\ldots)_3$ rounds to $(1212)_3 = 50_{10}$. (Note that $\beta^n$ is an odd number here.)

Assume we want to correctly round a real number, whose binary expansion is $2^e \cdot 0.1b_2 \ldots b_n b_{n+1} \ldots$, to $n$ bits. It is enough to know the values of $r = b_{n+1}$ — called the *round bit* — and that of the *sticky bit* $s$, which is 0 when $b_{n+2} b_{n+3} \ldots$ is identically zero, and 1 otherwise. Table 3.1 shows how to correctly round given $r$, $s$, and the given rounding mode; rounding to $\pm\infty$ being converted to rounding towards zero or away from zero, according to the sign of the number. The entry "$b_n$" is for round to nearest in the case of a tie: if $b_n = 0$ it will be unchanged, but if $b_n = 1$ we add 1 (thus changing $b_n$ to 0).

| $r$ | $s$ | towards zero | to nearest | away from zero |
|-----|-----|--------------|------------|----------------|
| 0 | 0 | 0 | 0 | 0 |
| 0 | 1 | 0 | 0 | 1 |
| 1 | 0 | 0 | $b_n$ | 1 |
| 1 | 1 | 0 | 1 | 1 |

Table 3.1: Rounding rules according to the round bit $r$ and the sticky bit $s$: a "0" entry means truncate (round towards zero), a "1" means round away from zero (add 1 to the truncated significand).

In general, we do not have an infinite expansion, but a finite approximation $y$ of an unknown real value $x$. For example, $y$ might be the result of an arithmetic operation such as division, or an approximation to the value of a transcendental function such as exp. The following problem arises: given the approximation $y$, and a bound on the error $|y - x|$, is it possible to determine the correct rounding of $x$? Algorithm **RoundingPossible** returns *true* if and only if it is possible.

**Proof.** Since rounding is monotonic, it is possible to determine $\circ(x)$ exactly

---

**Algorithm 3.1** RoundingPossible

---

**Input:** a floating-point number $y = 0.1y_2 \ldots y_m$, a precision $n \leq m$, an error
 bound $\varepsilon = 2^{-k}$, a rounding mode $\circ$

**Output:** *true* when $\circ_n(x)$ can be determined for $|y - x| \leq \varepsilon$

    **if** $k \leq n + 1$ **then** return *false*

    **if** $\circ$ is to nearest **then** $r \leftarrow 1$ **else** $r \leftarrow 0$

    **if** $y_{n+1} = r$ and $y_{n+2} = \cdots = y_k = 0$ **then** $s \leftarrow 0$ **else** $s \leftarrow 1$

    **if** $s = 1$ **then** return *true* **else** return *false*.

---

when $\circ(y - 2^{-k}) = \circ(y + 2^{-k})$, or in other words when the interval $[y - 2^{-k}, y + 2^{-k}]$ contains no rounding boundary (or only one as $y - 2^{-k}$ or $y + 2^{-k}$).

If $k \leq n + 1$, then the interval $[-2^{-k}, 2^{-k}]$ has width at least $2^{-n}$, thus contains at least one rounding boundary in its interior, or two rounding boundaries, and it is not possible to round correctly. In the case of directed rounding (resp. rounding to nearest), if $s = 0$ the approximation $y$ is representable (resp. the middle of two representable numbers) in precision $n$, and it is clearly not possible to round correctly; if $s = 1$ the interval $[y - 2^{-k}, y + 2^{-k}]$ contains at most one rounding boundary, and if so it is one of the bounds, thus it is possible to round correctly.     □

## The Double Rounding Problem

When a given real value $x$ is first rounded to precision $m$, then to precision $n < m$, we say that a "double rounding" occurs. The "double rounding problem" happens when this latter value differs from the direct rounding of $x$ to the smaller precision $n$, assuming the same rounding mode is used in all cases, i.e., when:

$$\circ_n(\circ_m(x)) \neq \circ_n(x).$$

The double rounding problem does not occur for directed rounding modes. For these rounding modes, the rounding boundaries at the larger precision $m$ refine those at the smaller precision $n$, thus all real values $x$ that round to the same value $y$ at precision $m$ also round to the same value at precision $n$, namely $\circ_n(y)$.

Consider the decimal value $x = 3.14251$. Rounding to nearest to 5 digits, we get $y = 3.1425$; rounding $y$ to nearest-even to 4 digits, we get $3.142$, whereas direct rounding of $x$ would give $3.143$.

With rounding to nearest mode, the double rounding problem only occurs when the second rounding involves the even-rule, i.e., the value $y = \circ_m(x)$ is a rounding boundary at precision $n$. Otherwise $y$ has distance at least one ulp (in precision $m$) from a rounding boundary at precision $n$, and since $|y - x|$ is bounded by half an ulp (in precision $m$), all possible values for $x$ round to the same value in precision $n$.

Note that the double rounding problem does not occur with all ways of breaking ties for rounding to nearest (Exercise 3.2).

### 3.1.10   Strategies

To determine the correct rounding of $f(x)$ with $n$ bits of precision, the best strategy is usually to first compute an approximation $y$ to $f(x)$ with a working precision of $m = n + h$ bits, with $h$ relatively small. Several strategies are possible in Ziv's algorithm (§3.1.8) when this first approximation $y$ is not accurate enough, or too close to a rounding boundary:

- compute the exact value of $f(x)$, and round it to the target precision $n$. This is possible for a basic operation, for example $f(x) = x^2$, or more generally $f(x, y) = x + y$ or $x \times y$. Some elementary functions may yield an exactly representable output too, for example $\sqrt{2.25} = 1.5$. An "exact result" test after the first approximation avoids possibly unnecessary further computations;

- repeat the computation with a larger working precision $m' = n + h'$. Assuming that the digits of $f(x)$ behave "randomly" and that $|f'(x)/f(x)|$ is not too large, using $h' \approx \lg n$ is enough to guarantee that rounding is possible with probability $1 - O(1/n)$. If rounding is still not possible, because the $h'$ last digits of the approximation encode 0 or $2^{h'} - 1$, one can increase the working precision and try again. A check for exact results guarantees that this process will eventually terminate, provided the algorithm used has the property that it gives the exact result if this result is representable and the working precision is high enough. For example, the square root algorithm should return the exact result if it is representable (see Algorithm **FPSqrt** in §3.5, and also Exercise 3.3).

# 3.2 Addition, Subtraction, Comparison

Addition and subtraction of floating-point numbers operate from the most significant digits, whereas integer addition and subtraction start from the least significant digits. Thus completely different algorithms are involved. Also, in the floating-point case, part or all of the inputs might have no impact on the output, except in the rounding phase.

In summary, floating-point addition and subtraction are more difficult to implement than integer addition/subtraction for two reasons:

- scaling due to the exponents requires shifting the significands before adding or subtracting them. In principle one could perform all operations using only integer operations, but this might require huge integers, for example when adding 1 and $2^{-1000}$;

- as the carries are propagated from least to most significant digits, one may have to look at arbitrarily low input digits to guarantee correct rounding.

In this section, we distinguish between "addition", where both operands to be added have the same sign, and "subtraction", where the operands to be added have different signs (we assume a sign-magnitude representation). The case of one or both operands zero is treated separately; in the description below we assume that all operands are nonzero.

## 3.2.1 Floating-Point Addition

Algorithm **FPadd** adds two binary floating-point numbers $b$ and $c$ of the same sign. More precisely, it computes the correct rounding of $b + c$, with respect to the given rounding mode $\circ$. For the sake of simplicity, we assume $b$ and $c$ are positive, $b \geq c > 0$. It will also be convenient to scale $b$ and $c$ so that $2^{n-1} \leq b < 2^n$ and $2^{m-1} \leq c < 2^m$, where $n$ is the desired precision of the output, and $m \leq n$. Of course, if the inputs $b$ and $c$ to Algorithm **FPadd** are scaled by $2^k$, then to compensate for this the output must be scaled by $2^{-k}$. We assume that the rounding mode is to nearest, towards zero, or away from zero (rounding to $\pm\infty$ reduces to rounding towards zero or away from zero, depending on the sign of the operands).

The values of $\text{round}(\circ, r, s)$ and $\text{round2}(\circ, a, t)$ are given in Table 3.2. We have simplified some of the expressions given in Table 3.2. For example, in

---

**Algorithm 3.2** FPadd

---

**Input:** $b \geq c > 0$ two binary floating-point numbers, a precision $n$ such that $2^{n-1} \leq b < 2^n$, and a rounding mode $\circ$

**Output:** a floating-point number $a$ of precision $n$ and scale $e$ such that $a \cdot 2^e = \circ(b + c)$

1: split $b$ into $b_h + b_\ell$ where $b_h$ contains the $n$ most significant bits of $b$.
2: split $c$ into $c_h + c_\ell$ where $c_h$ contains the most significant bits of $c$, and $\mathrm{ulp}(c_h) = \mathrm{ulp}(b_h) = 1$                                          $\triangleright$ $c_h$ might be zero
3: $a_h \leftarrow b_h + c_h, \quad e \leftarrow 0$
4: $(c, r, s) \leftarrow b_\ell + c_\ell$                                          $\triangleright$ see the text
5: $(a, t) \leftarrow (a_h + c + \mathrm{round}(\circ, r, s), \textit{etc.})$          $\triangleright$ for $t$ see Table 3.2 (upper)
6: **if** $a \geq 2^n$ **then**
7: $\quad (a, e) \leftarrow (\mathrm{round2}(\circ, a, t), e + 1)$                   $\triangleright$ see Table 3.2 (lower)
8: $\quad$ **if** $a = 2^n$ **then** $(a, e) \leftarrow (a/2, e + 1)$
9: return $(a, e)$.

---

the upper half of the table, $r \vee s$ means 0 if $r = s = 0$, and 1 otherwise. In the lower half of the table, $2\lfloor (a+1)/4 \rfloor$ is $(a-1)/2$ if $a = 1 \bmod 4$, and $(a+1)/2$ if $a = 3 \bmod 4$.

At step 4 of Algorithm **FPadd**, the notation $(c, r, s) \leftarrow b_\ell + c_\ell$ means that $c$ is the carry bit of $b_\ell + c_\ell$, $r$ the round bit, and $s$ the sticky bit; $c, r, s \in \{0, 1\}$. For rounding to nearest, $t = \mathrm{sign}(b + c - a)$ is a ternary value which is respectively positive, zero, or negative when $a$ is smaller than, equal to, or larger than the exact sum $b + c$.

**Theorem 3.2.1** *Algorithm* **FPadd** *is correct.*

**Proof.** We have $2^{n-1} \leq b < 2^n$ and $2^{m-1} \leq c < 2^m$, with $m \leq n$. Thus $b_h$ and $c_h$ are the integer parts of $b$ and $c$, $b_\ell$ and $c_\ell$ their fractional parts. Since $b \geq c$, we have $c_h \leq b_h$ and $2^{n-1} \leq b_h \leq 2^n - 1$, thus $2^{n-1} \leq a_h \leq 2^{n+1} - 2$, and at step 5, $2^{n-1} \leq a \leq 2^{n+1}$. If $a < 2^n$, $a$ is the correct rounding of $b + c$. Otherwise, we face the "double rounding" problem: rounding $a$ down to $n$ bits will give the correct result, except when $a$ is odd and rounding is to nearest. In that case, we need to know if the first rounding was exact, and if not in which direction it was rounded; this information is encoded in the ternary value $t$. After the second rounding, we have $2^{n-1} \leq a \leq 2^n$.  □

Note that the exponent $e_a$ of the result lies between $e_b$ (the exponent of $b$, here we considered the case $e_b = n$) and $e_b + 2$. Thus no underflow can occur in an addition. The case $e_a = e_b + 2$ can occur only when the destination precision is less than that of the operands.

| $\circ$ | $r$ | $s$ | round($\circ, r, s$) | $t$ |
|:---:|:---:|:---:|:---:|:---:|
| towards 0 | any | any | 0 | – |
| away from 0 | any | any | $r \vee s$ | – |
| to nearest | 0 | any | 0 | $s$ |
| to nearest | 1 | 0 | 0/1 (even rounding) | $+1/-1$ |
| to nearest | 1 | $\neq 0$ | 1 | $-1$ |

| $\circ$ | $a \bmod 2$ | $t$ | round2($\circ, a, t$) |
|:---:|:---:|:---:|:---:|
| any | 0 | any | $a/2$ |
| towards 0 | 1 | any | $(a-1)/2$ |
| away from 0 | 1 | any | $(a+1)/2$ |
| to nearest | 1 | 0 | $2\lfloor (a+1)/4 \rfloor$ |
| to nearest | 1 | $\pm 1$ | $(a+t)/2$ |

Table 3.2: Rounding rules for addition.

## 3.2.2 Floating-Point Subtraction

Floating-point subtraction (of positive operands) is very similar to addition, with the difference that *cancellation* can occur. Consider for example the subtraction $6.77823 - 5.98771$. The most significant digit of both operands disappeared in the result $0.79052$. This cancellation can be dramatic, as in $6.7782357934 - 6.7782298731 = 0.0000059203$, where six digits were cancelled.

Two approaches are possible, assuming $n$ result digits are wanted, and the exponent difference between the inputs is $d$:

- subtract the $n - d$ most-significant digits of the smaller operand from the $n$ most-significant digits of the larger operand. If the result has $n-e$ digits with $e > 0$, restart with $n + e$ digits from the larger operand and $(n + e) - d$ from the smaller operand;

- alternatively, predict the number $e$ of cancelled digits in the subtraction, and directly subtract the $(n + e) - d$ most-significant digits of the smaller operand from the $n + e$ most-significant digits of the larger one.

Note that in the first approach, we might have $e = n$ if all most-significant digits cancel, thus the process might need to be repeated several times.

The first step in the second approach is usually called *leading zero detection*. Note that the number $e$ of cancelled digits might depend on the rounding mode. For example, $6.778 - 5.7781$ with a 3-digit result yields $0.999$ with rounding toward zero, and $1.00$ with rounding to nearest. Therefore, in a real implementation, the definition of $e$ has to be made precise.

In practice we might consider $n + g$ and $(n + g) - d$ digits instead of $n$ and $n - d$, where the $g$ "guard digits" would prove useful (i) to decide the final rounding, and/or (ii) to avoid another loop in case $e \leq g$.

### Sterbenz's Theorem

Sterbenz's Theorem is an important result concerning floating-point subtraction (of operands of the same sign). It states that the rounding error is zero in some common cases. More precisely:

**Theorem 3.2.2 (Sterbenz)** If $x$ and $y$ are two floating-point numbers of same precision $n$, such that $y$ lies in the interval $[x/2, 2x] \cup [2x, x/2]$, then $y - x$ is exactly representable in precision $n$, if there is no underflow.

**Proof.** The case $x = y = 0$ is trivial, so assume that $x \neq 0$. Since $y \in [x/2, 2x] \cup [2x, x/2]$, $x$ and $y$ must have the same sign. We assume without loss of generality that $x$ and $y$ are positive, so $y \in [x/2, 2x]$.

Assume $x \leq y \leq 2x$ (the same reasoning applies for $x/2 \leq y \leq x$, i.e., $y \leq x \leq 2y$, by interchanging $x$ and $y$). Since $x \leq y$, we have $\mathrm{ulp}(x) \leq \mathrm{ulp}(y)$, thus $y$ is an integer multiple of $\mathrm{ulp}(x)$. It follows that $y - x$ is an integer multiple of $\mathrm{ulp}(x)$. Since $0 \leq y - x \leq x$, $y - x$ is necessarily representable with the precision of $x$. $\qquad\square$

It is important to note that Sterbenz's Theorem applies for any radix $\beta$; the constant 2 in $[x/2, 2x]$ has nothing to do with the radix.

## 3.3 Multiplication

Multiplication of floating-point numbers is called a *short product*. This reflects the fact that, in some cases, the low part of the full product of the significands has no impact — except perhaps for the rounding — on the final result. Consider the multiplication $x \times y$, where $x = \ell\beta^e$ and $y = m\beta^f$. Then $\circ(xy) = \circ(\ell m)\beta^{e+f}$, thus it suffices to consider the case that $x = \ell$ and $y = m$ are integers, and the product is rounded at some weight $\beta^g$ for $g \geq 0$. Either the integer product $\ell \times m$ is computed exactly, using one of the algorithms from Chapter 1, and then rounded; or the upper part is computed directly using a "short product algorithm", with correct rounding. The different cases that can occur are depicted in Figure 3.1.

An interesting question is: how many consecutive identical bits can occur after the round bit? Without loss of generality, we can rephrase this question as follows. Given two odd integers of at most $n$ bits, what is the longest run of identical bits in their product? (In the case of an even significand, one might write it $m = \ell 2^e$ with $\ell$ odd.) There is no *a priori* bound except the trivial one of $2n - 2$ for the number of zeros, and $2n - 1$ for the number of ones. For example, with a precision 5 bits, $27 \times 19 = (1\,000\,000\,001)_2$. More generally, such a case corresponds to a factorisation of $2^{2n-1} + 1$ into two integers of $n$ bits, for example $258\,513 \times 132\,913 = 2^{35} + 1$. $2n$ consecutive ones are not possible since $2^{2n} - 1$ can not factor into two integers of at most $n$ bits. Therefore the maximal runs have $2n - 1$ ones, for example $217 \times 151 = (111\,111\,111\,111\,111)_2$ for $n = 8$. A larger example is $849\,583 \times 647\,089 = 2^{39} - 1$.

The exact product of two floating-point numbers $m\beta^e$ and $m'\beta^{e'}$ is $(mm')\beta^{e+e'}$. Therefore, if no underflow or overflow occurs, the problem reduces to the multiplication of the significands $m$ and $m'$. See Algorithm **FPmultiply**.

The product at step 1 of **FPmultiply** is a *short product*, i.e., a product whose most significant part only is wanted, as discussed at the start of this section. In the quadratic range, it can be computed in about half the time of a full product. In the Karatsuba and Toom-Cook ranges, Mulders' algorithm can gain 10% to 20%; however, due to carries, implementing this algorithm for floating-point computations is tricky. In the FFT range, no better algorithm is known than computing the full product $mm'$ and then rounding it.

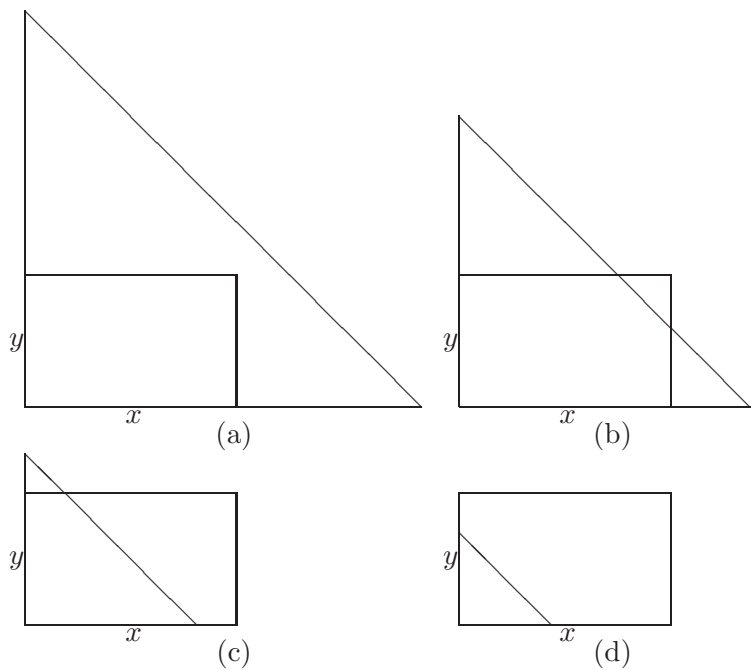

Figure 3.1: Different multiplication scenarios, according to the input and output precisions. The rectangle corresponds to the full product of the inputs $x$ and $y$ (most significant digits bottom left), the triangle to the wanted short product. Case (a): no rounding is necessary, the product being exact; case (b): the full product needs to be rounded, but the inputs should not be; case (c): the input $x$ with the larger precision might be truncated before performing a short product; case (d): both inputs might be truncated.

Hence our advice is to perform a full product of $m$ and $m'$, possibly after truncating them to $n + g$ digits if they have more than $n + g$ digits. Here $g$ (the number of *guard digits*) should be positive (see Exercise 3.4).

It seems wasteful to multiply $n$-bit operands, producing a $2n$-bit product, only to discard the low-order $n$ bits. Algorithm **ShortProduct** computes an approximation to the short product without computing the $2n$-bit full product. It uses a threshold $n_0 \geq 1$, which should be optimized for the given code base.

---

**Algorithm 3.3** FPmultiply

---

**Input:** $x = m \cdot \beta^e$, $x' = m' \cdot \beta^{e'}$, a precision $n$, a rounding mode $\circ$

**Output:** $\circ(xx')$ rounded to precision $n$

  1: $m'' \leftarrow \circ(mm')$ rounded to precision $n$

  2: return $m'' \cdot \beta^{e+e'}$.

---

**Error analysis of the short product.** Consider two $n$-word normalised significands $A$ and $B$ that we multiply using a short product algorithm, where the notation **FullProduct**$(A, B)$ means the full integer product $A \cdot B$.

---

**Algorithm 3.4** ShortProduct

---

**Input:** integers $A, B$, and $n$, with $0 \leq A, B < \beta^n$

**Output:** an approximation to $AB$ div $\beta^n$

**Require:** a threshold $n_0$

  **if** $n \leq n_0$ **then** return **FullProduct**$(A, B)$ div $\beta^n$

  choose $k \geq n/2$, $\ell \leftarrow n - k$

  $C_1 \leftarrow$ **FullProduct**$(A$ div $\beta^\ell, B$ div $\beta^\ell)$ div $\beta^{k-\ell}$

  $C_2 \leftarrow$ **ShortProduct**$(A \bmod \beta^\ell, B$ div $\beta^k, \ell)$

  $C_3 \leftarrow$ **ShortProduct**$(A$ div $\beta^k, B \bmod \beta^\ell, \ell)$

  return $C_1 + C_2 + C_3$.

---

**Theorem 3.3.1** *The value $C'$ returned by Algorithm* **ShortProduct** *differs from the exact short product $C = AB$ div $\beta^n$ by at most $3(n-1)$:*

$$C' \leq C \leq C' + 3(n-1).$$

**Proof.** First, since $A, B$ are nonnegative, and all roundings are truncations, the inequality $C' \leq C$ follows.

Let $A = \sum_i a_i \beta^i$ and $B = \sum_j b_j \beta^j$, where $0 \leq a_i, b_j < \beta$. The possible errors come from: (i) the neglected $a_i b_j$ terms, i.e., parts $C'_2, C'_3, C_4$ of Figure 3.2; (ii) the truncation while computing $C_1$; (iii) the error in the recursive calls for $C_2$ and $C_3$.

We first prove that the algorithm accumulates all products $a_i b_j$ with $i + j \geq n - 1$. This corresponds to all terms on and below the diagonal in Figure 3.2. The most significant neglected terms are the bottom-left terms from $C'_2$ and $C'_3$, respectively $a_{\ell-1} b_{k-1}$ and $a_{k-1} b_{\ell-1}$. Their contribution is at most $2(\beta-1)^2\beta^{n-2}$. The neglected terms from the next diagonal contribute at most $4(\beta-1)^2\beta^{n-3}$, and so on. The total contribution of neglected terms is thus bounded by:

$$(\beta-1)^2\beta^n[2\beta^{-2}+4\beta^{-3}+6\beta^{-4}+\cdots] < 2\beta^n$$

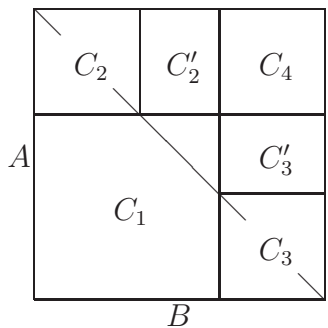

Figure 3.2: Graphical view of Algorithm **ShortProduct**: the computed parts are $C_1, C_2, C_3$, and the neglected parts are $C_2', C_3', C_4$ (most significant part bottom left).

(the inequality is strict since the sum is finite).

The truncation error in $C_1$ is at most $\beta^n$, thus the maximal difference $\varepsilon(n)$ between $C$ and $C'$ satisfies:

$$\varepsilon(n) < 3 + 2\varepsilon(\lfloor n/2 \rfloor),$$

which gives $\varepsilon(n) < 3(n-1)$, since $\varepsilon(1) = 0$.          □

REMARK: if one of the operands was truncated before applying Algorithm **ShortProduct**, simply add one unit to the upper bound (the truncated part is less than 1, thus its product by the other operand is bounded by $\beta^n$).

The complexity $S(n)$ of Algorithm **ShortProduct** satifies the recurrence $S(n) = M(k) + 2S(n-k)$. The optimal choice of $k$ depends on the underlying multiplication algorithm. Assuming $M(n) \approx n^\alpha$ for $\alpha > 1$ and $k = \gamma n$, we get

$$S(n) = \frac{\gamma^\alpha}{1 - 2(1-\gamma)^\alpha} M(n),$$

where the optimal value is $\gamma = 1/2$ in the quadratic range, $\gamma \approx 0.694$ in the Karatsuba range, and $\gamma \approx 0.775$ in the Toom-Cook 3-way range, giving respectively $S(n) \sim 0.5M(n)$, $S(n) \sim 0.808M(n)$, and $S(n) \sim 0.888M(n)$. The ratio $S(n)/M(n) \to 1$ as $r \to \infty$ for Toom-Cook $r$-way. In the FFT range, Algorithm **ShortProduct** is not any faster than a full product.

### 3.3.1 Integer Multiplication via Complex FFT

To multiply $n$-bit integers, it may be advantageous to use the Fast Fourier Tranform (FFT for short, see §1.3.4, §2.3). Note that three FFTs give the cyclic convolution $z = x * y$ defined by

$$z_k = \sum_{0 \le j < N} x_j y_{k-j \bmod N} \text{ for } 0 \le k < N.$$

In order to use the FFT for integer multiplication, we have to pad the input vectors with zeros, thus increasing the length of the transform from $N$ to $2N$.

FFT algorithms fall into two classes: those using number theoretical properties (typically working over a finite ring, as in §2.3.3), and those based on complex floating-point computations. The latter, while not having the best asymptotic complexity, exhibit good practical behaviour, because they take advantage of the efficiency of floating-point hardware. The drawback of the complex floating-point FFT (complex FFT for short) is that, being based on floating-point computations, it requires a rigorous error analysis. However, in some contexts where occasional errors are not disastrous, one may accept a small probability of error if this speeds up the computation. For example, in the context of integer factorisation, a small probability of error is acceptable because the result (a purported factorisation) can easily be checked and discarded if incorrect.

The following theorem provides a tight error analysis:

**Theorem 3.3.2** *The complex FFT allows computation of the cyclic convolution $z = x * y$ of two vectors of length $N = 2^n$ of complex values such that*

$$||z' - z||_\infty \le ||x|| \cdot ||y|| \cdot ((1+\varepsilon)^{3n}(1+\varepsilon\sqrt{5})^{3n+1}(1+\mu)^{3n} - 1), \quad (3.2)$$

*where $||\cdot||$ and $||\cdot||_\infty$ denote the Euclidean and infinity norms respectively, $\varepsilon$ is such that $|(a \pm b)' - (a \pm b)| \le \varepsilon|a \pm b|$, $|(ab)' - (ab)| \le \varepsilon|ab|$ for all machine floats $a$, $b$. Here $\mu \ge |(w^k)' - (w^k)|$, $0 \le k < N$, $w = e^{2\pi i/N}$, and $(\cdot)'$ refers to the computed (stored) value of $(\cdot)$ for each expression.*

| $n$ | $b$ | $m$ |
|-----|-----|-----|
| 1   | 25  | 25  |
| 2   | 24  | 48  |
| 3   | 23  | 92  |
| 4   | 22  | 176 |
| 5   | 22  | 352 |
| 6   | 21  | 672 |
| 7   | 20  | 1280 |
| 8   | 20  | 2560 |
| 9   | 19  | 4864 |
| 10  | 19  | 9728 |

| $n$ | $b$ | $m$ |
|-----|-----|-----|
| 11  | 18  | 18432 |
| 12  | 17  | 34816 |
| 13  | 17  | 69632 |
| 14  | 16  | 131072 |
| 15  | 16  | 262144 |
| 16  | 15  | 491520 |
| 17  | 15  | 983040 |
| 18  | 14  | 1835008 |
| 19  | 14  | 3670016 |
| 20  | 13  | 6815744 |

Table 3.3: Maximal number $b$ of bits per IEEE 754 double-precision floating-point number `binary64` (53-bit significand), and maximal $m$ for a plain $m \times m$ bit integer product, for a given FFT size $2^n$, with signed components.

For the IEEE 754 double-precision format, with rounding to nearest, we have $\varepsilon = 2^{-53}$, and if the $w^k$ are correctly rounded, we can take $\mu = \varepsilon/\sqrt{2}$. For a fixed FFT size $N = 2^n$, the inequality (3.2) enables us to compute a bound $B$ on the components of $x$ and $y$ that guarantees $||z' - z||_\infty < 1/2$. If we know that the exact result $z \in \mathbb{Z}^N$, this enables us to uniquely round the components of $z'$ to $z$. Table 3.3 gives $b = \lg B$, the number of bits that can be used in a 64-bit floating-point word, if we wish to perform $m$-bit multiplication exactly (here $m = 2^{n-1}b$). It is assumed that the FFT is performed with signed components in $\mathbb{Z} \cap [-2^{b-1}, +2^{b-1})$, see for example [80, p. 161].

Note that Theorem 3.3.2 is a worst-case result; with rounding to nearest we expect the error to be smaller due to cancellation – see Exercise 3.9.

Since 64-bit floating-point numbers have bounded precision, we can not compute arbitrarily large convolutions by this method — the limit is about $n = 43$. However, this corresponds to vectors of size $N = 2^n = 2^{43} > 10^{12}$, which is more than enough for practical purposes. (See also Exercise 3.11.)

### 3.3.2 The Middle Product

Given two integers of $2n$ and $n$ bits respectively, their "middle product" consists of the middle $n$ bits of their $3n$-bit product (see Fig. 3.3). The middle product might be computed using two short products, one (low) short product between $x$ and the high part of $y$, and one (high) short product between $x$ and the low part of $y$. However there are algorithms to compute a $2n \times n$ middle product with the same $\sim M(n)$ complexity as an $n \times n$ full product (see §3.8).

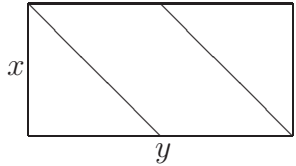

Figure 3.3: The middle product of $x$ of $n$ bits and $y$ of $2n$ bits corresponds to the middle region (most significant bits bottom left).

Several applications benefit from an efficient middle product. One of these applications is Newton's method (§4.2). Consider, for example, the reciprocal iteration (§4.2.2): $x_{j+1} = x_j + x_j(1 - x_j y)$. If $x_j$ has $n$ bits, one has to consider $2n$ bits from $y$ in order to get $2n$ accurate bits in $x_{j+1}$. The product $x_j y$ has $3n$ bits, but if $x_j$ is accurate to $n$ bits, the $n$ most significant bits of $x_j y$ cancel with 1, and the $n$ least significant bits can be ignored as they only contribute noise. Thus, the middle product of $x_j$ and $y$ is exactly what is needed.

#### Payne and Hanek Argument Reduction

Another application of the middle product is Payne and Hanek argument reduction. Assume $x = m \cdot 2^e$ is a floating-point number with a significand $0.5 \le m < 1$ of $n$ bits and a large exponent $e$ (say $n = 53$ and $e = 1024$ to fix the ideas). We want to compute $\sin x$ with a precision of $n$ bits. The classical argument reduction works as follows: first compute $k = \lfloor x/\pi \rfloor$, then compute the reduced argument

$$x' = x - k\pi. \tag{3.3}$$

About $e$ bits will be cancelled in the subtraction $x - (k\pi)$, thus we need to compute $k\pi$ with a precision of at least $e + n$ bits to get an accuracy of at least $n$ bits for $x'$. Of course, this assumes that $x$ is known exactly – otherwise there is no point in trying to compute $\sin x$. Assuming $1/\pi$ has been precomputed to precision $e$, the computation of $k$ costs $M(e, n)$, and the multiplication $k \times \pi$ costs $M(e, e + n)$, thus the total cost is about $M(e)$ when $e \gg n$.

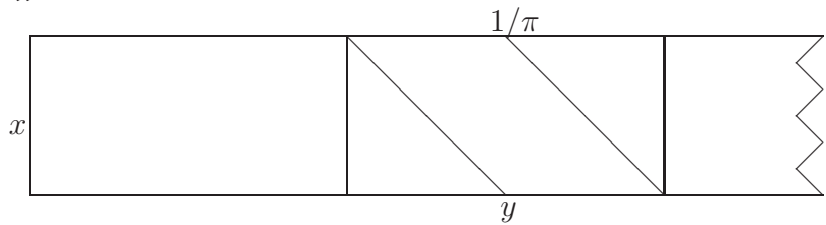

Figure 3.4: A graphical view of Payne and Hanek algorithm.

The key idea of the Payne and Hanek algorithm is to rewrite Eqn. (3.3) as

$$x' = \pi \left( \frac{x}{\pi} - k \right). \tag{3.4}$$

If the significand of $x$ has $n < e$ bits, only about $2n$ bits from the expansion of $1/\pi$ will effectively contribute to the $n$ most significant bits of $x'$, namely the bits of weight $2^{-e-n}$ to $2^{-e+n}$. Let $y$ be the corresponding $2n$-bit part of $1/\pi$. Payne and Hanek's algorithm works as follows: first multiply the $n$-bit significand of $x$ by $y$, keep the $n$ middle bits, and multiply by an $n$-bit approximation of $\pi$. The total cost is $\sim (M(2n, n) + M(n))$, or even $\sim 2M(n)$ if the middle product is performed in time $M(n)$, thus independent of $e$.

## 3.4  Reciprocal and Division

As for integer operations (§1.4), one should try as far as possible to trade floating-point divisions for multiplications, since the cost of a floating-point multiplication is theoretically smaller than the cost of a division by a constant factor (usually from 2 to 5, depending on the algorithm used). In practice, the ratio might not even be constant unless care is taken in implementing division. Some implementations provide division with cost $\Theta(M(n) \log n)$ or $\Theta(n^2)$.

When several divisions have to be performed with the same divisor, a well-known trick is to first compute the reciprocal of the divisor (§3.4.1); then each division reduces to a multiplications by the reciprocal. A small drawback is that each division incurs two rounding errors (one for the reciprocal and one for multiplication by the reciprocal) instead of one, so we can no longer guarantee a correctly rounded result. For example, in base ten with six digits, $3.0/3.0$ might evaluate to $0.999\,999 = 3.0 \times 0.333\,333$.

The cases of a single division, or several divisions with a varying divisor, are considered in §3.4.2.

### 3.4.1   Reciprocal

Here we describe algorithms that compute an approximate reciprocal of a positive floating-point number $a$, using integer-only operations (see Chapter 1). The integer operations simulate floating-point computations, but all roundings are made explicit. The number $a$ is represented by an integer $A$ of $n$ words in radix $\beta$: $a = \beta^{-n}A$, and we assume $\beta^n/2 \leq A$, thus requiring $1/2 \leq a < 1$. (This does not cover all cases for $\beta \geq 3$, but if $\beta^{n-1} \leq A < \beta^n/2$, multiplying $A$ by some appropriate integer $k < \beta$ will reduce to the case $\beta^n/2 \leq A$, then it suffices to multiply the reciprocal of $ka$ by $k$.)

We first perform an error analysis of Newton's method (§4.2) assuming all computations are done with infinite precision, thus neglecting roundoff errors.

**Lemma 3.4.1** *Let $1/2 \leq a < 1$, $\rho = 1/a$, $x > 0$, and $x' = x + x(1 - ax)$. Then:*

$$0 \leq \rho - x' \leq \frac{x^2}{\theta^3}(\rho - x)^2,$$

*for some $\theta \in [\min(x, \rho), \max(x, \rho)]$.*

**Proof.** Newton's iteration is based on approximating the function by its tangent. Let $f(t) = a - 1/t$, with $\rho$ the root of $f$. The second-order expansion of $f$ at $t = \rho$ with explicit remainder is:

$$f(\rho) = f(x) + (\rho - x)f'(x) + \frac{(\rho - x)^2}{2}f''(\theta),$$

for some $\theta \in [\min(x, \rho), \max(x, \rho)]$. Since $f(\rho) = 0$, this simplifies to

$$\rho = x - \frac{f(x)}{f'(x)} - \frac{(\rho - x)^2}{2}\frac{f''(\theta)}{f'(x)}. \tag{3.5}$$

Substituting $f(t) = a - 1/t$, $f'(t) = 1/t^2$ and $f''(t) = -2/t^3$, it follows that:

$$\rho = x + x(1 - ax) + \frac{x^2}{\theta^3}(\rho - x)^2,$$

which proves the claim.                                                           □

Algorithm **ApproximateReciprocal** computes an approximate reciprocal. The input $A$ is assumed to be normalised, i.e., $\beta^n/2 \leq A < \beta^n$. The output integer $X$ is an approximation to $\beta^{2n}/A$.

---

**Algorithm 3.5** ApproximateReciprocal

---

**Input:** $A = \sum_{i=0}^{n-1} a_i \beta^i$, with $0 \leq a_i < \beta$ and $\beta/2 \leq a_{n-1}$
**Output:** $X = \beta^n + \sum_{i=0}^{n-1} x_i \beta^i$ with $0 \leq x_i < \beta$
1: **if** $n \leq 2$ **then** return $\lceil \beta^{2n}/A \rceil - 1$
2: $\ell \leftarrow \lfloor (n-1)/2 \rfloor$, $h \leftarrow n - \ell$
3: $A_h \leftarrow \sum_{i=0}^{h-1} a_{\ell+i} \beta^i$
4: $X_h \leftarrow$ **ApproximateReciprocal**$(A_h)$
5: $T \leftarrow AX_h$
6: **while** $T \geq \beta^{n+h}$ **do**
7:     $(X_h, T) \leftarrow (X_h - 1, T - A)$
8: $T \leftarrow \beta^{n+h} - T$
9: $T_m \leftarrow \lfloor T\beta^{-\ell} \rfloor$
10: $U \leftarrow T_m X_h$
11: return $X_h \beta^\ell + \lfloor U\beta^{\ell-2h} \rfloor$.

---

**Lemma 3.4.2** *If $\beta$ is a power of two satisfying $\beta \geq 8$, and $\beta^n/2 \leq A < \beta^n$, then the output $X$ of Algorithm* **ApproximateReciprocal** *satisfies:*

$$AX < \beta^{2n} < A(X + 2).$$

**Proof.** For $n \leq 2$ the algorithm returns $X = \lfloor \beta^{2n}/A \rfloor$, unless $A = \beta^n/2$ when it returns $X = 2\beta^n - 1$. In both cases we have $AX < \beta^{2n} \leq A(X+1)$, thus the lemma holds for $n \leq 2$.

Now consider $n \geq 3$. We have $\ell = \lfloor (n-1)/2 \rfloor$ and $h = n-\ell$, thus $n = h+\ell$ and $h > \ell$. The algorithm first computes an approximate reciprocal of the upper $h$ words of $A$, and then updates it to $n$ words using Newton's iteration.

After the recursive call at line 4, we have by induction

$$A_h X_h < \beta^{2h} < A_h(X_h + 2). \tag{3.6}$$

After the product $T \leftarrow AX_h$ and the while-loop at steps 6–7, we still have $T = AX_h$, where $T$ and $X_h$ may have new values, and in addition $T < \beta^{n+h}$. We also have $\beta^{n+h} < T + 2A$; we prove this by distinguishing two cases. Either we entered the while-loop, then since the value of $T$ decreased by $A$ at each loop, the previous value $T + A$ was necessarily $\geq \beta^{n+h}$. If we did not enter the while-loop, the value of $T$ is still $AX_h$. Multiplying Eqn. (3.6) by $\beta^\ell$ gives: $\beta^{n+h} < A_h \beta^\ell (X_h + 2) \leq A(X_h + 2) = T + 2A$. Thus we have:

$$T < \beta^{n+h} < T + 2A.$$

It follows that $T > \beta^{n+h} - 2A > \beta^{n+h} - 2\beta^n$. As a consequence, the value of $\beta^{n+h} - T$ computed at step 8 can not exceed $2\beta^n - 1$. The last lines compute the product $T_m X_h$, where $T_m$ is the upper part of $T$, and put its $\ell$ most significant words in the low part $X_\ell$ of the result $X$.

Now let us perform the error analysis. Compared to Lemma 3.4.1, $x$ stands for $X_h \beta^{-h}$, $a$ stands for $A\beta^{-n}$, and $x'$ stands for $X\beta^{-n}$. The while-loop ensures that we start from an approximation $x < 1/a$, i.e., $AX_h < \beta^{n+h}$. Then Lemma 3.4.1 guarantees that $x \leq x' \leq 1/a$ if $x'$ is computed with infinite precision. Here we have $x \leq x'$, since $X = X_h \beta^h + X_\ell$, where $X_\ell \geq 0$. The only differences compared to infinite precision are:

- the low $\ell$ words from $1 - ax$ — here $T$ at line 8 — are neglected, and only its upper part $(1 - ax)_h$ — here $T_m$ — is considered;

- the low $2h - \ell$ words from $x(1 - ax)_h$ are neglected.

Those two approximations make the computed value of $x' \leq$ the value which would be computed with infinite precision. Thus, for the computed value $x'$, we have:

$$x \leq x' \leq 1/a.$$

From Lemma 3.4.1, the mathematical error is bounded by $x^2 \theta^{-3}(\rho - x)^2 < 4\beta^{-2h}$, since $x^2 \leq \theta^3$ and $|\rho - x| < 2\beta^{-h}$. The truncation from $1 - ax$, which is multiplied by $x < 2$, produces an error $< 2\beta^{-2h}$. Finally, the truncation of $x(1 - ax)_h$ produces an error $< \beta^{-n}$. The final result is thus:

$$x' \leq \rho < x' + 6\beta^{-2h} + \beta^{-n}.$$

Assuming $6\beta^{-2h} \leq \beta^{-n}$, which holds as soon as $\beta \geq 6$ since $2h > n$, this simplifies to:

$$x' \leq \rho < x' + 2\beta^{-n},$$

which gives with $x' = X\beta^{-n}$ and $\rho = \beta^n/A$:

$$X \leq \frac{\beta^{2n}}{A} < X + 2.$$

Since $\beta$ is assumed to be a power of two, equality can hold only when $A$ is itself a power of two, i.e., $A = \beta^n/2$. In this case there is only one value of $X_h$ that is possible for the recursive call, namely $X_h = 2\beta^h - 1$. In this case $T = \beta^{n+h} - \beta^n/2$ before the while-loop, which is not entered. Then $\beta^{n+h} - T = \beta^n/2$, which multiplied by $X_h$ gives (again) $\beta^{n+h} - \beta^n/2$, whose $h$ most significant words are $\beta - 1$. Thus $X_\ell = \beta^\ell - 1$, and $X = 2\beta^n - 1$. □

REMARK. Lemma 3.4.2 might be extended to the case $\beta^{n-1} \leq A < \beta^n$ or to a radix $\beta$ which is not a power of two. However, we prefer to state a restricted result with simple bounds.

COMPLEXITY ANALYSIS. Let $I(n)$ be the cost to invert an $n$-word number using Algorithm **ApproximateReciprocal**. If we neglect the linear costs, we have $I(n) \approx I(n/2) + M(n, n/2) + M(n/2)$, where $M(n, n/2)$ is the cost of an $n \times (n/2)$ product — the product $AX_h$ at step 5 — and $M(n/2)$ the cost of an $(n/2) \times (n/2)$ product — the product $T_m X_h$ at step 10. If the $n \times (n/2)$ product is performed via two $(n/2) \times (n/2)$ products, we have $I(n) \approx I(n/2) + 3M(n/2)$, which yields $I(n) \sim M(n)$ in the quadratic range, $\sim 1.5M(n)$ in the Karatsuba range, $\sim 1.704M(n)$ in the Toom-Cook 3-way range, and $\sim 3M(n)$ in the FFT range. In the FFT range, an $n \times (n/2)$ product might be directly computed by three FFTs of length $3n/2$ words, amounting to $\sim M(3n/4)$; in this case the complexity decreases to $\sim 2.5M(n)$ (see the comments at the end of §2.3.3, page 62).

THE WRAP-AROUND TRICK. We now describe a slight modification of Algorithm **ApproximateReciprocal** which yields a complexity $2M(n)$. In the product $AX_h$ at step 5, Eqn. (3.6) tells us that the result approaches $\beta^{n+h}$, or more precisely:

$$\beta^{n+h} - 2\beta^n < AX_h < \beta^{n+h} + 2\beta^n. \tag{3.7}$$

Assume we use an FFT-based algorithm such as the Schönhage-Strassen algorithm that computes products modulo $\beta^m + 1$, for some integer $m \in$ $(n, n+h)$. Let $AX_h = U\beta^m + V$ with $0 \leq V < \beta^m$. It follows from Eqn. (3.7) that $U = \beta^{n+h-m}$ or $U = \beta^{n+h-m}-1$. Let $T = AX_h \bmod (\beta^m+1)$ be the value computed by the algorithm. We have $T = V - U$ or $T = V - U + (\beta^m + 1)$. It follows that $AX_h = T + U(\beta^m + 1)$ or $AX_h = T + (U-1)(\beta^m + 1)$. Taking into account the two possible values of $U$, we have

$$AX_h = T + (\beta^{n+h-m} - \varepsilon)(\beta^m + 1),$$

where $\varepsilon \in \{0, 1, 2\}$. Since $\beta \geq 6$, $\beta^m > 4\beta^n$, thus only one value of $\varepsilon$ yields a value of $AX_h$ in the interval $(\beta^{n+h} - 2\beta^n, \beta^{n+h} + 2\beta^n)$.

Thus, we can replace step 5 in Algorithm **ApproximateReciprocal** by the following code:

> Compute $T = AX_h \bmod (\beta^m + 1)$ using FFTs with length $m > n$
> $T \leftarrow T + \beta^{n+h} + \beta^{n+h-m}$ $\qquad\qquad$ $\triangleright$ the case $\varepsilon = 0$
> **while** $T \geq \beta^{n+h} + 2\beta^n$ **do**
> $\quad T \leftarrow T - (\beta^m + 1)$

Assuming that one can take $m$ close to $n$, the cost of the product $AX_h$ is only about that of three FFTs of length $n$, that is $\sim M(n/2)$.

### 3.4.2  Division

In this section we consider the case where the divisor changes between successive operations, so no precomputation involving the divisor can be performed. We first show that the number of consecutive zeros in the result is bounded by the divisor length, then we consider the division algorithm and its complexity. Lemma 3.4.3 analyses the case where the division operands are truncated, because they have a larger precision than desired in the result. Finally we discuss "short division" and the error analysis of Barrett's algorithm.

A floating-point division reduces to an integer division as follows. Assume dividend $a = \ell \cdot \beta^e$ and divisor $d = m \cdot \beta^f$, where $\ell, m$ are integers. Then $a/d = (\ell/m)\beta^{e-f}$. If $k$ bits of the quotient are needed, we first determine a scaling factor $g$ such that $\beta^{k-1} \leq |\ell\beta^g/m| < \beta^k$, and we divide $\ell\beta^g$ — truncated if needed — by $m$. The following theorem gives a bound on the number of consecutive zeros after the integer part of the quotient of $\lfloor \ell\beta^g \rfloor$ by $m$.

**Theorem 3.4.1** *Assume we divide an $m$-digit positive integer by an $n$-digit positive integer in radix $\beta$, with $m \geq n$. Then the quotient is either exact, or its radix $\beta$ expansion admits at most $n-1$ consecutive zeros or ones after the digit of weight $\beta^0$.*

**Proof.** We first consider consecutive zeros. If the expansion of the quotient $q$ admits $n$ or more consecutive zeros after the binary point, we can write $q = q_1 + \beta^{-n} q_0$, where $q_1$ is an integer and $0 \leq q_0 < 1$. If $q_0 = 0$, then the quotient is exact. Otherwise, if $a$ is the dividend and $d$ is the divisor, one should have $a = q_1 d + \beta^{-n} q_0 d$. However, $a$ and $q_1 d$ are integers, and $0 < \beta^{-n} q_0 d < 1$, so $\beta^{-n} q_0 d$ can not be an integer, so we have a contradiction.

For consecutive ones, the proof is similar: write $q = q_1 - \beta^{-n} q_0$, with $0 \leq q_0 \leq 1$. Since $d < \beta_n$, we still have $0 \leq \beta^{-n} q_0 d < 1$.  □

Algorithm **DivideNewton** performs the division of two $n$-digit floating-point numbers. The key idea is to approximate the inverse of the divisor to half precision only, at the expense of additional steps. At step 4, **MiddleProduct**$(q_0, d)$ denotes the middle product of $q_0$ and $d$, i.e., the $n/2$ middle digits of that product. At step 2, $r$ is an approximation to $1/d_1$, and thus to $1/d$, with precision $n/2$ digits. Therefore at step 3, $q_0$ approximates $c/d$ to about $n/2$ digits, and the upper $n/2$ digits of $q_0 d$ at step 4 agree with those of $c$. The value $e$ computed at step 4 thus equals $q_0 d - c$ to precision $n/2$. It follows that $re \approx e/d$ agrees with $q_0 - c/d$ to precision $n/2$; hence the correction term (which is really a Newton correction) added in the last step.

---

**Algorithm 3.6** DivideNewton

---

**Input:** $n$-digit floating-point numbers $c$ and $d$, with $n$ even, $d$ normalised
**Output:** an approximation of $c/d$
1: write $d = d_1 \beta^{n/2} + d_0$ with $0 \leq d_1, d_0 < \beta^{n/2}$
2: $r \leftarrow$ **ApproximateReciprocal**$(d_1, n/2)$
3: $q_0 \leftarrow cr$ truncated to $n/2$ digits
4: $e \leftarrow$ **MiddleProduct**$(q_0, d)$
5: $q \leftarrow q_0 - re$.

---

In the FFT range, the cost of Algorithm **DivideNewton** is $\sim 2.5M(n)$: step 2 costs $\sim 2M(n/2) \sim M(n)$ with the wrap-around trick, and steps 3–5 each cost $\sim M(n/2)$ — using a fast middle product algorithm for step 4. By way of comparison, if we computed a full precision inverse as in Barrett's algorithm (see below), the cost would be $\sim 3.5M(n)$. (See §3.8 for improved asymptotic bounds on division.)

In the Karatsuba range, Algorithm **DivideNewton** costs $\sim 1.5M(n)$, and is useful provided the middle product of step 4 is performed with cost $\sim M(n/2)$. In the quadratic range, Algorithm **DivideNewton** costs $\sim 2M(n)$, and a classical division should be preferred.

When the requested precision for the output is smaller than that of the inputs of a division, one has to truncate the inputs, in order to avoid an unnecessarily expensive computation. Assume for example that we want to divide two numbers of $10,000$ bits, with a 10-bit quotient. To apply the following lemma, just replace $\mu$ by an appropriate value such that $A_1$ and $B_1$ have about $2n$ and $n$ digits respectively, where $n$ is the desired number of digits in the quotient; for example we might choose $\mu = \beta^k$ to truncate to $k$ words.

**Lemma 3.4.3** *Let $A, B, \mu \in \mathbb{N}^*$, $2 \le \mu \le B$. Let $Q = \lfloor A/B \rfloor$, $A_1 = \lfloor A/\mu \rfloor$, $B_1 = \lfloor B/\mu \rfloor$, $Q_1 = \lfloor A_1/B_1 \rfloor$. If $A/B \le 2B_1$, then*

$$Q \le Q_1 \le Q + 2.$$

The condition $A/B \le 2B_1$ is quite natural: it says that the truncated divisor $B_1$ should have essentially at least as many digits as the desired quotient.

**Proof.** Let $A_1 = Q_1 B_1 + R_1$. We have $A = A_1 \mu + A_0$, $B = B_1 \mu + B_0$, thus

$$\frac{A}{B} = \frac{A_1 \mu + A_0}{B_1 \mu + B_0} \le \frac{A_1 \mu + A_0}{B_1 \mu} = Q_1 + \frac{R_1 \mu + A_0}{B_1 \mu}.$$

Since $R_1 < B_1$ and $A_0 < \mu$, $R_1 \mu + A_0 < B_1 \mu$, thus $A/B < Q_1 + 1$. Taking the floor of each side proves, since $Q_1$ is an integer, that $Q \le Q_1$.

Now consider the second inequality. For given truncated parts $A_1$ and $B_1$, and thus given $Q_1$, the worst case is when $A$ is minimal, say $A = A_1 \mu$, and $B$ is maximal, say $B = B_1 \mu + (\mu - 1)$. In this case we have:

$$\left| \frac{A_1}{B_1} - \frac{A}{B} \right| = \left| \frac{A_1}{B_1} - \frac{A_1 \mu}{B_1 \mu + (\mu - 1)} \right| = \left| \frac{A_1(\mu - 1)}{B_1(B_1 \mu + \mu - 1)} \right|.$$

The numerator equals $A - A_1 \le A$, and the denominator equals $B_1 B$, thus the difference $A_1/B_1 - A/B$ is bounded by $A/(B_1 B) \le 2$, and so is the difference between $Q$ and $Q_1$. □

Algorithm **ShortDivision** is useful in the Karatsuba and Toom-Cook ranges. The key idea is that, when dividing a $2n$-digit number by an $n$-digit number, some work that is necessary for a full $2n$-digit division can be avoided (see Figure 3.5).

---

**Algorithm 3.7** ShortDivision

---

**Input:** $0 \leq A < \beta^{2n}$, $\beta^n/2 \leq B < \beta^n$
**Output:** an approximation of $A/B$
**Require:** a threshold $n_0$

1: **if** $n \leq n_0$ **then** return $\lfloor A/B \rfloor$
2: choose $k \geq n/2$, $\ell \leftarrow n - k$
3: $(A_1, A_0) \leftarrow (A \text{ div } \beta^{2\ell}, A \text{ mod } \beta^{2\ell})$
4: $(B_1, B_0) \leftarrow (B \text{ div } \beta^{\ell}, B \text{ mod } \beta^{\ell})$
5: $(Q_1, R_1) \leftarrow \mathbf{DivRem}(A_1, B_1)$
6: $A' \leftarrow R_1\beta^{2\ell} + A_0 - Q_1 B_0 \beta^{\ell}$
7: $Q_0 \leftarrow \mathbf{ShortDivision}(A' \text{ div } \beta^k, B \text{ div } \beta^k)$
8: return $Q_1\beta^{\ell} + Q_0$.

---

**Theorem 3.4.2** *The approximate quotient $Q'$ returned by* **ShortDivision** *differs at most by $2 \lg n$ from the exact quotient $Q = \lfloor A/B \rfloor$, more precisely:*

$$Q \leq Q' \leq Q + 2\lg n.$$

**Proof.** If $n \leq n_0$, $Q = Q'$ so the statement holds. Assume $n > n_0$. We have $A = A_1\beta^{2\ell} + A_0$ and $B = B_1\beta^{\ell} + B_0$, thus since $A_1 = Q_1 B_1 + R_1$, $A = (Q_1 B_1 + R_1)\beta^{2\ell} + A_0 = Q_1 B\beta^{\ell} + A'$, with $A' < \beta^{n+\ell}$. Let $A' = A'_1\beta^k + A'_0$, and $B = B'_1\beta^k + B'_0$, with $0 \leq A'_0, B'_0 < \beta^k$, and $A'_1 < \beta^{2\ell}$. From Lemma 3.4.3, the exact quotient of $A' \text{ div } \beta^k$ by $B \text{ div } \beta^k$ is greater or equal to that of $A'$ by $B$, thus by induction $Q_0 \geq A'/B$. Since $A/B = Q_1\beta^{\ell} + A'/B$, this proves that $Q' \geq Q$.

Now by induction $Q_0 \leq A'_1/B'_1 + 2\lg\ell$, and $A'_1/B'_1 \leq A'/B + 2$ (from Lemma 3.4.3 again, whose hypothesis $A'/B \leq 2B'_1$ is satisfied, since $A' < B_1\beta^{2\ell}$, thus $A'/B \leq \beta^{\ell} \leq 2B'_1$), so $Q_0 \leq A'/B + 2\lg n$, and $Q' \leq A/B + 2\lg n$. $\square$

As shown at the right of Figure 3.5, we can use a short product to compute $Q_1 B_0$ at step 6. Indeed, we need only the upper $\ell$ words of $A'$, thus only the upper $\ell$ words of $Q_1 B_0$. The complexity of Algorithm **ShortDivision** satisfies $D^*(n) = D(k) + M^*(n-k) + D^*(n-k)$ with $k \geq n/2$, where $D(n)$ denotes the cost of a division with remainder, and $M^*(n)$ the cost of a short product. In the Karatsuba range we have $D(n) \sim 2M(n)$, $M^*(n) \sim 0.808M(n)$, and the best possible value of $k$ is $k \approx 0.542n$, with corresponding cost $D^*(n) \sim 1.397M(n)$. In the Toom-Cook 3-way range, $k \approx 0.548n$ is optimal, and gives $D^*(n) \sim 1.988M(n)$.

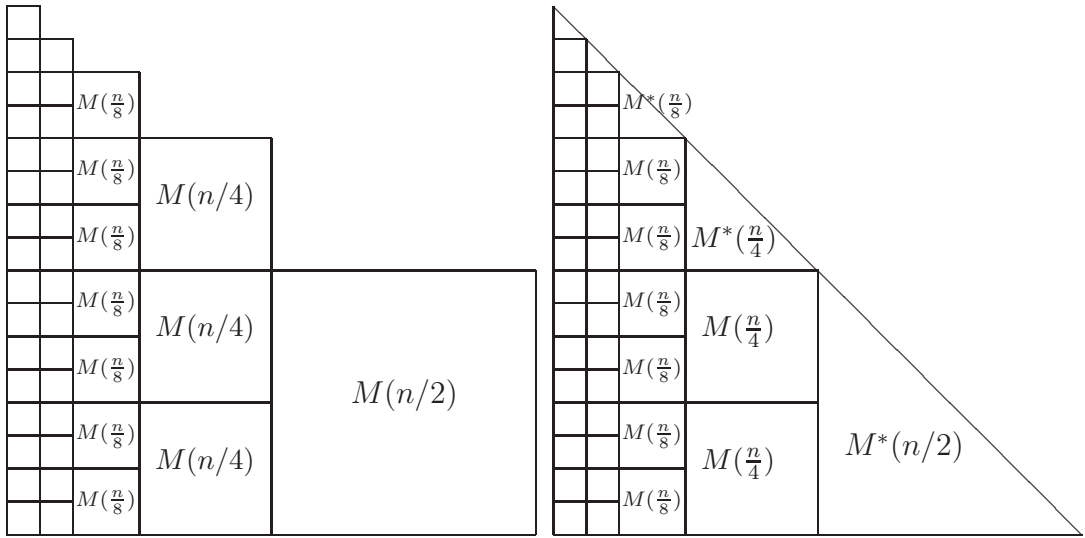

Figure 3.5: Divide and conquer short division: a graphical view. Left: with plain multiplication; right: with short multiplication. See also Figure 1.3.

**Barrett's floating-point division algorithm**

Here we consider floating-point division using Barrett's algorithm and provide a rigorous error bound (see §2.4.1 for an exact integer version). The algorithm is useful when the same divisor is used several times; otherwise Algorithm **DivideNewton** is faster (see Exercise 3.13). Assume we want to divide $a$ by $b$ of $n$ bits, each with a quotient of $n$ bits. Barrett's algorithm is as follows:

1. Compute the reciprocal $r$ of $b$ to $n$ bits [rounding to nearest]

2. $q \leftarrow \circ_n(a \times r)$ [rounding to nearest]

The cost of the algorithm in the FFT range is $\sim 3M(n)$: $\sim 2M(n)$ to compute the reciprocal with the wrap-around trick, and $M(n)$ for the product $a \times r$.

**Lemma 3.4.4** *At step 2 of Barrett's algorithm, we have $|a - bq| \leq 3|b|/2$.*

**Proof.** By scaling $a$ and $b$, we can assume that $b$ and $q$ are integers, that $2^{n-1} \leq b, q < 2^n$, thus $a < 2^{2n}$. We have $r = 1/b + \varepsilon$ with $|\varepsilon| \leq \text{ulp}(2^{-n}/2) = 2^{-2n}$. Also $q = ar + \varepsilon'$ with $|\varepsilon'| \leq \text{ulp}(q)/2 = 1/2$ since $q$ has $n$ bits. Thus $q = a(1/b + \varepsilon) + \varepsilon' = a/b + a\varepsilon + \varepsilon'$, and $|bq - a| = |b||a\varepsilon + \varepsilon'| \leq 3|b|/2$. $\qquad \square$

As a consequence, $q$ differs by at most one unit in last place from the $n$-bit quotient of $a$ and $b$, rounded to nearest.

Lemma 3.4.4 can be applied as follows: to perform several divisions with a precision of $n$ bits with the same divisor, precompute a reciprocal with $n + g$ bits, and use the above algorithm with a working precision of $n + g$ bits. If the last $g$ bits of $q$ are neither $000 \ldots 00x$ nor $111 \ldots 11x$ (where $x$ stands for 0 or 1), then rounding $q$ down to $n$ bits will yield $\circ_n(a/b)$ for a directed rounding mode.

### Which Algorithm to Use?

In this section, we described three algorithms to compute $x/y$: **Divide-Newton** uses Newton's method for $1/y$ and incorporates the dividend $x$ at the last iteration, **ShortDivision** is a recursive algorithm using division with remainder and short products, and Barrett's algorithm assumes we have precomputed an approximation to $1/y$. When the same divisor $y$ is used several times, clearly Barrett's algorithm is better, since each division costs only a short product. Otherwise **ShortDivision** is theoretically faster than **DivideNewton** in the schoolbook and Karatsuba ranges, and taking $k = n/2$ as parameter in **ShortDivision** is close to optimal. In the FFT range, **DivideNewton** should be preferred.

## 3.5  Square Root

Algorithm **FPSqrt** computes a floating-point square root, using as subroutine Algorithm **SqrtRem** (§1.5.1 to determine an integer square root (with remainder). It assumes an integer significand $m$, and a directed rounding mode (see Exercise 3.14 for rounding to nearest).

---

**Algorithm 3.8** FPSqrt

---

**Input:** $x = m \cdot 2^e$, a target precision $n$, a directed rounding mode $\circ$

**Output:** $y = \circ_n(\sqrt{x})$

    **if** $e$ is odd **then** $(m', f) \leftarrow (2m, e-1)$ **else** $(m', f) \leftarrow (m, e)$

    define $m' := m_1 2^{2k} + m_0$, $m_1$ integer of $2n$ or $2n-1$ bits, $0 \le m_0 < 2^{2k}$

    $(s, r) \leftarrow \textbf{SqrtRem}(m_1)$

    **if** ($\circ$ is round towards zero or down) or ($r = m_0 = 0$)

        **then** return $s \cdot 2^{k+f/2}$ **else** return $(s+1) \cdot 2^{k+f/2}$.

---

**Theorem 3.5.1** *Algorithm* **FPSqrt** *returns the correctly rounded square root of* $x$.

**Proof.** Since $m_1$ has $2n$ or $2n-1$ bits, $s$ has exactly $n$ bits, and we have $x \ge s^2 2^{2k+f}$, thus $\sqrt{x} \ge s 2^{k+f/2}$. On the other hand, **SqrtRem** ensures that $r \le 2s$, thus $x2^{-f} = (s^2 + r)2^{2k} + m_0 < (s^2 + r + 1)2^{2k} \le (s+1)^2 2^{2k}$. Since $y := s \cdot 2^{k+f/2}$ and $y^+ = (s+1) \cdot 2^{k+f/2}$ are two consecutive $n$-bit floating-point numbers, this concludes the proof. □

Note: in the case $s = 2^n - 1$, $s + 1 = 2^n$ is still representable in $n$ bits.

A different method is to use an initial approximation to the reciprocal square root $x^{-1/2}$ (§3.5.1), see Exercise 3.15. Faster algorithms are mentioned in §3.8.

### 3.5.1 Reciprocal Square Root

In this section we describe an algorithm to compute the reciprocal square root $a^{-1/2}$ of a floating-point number $a$, with a rigorous error bound.

**Lemma 3.5.1** *Let* $a, x > 0$, $\rho = a^{-1/2}$, *and* $x' = x + (x/2)(1 - ax^2)$. *Then*

$$0 \le \rho - x' \le \frac{3x^3}{2\theta^4}(\rho - x)^2,$$

*for some* $\theta \in [\min(x, \rho), \max(x, \rho)]$.

**Proof.** The proof is very similar to that of Lemma 3.4.1. Here we use $f(t) = a - 1/t^2$, with $\rho$ the root of $f$. Eqn. (3.5) translates to:

$$\rho = x + \frac{x}{2}(1 - ax^2) + \frac{3x^3}{2\theta^4}(\rho - x)^2,$$

which proves the Lemma. □

---

**Algorithm 3.9** ApproximateRecSquareRoot

---

**Input:** integer $A$ with $\beta^n \leq A < 4\beta^n$, $\beta \geq 38$

**Output:** integer $X$, $\beta^n/2 \leq X < \beta^n$ satisfying Lemma 3.5.2

 1: **if** $n \leq 2$ **then** return $\min(\beta^n - 1, \lfloor \beta^n/\sqrt{A\beta^{-n}} \rfloor)$
 2: $\ell \leftarrow \lfloor (n-1)/2 \rfloor$, $h \leftarrow n - \ell$
 3: $A_h \leftarrow \lfloor A\beta^{-\ell} \rfloor$
 4: $X_h \leftarrow$ **ApproximateRecSquareRoot**$(A_h)$
 5: $T \leftarrow AX_h^2$
 6: $T_h \leftarrow \lfloor T\beta^{-n} \rfloor$
 7: $T_\ell \leftarrow \beta^{2h} - T_h$
 8: $U \leftarrow T_\ell X_h$
 9: return $\min(\beta^n - 1, X_h\beta^\ell + \lfloor U\beta^{\ell-2h}/2 \rfloor)$.

---

**Lemma 3.5.2** *Provided that $\beta \geq 38$, if $X$ is the value returned by Algorithm* **ApproximateRecSquareRoot***, $a = A\beta^{-n}$, $x = X\beta^{-n}$, then $1/2 \leq x < 1$ and*

$$|x - a^{-1/2}| \leq 2\beta^{-n}.$$

**Proof.** We have $1 \leq a < 4$. Since $X$ is bounded by $\beta^n - 1$ at lines 1 and 9, we have $x, x_h < 1$, with $x_h = X_h\beta^{-h}$. We prove the statement by induction on $n$. It is true for $n \leq 2$. Now assume the value $X_h$ at step 4 satisfies:

$$|x_h - a_h^{-1/2}| \leq \beta^{-h},$$

where $a_h = A_h\beta^{-h}$. We have three sources of error, that we will bound separately:

1. the rounding errors in steps 6 and 9;

2. the mathematical error given by Lemma 3.5.1, which would occur even if all computations were exact;

3. the error coming from the fact we use $A_h$ instead of $A$ in the recursive call at step 4.

At step 5 we have exactly:

$$t := T\beta^{-n-2h} = ax_h^2,$$

which gives $|t_h - ax_h^2| < \beta^{-2h}$ with $t_h := T_h \beta^{-2h}$, and in turn $|t_\ell - (1 - ax_h^2)| < \beta^{-2h}$ with $t_\ell := T_\ell \beta^{-2h}$. At step 8, it follows $|u - x_h(1 - ax_h^2)| < \beta^{-2h}$, where $u = U\beta^{-3h}$. Thus, finally $|x - [x_h + x_h(1 - ax_h^2)/2]| < (\beta^{-2h} + \beta^{-n})/2$, after taking into account the rounding error in the last step.

Now we apply Lemma 3.5.1 to $x \to x_h$, $x' \to x$, to bound the mathematical error, assuming no rounding error occurs:

$$0 \le a^{-1/2} - x \le \frac{3x_h^3}{2\theta^4}(a^{-1/2} - x_h)^2,$$

which gives[1] $|a^{-1/2} - x| \le 3.04(a^{-1/2} - x_h)^2$. Now $|a^{-1/2} - a_h^{-1/2}| \le |a - a_h|\nu^{-3/2}/2$ for $\nu \in [\min(a_h, a), \max(a_h, a)]$, thus $|a^{-1/2} - a_h^{-1/2}| \le \beta^{-h}/2$. Together with the induction hypothesis $|x_h - a_h^{-1/2}| \le 2\beta^{-h}$, it follows that $|a^{-1/2} - x_h| \le 2.5\beta^{-h}$. Thus $|a^{-1/2} - x| \le 19\beta^{-2h}$.

The total error is thus bounded by:

$$|a^{-1/2} - x| \le \frac{3}{2}\beta^{-n} + 19\beta^{-2h}.$$

Since $2h \ge n + 1$, we see that $19\beta^{-2h} \le \beta^{-n}/2$ for $\beta \ge 38$, and the lemma follows.  □

NOTE: if $A_h X_h^2 \le \beta^{3h}$ at step 4 of Algorithm **ApproximateRecSquare-Root**, we could have $AX_h^2 > \beta^{n+2h}$ at step 5, which might cause $T_\ell$ to be negative.

Let $R(n)$ be the cost of **ApproximateRecSquareRoot** for an $n$-digit input. We have $h, \ell \approx n/2$, thus the recursive call costs $R(n/2)$, step 5 costs $M(n/2)$ to compute $X_h^2$, and $M(n)$ for the product $AX_h^2$ (or $M(3n/4)$ in the FFT range using the wrap-around trick described in §3.4.1, since we know the upper $n/2$ digits of the product give 1), and again $M(n/2)$ for step 8. We get $R(n) = R(n/2) + 2M(n)$ (or $R(n/2) + 7M(n)/4$ in the FFT range), which yields $R(n) \sim 4M(n)$ (or $R(n) \sim 3.5M(n)$ in the FFT range).

This algorithm is not optimal in the FFT range, especially when using an FFT algorithm with cheap point-wise products (such as the complex FFT,

---

[1] Since $\theta \in [x_h, a^{-1/2}]$ and $|x_h - a^{-1/2}| \le 2.5\beta^{-h}$, we have $\theta \ge x_h - 2.5\beta^{-h}$, thus $x_h/\theta \le 1 + 2.5\beta^{-h}/\theta \le 1 + 5\beta^{-h}$ (remember $\theta \in [x_h, a^{-1/2}]$), and it follows that $\theta \ge 1/2$. For $\beta \ge 38$, since $h \ge 2$, we have $1 + 5\beta^{-h} \le 1.0035$, thus $1.5x_h^3/\theta^4 \le (1.5/\theta)(1.0035)^3 \le 3.04$.

see §3.3.1). Indeed, Algorithm **ApproximateRecSquareRoot** uses the following form of Newton's iteration:

$$x' = x + \frac{x}{2}(1 - ax^2).$$

It might be better to write:

$$x' = x + \frac{1}{2}(x - ax^3).$$

Here, the product $x^3$ might be computed with a *single* FFT transform of length $3n/2$, replacing the point-wise products $\widehat{x}_i^2$ by $\widehat{x}_i^3$, with a total cost $\sim 0.75M(n)$. Moreover, the same idea can be used for the full product $ax^3$ of $5n/2$ bits, whose upper $n/2$ bits match those of $x$. Thus, using the wrap-around trick, a transform of length $2n$ is enough, with a cost of $\sim M(n)$ for the last iteration, and a total cost of $\sim 2M(n)$ for the reciprocal square root. With this improvement, the algorithm of Exercise 3.15 costs only $\sim 2.25M(n)$.

## 3.6   Conversion

Since most software tools work in radix 2 or $2^k$, and humans usually enter or read floating-point numbers in radix 10 or $10^k$, conversions are needed from one radix to the other one. Most applications perform very few conversions, in comparison to other arithmetic operations, thus the efficiency of the conversions is rarely critical.[2] The main issue here is more correctness than efficiency. Correctness of floating-point conversions is not an easy task, as can be seen from the history of bugs in Microsoft Excel.[3]

The algorithms described in this section use as subroutines the integer-conversion algorithms from Chapter 1. As a consequence, their efficiency depends on the efficiency of the integer-conversion algorithms.

### 3.6.1   Floating-Point Output

In this section we follow the convention of using lower-case letters for parameters related to the internal radix $b$, and upper-case for parameters related to the external radix $B$. Consider the problem of printing a floating-point number, represented internally in radix $b$ (say $b = 2$) in an external radix $B$ (say $B = 10$). We distinguish here two kinds of floating-point output:

- fixed-format output, where the output precision is given by the user, and we want the output value to be correctly rounded according to the given rounding mode. This is the usual method when values are to be used by humans, for example to fill a table of results. The input and output precisions may be very different: for example one may want to print 1000 digits of $2/3$, which uses only one digit internally in radix 3. Conversely, one may want to print only a few digits of a number accurate to 1000 bits.

- free-format output, where we want the output value, when read with correct rounding (usually to nearest), to give *exactly* the initial number. Here the minimal number of printed digits may depend on the input number. This kind of output is useful when storing data in a file, while guaranteeing that reading the data back will produce exactly the same internal numbers, or for exchanging data between different programs.

---

[2] An important exception is the computation of billions of digits of constants like $\pi$, $\log 2$, where a quadratic conversion routine would be far too slow.

[3] In Excel 2007, the product $850 \times 77.1$ prints as $100,000$ instead of $65,535$; this is really an output bug, since if one multiplies "$100,000$" by 2, one gets $131,070$. An input bug occurred in Excel 3.0 to 7.0, where the input $1.40737488355328$ gave $0.64$.

In other words, if $x$ is the number that we want to print, and $X$ is the printed value, the fixed-format output requires $|x - X| < \text{ulp}(X)$, and the free-format output requires $|x - X| < \text{ulp}(x)$ for directed rounding. Replace $< \text{ulp}(\cdot)$ by $\leq \text{ulp}(\cdot)/2$ for rounding to nearest.

---

**Algorithm 3.10** PrintFixed

---

**Input:** $x = f \cdot b^{e-p}$ with $f, e, p$ integers, $b^{p-1} \leq |f| < b^p$, external radix $B$ and precision $P$, rounding mode $\circ$

**Output:** $X = F \cdot B^{E-P}$ with $F, E$ integers, $B^{P-1} \leq |F| < B^P$, such that $X = \circ(x)$ in radix $B$ and precision $P$

1: $\lambda \leftarrow \circ(\log b / \log B)$
2: $E \leftarrow 1 + \lfloor (e-1)\lambda \rfloor$
3: $q \leftarrow \lceil P/\lambda \rceil$
4: $y \leftarrow \circ(xB^{P-E})$ with precision $q$
5: **if** one can not round $y$ to an integer **then** increase $q$ and go to step 4
6: $F \leftarrow \text{Integer}(y, \circ)$.　　　　　　　　　　　　　　　　$\triangleright$ see §1.7
7: **if** $|F| \geq B^P$ **then** $E \leftarrow E + 1$ and go to step 4.
8: return $F, E$.

---

Some comments on Algorithm **PrintFixed**:

- it assumes that we have precomputed values of $\lambda_B = \circ(\log b / \log B)$ for any possible external radix $B$ (the internal radix $b$ is assumed to be fixed for a given implementation). Assuming the input exponent $e$ is bounded, it is possible — see Exercise 3.17 — to choose these values precisely enough that

$$E = 1 + \left\lfloor (e-1)\frac{\log b}{\log B} \right\rfloor, \qquad (3.8)$$

  thus the value of $\lambda$ at step 1 is simply read from a table;

- the difficult part is step 4, where one has to perform the exponentiation $B^{P-E}$ — remember all computations are done in the internal radix $b$ — and multiply the result by $x$. Since we expect an integer of $q$ digits in step 6, there is no need to use a precision of more than $q$ digits in these computations, but a rigorous bound on the rounding errors is required, so as to be able to correctly round $y$;

- in step 5, "one can round $y$ to an integer" means that the interval containing all possible values of $xB^{P-E}$ — including the rounding errors while approaching $xB^{P-E}$, and the error while rounding to precision $q$ — contains no rounding boundary (if $\circ$ is a directed rounding, it should contain no integer; if $\circ$ is rounding to nearest, it should contain no half-integer).

**Theorem 3.6.1** *Algorithm* **PrintFixed** *is correct.*

**Proof.** First assume that the algorithm finishes. Eqn. (3.8) implies $B^{E-1} \leq b^{e-1}$, thus $|x|B^{P-E} \geq B^{P-1}$, which implies that $|F| \geq B^{P-1}$ at step 6. Thus $B^{P-1} \leq |F| < B^P$ at the end of the algorithm. Now, printing $x$ gives $F \cdot B^a$ iff printing $xB^k$ gives $F \cdot B^{a+k}$ for any integer $k$. Thus it suffices to check that printing $xB^{P-E}$ gives $F$, which is clear by construction.

The algorithm terminates because at step 4, $xB^{P-E}$, if not an integer, can not be arbitrarily close to an integer. If $P - E \geq 0$, let $k$ be the number of digits of $B^{P-E}$ in radix $b$, then $xB^{P-E}$ can be represented exactly with $p + k$ digits. If $P - E < 0$, let $g = B^{E-P}$, of $k$ digits in radix $b$. Assume $f/g = n + \varepsilon$ with $n$ integer; then $f - gn = g\varepsilon$. If $\varepsilon$ is not zero, $g\varepsilon$ is a non-zero integer, thus $|\varepsilon| \geq 1/g \geq 2^{-k}$.

The case $|F| \geq B^P$ at step 7 can occur for two reasons: either $|x|B^{P-E} \geq B^P$, thus its rounding also satisfies this inequality; or $|x|B^{P-E} < B^P$, but its rounding equals $B^P$ (this can only occur for rounding away from zero or to nearest). In the former case we have $|x|B^{P-E} \geq B^{P-1}$ at the next pass in step 4, while in the latter case the rounded value $F$ equals $B^{P-1}$ and the algorithm terminates. □

Now consider free-format output. For a directed rounding mode we want $|x - X| < \mathrm{ulp}(x)$ knowing $|x - X| < \mathrm{ulp}(X)$. Similarly for rounding to nearest, if we replace ulp by ulp $/2$.

It is easy to see that a sufficient condition is that $\mathrm{ulp}(X) \leq \mathrm{ulp}(x)$, or equivalently $B^{E-P} \leq b^{e-p}$ in Algorithm **PrintFixed** (with $P$ not fixed at input, which explain the "free-format" name). To summarise, we have

$$b^{e-1} \leq |x| < b^e, \quad B^{E-1} \leq |X| < B^E.$$

Since $|x| < b^e$, and $X$ is the rounding of $x$, it suffices to have $B^{E-1} \leq b^e$. It follows that $B^{E-P} \leq b^e B^{1-P}$, and the above sufficient condition becomes:

$$P \geq 1 + p \, \frac{\log b}{\log B}.$$

For example, with $b = 2$ and $B = 10$, $p = 53$ gives $P \geq 17$, and $p = 24$ gives $P \geq 9$. As a consequence, if a double-precision IEEE 754 binary floating-point number is printed with at least 17 significant decimal digits, it can be read back without any discrepancy, assuming input and output are performed with correct rounding to nearest (or directed rounding, with appropriately chosen directions).

## 3.6.2 Floating-Point Input

The problem of floating-point input is the following. Given a floating-point number $X$ with a significand of $P$ digits in some radix $B$ (say $B = 10$), a precision $p$ and a given rounding mode, we want to correctly round $X$ to a floating-point number $x$ with $p$ digits in the internal radix $b$ (say $b = 2$).

At first glance, this problem looks very similar to the floating-point output problem, and one might think it suffices to apply Algorithm **PrintFixed**, simply exchanging $(b, p, e, f)$ and $(B, P, E, F)$. Unfortunately, this is not the case. The difficulty is that, in Algorithm **PrintFixed**, all arithmetic operations are performed in the *internal* radix $b$, and we do not have such operations in radix $B$ (see however Exercise 1.37).

# 3.7   Exercises

**Exercise 3.1** In §3.1.5 we described a trick to get the next floating-point number in the direction away from zero. Determine for which IEEE 754 double-precision numbers the trick works.

**Exercise 3.2 (Kidder, Boldo)** Assume a binary representation. The "rounding to odd" mode [42, 149, 221] is defined as follows: in case the exact value is not representable, it rounds to the unique adjacent number with an odd significand. ("Von Neumann rounding" [42] omits the test for the exact value being representable or not, and rounds to odd in all nonzero cases.) Note that overflow never occurs during rounding to odd. Prove that if $y = \text{round}(x, p + k, \text{odd})$ and $z = \text{round}(y, p, \text{nearest\_even})$, and $k > 1$, then $z = \text{round}(x, p, \text{nearest\_even})$, i.e., the double-rounding problem does not occur.

**Exercise 3.3** Show that, if $\sqrt{a}$ is computed using Newton's iteration for $a^{-1/2}$:

$$x' = x + \frac{3}{2}(1 - ax^2)$$

(see §3.5.1), and the identity $\sqrt{a} = a \times a^{-1/2}$, with rounding mode "round towards zero", then it might never be possible to determine the correctly rounded value of $\sqrt{a}$, regardless of the number of additional guard digits used in the computation.

**Exercise 3.4** How does truncating the operands of a multiplication to $n + g$ digits (as suggested in §3.3) affect the accuracy of the result? Considering the cases $g = 1$ and $g > 1$ separately, what could happen if the same strategy were used for subtraction?

**Exercise 3.5** Is the bound of Theorem 3.3.1 optimal?

**Exercise 3.6** Adapt Mulders' short product algorithm [174] to floating-point numbers. In case the first rounding fails, can you compute additional digits without starting again from scratch?

**Exercise 3.7** Show that, if a balanced ternary system is used (radix 3 with digits $\{0, \pm 1\}$), then "round to nearest" is equivalent to truncation.

**Exercise 3.8 (Percival)** Suppose we compute the product of two complex floating-point numbers $z_0 = a_0 + ib_0$ and $z_1 = a_1 + ib_1$ in the following way: $x_a = \circ(a_0 a_1)$, $x_b = \circ(b_0 b_1)$, $y_a = \circ(a_0 b_1)$, $y_b = \circ(a_1 b_0)$, $z = \circ(x_a - x_b) + i \circ(y_a + y_b)$. All computations are done in precision $n$, with rounding to nearest. Compute an error bound of the form $|z - z_0 z_1| \le c2^{-n}|z_0 z_1|$. What is the best possible constant $c$?

**Exercise 3.9** Show that, if $\mu = O(\varepsilon)$ and $n\varepsilon < 1$, the bound in Theorem 3.3.2 simplifies to

$$||z' - z||_\infty = O(|x| \cdot |y| \cdot n\varepsilon).$$

If the rounding errors cancel we expect the error in each component of $z'$ to be $O(|x| \cdot |y| \cdot n^{1/2}\varepsilon)$. The error $||z' - z||_\infty$ could be larger since it is a maximum of $N = 2^n$ component errors. Using your favourite implementation of the FFT, compare the worst-case error bound given by Theorem 3.3.2 with the error $||z' - z||_\infty$ that occurs in practice.

**Exercise 3.10 (Enge)** Design an algorithm that correctly rounds the product of two complex floating-point numbers with 3 multiplications only. [Hint: assume all operands and the result have $n$-bit significand.]

**Exercise 3.11** Write a computer program to check the entries of Table 3.3 are correct and optimal, given Theorem 3.3.2.

**Exercise 3.12 (Bodrato)** Assuming one uses an FFT modulo $\beta^m - 1$ in the wrap-around trick, how should one modify step 5 of **ApproximateReciprocal**?

**Exercise 3.13** To perform $k$ divisions with the same divisor, which of Algorithm **DivideNewton** and Barrett's algorithm is faster?

**Exercise 3.14** Adapt Algorithm **FPSqrt** to the rounding to nearest mode.

**Exercise 3.15** Devise an algorithm similar to Algorithm **FPSqrt** but using Algorithm **ApproximateRecSquareRoot** to compute an $n/2$-bit approximation to $x^{-1/2}$, and doing one Newton-like correction to return an $n$-bit approximation to $x^{1/2}$. In the FFT range, your algorithm should take time $\sim 3M(n)$ (or better).

**Exercise 3.16** Prove that for any $n$-bit floating-point numbers $(x, y) \neq (0, 0)$, and if all computations are correctly rounded, with the same rounding mode, the result of $x/\sqrt{x^2 + y^2}$ lies in $[-1, 1]$, except in a special case. What is this special case and for what rounding mode does it occur?

**Exercise 3.17** Show that the computation of $E$ in Algorithm **PrintFixed**, step 2, is correct — i.e., $E = 1 + \lfloor (e-1) \log b / \log B \rfloor$ — as long as there is no integer $n$ such that $|n/(e-1) \log B / \log b - 1| < \varepsilon$, where $\varepsilon$ is the relative precision when computing $\lambda$: $\lambda = \log B / \log b(1 + \theta)$ with $|\theta| \leq \varepsilon$. For a fixed range of exponents $-e_{\max} \leq e \leq e_{\max}$, deduce a working precision $\varepsilon$. Application: for $b = 2$, and $e_{\max} = 2^{31}$, compute the required precision for $3 \leq B \leq 36$.

**Exercise 3.18 (Lefèvre)** The IEEE 754-1985 standard required binary to decimal conversions to be correctly rounded in the range $m \cdot 10^n$ for $|m| \leq 10^{17} - 1$ and $|n| \leq 27$ in double precision. Find the hardest-to-print double-precision number in this range (with rounding to nearest, for example). Write a C program that outputs double-precision numbers in this range, and compare it to the `sprintf` C-language function of your system. Similarly for a conversion from the IEEE 754-2008 `binary64` format (significand of 53 bits, $2^{-1074} \leq |x| < 2^{1024}$) to the `decimal64` format (significand of 16 decimal digits).

**Exercise 3.19** The same question as in Exercise 3.18, but for decimal to binary conversion, and the `atof` C-language function.

## 3.8   Notes and References

In her PhD thesis [163, Chapter V], Valérie Ménissier-Morain discusses continued fractions and redundant representations as alternatives to the classical non-redundant representation considered here. She also considers [163, Chapter III] the theory of computable reals, their representation by $B$-adic numbers, and the computation of algebraic or transcendental functions.

Other representations were designed to increase the range of representable values; in particular Clenshaw and Olver [70] invented *level-index arithmetic*, where for example 2009 is approximated by 3.7075, since $2009 \approx \exp(\exp(\exp(0.7075)))$, and the leading 3 indicates the number of iterated exponentials. The obvious drawback is that it is expensive to perform arithmetic operations such as addition on numbers in the level-index representation.

Clenshaw and Olver [69] also introduced the idea of *unrestricted algorithm* (meaning no restrictions on the precision or exponent range). Several such algorithms were described in [48].

Nowadays most computers use radix two, but other choices (for example radix 16) were popular in the past, before the widespread adoption of the IEEE 754 standard. A discussion of the best choice of radix is given in [42].

For a general discussion of floating-point addition, rounding modes, the sticky bit, *etc.*, see Hennessy, Patterson and Goldberg [120, Appendix A.4].[4]

The main reference for floating-point arithmetic is the IEEE 754 standard [5], which defines four binary formats: single precision, single extended (deprecated), double precision, and double extended. The IEEE 854 standard [72] defines radix-independent arithmetic, and mainly decimal arithmetic. Both standards were replaced by the revision of IEEE 754 (approved by the IEEE Standards Committee on June 12, 2008).

---

[4]We refer to the first edition as later editions may not include the relevant Appendix by Goldberg.

We have not found the source of Theorem 3.1.1 — it seems to be "folklore". The rule regarding the precision of a result given possibly differing precisions of the operands was considered by Brent [49] and Hull [127].

Floating-point expansions were introduced by Priest [187]. They are mainly useful for a small number of summands, typically two or three, and when the main operations are additions or subtractions. For a larger number of summands the combinatorial logic becomes complex, even for addition. Also, except in simple cases, it seems difficult to obtain correct rounding with expansions.

Some good references on error analysis of floating-point algorithms are the books by Higham [121] and Muller [175]. Older references include Wilkinson's classics [229, 230].

Collins and Krandick [74], and Lefèvre [154], proposed algorithms for multiple-precision floating-point addition.

The problem of leading zero anticipation and detection in hardware is classical; see [195] for a comparison of different methods. Sterbenz's theorem may be found in his book [211].

The idea of having a "short product" together with correct rounding was studied by Krandick and Johnson [146]. They attributed the term "short product" to Knuth. They considered both the schoolbook and the Karatsuba domains. Algorithms **ShortProduct** and **ShortDivision** are due to Mulders [174]. The problem of consecutive zeros or ones — also called *runs* of zeros or ones — has been studied by several authors in the context of computer arithmetic: Iordache and Matula [129] studied division (Theorem 3.4.1), square root, and reciprocal square root. Muller and Lang [152] generalised their results to algebraic functions.

The Fast Fourier Transform (FFT) using complex floating-point numbers and the Schönhage-Strassen algorithm are described in Knuth [143]. Many variations of the FFT are discussed in the books by Crandall [79, 80]. For further references, see §2.9.

Theorem 3.3.2 is from Percival [184]; previous rigorous error analyses of complex FFT gave very pessimistic bounds. Note that [55] corrects the erroneous proof given in [184] (see also Exercise 3.8).

The concept of "middle product" for power series is discussed in Hanrot *et al.* [111]. Bostan, Lecerf and Schost [40] have shown that it can be seen as a special case of "Tellegen's principle", and have generalised it to operations other than multiplication. The link between usual multiplication and the middle product using trilinear forms was mentioned by Victor Pan [182] for the multiplication of two complex numbers: "*The duality technique enables us to extend any successful bilinear algorithms to two new ones for the new problems, sometimes quite* *different from the original problem* $\cdots$ " David Harvey [115] has shown how to efficiently implement the middle product for integers. A detailed and comprehensive description of the Payne and Hanek argument reduction method can be found in Muller [175].

In this section we drop the "$\sim$" that strictly should be included in the complexity bounds. The $2M(n)$ reciprocal algorithm of §3.4.1 — with the wrap-around trick — is due to Schönhage, Grotefeld and Vetter [199]. It can be improved, as noticed by Dan Bernstein [20]. If we keep the FFT-transform of $x$, we can save $M(n)/3$ (assuming the term-to-term products have negligible cost), which gives $5M(n)/3$. Bernstein also proposes a "messy" $3M(n)/2$ algorithm [20]. Schönhage's $3M(n)/2$ algorithm is simpler [198]. The idea is to write Newton's iteration as $x' = 2x - ax^2$. If $x$ is accurate to $n/2$ bits, then $ax^2$ has (in theory) $2n$ bits, but we know the upper $n/2$ bits cancel with $x$, and we are not interested in the low $n$ bits. Thus we can perform modular FFTs of size $3n/2$, with cost $M(3n/4)$ for the last iteration, and $1.5M(n)$ overall. This $1.5M(n)$ bound for the reciprocal was improved to $1.444M(n)$ by Harvey [116]. See also [78] for the roundoff error analysis when using a floating-point multiplier.

The idea of incorporating the dividend in Algorithm **DivideNewton** is due to Karp and Markstein [138], and is usually known as the Karp-Markstein trick; we already used it in Algorithm **ExactDivision** in Chapter 1. The asymptotic complexity $5M(n)/2$ of floating-point division can be improved to $5M(n)/3$, as shown by van der Hoeven in [125]. Another well-known method to perform a floating-point division is Goldschmidt's iteration: starting from $a/b$, first find $c$ such that $b_1 = cb$ is close to 1, and $a/b = a_1/b_1$ with $a_1 = ca$. At step $k$, assuming $a/b = a_k/b_k$, we multiply both $a_k$ and $b_k$ by $2 - b_k$, giving $a_{k+1}$ and $b_{k+1}$. The sequence $(b_k)$ converges to 1, and $(a_k)$ converges to $a/b$. Goldschmidt's iteration works because, if $b_k = 1 + \varepsilon_k$ with $\varepsilon_k$ small, then $b_{k+1} = (1 + \varepsilon_k)(1 - \varepsilon_k)$ $= 1 - \varepsilon_k^2$. Goldschmidt's iteration admits quadratic convergence like Newton's method. However, unlike Newton's method, Goldschmidt's iteration is not self-correcting. Thus, it yields an arbitrary precision division with cost $\Theta(M(n)\log n)$. For this reason, Goldschmidt's iteration should only be used for small, fixed precision. A detailed analysis of Goldschmidt's algorithms for division and square root, and a comparison with Newton's method, is given in Markstein [159].

Bernstein [20] obtained faster square root algorithms in the FFT domain, by caching some Fourier transforms. More precisely, he obtained $11M(n)/6$ for the square root, and $5M(n)/2$ for the simultaneous computation of $x^{1/2}$ and $x^{-1/2}$. The bound for the square root was reduced to $4M(n)/3$ by Harvey [116].

Classical floating-point conversion algorithms are due to Steele and White [208], Gay [103], and Clinger [71]; most of these authors assume fixed precision. Cowlishaw maintains an extensive bibliography of conversion to and from deci-mal formats (see §5.3). What we call "free-format" output is called "idempotent conversion" by Kahan [133]; see also Knuth [143, exercise 4.4-18]. Another useful reference on binary to decimal conversion is Cornea *et al.* [77].

Bürgisser, Clausen and Shokrollahi [59] is an excellent book on topics such as lower bounds, fast multiplication of numbers and polynomials, Strassen-like algorithms for matrix multiplication, and the tensor rank problem.

There is a large literature on interval arithmetic, which is outside the scope of this chapter. A recent book is Kulisch [150], and a good entry point is the Interval Computations web page (see Chapter 5).

In this chapter we did not consider complex arithmetic, except where relevant for its use in the FFT. An algorithm for the complex (floating-point) square root, which allows correct rounding, is given in [91]. See also the comments on Friedland's algorithm in §4.12.

# Chapter 4

# Elementary and Special Function Evaluation

Here we consider various applications of Newton's method, which can be used to compute reciprocals, square roots, and more generally algebraic and functional inverse functions. We then consider unrestricted algorithms for computing elementary and special functions. The algorithms of this chapter are presented at a higher level than in Chapter 3. A full and detailed analysis of one special function might be the subject of an entire chapter!

## 4.1   Introduction

This chapter is concerned with algorithms for computing elementary and special functions, although the methods apply more generally. First we consider Newton's method, which is useful for computing inverse functions. For example, if we have an algorithm for computing $y = \ln x$, then Newton's method can be used to compute $x = \exp y$ (see §4.2.5). However, Newton's method has many other applications. In fact we already mentioned Newton's method in Chapters 1–3, but here we consider it in more detail.

After considering Newton's method, we go on to consider various methods for computing elementary and special functions. These methods include power series (§4.4), asymptotic expansions (§4.5), continued fractions (§4.6), recurrence relations (§4.7), the arithmetic-geometric mean (§4.8), binary splitting (§4.9), and contour integration (§4.10). The methods that we consider are *unrestricted* in the sense that there is no restriction on the at-tainable precision — in particular, it is not limited to the precision of IEEE standard 32-bit or 64-bit floating-point arithmetic. Of course, this depends on the availability of a suitable software package for performing floating-point arithmetic on operands of arbitrary precision, as discussed in Chapter 3.

Unless stated explicitly, we do not consider rounding issues in this chapter; it is assumed that methods described in Chapter 3 are used. Also, to simplify the exposition, we assume a binary radix ($\beta = 2$), although most of the content could be extended to any radix. We recall that $n$ denotes the relative precision (in bits here) of the desired approximation; if the absolute computed value is close to 1, then we want an approximation to within $2^{-n}$.

## 4.2   Newton's Method

Newton's method is a major tool in arbitrary-precision arithmetic. We have already seen it or its $p$-adic counterpart, namely Hensel lifting, in previous chapters (see for example Algorithm **ExactDivision** in §1.4.5, or the iteration (2.3) to compute a modular inverse in §2.5). Newton's method is also useful in small precision: most modern processors only implement addition and multiplication in hardware; division and square root are microcoded, using either Newton's method if a fused multiply-add instruction is available, or the SRT algorithm. See the algorithms to compute a floating-point reciprocal or reciprocal square root in §3.4.1 and §3.5.1.

This section discusses Newton's method is more detail, in the context of floating-point computations, for the computation of inverse roots (§4.2.1), reciprocals (§4.2.2), reciprocal square roots (§4.2.3), formal power series (§4.2.4), and functional inverses (§4.2.5). We also discuss higher order Newton-like methods (§4.2.6).

### Newton's Method via Linearisation

Recall that a function $f$ of a real variable is said to have a *zero* $\zeta$ if $f(\zeta) = 0$. If $f$ is differentiable in a neighbourhood of $\zeta$, and $f'(\zeta) \neq 0$, then $\zeta$ is said to be a *simple* zero. Similarly for functions of several real (or complex) variables. In the case of several variables, $\zeta$ is a simple zero if the Jacobian matrix evaluated at $\zeta$ is nonsingular.

*Newton's method* for approximating a simple zero $\zeta$ of $f$ is based on the idea of making successive linear approximations to $f(x)$ in a neighbourhood of $\zeta$. Suppose that $x_0$ is an initial approximation, and that $f(x)$ has two continuous derivatives in the region of interest. From Taylor's theorem,[1]

$$f(\zeta) = f(x_0) + (\zeta - x_0)f'(x_0) + \frac{(\zeta - x_0)^2}{2} f''(\xi) \qquad (4.1)$$

for some point $\xi$ in an interval including $\{\zeta, x_0\}$. Since $f(\zeta) = 0$, we see that

$$x_1 = x_0 - f(x_0)/f'(x_0)$$

is an approximation to $\zeta$, and

$$x_1 - \zeta = O\left(|x_0 - \zeta|^2\right).$$

Provided $x_0$ is sufficiently close to $\zeta$, we will have

$$|x_1 - \zeta| \le |x_0 - \zeta|/2 < 1.$$

This motivates the definition of *Newton's method* as the iteration

$$x_{j+1} = x_j - \frac{f(x_j)}{f'(x_j)}, \quad j = 0, 1, \ldots \qquad (4.2)$$

Provided $|x_0 - \zeta|$ is sufficiently small, we expect $x_n$ to converge to $\zeta$. The *order of convergence* will be at least two, that is

$$|e_{n+1}| \le K|e_n|^2$$

for some constant $K$ independent of $n$, where $e_n = x_n - \zeta$ is the error after $n$ iterations.

A more careful analysis shows that

$$e_{n+1} = \frac{f''(\zeta)}{2f'(\zeta)} e_n^2 + O\left(|e_n^3|\right), \qquad (4.3)$$

provided $f \in C^3$ near $\zeta$. Thus, the order of convergence is exactly two if $f''(\zeta) \ne 0$ and $e_0$ is sufficiently small but nonzero. (Such an iteration is also said to be *quadratically convergent.*)

---

[1] Here we use Taylor's theorem at $x_0$, since this yields a formula in terms of derivatives at $x_0$, which is known, instead of at $\zeta$, which is unknown. Sometimes (for example in the derivation of (4.3)), it is preferable to use Taylor's theorem at the (unknown) zero $\zeta$.

## 4.2.1 Newton's Method for Inverse Roots

Consider applying Newton's method to the function

$$f(x) = y - x^{-m},$$

where $m$ is a positive integer constant, and (for the moment) $y$ is a positive constant. Since $f'(x) = mx^{-(m+1)}$, Newton's iteration simplifies to

$$x_{j+1} = x_j + x_j(1 - x_j^m y)/m. \qquad (4.4)$$

This iteration converges to $\zeta = y^{-1/m}$ provided the initial approximation $x_0$ is sufficiently close to $\zeta$. It is perhaps surprising that (4.4) does not involve divisions, except for a division by the integer constant $m$. In particular, we can easily compute reciprocals (the case $m = 1$) and reciprocal square roots (the case $m = 2$) by Newton's method. These cases are sufficiently important that we discuss them separately in the following subsections.

## 4.2.2 Newton's Method for Reciprocals

Taking $m = 1$ in (4.4), we obtain the iteration

$$x_{j+1} = x_j + x_j(1 - x_j y) \qquad (4.5)$$

which we expect to converge to $1/y$ provided $x_0$ is a sufficiently good approximation. (See §3.4.1 for a concrete algorithm with error analysis.) To see what "sufficiently good" means, define

$$u_j = 1 - x_j y.$$

Note that $u_j \to 0$ if and only if $x_j \to 1/y$. Multiplying each side of (4.5) by $y$, we get

$$1 - u_{j+1} = (1 - u_j)(1 + u_j),$$

which simplifies to

$$u_{j+1} = u_j^2. \qquad (4.6)$$

Thus

$$u_j = (u_0)^{2^j}. \qquad (4.7)$$

We see that the iteration converges if and only if $|u_0| < 1$, which (for real $x_0$ and $y$) is equivalent to the condition $x_0 y \in (0, 2)$. Second-order convergence is reflected in the double exponential with exponent 2 on the right-hand-side of (4.7).

The iteration (4.5) is sometimes implemented in hardware to compute reciprocals of floating-point numbers (see §4.12). The sign and exponent of the floating-point number are easily handled, so we can assume that $y \in [0.5, 1.0)$ (recall we assume a binary radix in this chapter). The initial approximation $x_0$ is found by table lookup, where the table is indexed by the first few bits of $y$. Since the order of convergence is two, the number of correct bits approximately doubles at each iteration. Thus, we can predict in advance how many iterations are required. Of course, this assumes that the table is initialised correctly.[2]

**Computational Issues**

At first glance, it seems better to replace Eqn. (4.5) by

$$x_{j+1} = x_j(2 - x_j y), \tag{4.8}$$

which looks simpler. However, although those two forms are mathematically equivalent, they are not computationally equivalent. Indeed, in Eqn. (4.5), if $x_j$ approximates $1/y$ to within $n/2$ bits, then $1 - x_j y = O(2^{-n/2})$, and the product of $x_j$ by $1 - x_j y$ might be computed with a precision of only $n/2$ bits. In the apparently simpler form (4.8), $2 - x_j y = 1 + O(2^{-n/2})$, thus the product of $x_j$ by $2 - x_j y$ has to be performed with a full precision of $n$ bits, to get $x_{j+1}$ accurate to within $n$ bits.

As a general rule, it is best to separate the terms of different order in Newton's iteration, and not try to factor common expressions. For an exception, see the discussion of Schönhage's $3M(n)/2$ reciprocal algorithm in §3.8.

### 4.2.3   Newton's Method for (Reciprocal) Square Roots

Taking $m = 2$ in (4.4), we obtain the iteration

$$x_{j+1} = x_j + x_j(1 - x_j^2 y)/2, \tag{4.9}$$

---

[2] In the case of the infamous *Pentium* `fdiv` *bug* [109, 176], a lookup table used for division was initialised incorrectly, and the division was occasionally inaccurate. In this case division used the SRT algorithm, but the moral is the same – tables must be initialised correctly.

which we expect to converge to $y^{-1/2}$ provided $x_0$ is a sufficiently good approximation.

If we want to compute $y^{1/2}$, we can do this in one multiplication after first computing $y^{-1/2}$, since

$$y^{1/2} = y \times y^{-1/2}.$$

This method does not involve any divisions (except by 2, see Ex. 3.15). In contrast, if we apply Newton's method to the function $f(x) = x^2 - y$, we obtain Heron's[3] iteration (see Algorithm **SqrtInt** in §1.5.1) for the square root of $y$:

$$x_{j+1} = \frac{1}{2} \left( x_j + \frac{y}{x_j} \right). \tag{4.10}$$

This requires a division by $x_j$ at iteration $j$, so it is essentially different from the iteration (4.9). Although both iterations have second-order convergence, we expect (4.9) to be more efficient (however this depends on the relative cost of division compared to multiplication). See also §3.5.1 and, for various optimisations, §3.8.

## 4.2.4   Newton's Method for Formal Power Series

This section is not required for function evaluation, however it gives a complementary point of view on Newton's method, and has applications to computing constants such as Bernoulli numbers (see Exercises 4.41–4.42).

Newton's method can be applied to find roots of functions defined by formal power series as well as of functions of a real or complex variable. For simplicity we consider formal power series of the form

$$A(z) = a_0 + a_1 z + a_2 z^2 + \cdots$$

where $a_i \in \mathbb{R}$ (or any field of characteristic zero) and ord$(A) = 0$, i.e., $a_0 \neq 0$.

For example, if we replace $y$ in (4.5) by $1 - z$, and take initial approximation $x_0 = 1$, we obtain a quadratically-convergent iteration for the formal power series

$$(1 - z)^{-1} = \sum_{n=0}^{\infty} z^n.$$

_________________

[3]Heron of Alexandria, *circa* 10–75 AD.

In the case of formal power series, "quadratically convergent" means that $\mathrm{ord}(e_j) \to +\infty$ like $2^j$, where $e_j$ is the difference between the desired result and the $j$th approximation. In our example, with the notation of §4.2.2, $u_0 = 1 - x_0 y = z$, so $u_j = z^{2^j}$ and

$$x_j = \frac{1 - u_j}{1 - z} = \frac{1}{1 - z} + O\left(z^{2^j}\right).$$

Given a formal power series $A(z) = \sum_{j \geq 0} a_j z^j$, we can define the formal *derivative*

$$A'(z) = \sum_{j > 0} j a_j z^{j-1} = a_1 + 2 a_2 z + 3 a_3 z^2 + \cdots,$$

and the *integral*

$$\sum_{j \geq 0} \frac{a_j}{j+1} \, z^{j+1},$$

but there is no useful analogue for multiple-precision integers $\sum_{j=0}^n a_j \beta^j$. This means that some fast algorithms for operations on power series have no analogue for operations on integers (see for example Exercise 4.1).

### 4.2.5   Newton's Method for Functional Inverses

Given a function $g(x)$, its functional inverse $h(x)$ satisfies $g(h(x)) = x$, and is denoted by $h(x) := g^{(-1)}(x)$. For example, $g(x) = \ln x$ and $h(x) = \exp x$ are functional inverses, as are $g(x) = \tan x$ and $h(x) = \arctan x$. Using the function $f(x) = y - g(x)$ in (4.2), one gets a root $\zeta$ of $f$, i.e., a value such that $g(\zeta) = y$, or $\zeta = g^{(-1)}(y)$:

$$x_{j+1} = x_j + \frac{y - g(x_j)}{g'(x_j)}.$$

Since this iteration only involves $g$ and $g'$, it provides an efficient way to evaluate $h(y)$, assuming that $g(x_j)$ and $g'(x_j)$ can be efficiently computed. Moreover, if the complexity of evaluating $g'$ — and of division — is no greater than that of $g$, we get a means to evaluate the functional inverse $h$ of $g$ with the same order of complexity as that of $g$.

As an example, if one has an efficient implementation of the logarithm, a similarly efficient implementation of the exponential is deduced as follows. Consider the root $e^y$ of the function $f(x) = y - \ln x$, which yields the iteration:

$$x_{j+1} = x_j + x_j(y - \ln x_j), \tag{4.11}$$

and in turn Algorithm **LiftExp** (for the sake of simplicity, we consider here only one Newton iteration).

---

**Algorithm 4.1** LiftExp

**Input:** $x_j$, $(n/2)$-bit approximation to $\exp(y)$

**Output:** $x_{j+1}$, $n$-bit approximation to $\exp(y)$

$\quad t \leftarrow \ln x_j \qquad\qquad\qquad \triangleright t$ computed to $n$-bit accuracy

$\quad u \leftarrow y - t \qquad\qquad\qquad \triangleright u$ computed to $(n/2)$-bit accuracy

$\quad v \leftarrow x_j u \qquad\qquad\qquad \triangleright v$ computed to $(n/2)$-bit accuracy

$\quad x_{j+1} \leftarrow x_j + v.$

---

## 4.2.6   Higher Order Newton-like Methods

The classical Newton's method is based on a linear approximation of $f(x)$ near $x_0$. If we use a higher-order approximation, we can get a higher-order method. Consider for example a second-order approximation. Equation (4.1) becomes:

$$f(\zeta) = f(x_0) + (\zeta - x_0)f'(x_0) + \frac{(\zeta - x_0)^2}{2}f''(x_0) + \frac{(\zeta - x_0)^3}{6}f'''(\xi).$$

Since $f(\zeta) = 0$, we have

$$\zeta = x_0 - \frac{f(x_0)}{f'(x_0)} - \frac{(\zeta - x_0)^2}{2}\frac{f''(x_0)}{f'(x_0)} + O((\zeta - x_0)^3). \qquad (4.12)$$

A difficulty here is that the right-hand-side of (4.12) involves the unknown $\zeta$. Let $\zeta = x_0 - f(x_0)/f'(x_0) + \nu$, where $\nu$ is a second-order term. Substituting this in the right-hand-side of (4.12) and neglecting terms of order $(\zeta - x_0)^3$ yields the cubic iteration:

$$x_{j+1} = x_j - \frac{f(x_j)}{f'(x_j)} - \frac{f(x_j)^2 f''(x_j)}{2f'(x_j)^3}.$$

For the computation of the reciprocal (§4.2.2) with $f(x) = y - 1/x$, this yields

$$x_{j+1} = x_j + x_j(1 - x_j y) + x_j(1 - x_j y)^2. \qquad (4.13)$$

For the computation of $\exp y$ using functional inversion (§4.2.5), one gets:

$$x_{j+1} = x_j + x_j(y - \ln x_j) + \frac{1}{2}x_j(y - \ln x_j)^2. \qquad (4.14)$$

These iterations can be obtained in a more systematic way that generalises to give iterations of arbitrarily high order. For the computation of the reciprocal, let $\varepsilon_j = 1 - x_j y$, so $x_j y = 1 - \varepsilon_j$ and (assuming $|\varepsilon_j| < 1$),

$$1/y = x_j/(1 - \varepsilon_j) = x_j(1 + \varepsilon_j + \varepsilon_j^2 + \cdots).$$

Truncating after the term $\varepsilon_j^{k-1}$ gives a $k$-th order iteration

$$x_{j+1} = x_j(1 + \varepsilon_j + \varepsilon_j^2 + \cdots + \varepsilon_j^{k-1}) \qquad (4.15)$$

for the reciprocal. The case $k = 2$ corresponds to Newton's method, and the case $k = 3$ is just the iteration (4.13) that we derived above.

Similarly, for the exponential we take $\varepsilon_j = y - \ln x_j = \ln(x/x_j)$, so

$$x/x_j = \exp \varepsilon_j = \sum_{m=0}^{\infty} \frac{\varepsilon_j^m}{m!}.$$

Truncating after $k$ terms gives a $k$-th order iteration

$$x_{j+1} = x_j \left( \sum_{m=0}^{k-1} \frac{\varepsilon_j^m}{m!} \right) \qquad (4.16)$$

for the exponential function. The case $k = 2$ corresponds to the Newton iteration, the case $k = 3$ is the iteration (4.14) that we derived above, and the cases $k > 3$ give higher-order Newton-like iterations. For a generalisation to other functions, see Exercises 4.3, 4.6.

## 4.3 Argument Reduction

*Argument reduction* is a classical method to improve the efficiency of the evaluation of mathematical functions. The key idea is to reduce the initial problem to a domain where the function is easier to evaluate. More precisely, given $f$ to evaluate at $x$, one proceeds in three steps:

- *argument reduction*: $x$ is transformed into a *reduced argument $x'$*;

- *evaluation*: $f$ is evaluated at $x'$;

- *reconstruction*: $f(x)$ is computed from $f(x')$ using a functional identity.

In some cases the argument reduction or the reconstruction is trivial, for example $x' = x/2$ in radix 2, or $f(x) = \pm f(x')$ (some examples illustrate this below). It might also be that the evaluation step uses a different function $g$ instead of $f$; for example $\sin(x + \pi/2) = \cos(x)$.

Unfortunately, argument reduction formulæ do not exist for every function; for example, no argument reduction is known for the error function. Argument reduction is only possible when a functional identity relates $f(x)$ and $f(x')$ (or $g(x)$ and $g(x')$). The elementary functions have *addition formulae* such as

$$
\begin{aligned}
\exp(x+y) &= \exp(x)\exp(y), \\
\log(xy) &= \log(x) + \log(y), \\
\sin(x+y) &= \sin(x)\cos(y) + \cos(x)\sin(y), \\
\tan(x+y) &= \frac{\tan(x) + \tan(y)}{1 - \tan(x)\tan(y)}.
\end{aligned}
\tag{4.17}
$$

We use these formulæ to reduce the argument so that power series converge more rapidly. Usually we take $x = y$ to get *doubling formulae* such as

$$
\exp(2x) = \exp(x)^2,
\tag{4.18}
$$

though occasionally *tripling formulae* such as

$$
\sin(3x) = 3\sin(x) - 4\sin^3(x)
$$

might be useful. This tripling formula only involves one function (sin), whereas the doubling formula $\sin(2x) = 2\sin x \cos x$ involves two functions (sin and cos), but this problem can be overcome: see §4.3.4 and §4.9.1.

We usually distinguish two kinds of argument reduction:

- *additive argument reduction*, where $x' = x - kc$, for some real constant $c$ and some integer $k$. This occurs in particular when $f(x)$ is periodic, for example for the sine and cosine functions with $c = 2\pi$;

- *multiplicative argument reduction*, where $x' = x/c^k$ for some real constant $c$ and some integer $k$. This occurs with $c = 2$ in the computation of $\exp x$ when using the doubling formula (4.18): see §4.3.1.

Note that, for a given function, both kinds of argument reduction might be available. For example, for $\sin x$, one might either use the tripling formula $\sin(3x) = 3\sin x - 4\sin^3 x$, or the additive reduction $\sin(x + 2k\pi) = \sin x$ that arises from the periodicity of $\sin$.

Sometime "reduction" is not quite the right word, since a functional identity is used to *increase* rather than to *decrease* the argument. For example, the Gamma function $\Gamma(x)$ satisfies an identity

$$x\Gamma(x) = \Gamma(x + 1),$$

that can be used repeatedly to *increase* the argument until we reach the region where Stirling's asymptotic expansion is sufficiently accurate, see §4.5.

## 4.3.1   Repeated Use of a Doubling Formula

If we apply the doubling formula (4.18) for the exponential function $k$ times, we get

$$\exp(x) = \exp(x/2^k)^{2^k}.$$

Thus, if $|x| = \Theta(1)$, we can reduce the problem of evaluating $\exp(x)$ to that of evaluating $\exp(x/2^k)$, where the argument is now $O(2^{-k})$. This is better since the power series converges more quickly for $x/2^k$. The cost is the $k$ squarings that we need to reconstruct the final result from $\exp(x/2^k)$.

There is a trade-off here, and $k$ should be chosen to minimise the total time. If the obvious method for power series evaluation is used, then the optimal $k$ is of order $\sqrt{n}$ and the overall time is $O(n^{1/2}M(n))$. We shall see in §4.4.3 that there are faster ways to evaluate power series, so this is not the best possible result.

We assumed here that $|x| = \Theta(1)$. A more careful analysis shows that the optimal $k$ depends on the order of magnitude of $x$ (see Exercise 4.5).

## 4.3.2   Loss of Precision

For some power series, especially those with alternating signs, a loss of precision might occur due to a cancellation between successive terms. A typical example is the series for $\exp(x)$ when $x < 0$. Assume for example that we want 10 significant digits of $\exp(-10)$. The first ten terms $x^k/k!$ for $x = -10$ are approximately:

$$1., -10., 50., -166.6666667, 416.6666667, -833.3333333, 1388.888889,$$
$$-1984.126984, 2480.158730, -2755.731922.$$

Note that these terms alternate in sign and initially *increase* in magnitude. They only start to decrease in magnitude for $k > |x|$. If we add the first 51 terms with a working precision of 10 decimal digits, we get an approximation to $\exp(-10)$ that is only accurate to about 3 digits!

A much better approach is to use the identity

$$\exp(x) = 1/\exp(-x)$$

to avoid cancellation in the power series summation. In other cases a different power series without sign changes might exist for a closely related function: for example, compare the series (4.22) and (4.23) for computation of the error function $\text{erf}(x)$. See also Exercises 4.19–4.20.

### 4.3.3   Guard Digits

*Guard digits* are digits in excess of the number of digits that are required in the final answer. Generally, it is necessary to use some guard digits during a computation in order to obtain an accurate result (one that is correctly rounded or differs from the correctly rounded result by a small number of units in the last place). Of course, it is expensive to use too many guard digits. Thus, care has to be taken to use the right number of guard digits, that is the right working precision. Here and below, we use the generic term "guard digits", even for radix $\beta = 2$.

Consider once again the example of $\exp x$, with reduced argument $x/2^k$ and $x = \Theta(1)$. Since $x/2^k$ is $O(2^{-k})$, when we sum the power series $1 + x/2^k + \cdots$ from left to right (forward summation), we "lose" about $k$ bits of precision. More precisely, if $x/2^k$ is accurate to $n$ bits, then $1 + x/2^k$ is accurate to $n + k$ bits, but if we use the same working precision $n$, we obtain only $n$ correct bits. After squaring $k$ times in the reconstruction step, about $k$ bits will be lost (each squaring loses about one bit), so the final accuracy will be only $n-k$ bits. If we summed the power series in reverse order instead (backward summation), and used a working precision of $n + k$ when adding 1 and $x/2^k + \cdots$ and during the squarings, we would obtain an accuracy of $n+k$ bits before the $k$ squarings, and an accuracy of $n$ bits in the final result.

Another way to avoid loss of precision is to evaluate $\text{expm1}(x/2^k)$, where the function expm1 is defined by

$$\text{expm1}(x) = \exp(x) - 1$$

and has a doubling formula that avoids loss of significance when $|x|$ is small. See Exercises 4.7–4.9.

### 4.3.4  Doubling versus Tripling

Suppose we want to compute the function $\sinh(x) = (e^x - e^{-x})/2$. The obvious doubling formula for sinh,

$$\sinh(2x) = 2\sinh(x)\cosh(x),$$

involves the auxiliary function $\cosh(x) = (e^x + e^{-x})/2$. Since $\cosh^2(x) - \sinh^2(x) = 1$, we could use the doubling formula

$$\sinh(2x) = 2\sinh(x)\sqrt{1 + \sinh^2(x)},$$

but this involves the overhead of computing a square root. This suggests using the tripling formula

$$\sinh(3x) = \sinh(x)(3 + 4\sinh^2(x)). \tag{4.19}$$

However, it is usually more efficient to do argument reduction via the doubling formula (4.18) for exp, because it takes one multiplication and one squaring to apply the tripling formula, but only two squarings to apply the doubling formula twice (and $3 < 2^2$). A drawback is loss of precision, caused by cancellation in the computation of $\exp(x) - \exp(-x)$, when $|x|$ is small. In this case it is better to use (see Exercise 4.10)

$$\sinh(x) = (\text{expm1}(x) - \text{expm1}(-x))/2. \tag{4.20}$$

See §4.12 for further comments on doubling versus tripling, especially in the FFT range.

# 4.4   Power Series

Once argument reduction has been applied, where possible (§4.3), one is usually faced with the evaluation of a power series. The elementary and special functions have power series expansions such as:

$$\exp x \;=\; \sum_{j \geq 0} \frac{x^j}{j!}, \qquad \ln(1+x) = \sum_{j \geq 0} \frac{(-1)^j x^{j+1}}{j+1},$$

$$\arctan x \;=\; \sum_{j \geq 0} \frac{(-1)^j x^{2j+1}}{2j+1}, \quad \sinh x = \sum_{j \geq 0} \frac{x^{2j+1}}{(2j+1)!}, \;\; etc.$$

This section discusses several techniques to recommend or to avoid. We use the following notations: $x$ is the evaluation point, $n$ is the desired precision, and $d$ is the number of terms retained in the power series, or $d-1$ is the degree of the corresponding polynomial $\sum_{0 \leq j < d} a_j x^j$.

If $f(x)$ is analytic in a neighbourhood of some point $c$, an obvious method to consider for the evaluation of $f(x)$ is summation of the Taylor series

$$f(x) = \sum_{j=0}^{d-1} (x-c)^j \, \frac{f^{(j)}(c)}{j!} + R_d(x, c).$$

As a simple but instructive example we consider the evaluation of $\exp(x)$ for $|x| \leq 1$, using

$$\exp(x) = \sum_{j=0}^{d-1} \frac{x^j}{j!} + R_d(x), \tag{4.21}$$

where $|R_d(x)| \leq |x|^d \exp(|x|)/d! \leq e/d!$.

Using Stirling's approximation for $d!$, we see that $d \geq K(n) \sim n/\lg n$ is sufficient to ensure that $|R_d(x)| = O(2^{-n})$. Thus, the time required to evaluate (4.21) with Horner's rule[4] is $O(nM(n)/\log n)$.

---

[4] By *Horner's rule* (with argument $x$) we mean evaluating the polynomial $s_0 = \sum_{0 \leq j \leq d} a_j x^j$ of degree $d$ (not $d-1$ in this footnote) by the recurrence $s_d = a_d$, $s_j = a_j + s_{j+1} x$ for $j = d-1, d-2, \ldots, 0$. Thus $s_k = \sum_{k \leq j \leq d} a_j x^{j-k}$. An evaluation by Horner's rule takes $d$ additions and $d$ multiplications, and is more efficient than explicitly evaluating the individual terms $a_j x^j$.

In practice it is convenient to sum the series in the forward direction $(j = 0, 1, \ldots, d - 1)$. The terms $t_j = x^j/j!$ and partial sums

$$S_j = \sum_{i=0}^{j} t_i$$

may be generated by the recurrence $t_j = xt_{j-1}/j$, $S_j = S_{j-1} + t_j$, and the summation terminated when $|t_d| < 2^{-n}/e$. Thus, it is not necessary to estimate $d$ in advance, as it would be if the series were summed by Horner's rule in the backward direction $(j = d-1, d-2, \ldots, 0)$ (see however Exercise 4.4).

We now consider the effect of rounding errors, under the assumption that floating-point operations are correctly rounded, i.e., satisfy

$$\circ(x \text{ op } y) = (x \text{ op } y)(1 + \delta),$$

where $|\delta| \leq \varepsilon$ and "op" = "+", "−", "×" or "/". Here $\varepsilon = 2^{-n}$ is the "machine precision" or "working precision". Let $\widehat{t_j}$ be the computed value of $t_j$, *etc.* Thus

$$|\widehat{t_j} - t_j| \, / \, |t_j| \; \leq \; 2j\varepsilon + O(\varepsilon^2)$$

and using $\sum_{j=0}^{d} t_j = S_d \leq e$:

$$
\begin{aligned}
|\widehat{S_d} - S_d| &\leq de\varepsilon + \sum_{j=1}^{d} 2j\varepsilon|t_j| + O(\varepsilon^2) \\
&\leq (d+2)e\varepsilon + O(\varepsilon^2) = O(n\varepsilon).
\end{aligned}
$$

Thus, to get $|\widehat{S_d} - S_d| = O(2^{-n})$, it is sufficient that $\varepsilon = O(2^{-n}/n)$. In other words, we need to work with about $\lg n$ guard digits. This is not a significant overhead if (as we assume) the number of digits may vary dynamically. We can sum with $j$ increasing (the *forward* direction) or decreasing (the *backward* direction). A slightly better error bound is obtainable for summation in the backward direction, but this method has the disadvantage that the number of terms $d$ has to be decided in advance (see however Exercise 4.4).

In practice it is inefficient to keep the working precision $\varepsilon$ fixed. We can profitably reduce it when computing $t_j$ from $t_{j-1}$ if $|t_{j-1}|$ is small, without significantly increasing the error bound. We can also vary the working precision when accumulating the sum, especially if it is computed in the backward direction (so the smallest terms are summed first).

It is instructive to consider the effect of relaxing our restriction that $|x| \leq 1$. First suppose that $x$ is large and positive. Since $|t_j| > |t_{j-1}|$ when $j < |x|$, it is clear that the number of terms required in the sum (4.21) is at least of order $|x|$. Thus, the method is slow for large $|x|$ (see §4.3 for faster methods in this case).

If $|x|$ is large and $x$ is negative, the situation is even worse. From Stirling's approximation we have

$$\max_{j \geq 0} |t_j| \simeq \frac{\exp |x|}{\sqrt{2\pi |x|}},$$

but the result is $\exp(-|x|)$, so about $2|x|/\log 2$ guard digits are required to compensate for what Lehmer called "catastrophic cancellation" [94]. Since $\exp(x) = 1/\exp(-x)$, this problem may easily be avoided, but the corresponding problem is not always so easily avoided for other analytic functions.

Here is a less trivial example. To compute the error function

$$\mathrm{erf}(x) = \frac{2}{\sqrt{\pi}} \int_0^x e^{-u^2}\, \mathrm{d}u,$$

we may use either the power series

$$\mathrm{erf}(x) = \frac{2x}{\sqrt{\pi}} \sum_{j=0}^{\infty} \frac{(-1)^j\, x^{2j}}{j!(2j+1)} \tag{4.22}$$

or the (mathematically, but not numerically) equivalent

$$\mathrm{erf}(x) = \frac{2x e^{-x^2}}{\sqrt{\pi}} \sum_{j=0}^{\infty} \frac{2^j\, x^{2j}}{1 \cdot 3 \cdot 5 \cdots (2j+1)} \,. \tag{4.23}$$

For small $|x|$, the series (4.22) is slightly faster than the series (4.23) because there is no need to compute an exponential. However, the series (4.23) is preferable to (4.22) for moderate $|x|$ because it involves no cancellation. For large $|x|$ neither series is satisfactory, because $\Omega(x^2)$ terms are required, and in this case it is preferable to use the asymptotic expansion for $\mathrm{erfc}(x) = 1 - \mathrm{erf}(x)$: see §4.5. In the borderline region use of the continued fraction (4.40) could be considered: see Exercise 4.31.

In the following subsections we consider different methods to evaluate power series. We generally ignore the effect of rounding errors, but the results obtained above are typical.

## Assumption about the Coefficients

We assume in this section that we have a power series $\sum_{j \geq 0} a_j x^j$ where $a_{j+\delta}/a_j$ is a rational function $R(j)$ of $j$, and hence it is easy to evaluate $a_0, a_1, a_2, \ldots$ sequentially. Here $\delta$ is a fixed positive constant, usually 1 or 2. For example, in the case of $\exp x$, we have $\delta = 1$ and

$$\frac{a_{j+1}}{a_j} = \frac{j!}{(j+1)!} = \frac{1}{j+1}\,.$$

Our assumptions cover the common case of hypergeometric functions. For the more general case of holonomic functions, see §4.9.2.

In common cases where our assumption is invalid, other good methods are available to evaluate the function. For example, $\tan x$ does not satisfy our assumption (the coefficients in its Taylor series are called *tangent numbers* and are related to Bernoulli numbers – see §4.7.2), but to evaluate $\tan x$ we can use Newton's method on the inverse function (arctan, which does satisfy our assumptions – see §4.2.5), or we can use $\tan x = \sin x / \cos x$.

## The Radius of Convergence

If the elementary function is an entire function (e.g., exp, sin) then the power series converges in the whole complex plane. In this case the degree of the denominator of $R(j) = a_{j+1}/a_j$ is greater than that of the numerator.

In other cases (such as $\ln$, arctan) the function is not entire. The power series only converges in a disk because the function has a singularity on the boundary of this disk. In fact $\ln(x)$ has a singularity at the origin, which is why we consider the power series for $\ln(1 + x)$. This power series has radius of convergence 1.

Similarly, the power series for $\arctan(x)$ has radius of convergence 1 because $\arctan(x)$ has singularities on the unit circle (at $\pm i$) even though it is uniformly bounded for all real $x$.

### 4.4.1   Direct Power Series Evaluation

Suppose that we want to evaluate a power series $\sum_{j \geq 0} a_j x^j$ at a given argument $x$. Using periodicity (in the cases of $\sin, \cos$) and/or argument reduction techniques (§4.3), we can often ensure that $|x|$ is sufficiently small. Thus, let us assume that $|x| \leq 1/2$ and that the radius of convergence of the series is at least 1.

As above, assume that $a_{j+\delta}/a_j$ is a rational function of $j$, and hence easy to evaluate. For simplicity we consider only the case $\delta = 1$. To sum the series with error $O(2^{-n})$ it is sufficient to take $n + O(1)$ terms, so the time required is $O(nM(n))$. If the function is entire, then the series converges faster and the time is reduced to $O(nM(n)/(\log n))$. However, we can do much better by carrying the argument reduction further, as demonstrated in the next section.

### 4.4.2   Power Series With Argument Reduction

Consider the evaluation of $\exp(x)$. By applying argument reduction $k + O(1)$ times, we can ensure that the argument $x$ satisfies $|x| < 2^{-k}$. Then, to obtain $n$-bit accuracy we only need to sum $O(n/k)$ terms of the power series. Assuming that a step of argument reduction is $O(M(n))$, which is true for the elementary functions, the total cost is $O((k + n/k)M(n))$. Indeed, the argument reduction and/or reconstruction requires $O(k)$ steps of $O(M(n))$, and the evaluation of the power series of order $n/k$ costs $(n/k)M(n)$; so choosing $k \sim n^{1/2}$ gives cost

$$O\left(n^{1/2}M(n)\right).$$

For example, our comments apply to the evaluation of $\exp(x)$ using

$$\exp(x) = \exp(x/2)^2,$$

to $\mathrm{log1p}(x) = \ln(1 + x)$ using

$$\mathrm{log1p}(x) = 2\,\mathrm{log1p}\left(\frac{x}{1 + \sqrt{1 + x}}\right),$$

and to $\arctan(x)$ using

$$\arctan x = 2\arctan\left(\frac{x}{1 + \sqrt{1 + x^2}}\right).$$

Note that in the last two cases each step of the argument reduction requires a square root, but this can be done with cost $O(M(n))$ by Newton's method (§3.5). Thus in all three cases the overall cost is $O(n^{1/2}M(n))$, although the implicit constant might be smaller for exp than for log1p or arctan. See Exercises 4.8–4.9.

**Using Symmetries**

A not-so-well-known idea is to evaluate $\ln(1+x)$ using the power series

$$\ln\left(\frac{1+y}{1-y}\right) = 2\sum_{j\geq 0}\frac{y^{2j+1}}{2j+1}$$

with $y$ defined by $(1+y)/(1-y) = 1+x$, i.e., $y = x/(2+x)$. This saves half the terms and also reduces the argument, since $y < x/2$ if $x > 0$. Unfortunately this nice idea can be applied only once. For a related example, see Exercise 4.11.

### 4.4.3   Rectangular Series Splitting

Once we determine how many terms in the power series are required for the desired accuracy, the problem reduces to evaluating a truncated power series, i.e., a polynomial.

Let $P(x) = \sum_{0\leq j<d}a_jx^j$ be the polynomial that we want to evaluate, $\deg(P) < d$. In the general case $x$ is a floating-point number of $n$ bits, and we aim at an accuracy of $n$ bits for $P(x)$. However the coefficients $a_j$, or their ratios $R(j) = a_{j+1}/a_j$, are usually small integers or rational numbers of $O(\log n)$ bits. A *scalar multiplication* involves one coefficient $a_j$ and the variable $x$ (or more generally an $n$-bit floating-point number), whereas a *nonscalar multiplication* involves two powers of $x$ (or more generally two $n$-bit floating-point numbers). Scalar multiplications are cheaper because the $a_j$ are small rationals of size $O(\log n)$, whereas $x$ and its powers generally have $\Theta(n)$ bits. It is possible to evaluate $P(x)$ with $O(\sqrt{n})$ nonscalar multiplications (plus $O(n)$ scalar multiplications and $O(n)$ additions, using $O(\sqrt{n})$ storage). The same idea applies, more generally, to evaluation of hypergeometric functions.

**Classical Splitting**

Suppose $d = jk$, define $y = x^k$, and write

$$P(x) = \sum_{\ell=0}^{j-1} y^\ell P_\ell(x) \quad \text{where} \quad P_\ell(x) = \sum_{m=0}^{k-1} a_{k\ell+m}\, x^m.$$

One first computes the powers $x^2, x^3, \ldots, x^{k-1}, x^k = y$; then the polynomials $P_\ell(x)$ are evaluated simply by multiplying $a_{k\ell+m}$ and the precomputed $x^m$ (it is important not to use Horner's rule here, since this would involve expensive nonscalar multiplications). Finally, $P(x)$ is computed from the $P_\ell(x)$ using Horner's rule with argument $y$. To see the idea geometrically, write $P(x)$ as

$$
\begin{array}{l}
y^0\ [a_0 \quad + \quad a_1 x \quad + \quad a_2 x^2 \quad + \cdots + \quad a_{k-1} x^{k-1}] \quad + \\
y^1\ [a_k \quad + \quad a_{k+1} x \quad + \quad a_{k+2} x^2 \quad + \cdots + \quad a_{2k-1} x^{k-1}] \quad + \\
y^2\ [a_{2k} \quad + \quad a_{2k+1} x \quad + \quad a_{2k+2} x^2 \quad + \cdots + \quad a_{3k-1} x^{k-1}] \quad + \\
\qquad\qquad \vdots \qquad\qquad\qquad \vdots \qquad\qquad\quad \vdots \\
y^{j-1}\ [a_{(j-1)k} \ + \ a_{(j-1)k+1} x \ + \ a_{(j-1)k+2} x^2 \ + \cdots + \ a_{jk-1} x^{k-1}]
\end{array}
$$

where $y = x^k$. The terms in square brackets are the polynomials $P_0(x)$, $P_1(x)$, $\ldots$, $P_{j-1}(x)$.

As an example, consider $d = 12$, with $j = 3$ and $k = 4$. This gives $P_0(x) = a_0 + a_1 x + a_2 x^2 + a_3 x^3$, $P_1(x) = a_4 + a_5 x + a_6 x^2 + a_7 x^3$, $P_2(x) = a_8 + a_9 x + a_{10} x^2 + a_{11} x^3$, then $P(x) = P_0(x) + y P_1(x) + y^2 P_2(x)$, where $y = x^4$. Here we need to compute $x^2$, $x^3$, $x^4$, which requires three nonscalar products — note that even powers like $x^4$ should be computed as $(x^2)^2$ to use squarings instead of multiplies — and we need two nonscalar products to evaluate $P(x)$, thus a total of five nonscalar products, instead of $d - 2 = 10$ with a naive application of Horner's rule to $P(x)$.[5]

**Modular Splitting**

An alternate splitting is the following, which may be obtained by transposing the matrix of coefficients above, swapping $j$ and $k$, and interchanging the powers of $x$ and $y$. It might also be viewed as a generalized odd-even scheme

---

[5]$P(x)$ has degree $d - 1$, so Horner's rule performs $d - 1$ products, but the first one $x \times a_{d-1}$ is a scalar product, hence there are $d - 2$ nonscalar products.

(§1.3.5). Suppose as before that $d = jk$, and write, with $y = x^j$:

$$P(x) = \sum_{\ell=0}^{j-1} x^\ell P_\ell(y) \quad \text{where} \quad P_\ell(y) = \sum_{m=0}^{k-1} a_{jm+\ell}\, y^m.$$

First compute $y = x^j, y^2, y^3, \ldots, y^{k-1}$. Now the polynomials $P_\ell(y)$ can be evaluated using only scalar multiplications of the form $a_{jm+\ell} \times y^m$.

To see the idea geometrically, write $P(x)$ as

$$
\begin{array}{llllll}
x^0 & [a_0 & + & a_j y & + & a_{2j} y^2 & + & \cdots] & + \\
x^1 & [a_1 & + & a_{j+1} y & + & a_{2j+1} y^2 & + & \cdots] & + \\
x^2 & [a_2 & + & a_{j+2} y & + & a_{2j+2} y^2 & + & \cdots] & + \\
 & \vdots & & \vdots & & \vdots & & \\
x^{j-1} & [a_{j-1} & + & a_{2j-1} y & + & a_{3j-1} y^2 & + & \cdots]
\end{array}
$$

where $y = x^j$. We traverse the first row of the array, then the second row, then the third, …, finally the $j$-th row, accumulating sums $S_0, S_1, \ldots, S_{j-1}$ (one for each row). At the end of this process $S_\ell = P_\ell(y)$ and we only have to evaluate

$$P(x) = \sum_{\ell=0}^{j-1} x^\ell S_\ell.$$

The complexity of each scheme is almost the same (see Exercise 4.12). With $d = 12$ ($j = 3$ and $k = 4$) we have $P_0(y) = a_0 + a_3 y + a_6 y^2 + a_9 y^3$, $P_1(y) = a_1 + a_4 y + a_7 y^2 + a_{10} y^3$, $P_2(y) = a_2 + a_5 y + a_8 y^2 + a_{11} y^3$. We first compute $y = x^3$, $y^2$ and $y^3$, then we evaluate $P_0(y)$ in three scalar multiplications $a_3 y$, $a_6 y^2$, and $a_9 y^3$ and three additions, similarly for $P_1$ and $P_2$, and finally we evaluate $P(x)$ using

$$P(x) = P_0(y) + x P_1(y) + x^2 P_2(y),$$

(here we might use Horner's rule). In this example, we have a total of six nonscalar multiplications: four to compute $y$ and its powers, and two to evaluate $P(x)$.

## Complexity of Rectangular Series Splitting

To evaluate a polynomial $P(x)$ of degree $d - 1 = jk - 1$, rectangular series splitting takes $O(j+k)$ nonscalar multiplications — each costing $O(M(n))$ — and $O(jk)$ scalar multiplications. The scalar multiplications involve multiplication and/or division of a multiple-precision number by small integers. Assume that these multiplications and/or divisions take time $c(d)n$ each (see Exercise 4.13 for a justification of this assumption). The function $c(d)$ accounts for the fact that the involved scalars (the coefficients $a_j$ or the ratios $a_{j+1}/a_j$) have a size depending on the degree $d$ of $P(x)$. In practice we can usually regard $c(d)$ as constant.

Choosing $j \sim k \sim d^{1/2}$ we get overall time

$$O(d^{1/2}M(n) + dn \cdot c(d)). \tag{4.24}$$

If $d$ is of the same order as the precision $n$ of $x$, this is not an improvement on the bound $O(n^{1/2}M(n))$ that we obtained already by argument reduction and power series evaluation (§4.4.2). However, we can do argument reduction before applying rectangular series splitting. Assuming that $c(n) = O(1)$ (see Exercise 4.14 for a detailed analysis), the total complexity is:

$$T(n) = O\left(\frac{n}{d}M(n) + d^{1/2}M(n) + dn\right),$$

where the extra $(n/d)M(n)$ term comes from argument reduction and/or reconstruction. Which term dominates? There are two cases:

1. $M(n) \gg n^{4/3}$. Here the minimum is obtained when the first two terms — argument reduction/reconstruction and nonscalar multiplications — are equal, i.e., for $d \sim n^{2/3}$, which yields $T(n) = O(n^{1/3}M(n))$. This case applies if we use classical or Karatsuba multiplication, since $\lg 3 > 4/3$, and similarly for Toom-Cook 3-, 4-, 5-, or 6-way multiplication (but not 7-way, since $\log_7 13 < 4/3$). In this case $T(n) \gg n^{5/3}$.

2. $M(n) \ll n^{4/3}$. Here the minimum is obtained when the first and the last terms — argument reduction/reconstruction and scalar multiplications — are equal. The optimal value of $d$ is then $\sqrt{M(n)}$, and we get an improved bound $\Theta(n\sqrt{M(n)}) \gg n^{3/2}$. We can not approach the $O(n^{1+\varepsilon})$ that is achievable with AGM-based methods (if applicable) – see §4.8.

# 4.5 Asymptotic Expansions

Often it is necessary to use different methods to evaluate a special function in different parts of its domain. For example, the exponential integral[6]

$$\mathrm{E}_1(x) = \int_x^\infty \frac{\exp(-u)}{u} \, du \qquad (4.25)$$

is defined for all $x > 0$. However, the power series

$$\mathrm{E}_1(x) + \gamma + \ln x = \sum_{j=1}^\infty \frac{(-1)^{j-1} x^j}{j! \, j} \qquad (4.26)$$

is unsatisfactory as a means of evaluating $\mathrm{E}_1(x)$ for large positive $x$, for the reasons discussed in §4.4 in connection with the power series (4.22) for $\mathrm{erf}(x)$, or the power series for $\exp(x)$ ($x$ negative). For sufficiently large positive $x$ it is preferable to use

$$e^x \, \mathrm{E}_1(x) = \sum_{j=1}^k \frac{(j-1)! \, (-1)^{j-1}}{x^j} + R_k(x), \qquad (4.27)$$

where

$$R_k(x) = k! \, (-1)^k \exp(x) \int_x^\infty \frac{\exp(-u)}{u^{k+1}} \, du. \qquad (4.28)$$

Note that

$$|R_k(x)| < \frac{k!}{x^{k+1}},$$

so

$$\lim_{x \to +\infty} R_k(x) = 0,$$

but $\lim_{k \to \infty} R_k(x)$ does not exist. In other words, the series

$$\sum_{j=1}^\infty \frac{(j-1)! \, (-1)^{j-1}}{x^j}$$

---

[6]The functions $\mathrm{E}_1(x)$ and $\mathrm{Ei}(x) = \mathrm{PV} \int_{-\infty}^x (\exp(t)/t) \, dt$ are both called "exponential integrals". Closely related is the "logarithmic integral" $\mathrm{li}(x) = \mathrm{Ei}(\ln x) = \mathrm{PV} \int_0^x (1/\ln t) \, dt$. Here the integrals $\mathrm{PV} \int \cdots$ should be interpreted as Cauchy principal values if there is a singularity in the range of integration. The power series (4.26) is valid for $x \in \mathbb{C}$ if $|\arg x| < \pi$ (see Exercise 4.16).

is divergent. In such cases we call this an *asymptotic series* and write

$$e^x \, \mathrm{E}_1(x) \sim \sum_{j>0} \frac{(j-1)!(-1)^{j-1}}{x^j}. \qquad (4.29)$$

Although they do not generally converge, asymptotic series are very useful. Often (though not always!) the error is bounded by the last term taken in the series (or by the first term omitted). Also, when the terms in the asymptotic series alternate in sign, it can often be shown that the true value lies between two consecutive approximations obtained by summing the series with (say) $k$ and $k+1$ terms. For example, this is true for the series (4.29) above, provided $x$ is real and positive.

When $x$ is large and positive, the relative error attainable by using (4.27) with $k = \lfloor x \rfloor$ is $O(x^{1/2} \exp(-x))$, because

$$|R_k(k)| \; \le \; k!/k^{k+1} \; = \; O(k^{-1/2} \exp(-k)) \qquad (4.30)$$

and the leading term on the right side of (4.27) is $1/x$. Thus, the asymptotic series may be used to evaluate $\mathrm{E}_1(x)$ to precision $n$ whenever $x > n \ln 2 + O(\ln n)$. More precise estimates can be obtained by using a version of Stirling's approximation with error bounds, for example

$$\left(\frac{k}{e}\right)^k \sqrt{2\pi k} < k! < \left(\frac{k}{e}\right)^k \sqrt{2\pi k} \, \exp\left(\frac{1}{12k}\right).$$

If $x$ is too small for the asymptotic approximation to be sufficiently accurate, we can avoid the problem of cancellation in the power series (4.26) by the technique of Exercise 4.19. However, the asymptotic approximation is faster and hence is preferable whenever it is sufficiently accurate.

Examples where asymptotic expansions are useful include the evaluation of $\mathrm{erfc}(x)$, $\Gamma(x)$, Bessel functions, *etc.* We discuss some of these below.

Asymptotic expansions often arise when the convergence of series is accelerated by the Euler-Maclaurin sum formula.[7] For example, Euler's constant

---

[7] The Euler-Maclaurin sum formula is a way of expressing the difference between a sum and an integral as an asymptotic expansion. For example, assuming that $a \in \mathbb{Z}$, $b \in \mathbb{Z}$, $a \le b$, and $f(x)$ satisfies certain conditions, one form of the formula is

$$\sum_{a \le k \le b} f(k) - \int_a^b f(x) \, \mathrm{d}x \sim \frac{f(a) + f(b)}{2} + \sum_{k \ge 1} \frac{B_{2k}}{(2k)!} \left( f^{(2k-1)}(b) - f^{(2k-1)}(a) \right).$$

Often we can let $b \to +\infty$ and omit the terms involving $b$ on the right-hand side. For more information see §4.12.

$\gamma$ is defined by

$$\gamma = \lim_{N \to \infty} \left( H_N - \ln N \right), \tag{4.31}$$

where $H_N = \sum_{1 \le j \le N} 1/j$ is a harmonic number. However, Eqn. (4.31) converges slowly, so to evaluate $\gamma$ accurately we need to accelerate the convergence. This can be done using the Euler-Maclaurin formula. The idea is to split the sum $H_N$ into two parts:

$$H_N = H_{p-1} + \sum_{j=p}^{N} \frac{1}{j}.$$

We approximate the second sum using the Euler-Maclaurin formula[7] with $a = p$, $b = N$, $f(x) = 1/x$, then let $N \to +\infty$. The result is

$$\gamma \sim H_p - \ln p + \sum_{k \ge 1} \frac{B_{2k}}{2k} p^{-2k}. \tag{4.32}$$

If $p$ and the number of terms in the asymptotic expansion are chosen judiciously, this gives a good algorithm for computing $\gamma$ (though not the best algorithm: see §4.12 for a faster algorithm that uses properties of Bessel functions).

Here is another example. The Riemann zeta-function $\zeta(s)$ is defined for $s \in \mathbb{C}$, $\Re(s) > 1$, by

$$\zeta(s) = \sum_{j=1}^{\infty} j^{-s}, \tag{4.33}$$

and by analytic continuation for other $s \ne 1$. $\zeta(s)$ may be evaluated to any desired precision if $m$ and $p$ are chosen large enough in the Euler-Maclaurin formula

$$\zeta(s) = \sum_{j=1}^{p-1} j^{-s} + \frac{p^{-s}}{2} + \frac{p^{1-s}}{s-1} + \sum_{k=1}^{m} T_{k,p}(s) + E_{m,p}(s), \tag{4.34}$$

where

$$T_{k,p}(s) = \frac{B_{2k}}{(2k)!} \, p^{1-s-2k} \prod_{j=0}^{2k-2} (s+j), \tag{4.35}$$

$$|E_{m,p}(s)| \; < \; |T_{m+1,p}(s) \, (s+2m+1)/(\sigma+2m+1)|, \tag{4.36}$$

$m \geq 0$, $p \geq 1$, $\sigma = \Re(s) > -(2m+1)$, and the $B_{2k}$ are Bernoulli numbers.

In arbitrary-precision computations we must be able to compute as many terms of an asymptotic expansion as are required to give the desired accuracy. It is easy to see that, if $m$ in (4.34) is bounded as the precision $n$ goes to $\infty$, then $p$ has to increase as an exponential function of $n$. To evaluate $\zeta(s)$ from (4.34) to precision $n$ in time polynomial in $n$, both $m$ and $p$ must tend to infinity with $n$. Thus, the Bernoulli numbers $B_2, \ldots, B_{2m}$ can not be stored in a table of fixed size,[8] but must be computed when needed (see §4.7). For this reason we can not use asymptotic expansions when the general form of the coefficients is unknown or the coefficients are too difficult to evaluate. Often there is a related expansion with known and relatively simple coefficients. For example, the asymptotic expansion (4.38) for $\ln \Gamma(x)$ has coefficients related to the Bernoulli numbers, like the expansion (4.34) for $\zeta(s)$, and thus is simpler to implement than Stirling's asymptotic expansion for $\Gamma(x)$ (see Exercise 4.42).

Consider the computation of the error function $\mathrm{erf}(x)$. As seen in §4.4, the series (4.22) and (4.23) are not satisfactory for large $|x|$, since they require $\Omega(x^2)$ terms. For example, to evaluate $\mathrm{erf}(1000)$ with an accuracy of six digits, Eqn. (4.22) requires at least $2\,718\,279$ terms! Instead, we may use an asymptotic expansion. The complementary error function $\mathrm{erfc}(x) = 1 - \mathrm{erf}(x)$ satisfies

$$\mathrm{erfc}(x) \sim \frac{e^{-x^2}}{x\sqrt{\pi}} \sum_{j=0}^{k} (-1)^j \frac{(2j)!}{j!} (2x)^{-2j}, \qquad (4.37)$$

with the error bounded in absolute value by the next term and of the same sign. In the case $x = 1000$, the term for $j = 1$ of the sum equals $-0.5 \times 10^{-6}$; thus $e^{-x^2}/(x\sqrt{\pi})$ is an approximation to $\mathrm{erfc}(x)$ with an accuracy of six digits. Because $\mathrm{erfc}(1000) \approx 1.86 \times 10^{-434\,298}$ is very small, this gives an *extremely* accurate approximation to $\mathrm{erf}(1000)$.

For a function like the error function where both a power series (at $x = 0$) and an asymptotic expansion (at $x = \infty$) are available, we might prefer to use the former or the latter, depending on the value of the argument and on the desired precision. We study here in some detail the case of the error function, since it is typical.

---

[8]In addition, we would have to store them as exact rationals, taking $\sim m^2 \lg m$ bits of storage, since a floating-point representation would not be convenient unless the target precision $n$ were known in advance. See §4.7.2 and Exercise 4.37.

The sum in (4.37) is divergent, since its $j$-th term is $\sim \sqrt{2}(j/ex^2)^j$. We need to show that the smallest term is $O(2^{-n})$ in order to be able to deduce an $n$-bit approximation to $\mathrm{erfc}(x)$. The terms decrease while $j < x^2 + 1/2$, so the minimum is obtained for $j \approx x^2$, and is of order $e^{-x^2}$, thus we need $x > \sqrt{n \ln 2}$. For example, for $n = 10^6$ bits this yields $x > 833$. However, since $\mathrm{erfc}(x)$ is small for large $x$, say $\mathrm{erfc}(x) \approx 2^{-\lambda}$, we need only $m = n - \lambda$ correct bits of $\mathrm{erfc}(x)$ to get $n$ correct bits of $\mathrm{erf}(x) = 1 - \mathrm{erfc}(x)$.

Consider $x$ fixed and $j$ varying in the terms in the sums (4.22) and (4.37). For $j < x^2$, $x^{2j}/j!$ is an *increasing* function of $j$, but $(2j)!/(j!(4x^2)^j)$ is a *decreasing* function of $j$. In this region the terms in Eqn. (4.37) are decreasing. Thus, comparing the series (4.22) and (4.37), we see that the latter should always be used if it can give sufficient accuracy. Similarly, (4.37) should if possible be used in preference to (4.23), as the magnitudes of corresponding terms in (4.22) and in (4.23) are similar.

---

**Algorithm 4.2** Erf

**Input:** positive floating-point number $x$, integer $n$
**Output:** an $n$-bit approximation to $\mathrm{erf}(x)$

$\quad m \leftarrow \lceil n - (x^2 + \ln x + (\ln \pi)/2)/(\ln 2) \rceil$
$\quad$ **if** $(m + 1/2)\ln(2) < x^2$ **then**
$\quad\quad t \leftarrow \mathrm{erfc}(x)$ with the asymptotic expansion (4.37) and precision $m$
$\quad\quad$ return $1 - t$ (in precision $n$)
$\quad$ **else if** $x < 1$ **then**
$\quad\quad$ compute $\mathrm{erf}(x)$ with the power series (4.22) in precision $n$
$\quad$ **else**
$\quad\quad$ compute $\mathrm{erf}(x)$ with the power series (4.23) in precision $n$.

---

Algorithm **Erf** computes $\mathrm{erf}(x)$ for real positive $x$ (for other real $x$, use the fact that $\mathrm{erf}(x)$ is an odd function, so $\mathrm{erf}(-x) = -\mathrm{erf}(x)$ and $\mathrm{erf}(0) = 0$). In Algorithm **Erf**, the number of terms needed if Eqn. (4.22) or Eqn. (4.23) is used is approximately the unique positive root $j_0$ (rounded up to the next integer) of

$$j(\ln j - 2\ln x - 1) = n \ln 2,$$

so $j_0 > ex^2$. On the other hand, if Eqn. (4.37) is used, then the number of terms $k < x^2 + 1/2$ (since otherwise the terms start increasing). The condition $(m + 1/2)\ln(2) < x^2$ in the algorithm ensures that the asymptotic expansion can give $m$-bit accuracy.

Here is an example: for $x = 800$ and a precision of one million bits, Equation (4.23) requires about $j_0 = 2\,339\,601$ terms. Eqn. (4.37) tells us that $\operatorname{erfc}(x) \approx 2^{-923\,335}$; thus we need only $m = 76\,665$ bits of precision for $\operatorname{erfc}(x)$; in this case Eqn. (4.37) requires only about $k = 10\,375$ terms. Note that using Eqn. (4.22) would be slower than using Eqn. (4.23), because we would have to compute about the same number of terms, but with higher precision, to compensate for cancellation. We recommend using Eqn. (4.22) only if $|x|$ is small enough that any cancellation is insignificant (for example, if $|x| < 1$).

Another example, closer to the boundary: for $x = 589$, still with $n = 10^6$, we have $m = 499\,489$, which gives $j_0 = 1\,497\,924$, and $k = 325\,092$. For somewhat smaller $x$ (or larger $n$) it might be desirable to use the continued fraction (4.40), see Exercise 4.31.

Occasionally an asymptotic expansion can be used to obtain arbitrarily high precision. For example, consider the computation of $\ln \Gamma(x)$. For large positive $x$, we can use Stirling's asymptotic expansion

$$\ln \Gamma(x) = \left(x - \frac{1}{2}\right) \ln x - x + \frac{\ln(2\pi)}{2} + \sum_{k=1}^{m-1} \frac{B_{2k}}{2k(2k-1)x^{2k-1}} + R_m(x), \quad (4.38)$$

where $R_m(x)$ is less in absolute value than the first term neglected, that is

$$\frac{B_{2m}}{2m(2m-1)x^{2m-1}},$$

and has the same sign.[9] The ratio of successive terms $t_k$ and $t_{k+1}$ of the sum is

$$\frac{t_{k+1}}{t_k} \approx -\left(\frac{k}{\pi x}\right)^2,$$

so the terms start to increase in absolute value for (approximately) $k > \pi x$. This gives a bound on the accuracy attainable, in fact

$$\ln |R_m(x)| > -2\pi x \ln(x) + O(x).$$

However, because $\Gamma(x)$ satisfies the functional equation $\Gamma(x+1) = x\Gamma(x)$, we can take $x' = x + \delta$ for some sufficiently large $\delta \in \mathbb{N}$, evaluate $\ln \Gamma(x')$ using the asymptotic expansion, and then compute $\ln \Gamma(x)$ from the functional equation. See Exercise 4.21.

---

[9]The asymptotic expansion is also valid for $x \in \mathbb{C}$, $|\arg x| < \pi$, $x \neq 0$, but the bound on the error term $R_m(x)$ in this case is more complicated. See for example [1, 6.1.42].

# 4.6   Continued Fractions

In §4.5 we considered the exponential integral $E_1(x)$. This can be computed using the *continued fraction*

$$e^x \, E_1(x) = \cfrac{1}{x + \cfrac{1}{1 + \cfrac{1}{x + \cfrac{2}{1 + \cfrac{2}{x + \cfrac{3}{1 + \cdots}}}}}}.$$

Writing continued fractions in this way takes a lot of space, so instead we use the shorthand notation

$$e^x \, E_1(x) = \frac{1}{x+} \, \frac{1}{1+} \, \frac{1}{x+} \, \frac{2}{1+} \, \frac{2}{x+} \, \frac{3}{1+} \, \cdots. \tag{4.39}$$

Another example is

$$\operatorname{erfc}(x) = \left( \frac{e^{-x^2}}{\sqrt{\pi}} \right) \frac{1}{x+} \, \frac{1/2}{x+} \, \frac{2/2}{x+} \, \frac{3/2}{x+} \, \frac{4/2}{x+} \, \frac{5/2}{x+} \, \cdots. \tag{4.40}$$

Formally, a continued fraction

$$f = b_0 + \frac{a_1}{b_1+} \, \frac{a_2}{b_2+} \, \frac{a_3}{b_3+} \, \cdots \in \widehat{\mathbb{C}}$$

is defined by two sequences $(a_j)_{j \in \mathbb{N}^*}$ and $(b_j)_{j \in \mathbb{N}}$, where $a_j, \ b_j \in \mathbb{C}$. Here $\widehat{\mathbb{C}} = \mathbb{C} \cup \{\infty\}$ is the set of *extended* complex numbers.[10] The expression $f$ is defined to be $\lim_{k \to \infty} f_k$, if the limit exists, where

$$f_k = b_0 + \frac{a_1}{b_1+} \, \frac{a_2}{b_2+} \, \frac{a_3}{b_3+} \, \cdots \, \frac{a_k}{b_k} \tag{4.41}$$

is the finite continued fraction — called the *k-th approximant* — obtained by truncating the infinite continued fraction after $k$ quotients.

---

[10]Arithmetic operations on $\mathbb{C}$ are extended to $\widehat{\mathbb{C}}$ in the obvious way, for example $1/0 = 1 + \infty = 1 \times \infty = \infty$, $1/\infty = 0$. Note that $0/0$, $0 \times \infty$ and $\infty \pm \infty$ are undefined.

Sometimes continued fractions are preferable, for computational purposes, to power series or asymptotic expansions. For example, Euler's continued fraction (4.39) converges for all real $x > 0$, and is better for computation of $E_1(x)$ than the power series (4.26) in the region where the power series suffers from catastrophic cancellation but the asymptotic expansion (4.27) is not sufficiently accurate. Convergence of (4.39) is slow if $x$ is small, so (4.39) is preferred for precision $n$ evaluation of $E_1(x)$ only when $x$ is in a certain interval, say $x \in (c_1 n, c_2 n)$, $c_1 \approx 0.1$, $c_2 = \ln 2 \approx 0.6931$ (see Exercise 4.24).

Continued fractions may be evaluated by either forward or backward recurrence relations. Consider the finite continued fraction

$$y = \frac{a_1}{b_1+} \frac{a_2}{b_2+} \frac{a_3}{b_3+} \cdots \frac{a_k}{b_k}. \tag{4.42}$$

The backward recurrence is $R_k = 1$, $R_{k-1} = b_k$,

$$R_j = b_{j+1} R_{j+1} + a_{j+2} R_{j+2} \qquad (j = k - 2, \ldots, 0), \tag{4.43}$$

and $y = a_1 R_1 / R_0$, with invariant

$$\frac{R_j}{R_{j-1}} = \frac{1}{b_j+} \frac{a_{j+1}}{b_{j+1}+} \cdots \frac{a_k}{b_k}.$$

The forward recurrence is $P_0 = 0$, $P_1 = a_1$, $Q_0 = 1$, $Q_1 = b_1$,

$$\left. \begin{array}{l} P_j = b_j P_{j-1} + a_j P_{j-2} \\ Q_j = b_j Q_{j-1} + a_j Q_{j-2} \end{array} \right\} \qquad (j = 2, \ldots, k), \tag{4.44}$$

and $y = P_k / Q_k$ (see Exercise 4.26).

The advantage of evaluating an infinite continued fraction such as (4.39) via the forward recurrence is that the cutoff $k$ need not be chosen in advance; we can stop when $|D_k|$ is sufficiently small, where

$$D_k = \frac{P_k}{Q_k} - \frac{P_{k-1}}{Q_{k-1}}. \tag{4.45}$$

The main disadvantage of the forward recurrence is that twice as many arithmetic operations are required as for the backward recurrence with the same value of $k$. Another disadvantage is that the forward recurrence may be less numerically stable than the backward recurrence.

If we are working with variable-precision floating-point arithmetic which is much more expensive than single-precision floating-point, then a useful strategy is to use the forward recurrence with single-precision arithmetic (scaled to avoid overflow/underflow) to estimate $k$, then use the backward recurrence with variable-precision arithmetic. One trick is needed: to evaluate $D_k$ using scaled single-precision we use the recurrence

$$\left.\begin{array}{l} D_1 = a_1/b_1, \\ D_j = -a_j Q_{j-2} D_{j-1}/Q_j \qquad (j = 2, 3, \ldots) \end{array}\right\} \tag{4.46}$$

which avoids the cancellation inherent in (4.45).

By analogy with the case of power series with decreasing terms that alternate in sign, there is one case in which it is possible to give a simple *a posteriori* bound for the error occurred in truncating a continued fraction. Let $f$ be a convergent continued fraction with approximants $f_k$ as in (4.41). Then

**Theorem 4.6.1** *If $a_j > 0$ and $b_j > 0$ for all $j \in \mathbb{N}^*$, then the sequence $(f_{2k})_{k \in \mathbb{N}}$ of even order approximants is strictly increasing, and the sequence $(f_{2k+1})_{k \in \mathbb{N}}$ of odd order approximants is strictly decreasing. Thus*

$$f_{2k} < f < f_{2k+1}$$

*and*

$$\left| f - \frac{f_{m-1} + f_m}{2} \right| < \left| \frac{f_m - f_{m-1}}{2} \right|$$

*for all $m \in \mathbb{N}^*$.*

In general, if the conditions of Theorem 4.6.1 are not satisfied, then it is difficult to give simple, sharp error bounds. Power series and asymptotic series are usually much easier to analyse than continued fractions.

## 4.7 Recurrence Relations

The evaluation of special functions by continued fractions is a special case of their evaluation by recurrence relations. To illustrate this, we consider the Bessel functions of the first kind, $J_\nu(x)$. Here $\nu$ and $x$ can in general be complex, but we restrict attention to the case $\nu \in \mathbb{Z}$, $x \in \mathbb{R}$. The functions $J_\nu(x)$ can be defined in several ways, for example by the generating function (elegant but only useful for $\nu \in \mathbb{Z}$):

$$\exp\left(\frac{x}{2}\left(t - \frac{1}{t}\right)\right) = \sum_{\nu=-\infty}^{+\infty} t^\nu J_\nu(x), \tag{4.47}$$

or by the power series (also valid if $\nu \notin \mathbb{Z}$):

$$J_\nu(x) = \left(\frac{x}{2}\right)^\nu \sum_{j=0}^\infty \frac{(-x^2/4)^j}{j!\,\Gamma(\nu+j+1)}\,. \tag{4.48}$$

We also need Bessel functions of the second kind (sometimes called Neumann functions or Weber functions) $Y_\nu(x)$, which may be defined by:

$$Y_\nu(x) = \lim_{\mu \to \nu} \frac{J_\mu(x)\cos(\pi\mu) - J_{-\mu}(x)}{\sin(\pi\mu)}\,. \tag{4.49}$$

Both $J_\nu(x)$ and $Y_\nu(x)$ are solutions of Bessel's differential equation

$$x^2 y'' + xy' + (x^2 - \nu^2)y = 0. \tag{4.50}$$

## 4.7.1   Evaluation of Bessel Functions

The Bessel functions $J_\nu(x)$ satisfy the recurrence relation

$$J_{\nu-1}(x) + J_{\nu+1}(x) = \frac{2\nu}{x}J_\nu(x). \tag{4.51}$$

Dividing both sides by $J_\nu(x)$, we see that

$$\frac{J_{\nu-1}(x)}{J_\nu(x)} = \frac{2\nu}{x} - 1 \left/ \frac{J_\nu(x)}{J_{\nu+1}(x)} \right.,$$

which gives a continued fraction for the ratio $J_\nu(x)/J_{\nu-1}(x)$ $(\nu \geq 1)$:

$$\frac{J_\nu(x)}{J_{\nu-1}(x)} = \frac{1}{2\nu/x-} \; \frac{1}{2(\nu+1)/x-} \; \frac{1}{2(\nu+2)/x-} \; \cdots. \tag{4.52}$$

However, (4.52) is not immediately useful for evaluating the Bessel functions $J_0(x)$ or $J_1(x)$, as it only gives their ratio.

The recurrence (4.51) may be evaluated backwards by *Miller's algorithm*. The idea is to start at some sufficiently large index $\nu'$, take $f_{\nu'+1} = 0$, $f_{\nu'} = 1$, and evaluate the recurrence

$$f_{\nu-1} + f_{\nu+1} = \frac{2\nu}{x}\,f_\nu \tag{4.53}$$

backwards to obtain $f_{\nu'-1}, \cdots, f_0$. However, (4.53) is the same recurrence as (4.51), so we expect to obtain $f_0 \approx cJ_0(x)$ where $c$ is some scale factor. We can use the identity

$$J_0(x) + 2 \sum_{\nu=1}^{\infty} J_{2\nu}(x) = 1 \qquad (4.54)$$

to determine $c$.

To understand why Miller's algorithm works, and why evaluation of the recurrence (4.51) in the forward direction is numerically unstable for $\nu > x$, we observe that the recurrence (4.53) has two independent solutions: the desired solution $J_\nu(x)$, and an undesired solution $Y_\nu(x)$, where $Y_\nu(x)$ is a Bessel function of the *second kind*, see Eqn. (4.49). The general solution of the recurrence (4.53) is a linear combination of the special solutions $J_\nu(x)$ and $Y_\nu(x)$. Due to rounding errors, the computed solution will also be a linear combination, say $aJ_\nu(x) + bY_\nu(x)$. Since $|Y_\nu(x)|$ increases exponentially with $\nu$ when $\nu > ex/2$, but $|J_\nu(x)|$ is bounded, the unwanted component will increase exponentially if we use the recurrence in the forward direction, but decrease if we use it in the backward direction.

More precisely, we have

$$J_\nu(x) \sim \frac{1}{\sqrt{2\pi\nu}} \left(\frac{ex}{2\nu}\right)^\nu \quad \text{and} \quad Y_\nu(x) \sim -\sqrt{\frac{2}{\pi\nu}} \left(\frac{2\nu}{ex}\right)^\nu \qquad (4.55)$$

as $\nu \to +\infty$ with $x$ fixed. Thus, when $\nu$ is large and greater than $ex/2$, $J_\nu(x)$ is small and $|Y_\nu(x)|$ is large.

Miller's algorithm seems to be the most effective method in the region where the power series (4.48) suffers from catastrophic cancellation but asymptotic expansions are not sufficiently accurate. For more on Miller's algorithm, see §4.12.

### 4.7.2 Evaluation of Bernoulli and Tangent numbers

In §4.5, Equations (4.35) and (4.38), the Bernoulli numbers $B_{2k}$ or scaled Bernoulli numbers $C_k = B_{2k}/(2k)!$ were required. These constants can be defined by the generating functions

$$\sum_{k=0}^{\infty} B_k \frac{x^k}{k!} = \frac{x}{e^x - 1}, \qquad (4.56)$$

$$\sum_{k=0}^{\infty} C_k x^{2k} = \frac{x}{e^x - 1} + \frac{x}{2} = \frac{x/2}{\tanh(x/2)} \, . \tag{4.57}$$

Multiplying both sides of (4.56) or (4.57) by $e^x - 1$, and equating coefficients, gives the recurrence relations

$$B_0 = 1, \ \sum_{j=0}^{k} \binom{k+1}{j} B_j = 0 \ \text{ for } \ k > 0, \tag{4.58}$$

and

$$\sum_{j=0}^{k} \frac{C_j}{(2k+1-2j)!} = \frac{1}{2\,(2k)!} \, . \tag{4.59}$$

These recurrences, or slight variants with similar numerical properties, have often been used to evaluate Bernoulli numbers.

In this chapter our philosophy is that the required precision is not known in advance, so it is not possible to precompute the Bernoulli numbers and store them in a table once and for all. Thus, we need a good algorithm for computing them at runtime.

Unfortunately, forward evaluation of the recurrence (4.58), or the corresponding recurrence (4.59) for the scaled Bernoulli numbers, is numerically unstable: using precision $n$ the relative error in the computed $B_{2k}$ or $C_k$ is of order $4^k 2^{-n}$: see Exercise 4.35.

Despite its numerical instability, use of (4.59) may give the $C_k$ to acceptable accuracy if they are only needed to generate coefficients in an Euler-Maclaurin expansion whose successive terms diminish by at least a factor of four (or if the $C_k$ are computed using exact rational arithmetic). If the $C_k$ are required to precision $n$, then (4.59) should be used with sufficient guard digits, or (better) a more stable recurrence should be used. If we multiply both sides of (4.57) by $\sinh(x/2)/x$ and equate coefficients, we get the recurrence

$$\sum_{j=0}^{k} \frac{C_j}{(2k+1-2j)!\,4^{k-j}} = \frac{1}{(2k)!\,4^k} \, . \tag{4.60}$$

If (4.60) is used to evaluate $C_k$, using precision $n$ arithmetic, the relative error is only $O(k^2 2^{-n})$. Thus, use of (4.60) gives a stable algorithm for evaluating the scaled Bernoulli numbers $C_k$ (and hence, if desired, the Bernoulli numbers).

An even better, and perfectly stable, way to compute Bernoulli numbers is to exploit their relationship with the *tangent numbers $T_j$*, defined by

$$\tan x = \sum_{j \geq 1} T_j \frac{x^{2j-1}}{(2j-1)!}.  \tag{4.61}$$

The tangent numbers are positive integers and can be expressed in terms of Bernoulli numbers:

$$T_j = (-1)^{j-1} 2^{2j} \left(2^{2j} - 1\right) \frac{B_{2j}}{2j}.  \tag{4.62}$$

Conversely, the Bernoulli numbers can be expressed in terms of tangent numbers:

$$B_j = \begin{cases} 1 & \text{if } j = 0, \\ -1/2 & \text{if } j = 1, \\ (-1)^{j/2-1} j T_{j/2}/(4^j - 2^j) & \text{if } j > 0 \text{ is even}, \\ 0 & \text{otherwise}. \end{cases}$$

Eqn. (4.62) shows that the odd primes in the denominator of the Bernoulli number $B_{2j}$ must be divisors of $2^{2j} - 1$. In fact, this is a consequence of Fermat's little theorem and the Von Staudt-Clausen theorem, which says that the primes $p$ dividing the denominator of $B_{2j}$ are precisely those for which $(p-1)|2j$ (see §4.12).

We now derive a recurrence that can be used to compute tangent numbers, using only integer arithmetic. For brevity write $t = \tan x$ and $D = \mathrm{d}/\mathrm{d}x$. Then $Dt = \sec^2 x = 1 + t^2$. It follows that $D(t^n) = nt^{n-1}(1 + t^2)$ for all $n \in \mathbb{N}^*$.

It is clear that $D^n t$ is a polynomial in $t$, say $P_n(t)$. For example, $P_0(t) = t$, $P_1(t) = 1 + t^2$, *etc.* Write $P_n(t) = \sum_{j \geq 0} p_{n,j} t^j$. From the recurrence $P_n(t) = DP_{n-1}(t)$, and the formula for $D(t^n)$ just noted, we see that $\deg(P_n) = n + 1$ and

$$\sum_{j \geq 0} p_{n,j} t^j = \sum_{j \geq 0} j p_{n-1,j} t^{j-1}(1 + t^2),$$

so

$$p_{n,j} = (j-1)p_{n-1,j-1} + (j+1)p_{n-1,j+1}  \tag{4.63}$$

for all $n \in \mathbb{N}^*$. Using (4.63) it is straightforward to compute the coefficients of the polynomials $P_1(t)$, $P_2(t)$, *etc.*

---

**Algorithm 4.3** TangentNumbers

---

**Input:** positive integer $m$

**Output:** Tangent numbers $T_1, \ldots, T_m$

$\quad T_1 \leftarrow 1$

$\quad$ **for** $k$ **from** 2 **to** $m$ **do**

$\quad\quad T_k \leftarrow (k-1)T_{k-1}$

$\quad$ **for** $k$ **from** 2 **to** $m$ **do**

$\quad\quad$ **for** $j$ **from** $k$ **to** $m$ **do**

$\quad\quad\quad T_j \leftarrow (j-k)T_{j-1} + (j-k+2)T_j$

$\quad$ return $T_1, T_2, \ldots, T_m$.

---

Observe that, since $\tan x$ is an odd function of $x$, the polynomials $P_{2k}(t)$ are odd, and the polynomials $P_{2k+1}(t)$ are even. Equivalently, $p_{n,j} = 0$ if $n + j$ is even.

We are interested in the tangent numbers $T_k = P_{2k-1}(0) = p_{2k-1,0}$. Using the recurrence (4.63) but avoiding computation of the coefficients that are known to vanish, we obtain Algorithm **TangentNumbers** for the in-place computation of tangent numbers. Note that this algorithm uses only arithmetic on non-negative integers. If implemented with single-precision integers, there may be problems with overflow as the tangent numbers grow rapidly. If implemented using floating-point arithmetic, it is numerically stable because there is no cancellation. An analogous algorithm **SecantNumbers** is the topic of Exercise 4.40.

The tangent numbers grow rapidly because the generating function $\tan x$ has poles at $x = \pm\pi/2$. Thus, we expect $T_k$ to grow roughly like $(2k-1)! \, (2/\pi)^{2k}$. More precisely,

$$\frac{T_k}{(2k-1)!} = \frac{2^{2k+1}(1-2^{-2k})\zeta(2k)}{\pi^{2k}}, \tag{4.64}$$

where $\zeta(s)$ is the usual Riemann zeta-function, and

$$(1-2^{-s})\zeta(s) = 1 + 3^{-s} + 5^{-s} + \cdots$$

is sometimes called the *odd* zeta-function.

The Bernoulli numbers also grow rapidly, but not quite as fast as the tangent numbers, because the singularities of the generating function (4.56) are further from the origin (at $\pm 2i\pi$ instead of $\pm\pi/2$). It is well-known that the Riemann zeta-function for even non-negative integer arguments can be expressed in terms of Bernoulli numbers – the relation is

$$(-1)^{k-1} \frac{B_{2k}}{(2k)!} = \frac{2\zeta(2k)}{(2\pi)^{2k}} \,. \tag{4.65}$$

Since $\zeta(2k) = 1 + O(4^{-k})$ as $k \to +\infty$, we see that

$$|B_{2k}| \sim \frac{2\,(2k)!}{(2\pi)^{2k}} \,. \tag{4.66}$$

It is easy to see that (4.64) and (4.65) are equivalent, in view of the relation (4.62).

An asymptotically fast way of computing Bernoulli numbers is the topic of Exercise 4.41. For yet another way of computing Bernoulli numbers, using very little space, see §4.10.

## 4.8  Arithmetic-Geometric Mean

The (theoretically) fastest known methods for very large precision $n$ use the arithmetic-geometric mean (AGM) iteration of Gauss and Legendre. The AGM is another nonlinear recurrence, important enough to treat separately. Its complexity is $O(M(n) \ln n)$; the implicit constant here can be quite large, so other methods are better for small $n$.

Given $(a_0, b_0)$, the AGM iteration is defined by

$$(a_{j+1}, b_{j+1}) = \left( \frac{a_j + b_j}{2}, \sqrt{a_j b_j} \right) .$$

For simplicity we only consider real, positive starting values $(a_0, b_0)$ here (for complex starting values, see §§4.8.5, 4.12). The AGM iteration converges *quadratically* to a limit which we denote by $\mathrm{AGM}(a_0, b_0)$.

The AGM is useful because:

1. it converges quadratically. Eventually the number of correct digits doubles at each iteration, so only $O(\log n)$ iterations are required;

2. each iteration takes time $O(M(n))$ because the square root can be computed in time $O(M(n))$ by Newton's method (see §3.5 and §4.2.3);

3. if we take suitable starting values $(a_0, b_0)$, the result $\text{AGM}(a_0, b_0)$ can be used to compute logarithms (directly) and other elementary functions (less directly), as well as constants such as $\pi$ and $\ln 2$.

## 4.8.1   Elliptic Integrals

The theory of the AGM iteration is intimately linked to the theory of elliptic integrals. The *complete elliptic integral of the first kind* is defined by

$$K(k) = \int_0^{\pi/2} \frac{d\theta}{\sqrt{1 - k^2 \sin^2 \theta}} = \int_0^1 \frac{dt}{\sqrt{(1 - t^2)(1 - k^2 t^2)}}, \qquad (4.67)$$

and the *complete elliptic integral of the second kind* is

$$E(k) = \int_0^{\pi/2} \sqrt{1 - k^2 \sin^2 \theta} \, d\theta = \int_0^1 \sqrt{\frac{1 - k^2 t^2}{1 - t^2}} \, dt,$$

where $k \in [0, 1]$ is called the *modulus* and $k' = \sqrt{1 - k^2}$ is the *complementary modulus*. It is traditional (though confusing as the prime does not denote differentiation) to write $K'(k)$ for $K(k')$ and $E'(k)$ for $E(k')$.

**The Connection With Elliptic Integrals.**   Gauss discovered that

$$\frac{1}{\text{AGM}(1, k)} = \frac{2}{\pi} K'(k). \qquad (4.68)$$

This identity can be used to compute the elliptic integral $K$ rapidly via the AGM iteration. We can also use it to compute logarithms. From the definition (4.67), we see that $K(k)$ has a series expansion that converges for $|k| < 1$ (in fact $K(k) = (\pi/2)F(1/2, 1/2; 1; k^2)$ is a hypergeometric function). For small $k$ we have

$$K(k) = \frac{\pi}{2} \left( 1 + \frac{k^2}{4} + O(k^4) \right). \qquad (4.69)$$

It can also be shown that

$$K'(k) = \frac{2}{\pi} \ln \left( \frac{4}{k} \right) K(k) - \frac{k^2}{4} + O(k^4). \qquad (4.70)$$

### 4.8.2    First AGM Algorithm for the Logarithm

From the formulæ (4.68), (4.69) and (4.70), we easily get

$$\frac{\pi/2}{\mathrm{AGM}(1, k)} = \ln\left(\frac{4}{k}\right)\left(1 + O(k^2)\right). \qquad (4.71)$$

Thus, if $x = 4/k$ is large, we have

$$\ln(x) = \frac{\pi/2}{\mathrm{AGM}(1, 4/x)}\left(1 + O\left(\frac{1}{x^2}\right)\right).$$

If $x \geq 2^{n/2}$, we can compute $\ln(x)$ to precision $n$ using the AGM iteration. It takes about $2\lg(n)$ iterations to converge if $x \in [2^{n/2}, 2^n]$.

Note that we need the constant $\pi$, which could be computed by using our formula twice with slightly different arguments $x_1$ and $x_2$, then taking differences to approximate $(\mathrm{d}\ln(x)/\mathrm{d}x)/\pi$ at $x_1$ (see Exercise 4.44). More efficient is to use the *Brent-Salamin* (or *Gauss-Legendre*) algorithm, which is based on the AGM and the Legendre relation

$$EK' + E'K - KK' = \frac{\pi}{2}. \qquad (4.72)$$

**Argument Expansion.**    If $x$ is not large enough, we can compute

$$\ln(2^{\ell}x) = \ell\,\ln 2 + \ln x$$

by the AGM method (assuming the constant $\ln 2$ is known). Alternatively, if $x > 1$, we can square $x$ enough times and compute

$$\ln\left(x^{2^{\ell}}\right) = 2^{\ell}\ln(x).$$

This method with $x = 2$ gives a way of computing $\ln 2$, assuming we already know $\pi$.

**The Error Term.**    The $O(k^2)$ error term in the formula (4.71) is a nuisance. A rigorous bound is

$$\left|\frac{\pi/2}{\mathrm{AGM}(1, k)} - \ln\left(\frac{4}{k}\right)\right| \leq 4k^2(8 - \ln k) \qquad (4.73)$$

for all $k \in (0, 1]$, and the bound can be sharpened to $0.37k^2(2.4 - \ln(k))$ if $k \in (0, 0.5]$.

The error $O(k^2 |\ln k|)$ makes it difficult to accelerate convergence by using a larger value of $k$ (i.e., a value of $x = 4/k$ smaller than $2^{n/2}$). There is an *exact* formula which is much more elegant and avoids this problem. Before giving this formula we need to define some *theta functions* and show how they can be used to parameterise the AGM iteration.

### 4.8.3   Theta Functions

We need the theta functions $\theta_2(q)$, $\theta_3(q)$ and $\theta_4(q)$, defined for $|q| < 1$ by:

$$\theta_2(q) = \sum_{n=-\infty}^{+\infty} q^{(n+1/2)^2} = 2q^{1/4} \sum_{n=0}^{+\infty} q^{n(n+1)}, \qquad (4.74)$$

$$\theta_3(q) = \sum_{n=-\infty}^{+\infty} q^{n^2} = 1 + 2 \sum_{n=1}^{+\infty} q^{n^2}, \qquad (4.75)$$

$$\theta_4(q) = \theta_3(-q) = 1 + 2 \sum_{n=1}^{+\infty} (-1)^n q^{n^2}. \qquad (4.76)$$

Note that the defining power series are sparse so it is easy to compute $\theta_2(q)$ and $\theta_3(q)$ for small $q$. Unfortunately, the rectangular splitting method of §4.4.3 does not help to speed up the computation.

The asymptotically fastest methods to compute theta functions use the AGM. However, we do not follow this trail because it would lead us in circles! We want to use theta functions to give starting values for the AGM iteration.

**Theta Function Identities.**   There are many classical identities involving theta functions. Two that are of interest to us are:

$$\frac{\theta_3^2(q) + \theta_4^2(q)}{2} = \theta_3^2(q^2) \quad \text{and} \quad \theta_3(q)\theta_4(q) = \theta_4^2(q^2).$$

The latter may be written as

$$\sqrt{\theta_3^2(q)\theta_4^2(q)} = \theta_4^2(q^2)$$

to show the connection with the AGM:

$$\begin{aligned}
\mathrm{AGM}(\theta_3^2(q), \theta_4^2(q)) &= \mathrm{AGM}(\theta_3^2(q^2), \theta_4^2(q^2)) = \cdots \\
&= \mathrm{AGM}(\theta_3^2(q^{2^k}), \theta_4^2(q^{2^k})) = \cdots = 1
\end{aligned}$$

for any $|q| < 1$. (The limit is 1 because $q^{2^k}$ converges to 0, thus both $\theta_3$ and $\theta_4$ converge to 1.) Apart from scaling, the AGM iteration is parameterised by $(\theta_3^2(q^{2^k}), \theta_4^2(q^{2^k}))$ for $k = 0, 1, 2, \ldots$

**The Scaling Factor.** Since $\mathrm{AGM}(\theta_3^2(q), \theta_4^2(q)) = 1$, and $\mathrm{AGM}(\lambda a, \lambda b) = \lambda \cdot \mathrm{AGM}(a, b)$, scaling gives $\mathrm{AGM}(1, k') = 1/\theta_3^2(q)$ if $k' = \theta_4^2(q)/\theta_3^2(q)$. Equivalently, since $\theta_2^4 + \theta_4^4 = \theta_3^4$ (Jacobi), $k = \theta_2^2(q)/\theta_3^2(q)$. However, we know (from (4.68) with $k \to k'$) that $1/\mathrm{AGM}(1, k') = 2K(k)/\pi$, so

$$K(k) = \frac{\pi}{2} \theta_3^2(q). \tag{4.77}$$

Thus, the theta functions are closely related to elliptic integrals. In the literature $q$ is usually called the *nome* associated with the modulus $k$.

**From $q$ to $k$ and $k$ to $q$.** We saw that $k = \theta_2^2(q)/\theta_3^2(q)$, which gives $k$ in terms of $q$. There is also a nice inverse formula which gives $q$ in terms of $k$: $q = \exp(-\pi K'(k)/K(k))$, or equivalently

$$\ln\left(\frac{1}{q}\right) = \frac{\pi K'(k)}{K(k)}. \tag{4.78}$$

**Sasaki and Kanada's Formula.** Substituting (4.68) and (4.77) with $k = \theta_2^2(q)/\theta_3^2(q)$ into (4.78) gives Sasaki and Kanada's elegant formula:

$$\ln\left(\frac{1}{q}\right) = \frac{\pi}{\mathrm{AGM}(\theta_2^2(q), \theta_3^2(q))}. \tag{4.79}$$

This leads to the following algorithm to compute $\ln x$.

### 4.8.4   Second AGM Algorithm for the Logarithm

Suppose $x$ is large. Let $q = 1/x$, compute $\theta_2(q^4)$ and $\theta_3(q^4)$ from their defining series (4.74) and (4.75), then compute $\mathrm{AGM}(\theta_2^2(q^4), \theta_3^2(q^4))$. Sasaki and Kanada's formula (with $q$ replaced by $q^4$ to avoid the $q^{1/4}$ term in the definition of $\theta_2(q)$) gives

$$\ln(x) = \frac{\pi/4}{\mathrm{AGM}(\theta_2^2(q^4), \theta_3^2(q^4))}\,.$$

There is a trade-off between increasing $x$ (by squaring or multiplication by a power of 2, see the paragraph on "Argument Expansion" in §4.8.2), and taking longer to compute $\theta_2(q^4)$ and $\theta_3(q^4)$ from their series. In practice it seems good to increase $x$ until $q = 1/x$ is small enough that $O(q^{36})$ terms are negligible. Then we can use

$$\theta_2(q^4) = 2\left(q + q^9 + q^{25} + O(q^{49})\right),$$

$$\theta_3(q^4) = 1 + 2\left(q^4 + q^{16} + O(q^{36})\right).$$

We need $x \geq 2^{n/36}$ which is much better than the requirement $x \geq 2^{n/2}$ for the first AGM algorithm. We save about four AGM iterations at the cost of a few multiplications.

**Implementation Notes.**   Since

$$\mathrm{AGM}(\theta_2^2, \theta_3^2) = \frac{\mathrm{AGM}(\theta_2^2 + \theta_3^2, 2\theta_2\theta_3)}{2}\,,$$

we can avoid the first square root in the AGM iteration. Also, it only takes two nonscalar multiplications to compute $2\theta_2\theta_3$ and $\theta_2^2 + \theta_3^2$ from $\theta_2$ and $\theta_3$: see Exercise 4.45. Another speedup is possible by trading the multiplications for squares, see §4.12.

**Drawbacks of the AGM.**   The AGM has three drawbacks:

1. the AGM iteration is *not* self-correcting, so we have to work with full precision (plus any necessary guard digits) throughout. In contrast, when using Newton's method or evaluating power series, many of the computations can be performed with reduced precision, which saves a $\log n$ factor (this amounts to using a *negative* number of guard digits);

2. the AGM with real arguments gives $\ln(x)$ directly. To obtain $\exp(x)$ we need to apply Newton's method (§4.2.5 and Exercise 4.6). To evaluate trigonometric functions such as $\sin(x)$, $\cos(x)$, $\arctan(x)$ we need to work with complex arguments, which increases the constant hidden in the "$O$" time bound. Alternatively, we can use Landen transformations for incomplete elliptic integrals, but this gives even larger constants;

3. because it converges so fast, it is difficult to speed up the AGM. At best we can save $O(1)$ iterations (see however §4.12).

### 4.8.5 The Complex AGM

In some cases the asymptotically fastest algorithms require the use of complex arithmetic to produce a real result. It would be nice to avoid this because complex arithmetic is significantly slower than real arithmetic. Examples where we seem to need complex arithmetic to get the asymptotically fastest algorithms are:

1. $\arctan(x)$, $\arcsin(x)$, $\arccos(x)$ via the AGM, using, for example,

$$\arctan(x) = \Im(\ln(1 + ix));$$

2. $\tan(x)$, $\sin(x)$, $\cos(x)$ using Newton's method and the above, or

$$\cos(x) + i\sin(x) = \exp(ix),$$

where the complex exponential is computed by Newton's method from the complex logarithm (see Eqn. (4.11)).

The theory that we outlined for the AGM iteration and AGM algorithms for $\ln(z)$ can be extended without problems to complex $z \notin (-\infty, 0]$, provided we always choose the square root with positive real part.

A complex multiplication takes three real multiplications (using Karatsuba's trick), and a complex squaring takes two real multiplications. We can do even better in the FFT domain, assuming that one multiplication of cost $M(n)$ is equivalent to three Fourier transforms. In this model a squaring costs $2M(n)/3$. A complex multiplication $(a + ib)(c + id) = (ac - bd) + i(ad + bc)$ requires four forward and two backward transforms, thus costs $2M(n)$. A complex squaring $(a + ib)^2 = (a + b)(a - b) + i(2ab)$ requires two forward and two backward transforms, thus costs $4M(n)/3$. Taking this into account, we get the asymptotic upper bounds relative to the cost of one multiplication given in Table 4.1 (0.666 should be interpreted as $\sim 2M(n)/3$, and so on). See §4.12 for details of the algorithms giving these constants.

| Operation | real | complex |
|-----------|------|---------|
| squaring | 0.666 | 1.333 |
| multiplication | 1.000 | 2.000 |
| reciprocal | 1.444 | 3.444 |
| division | 1.666 | 4.777 |
| square root | 1.333 | 5.333 |
| AGM iteration | 2.000 | 6.666 |
| log via AGM | $4.000 \lg n$ | $13.333 \lg n$ |

Table 4.1: Costs in the FFT domain

# 4.9   Binary Splitting

Since the asymptotically fastest algorithms for arctan, sin, cos, *etc.* have a large constant hidden in their time bound $O(M(n) \log n)$ (see "Drawbacks of the AGM", §4.8.4), page 176), it is interesting to look for other algorithms that may be competitive for a large range of precisions, even if not asymptotically optimal. One such algorithm (or class of algorithms) is based on *binary splitting* or the closely related *FEE method* (see §4.12). The time complexity of these algorithms is usually

$$O((\log n)^{\alpha} M(n))$$

for some constant $\alpha \geq 1$ depending on how fast the relevant power series converges, and also on the multiplication algorithm (classical, Karatsuba or quasi-linear).

**The Idea.**   Suppose we want to compute $\arctan(x)$ for rational $x = p/q$, where $p$ and $q$ are small integers and $|x| \leq 1/2$. The Taylor series gives

$$\arctan\left(\frac{p}{q}\right) \approx \sum_{0 \leq j \leq n/2} \frac{(-1)^j p^{2j+1}}{(2j+1)q^{2j+1}}.$$

The finite sum, if computed exactly, gives a rational approximation $P/Q$ to $\arctan(p/q)$, and

$$\log |Q| = O(n \log n).$$

(Note: the series for exp converges faster, so in this case we sum $\sim n/\ln n$ terms and get $\log |Q| = O(n)$.)

The finite sum can be computed by the "divide and conquer" strategy: sum the first half to get $P_1/Q_1$ say, and the second half to get $P_2/Q_2$, then

$$\frac{P}{Q} = \frac{P_1}{Q_1} + \frac{P_2}{Q_2} = \frac{P_1 Q_2 + P_2 Q_1}{Q_1 Q_2}.$$

The rationals $P_1/Q_1$ and $P_2/Q_2$ are computed by a recursive application of the same method, hence the term "binary splitting". If used with quadratic multiplication, this way of computing $P/Q$ does not help; however, fast multiplication speeds up the balanced products $P_1 Q_2$, $P_2 Q_1$, and $Q_1 Q_2$.

**Complexity.** The overall time complexity is

$$O\left(\sum_{k=1}^{\lceil \lg(n) \rceil} 2^k M(2^{-k} n \log n)\right) = O((\log n)^\alpha M(n)), \qquad (4.80)$$

where $\alpha = 2$ in the FFT range; in general $\alpha \le 2$ (see Exercise 4.47).

We can save a little by working to precision $n$ rather than $n \log n$ at the top levels; but we still have $\alpha = 2$ for quasi-linear multiplication.

In practice the multiplication algorithm would not be fixed but would depend on the size of the integers being multiplied. The complexity would depend on the algorithm(s) used at the top levels.

**Repeated Application of the Idea.** If $x \in (0, 0.25)$ and we want to compute $\arctan(x)$, we can approximate $x$ by a rational $p/q$ and compute $\arctan(p/q)$ as a first approximation to $\arctan(x)$, say $p/q \le x < (p+1)/q$. Now, from (4.17),

$$\tan(\arctan(x) - \arctan(p/q)) = \frac{x - p/q}{1 + px/q},$$

so

$$\arctan(x) = \arctan(p/q) + \arctan(\delta),$$

where

$$\delta = \frac{x - p/q}{1 + px/q} = \frac{qx - p}{q + px}.$$

We can apply the same idea to approximate $\arctan(\delta)$, until eventually we get a sufficiently accurate approximation to $\arctan(x)$. Note that $|\delta| < |x - p/q| < 1/q$, so it is easy to ensure that the process converges.

**Complexity of Repeated Application.**   If we use a sequence of about $\lg n$ rationals $p_1/q_1, p_2/q_2, \ldots$, where

$$q_i = 2^{2^i},$$

then the computation of each $\arctan(p_i/q_i)$ takes time $O((\log n)^\alpha M(n))$, and the overall time to compute $\arctan(x)$ is

$$O((\log n)^{\alpha+1} M(n)).$$

Indeed, we have $0 \le p_i < 2^{2^{i-1}}$, thus $p_i$ has at most $2^{i-1}$ bits, and $p_i/q_i$ as a rational has value $O(2^{-2^{i-1}})$ and size $O(2^i)$. The exponent $\alpha + 1$ is 2 or 3. Although this is not asymptotically as fast as AGM-based algorithms, the implicit constants for binary splitting are small and the idea is useful for quite large $n$ (at least $10^6$ decimal places).

**Generalisations.**   The idea of binary splitting can be generalised. For example, the Chudnovsky brothers gave a "bit-burst" algorithm which applies to fast evaluation of solutions of linear differential equations. This is described in §4.9.2.

## 4.9.1   A Binary Splitting Algorithm for $\sin$, $\cos$

In [45, Theorem 6.2], Brent claims an $O(M(n) \log^2 n)$ algorithm for $\exp x$ and $\sin x$, however the proof only covers the case of the exponential, and ends with "the proof of (6.28) is similar". He had in mind deducing $\sin x$ from a complex computation of $\exp(ix) = \cos x + i \sin x$. Algorithm **SinCos** is a variation of Brent's algorithm for $\exp x$ that computes $\sin x$ and $\cos x$ simultaneously, in a way that avoids computations with complex numbers. The simultaneous computation of $\sin x$ and $\cos x$ might be useful, for example, to compute $\tan x$ or a plane rotation through the angle $x$.

At step 2 of Algorithm **SinCos**, we have $x_j = y_j + x_{j+1}$, thus $\sin x_j = \sin y_j \cos x_{j+1} + \cos y_j \sin x_{j+1}$, and similarly for $\cos x_j$, explaining the formulæ used at step 6. Step 5 uses a binary splitting algorithm similar to the one

---

**Algorithm 4.4 SinCos**

**Input:** floating-point $0 < x < 1/2$, integer $n$

**Output:** an approximation of $\sin x$ and $\cos x$ with error $O(2^{-n})$

1: write $x \approx \sum_{i=0}^{k} p_i \cdot 2^{-2^{i+1}}$ where $0 \le p_i < 2^{2^i}$ and $k = \lceil \lg n \rceil - 1$

2: let $x_j = \sum_{i=j}^{k} p_i \cdot 2^{-2^{i+1}}$, with $x_{k+1} = 0$, and $y_j = p_j \cdot 2^{-2^{j+1}}$

3: $(S_{k+1}, C_{k+1}) \leftarrow (0, 1)$         ▷ $S_j$ is $\sin x_j$ and $C_j$ is $\cos x_j$

4: **for** $j$ **from** $k$ **downto** 0 **do**

5:     compute $\sin y_j$ and $\cos y_j$ using binary splitting

6:     $S_j \leftarrow \sin y_j \cdot C_{j+1} + \cos y_j \cdot S_{j+1}$,     $C_j \leftarrow \cos y_j \cdot C_{j+1} - \sin y_j \cdot S_{j+1}$

7: return $(S_0, C_0)$.

---

described above for $\arctan(p/q)$: $y_j$ is a small rational, or is small itself, so that all needed powers do not exceed $n$ bits in size. This algorithm has the same complexity $O(M(n) \log^2 n)$ as Brent's algorithm for $\exp x$.

## 4.9.2 The Bit-Burst Algorithm

The binary-splitting algorithms described above for $\arctan x$, $\exp x$, $\sin x$ rely on a functional equation: $\tan(x + y) = (\tan x + \tan y)/(1 - \tan x \tan y)$, $\exp(x + y) = \exp(x)\exp(y)$, $\sin(x + y) = \sin x \cos y + \sin y \cos x$. We describe here a more general algorithm, known as the "bit-burst" algorithm, which does not require such a functional equation. This algorithm applies to a class of functions known as *holonomic* functions. Other names are *differentiably finite* and *D-finite*.

A function $f(x)$ is said to be *holonomic* iff it satisfies a linear homogeneous differential equation with polynomial coefficients in $x$. Equivalently, the Taylor coefficients $u_k$ of $f$ satisfy a linear recurrence with coefficients polynomial in $k$. The set of holonomic functions is closed under the operations of addition and multiplication, but not necessarily under division. For example, the $\exp, \ln, \sin, \cos$ functions are holonomic, but $\tan$ is not.

An important subclass of holonomic functions is the hypergeometric functions, whose Taylor coefficients satisfy a recurrence $u_{k+1}/u_k = R(k)$, where $R(k)$ is a rational function of $k$ (see §4.4). This matches the second definition above, because we can write it as $u_{k+1}Q(k) - u_k P(k) = 0$ if $R(k) = P(k)/Q(k)$. Holonomic functions are much more general than hypergeomet-ric functions (see Exercise 4.48); in particular the ratio of two consecutive terms in a hypergeometric series has size $O(\log k)$ (as a rational number), but can be much larger for holonomic functions.

**Theorem 4.9.1** *If $f$ is holonomic and has no singularities on a finite, closed interval $[A, B]$, where $A < 0 < B$ and $f(0) = 0$, then $f(x)$ can be computed to an (absolute) accuracy of $n$ bits, for any $n$-bit floating-point number $x \in (A, B)$, in time $O(M(n) \log^3 n)$.*

NOTES: For a sharper result, see Exercise 4.49. The condition $f(0) = 0$ is just a technical condition to simplify the proof of the theorem; $f(0)$ can be any value that can be computed to $n$ bits in time $O(M(n) \log^3 n)$.

**Proof.** Without loss of generality, we assume $0 \leq x < 1 < B$; the binary expansion of $x$ can then be written $x = 0.b_1 b_2 \ldots b_n$. Define $r_1 = 0.b_1$, $r_2 = 0.0 b_2 b_3$, $r_3 = 0.000 b_4 b_5 b_6 b_7$ (the same decomposition was already used in Algorithm **SinCos**): $r_1$ consists of the first bit of the binary expansion of $x$, $r_2$ consists of the next two bits, $r_3$ the next four bits, and so on. We thus have $x = r_1 + r_2 + \ldots + r_k$ where $2^{k-1} \leq n < 2^k$.

Define $x_i = r_1 + \cdots + r_i$ with $x_0 = 0$. The idea of the algorithm is to translate the Taylor series of $f$ from $x_i$ to $x_{i+1}$; since $f$ is holonomic, this reduces to translating the recurrence on the corresponding coefficients. The condition that $f$ has no singularity in $[0, x] \subset [A, B]$ ensures that the translated recurrence is well-defined. We define $f_0(t) = f(t)$, $f_1(t) = f_0(r_1 + t)$, $f_2(t) = f_1(r_2 + t)$, ..., $f_i(t) = f_{i-1}(r_i + t)$ for $i \leq k$. We have $f_i(t) = f(x_i + t)$, and $f_k(t) = f(x + t)$ since $x_k = x$. Thus we are looking for $f_k(0) = f(x)$.

Let $f_i^*(t) = f_i(t) - f_i(0)$ be the non-constant part of the Taylor expansion of $f_i$. We have $f_i^*(r_{i+1}) = f_i(r_{i+1}) - f_i(0) = f_{i+1}(0) - f_i(0)$ because $f_{i+1}(t) = f_i(r_{i+1} + t)$. Thus:

$$
\begin{aligned}
f_0^*(r_1) + \cdots + f_{k-1}^*(r_k) &= (f_1(0) - f_0(0)) + \cdots + (f_k(0) - f_{k-1}(0)) \\
&= f_k(0) - f_0(0) = f(x) - f(0).
\end{aligned}
$$

Since $f(0) = 0$, this gives:

$$
f(x) = \sum_{i=0}^{k-1} f_i^*(r_{i+1}).
$$

To conclude the proof, we will show that each term $f_i^*(r_{i+1})$ can be evaluated to $n$ bits in time $O(M(n)\log^2 n)$. The rational $r_{i+1}$ has a numerator of at most $2^i$ bits, and

$$0 \leq r_{i+1} < 2^{1-2^i}.$$

Thus, to evaluate $f_i^*(r_{i+1})$ to $n$ bits, $n/2^i + O(\log n)$ terms of the Taylor expansion of $f_i^*(t)$ are enough. We now use the fact that $f$ is holonomic. Assume $f$ satisfies the following homogeneous linear[11] differential equation with polynomial coefficients:

$$c_m(t)f^{(m)}(t) + \cdots + c_1(t)f'(t) + c_0(t)f(t) = 0.$$

Substituting $x_i + t$ for $t$, we obtain a differential equation for $f_i$:

$$c_m(x_i+t)f_i^{(m)}(t) + \cdots + c_1(x_i+t)f_i'(t) + c_0(x_i+t)f_i(t) = 0.$$

From this equation we deduce (see §4.12) a linear recurrence for the Taylor coefficients of $f_i(t)$, of the same order as that for $f(t)$. The coefficients in the recurrence for $f_i(t)$ have $O(2^i)$ bits, since $x_i = r_1 + \cdots + r_i$ has $O(2^i)$ bits. It follows that the $\ell$-th Taylor coefficient of $f_i(t)$ has size $O(\ell(2^i + \log \ell))$. The $\ell \log \ell$ term comes from the polynomials in $\ell$ in the recurrence. Since $\ell \leq n/2^i + O(\log n)$, this is $O(n\log n)$.

However, we do not want to evaluate the $\ell$-th Taylor coefficient $u_\ell$ of $f_i(t)$, but the series

$$s_\ell = \sum_{j=1}^{\ell} u_j r_{i+1}^j \approx f_i^*(r_{i+1}).$$

Noting that $u_\ell = (s_\ell - s_{\ell-1})/r_{i+1}^\ell$, and substituting this value in the recurrence for $(u_\ell)$, say of order $d$, we obtain a recurrence of order $d+1$ for $(s_\ell)$. Putting this latter recurrence in matrix form $S_\ell = M_\ell S_{\ell-1}$, where $S_\ell$ is the vector $(s_\ell, s_{\ell-1}, \ldots, s_{\ell-d})$, we obtain

$$S_\ell = M_\ell M_{\ell-1} \cdots M_{d+1} S_d, \tag{4.81}$$

where the matrix product $M_\ell M_{\ell-1} \cdots M_{d+1}$ can be evaluated in time $O(M(n)\log^2 n)$ using binary splitting. □

---

[11]If $f$ satisfies a non-homogeneous differential equation, say $E(t, f(t), f'(t), \ldots, f^{(k)}(t)) = b(t)$, where $b(t)$ is polynomial in $t$, differentiating it yields $F(t, f(t), f'(t), \ldots, f^{(k+1)}(t)) = b'(t)$, and $b'(t)E(\cdot) - b(t)F(\cdot)$ is homogeneous.

We illustrate Theorem 4.9.1 with the arc-tangent function, which satisfies the differential equation $f'(t)(1 + t^2) = 1$. This equation evaluates at $x_i + t$ to $f_i'(t)(1 + (x_i + t)^2) = 1$, where $f_i(t) = f(x_i + t)$. This gives the recurrence

$$(1 + x_i^2)\ell u_\ell + 2x_i(\ell - 1)u_{\ell-1} + (\ell - 2)u_{\ell-2} = 0$$

for the Taylor coefficients $u_\ell$ of $f_i$. This recurrence translates to

$$(1 + x_i^2)\ell v_\ell + 2x_i r_{i+1}(\ell - 1)v_{\ell-1} + r_{i+1}^2(\ell - 2)v_{\ell-2} = 0$$

for $v_\ell = u_\ell r_{i+1}^\ell$, and to

$$(1 + x_i^2)\ell(s_\ell - s_{\ell-1}) + 2x_i r_{i+1}(\ell - 1)(s_{\ell-1} - s_{\ell-2}) + r_{i+1}^2(\ell - 2)(s_{\ell-2} - s_{\ell-3}) = 0$$

for $s_\ell = \sum_{j=1}^\ell v_j$. This recurrence of order 3 can be written in matrix form, and Eqn. (4.81) enables one to efficiently compute $s_\ell \approx f_i(r_i + 1) - f_i(0)$ using multiplication of $3 \times 3$ matrices and fast integer multiplication.

## 4.10 Contour Integration

In this section we assume that facilities for arbitrary-precision complex arithmetic are available. These can be built on top of an arbitrary-precision real arithmetic package (see Chapters 3 and 5).

Let $f(z)$ be holomorphic in the disc $|z| < R$, $R > 1$, and let the power series for $f$ be

$$f(z) = \sum_{j=0}^\infty a_j z^j. \tag{4.82}$$

From Cauchy's theorem [122, Ch. 7] we have

$$a_j = \frac{1}{2\pi i} \int_C \frac{f(z)}{z^{j+1}} \, dz, \tag{4.83}$$

where $C$ is the unit circle. The contour integral in (4.83) may be approximated numerically by sums

$$S_{j,k} = \frac{1}{k} \sum_{m=0}^{k-1} f(e^{2\pi im/k}) e^{-2\pi ijm/k}. \tag{4.84}$$

Let $C'$ be a circle with centre at the origin and radius $\rho \in (1, R)$. From Cauchy's theorem, assuming that $j < k$, we have (see Exercise 4.50):

$$S_{j,k} - a_j = \frac{1}{2\pi i} \int_{C'} \frac{f(z)}{(z^k - 1)z^{j+1}} \, \mathrm{d}z = a_{j+k} + a_{j+2k} + \cdots, \qquad (4.85)$$

so $|S_{j,k} - a_j| = O((R - \delta)^{-(j+k)})$ as $k \to \infty$, for any $\delta > 0$. For example, let

$$f(z) = \frac{z}{e^z - 1} + \frac{z}{2} \qquad (4.86)$$

be the generating function for the Bernoulli numbers as in (4.57), so $a_{2j} = C_j = B_{2j}/(2j)!$ and $R = 2\pi$ (because of the poles at $\pm 2\pi i$). Then

$$S_{2j,k} - \frac{B_{2j}}{(2j)!} = \frac{B_{2j+k}}{(2j+k)!} + \frac{B_{2j+2k}}{(2j+2k)!} + \cdots, \qquad (4.87)$$

so we can evaluate $B_{2j}$ with relative error $O((2\pi)^{-k})$ by evaluating $f(z)$ at $k$ points on the unit circle.

There is some cancellation when using (4.84) to evaluate $S_{2j,k}$ because the terms in the sum are of order unity but the result is of order $(2\pi)^{-2j}$. Thus $O(j)$ guard digits are needed. In the following we assume $j = O(n)$.

If $\exp(-2\pi i j m/k)$ is computed efficiently from $\exp(-2\pi i/k)$ in the obvious way, the time required to evaluate $B_2, \ldots, B_{2j}$ to precision $n$ is $O(jnM(n))$, and the space required is $O(n)$. We assume here that we need all Bernoulli numbers up to index $2j$, but we do not need to store all of them simultaneously. This is the case if we are using the Bernoulli numbers as coefficients in a sum such as (4.38).

The recurrence relation method of §4.7.2 is faster but requires space $\Theta(jn)$. Thus, the method of contour integration has advantages if space is critical.

For comments on other forms of numerical quadrature, see §4.12.

## 4.11 Exercises

**Exercise 4.1** If $A(x) = \sum_{j \geq 0} a_j x^j$ is a formal power series over $\mathbb{R}$ with $a_0 = 1$, show that $\ln(A(x))$ can be computed with error $O(x^n)$ in time $O(M(n))$, where $M(n)$ is the time required to multiply two polynomials of degree $n - 1$. Assume a reasonable smoothness condition on the growth of $M(n)$ as a function of $n$. [Hint: $(\mathrm{d}/\mathrm{d}x)\ln(A(x)) = A'(x)/A(x)$.] Does a similar result hold for $n$-bit numbers if $x$ is replaced by $1/2$?

**Exercise 4.2 (Schönhage [198] and Schost)** Assume one wants to compute $1/s(x) \bmod x^n$, for $s(x)$ a power series. Design an algorithm using an odd-even scheme (§1.3.5), and estimate its complexity in the FFT range.

**Exercise 4.3** Suppose that $g$ and $h$ are sufficiently smooth functions satisfying $g(h(x)) = x$ on some interval. Let $y_j = h(x_j)$. Show that the iteration

$$x_{j+1} = x_j + \sum_{m=1}^{k-1} (y - y_j)^m \frac{g^{(m)}(y_j)}{m!}$$

is a $k$-th order iteration that (under suitable conditions) will converge to $x = g(y)$. [Hint: generalise the argument leading to (4.16).]

**Exercise 4.4** Design a Horner-like algorithm for evaluating a series $\sum_{j=0}^{k} a_j x^j$ in the forward direction, while deciding dynamically where to stop. For the stopping criterion, assume that the $|a_j|$ are monotonic decreasing and that $|x| < 1/2$. [Hint: use $y = 1/x$.]

**Exercise 4.5** Assume one wants $n$ bits of $\exp x$ for $x$ of order $2^j$, with the repeated use of the doubling formula (§4.3.1), and the naive method to evaluate power series. What is the best reduced argument $x/2^k$ in terms of $n$ and $j$? [Consider both cases $j \geq 0$ and $j < 0$.]

**Exercise 4.6** Assuming one can compute an $n$-bit approximation to $\ln x$ in time $T(n)$, where $n \ll M(n) = o(T(n))$, show how to compute an $n$-bit approximation to $\exp x$ in time $\sim T(n)$. Assume that $T(n)$ and $M(n)$ satisfy reasonable smoothness conditions.

**Exercise 4.7** Care has to be taken to use enough guard digits when computing $\exp(x)$ by argument reduction followed by the power series (4.21). If $x$ is of order unity and $k$ steps of argument reduction are used to compute $\exp(x)$ via

$$\exp(x) = \left( \exp(x/2^k) \right)^{2^k},$$

show that about $k$ bits of precision will be lost (so it is necessary to use about $k$ guard bits).

**Exercise 4.8** Show that the problem analysed in Exercise 4.7 can be avoided if we work with the function

$$\text{expm1}(x) = \exp(x) - 1 = \sum_{j=1}^{\infty} \frac{x^j}{j!}$$

which satisfies the doubling formula $\text{expm1}(2x) = \text{expm1}(x)(2 + \text{expm1}(x))$.

**Exercise 4.9** For $x > -1$, prove the reduction formula

$$\mathrm{log1p}(x) = 2\,\mathrm{log1p}\left(\frac{x}{1 + \sqrt{1+x}}\right)$$

where the function $\mathrm{log1p}(x)$ is defined by $\mathrm{log1p}(x) = \ln(1+x)$, as in §4.4.2. Explain why it might be desirable to work with $\mathrm{log1p}$ instead of $\ln$ in order to avoid loss of precision (in the argument reduction, rather than in the reconstruction as in Exercise 4.7). Note however that argument reduction for $\mathrm{log1p}$ is more expensive than that for $\mathrm{expm1}$, because of the square root.

**Exercise 4.10** Give a numerically stable way of computing $\sinh(x)$ using one evaluation of $\mathrm{expm1}(|x|)$ and a small number of additional operations (compare Eqn. (4.20)).

**Exercise 4.11 (White)** Show that $\exp(x)$ can be computed via $\sinh(x)$ using the formula

$$\exp(x) = \sinh(x) + \sqrt{1 + \sinh^2(x)}.$$

Since

$$\sinh(x) = \frac{e^x - e^{-x}}{2} = \sum_{k \geq 0} \frac{x^{2k+1}}{(2k+1)!},$$

this saves computing about half the terms in the power series for $\exp(x)$ at the expense of one square root. How would you modify this method to preserve numerical stability for negative arguments $x$? Can this idea be used for other functions than $\exp(x)$?

**Exercise 4.12** Count precisely the number of nonscalar products necessary for the two variants of rectangular series splitting (§4.4.3).

**Exercise 4.13** A drawback of rectangular series splitting as presented in §4.4.3 is that the coefficients ($a_{k\ell+m}$ in the classical splitting, or $a_{jm+\ell}$ in the modular splitting) involved in the scalar multiplications might become large. Indeed, they are typically a product of factorials, and thus have size $O(d \log d)$. Assuming that the ratios $a_{i+1}/a_i$ are small rationals, propose an alternate way of evaluating $P(x)$.

**Exercise 4.14** Make explicit the cost of the slowly growing function $c(d)$ (§4.4.3).

**Exercise 4.15** Prove the remainder term (4.28) in the expansion (4.27) for $\mathrm{E}_1(x)$. [Hint: prove the result by induction on $k$, using integration by parts in the formula (4.28).]

**Exercise 4.16** Show that we can avoid using Cauchy principal value integrals by defining $\mathrm{Ei}(z)$ and $\mathrm{E}_1(z)$ in terms of the entire function

$$\mathrm{Ein}(z) = \int_0^z \frac{1 - \exp(-t)}{t}\, \mathrm{d}t = \sum_{j=1}^\infty \frac{(-1)^{j-1} z^j}{j!\, j}\,.$$

**Exercise 4.17** Let $\mathrm{E}_1(x)$ be defined by (4.25) for real $x > 0$. Using (4.27), show that

$$\frac{1}{x} - \frac{1}{x^2} < e^x\, \mathrm{E}_1(x) < \frac{1}{x}\,.$$

**Exercise 4.18** In this exercise the series are purely formal, so ignore any questions of convergence. Applications are given in Exercises 4.19–4.20.

Suppose that $(a_j)_{j \in \mathbb{N}}$ is a sequence with exponential generating function $s(z) = \sum_{j=0}^\infty a_j z^j / j!$. Suppose that $A_n = \sum_{j=0}^n \binom{n}{j} a_j$, and let $S(z) = \sum_{j=0}^\infty A_j z^j / j!$ be the exponential generating function of the sequence $(A_n)_{n \in \mathbb{N}}$. Show that

$$S(z) = \exp(z) s(z).$$

**Exercise 4.19** The power series for $\mathrm{Ein}(z)$ given in Exercise 4.16 suffers from catastrophic cancellation when $z$ is large and positive (like the series for $\exp(-z)$). Use Exercise 4.18 to show that this problem can be avoided by using the power series (where $H_n$ denotes the $n$-th harmonic number)

$$e^z\, \mathrm{Ein}(z) = \sum_{j=1}^\infty \frac{H_j z^j}{j!}\,.$$

**Exercise 4.20** Show that Eqn. (4.23) for $\mathrm{erf}(x)$ follows from Eqn. (4.22). [Hint: this is similar to Exercise 4.19.]

**Exercise 4.21** Give an algorithm to evaluate $\Gamma(x)$ for real $x \geq 1/2$, with guaranteed relative error $O(2^{-n})$. Use the method sketched in §4.5 for $\ln\Gamma(x)$. What can you say about the complexity of the algorithm?

**Exercise 4.22** Extend your solution to Exercise 4.21 to give an algorithm to evaluate $1/\Gamma(z)$ for $z \in \mathbb{C}$, with guaranteed relative error $O(2^{-n})$. *Note:* $\Gamma(z)$ has poles at zero and the negative integers (that is, for $-z \in \mathbb{N}$), but we overcome this difficulty by computing the entire function $1/\Gamma(z)$. *Warning:* $|\Gamma(z)|$ can be very small if $\Im(z)$ is large. This follows from Stirling's asymptotic expansion. In the particular case of $z = iy$ on the imaginary axis we have

$$2\ln|\Gamma(iy)| = \ln\left(\frac{\pi}{y\sinh(\pi y)}\right) \approx -\pi|y|.$$

More generally,

$$|\Gamma(x+iy)|^2 \approx 2\pi|y|^{2x-1}\exp(-\pi|y|)$$

for $x, y \in \mathbb{R}$ and $|y|$ large.

**Exercise 4.23** The usual form (4.38) of Stirling's approximation for $\ln(\Gamma(z))$ involves a divergent series. It is possible to give a version of Stirling's approximation where the series is convergent:

$$\ln\Gamma(z) = \left(z - \frac{1}{2}\right)\ln z - z + \frac{\ln(2\pi)}{2} + \sum_{k=1}^{\infty}\frac{c_k}{(z+1)(z+2)\cdots(z+k)}, \quad (4.88)$$

where the constants $c_k$ can be expressed in terms of *Stirling numbers of the first kind*, $s(n, k)$, defined by the generating function

$$\sum_{k=0}^{n} s(n,k)x^k = x(x-1)\cdots(x-n+1).$$

In fact

$$c_k = \frac{1}{2k}\sum_{j=1}^{k}\frac{j|s(n,j)|}{(j+1)(j+2)}.$$

The Stirling numbers $s(n, k)$ can be computed easily from a three-term recurrence, so this gives a feasible alternative to the usual form of Stirling's approximation with coefficients related to Bernoulli numbers.

Show, experimentally and/or theoretically, that the convergent form of Stirling's approximation is *not* an improvement over the usual form as used in Exercise 4.21.

**Exercise 4.24** Implement procedures to evaluate $\mathrm{E}_1(x)$ to high precision for real positive $x$, using (a) the power series (4.26), (b) the asymptotic expansion (4.27) (if sufficiently accurate), (c) the method of Exercise 4.19, and (d) the continued fraction (4.39) using the backward and forward recurrences as suggested in §4.6. Determine empirically the regions where each method is the fastest.

**Exercise 4.25** Prove the backward recurrence (4.43).

**Exercise 4.26** Prove the forward recurrence (4.44).
[Hint: let
$$y_k(x) = \frac{a_1}{b_1+} \ \cdots \ \frac{a_{k-1}}{b_{k-1}+} \ \frac{a_k}{b_k + x} \,.$$
Show, by induction on $k \geq 1$, that
$$y_k(x) = \frac{P_k + P_{k-1} x}{Q_k + Q_{k-1} x} \,. \ ]$$

**Exercise 4.27** For the forward recurrence (4.44), show that
$$\left( \begin{array}{cc} Q_k & Q_{k-1} \\ P_k & P_{k-1} \end{array} \right) = \left( \begin{array}{cc} b_1 & 1 \\ a_1 & 0 \end{array} \right) \left( \begin{array}{cc} b_2 & 1 \\ a_2 & 0 \end{array} \right) \cdots \left( \begin{array}{cc} b_k & 1 \\ a_k & 0 \end{array} \right)$$
holds for $k > 0$ (and for $k = 0$ if we define $P_{-1}$, $Q_{-1}$ appropriately).
*Remark.* This gives a way to use parallelism when evaluating continued fractions.

**Exercise 4.28** For the forward recurrence (4.44), show that
$$\left| \begin{array}{cc} Q_k & Q_{k-1} \\ P_k & P_{k-1} \end{array} \right| = (-1)^k a_1 a_2 \cdots a_k.$$

**Exercise 4.29** Prove the identity (4.46).

**Exercise 4.30** Prove Theorem 4.6.1.

**Exercise 4.31** Investigate using the continued fraction (4.40) for evaluating the complementary error function $\mathrm{erfc}(x)$ or the error function $\mathrm{erf}(x) = 1 - \mathrm{erfc}(x)$. Is there a region where the continued fraction is preferable to any of the methods used in Algorithm **Erf** of §4.6?

**Exercise 4.32** Show that the continued fraction (4.41) can be evaluated in time $O(M(k) \log k)$ if the $a_j$ and $b_j$ are bounded integers (or rational numbers with bounded numerators and denominators). [Hint: use Exercise 4.27.]

**Exercise 4.33** Instead of (4.54), a different normalisation condition
$$J_0(x)^2 + 2 \sum_{\nu=1}^{\infty} J_\nu(x)^2 = 1 \tag{4.89}$$
could be used in Miller's algorithm. Which of these normalisation conditions is preferable?

**Exercise 4.34** Consider the recurrence $f_{\nu-1} + f_{\nu+1} = 2Kf_{\nu}$, where $K > 0$ is a fixed real constant. We can expect the solution to this recurrence to give some insight into the behaviour of the recurrence (4.53) in the region $\nu \approx Kx$. Assume for simplicity that $K \neq 1$. Show that the general solution has the form

$$f_{\nu} = A\lambda^{\nu} + B\mu^{\nu},$$

where $\lambda$ and $\mu$ are the roots of the quadratic equation $x^2 - 2Kx + 1 = 0$, and $A$ and $B$ are constants determined by the initial conditions. Show that there are two cases: if $K < 1$ then $\lambda$ and $\mu$ are complex conjugates on the unit circle, so $|\lambda| = |\mu| = 1$; if $K > 1$ then there are two real roots satisfying $\lambda\mu = 1$.

**Exercise 4.35** Prove (or give a plausibility argument for) the statements made in §4.7 that: (a) if a recurrence based on (4.59) is used to evaluate the scaled Bernoulli number $C_k$, using precision $n$ arithmetic, then the relative error is of order $4^k 2^{-n}$; and (b) if a recurrence based on (4.60) is used, then the relative error is $O(k^2 2^{-n})$.

**Exercise 4.36** Starting from the definition (4.56), prove Eqn. (4.57). Deduce the relation (4.62) connecting tangent numbers and Bernoulli numbers.

**Exercise 4.37** (a) Show that the number of bits required to represent the tangent number $T_k$ exactly is $\sim 2k \lg k$ as $k \to \infty$. (b) Show that the same applies for the exact representation of the Bernoulli number $B_{2k}$ as a rational number.

**Exercise 4.38** Explain how the correctness of Algorithm **TangentNumbers** (§4.7.2) follows from the recurrence (4.63).

---

**Algorithm 4.5** SecantNumbers

---

**Input:** positive integer $m$
**Output:** Secant numbers $S_0, S_1, \ldots, S_m$
    $S_0 \leftarrow 1$
    **for** $k$ **from** 1 **to** $m$ **do**
        $S_k \leftarrow kS_{k-1}$
    **for** $k$ **from** 1 **to** $m$ **do**
        **for** $j$ **from** $k+1$ **to** $m$ **do**
            $S_j \leftarrow (j-k)S_{j-1} + (j-k+1)S_j$
    **return** $S_0, S_1, \ldots, S_m$.

---

**Exercise 4.39** Show that the complexity of computing the tangent numbers $T_1, \ldots, T_m$ by Algorithm **TangentNumbers** (§4.7.2) is $O(m^3 \log m)$. Assume that the multiplications of tangent numbers $T_j$ by small integers take time $O(\log T_j)$. [Hint: use the result of Exercise 4.37.]

**Exercise 4.40** Verify that Algorithm **SecantNumbers** computes in-place the *Secant numbers* $S_k$, defined by the generating function

$$\sum_{k \geq 0} S_k \frac{x^{2k}}{(2k)!} = \sec x = \frac{1}{\cos x} \,,$$

in much the same way that Algorithm **TangentNumbers** (§4.7.2) computes the Tangent numbers.

**Exercise 4.41 (Harvey)** The generating function (4.56) for Bernoulli numbers can be written as

$$\sum_{k \geq 0} B_k \frac{x^k}{k!} = 1 \left/ \sum_{k \geq 0} \frac{x^k}{(k+1)!} \right. \,,$$

and we can use an asymptotically fast algorithm to compute the first $n+1$ terms in the reciprocal of the power series. This should be asymptotically faster than using the recurrences given in §4.7.2. Give an algorithm using this idea to compute the Bernoulli numbers $B_0, B_1, \ldots, B_n$ in time $O(n^2 (\log n)^{2+\varepsilon})$. Implement your algorithm and see how large $n$ needs to be for it to be faster than the algorithms discussed in §4.7.2.

---

**Algorithm 4.6** SeriesExponential

**Input:** positive integer $m$ and real numbers $a_1, a_2, \ldots, a_m$
**Output:** real numbers $b_0, b_1, \ldots, b_m$ such that
$$b_0 + b_1 x + \cdots + b_m x^m = \exp(a_1 x + \cdots + a_m x^m) + O(x^{m+1})$$
$\quad b_0 \leftarrow 1$
$\quad$**for** $k$ **from** 1 **to** $m$ **do**
$\qquad b_k \leftarrow \left( \sum_{j=1}^{k} j a_j b_{k-j} \right) \big/ k$
$\quad$return $b_0, b_1, \ldots, b_m$.

---

**Exercise 4.42** (a) Show that Algorithm **SeriesExponential** computes $B(x) = \exp(A(x))$ up to terms of order $x^{m+1}$, where $A(x) = a_1 x + a_2 x^2 + \cdots + a_m x^m$ is input data and $B(x) = b_0 + b_1 x + \cdots + b_m x^m$ is the output. [Hint: compare Exercise 4.1.]

(b) Apply this to give an algorithm to compute the coefficients $b_k$ in Stirling's approximation for $n!$ (or $\Gamma(n + 1)$):

$$n! \sim \left(\frac{n}{e}\right) \sqrt{2\pi n} \sum_{k \geq 0} \frac{b_k}{n^k}.$$

[Hint: we know the coefficients in Stirling's approximation (4.38) for $\ln \Gamma(z)$ in terms of Bernoulli numbers.]

(c) Is this likely to be useful for high-precision computation of $\Gamma(x)$ for real positive $x$?

**Exercise 4.43** Deduce from Eqn. (4.69) and (4.70) an expansion of $\ln(4/k)$ with error term $O(k^4 \log(4/k))$. Use any means to figure out an effective bound on the $O()$ term. Deduce an algorithm requiring $x \geq 2^{n/4}$ only to get $n$ bits of $\ln x$.

**Exercise 4.44** Show how both $\pi$ and $\ln 2$ can be evaluated using Eqn. (4.71).

**Exercise 4.45** In §4.8.4 we mentioned that $2\theta_2\theta_3$ and $\theta_2^2 + \theta_3^2$ can be computed using two nonscalar multiplications. For example, we could (A) compute $u = (\theta_2 + \theta_3)^2$ and $v = \theta_2\theta_3$; then the desired values are $2v$ and $u - 2v$. Alternatively, we could (B) compute $u$ and $w = (\theta_2 - \theta_3)^2$; then the desired values are $(u \pm w)/2$. Which method (A) or (B) is preferable?

**Exercise 4.46** Improve the constants in Table 4.1.

**Exercise 4.47** Justify Eqn. (4.80) and give an upper bound on the constant $\alpha$ if the multiplication algorithm satisfies $M(n) = \Theta(n^c)$ for some $c \in (1, 2]$.

**Exercise 4.48 (Salvy)** Is the function $\exp(x^2) + x/(1 - x^2)$ holonomic?

**Exercise 4.49 (van der Hoeven, Mezzarobba)** Improve to $O(M(n) \log^2 n)$ the complexity given in Theorem 4.9.1.

**Exercise 4.50** If $w = e^{2\pi i/k}$, show that

$$\frac{1}{z^k - 1} = \frac{1}{k} \sum_{m=0}^{k-1} \frac{w^m}{z - w^m}.$$

Deduce that $S_{j,k}$, defined by Eqn. (4.84), satisfies

$$S_{j,k} = \frac{1}{2\pi i} \int_{C'} \frac{z^{k-j-1}}{z^k - 1} f(z) \, dz$$

for $j < k$, where the contour $C'$ is as in §4.10. Deduce Eqn. (4.85).

*Remark.* Eqn. (4.85) illustrates the phenomenon of *aliasing*: observations at $k$ points can not distinguish between the Fourier coefficients $a_j$, $a_{j+k}$, $a_{j+2k}$, *etc.*

**Exercise 4.51** Show that the sum $S_{2j,k}$ of §4.10 can be computed with (essentially) only about $k/4$ evaluations of $f$ if $k$ is even. Similarly, show that about $k/2$ evaluations of $f$ suffice if $k$ is odd. On the other hand, show that the error bound $O((2\pi)^{-k})$ following Eqn. (4.87) can be improved if $k$ is odd.

## 4.12 Notes and References

One of the main references for special functions is the "Handbook of Mathematical Functions" by Abramowitz and Stegun [1], which gives many useful results but no proofs. A more recent book is that of Nico Temme [215], and a comprehensive reference is Andrews *et al.* [4]. A large part of the content of this chapter comes from [48], and was implemented in the MP package [47]. In the context of floating-point computations, the "Handbook of Floating-Point Arithmetic" [57] is a useful reference, especially Chapter 11.

The SRT algorithm for division is named after Sweeney, Robertson [190] and Tocher [217]. Original papers on Booth recoding, SRT division, *etc.*, are reprinted in the book by Swartzlander [213]. SRT division is similar to non-restoring division, but uses a lookup table based on the dividend and the divisor to determine each quotient digit. The Intel Pentium `fdiv` bug was caused by an incorrectly initialised lookup table.

Basic material on Newton's method may be found in many references, for example the books by Brent [41, Ch. 3], Householder [126] or Traub [219]. Some details on the use of Newton's method in modern processors can be found in [128]. The idea of first computing $y^{-1/2}$, then multiplying by $y$ to get $y^{1/2}$ (§4.2.3) was pushed further by Karp and Markstein [138], who perform this at the penultimate iteration, and modify the last iteration of Newton's method for $y^{-1/2}$ to directly get $y^{1/2}$ (see §1.4.5 for an example of the Karp-Markstein trick for division). For more on Newton's method for power series, we refer to [43, 52, 56, 143, 151, 203].

Some good references on error analysis of floating-point algorithms are the books by Higham [121] and Muller [175]. Older references include Wilkinson's classics [229, 230].

Regarding doubling versus tripling: in §4.3.4 we assumed that one multiplication and one squaring were required to apply the tripling formula (4.19). However, one might use the form $\sinh(3x) = 3\sinh(x) + 4\sinh^3(x)$, which requires only one cubing. Assuming a cubing costs 50% more than a squaring — in the FFT range — the ratio would be $1.5\log_3 2 \approx 0.946$. Thus, if a specialised cubing routine is available, tripling may sometimes be slightly faster than doubling.

For an example of a detailed error analysis of an unrestricted algorithm, see [69].

The idea of rectangular series splitting to evaluate a power series with $O(\sqrt{n})$ nonscalar multiplications (§4.4.3) was first published in 1973 by Paterson and Stockmeyer [183]. It was rediscovered in the context of multiple-precision evaluation of elementary functions by Smith [205, §8.7] in 1991. Smith gave it the name "concurrent series". Smith proposed modular splitting of the series, but classical splitting seems slightly better. Smith noticed that the simultaneous use of this fast technique and argument reduction yields $O(n^{1/3}M(n))$ algorithms. Earlier, in 1960, Estrin [92] had found a similar technique with $n/2$ nonscalar multiplications, but $O(\log n)$ parallel complexity.

There are several variants of the Euler-Maclaurin sum formula, with and without bounds on the remainder. See for example Abramowitz and Stegun [1, Ch. 23], and Apostol [6].

Most of the asymptotic expansions that we have given in §4.5 may be found in Abramowitz and Stegun [1]. For more background on asymptotic expansions of special functions, see for example the books by de Bruijn [84], Olver [181] and Wong [232]. We have omitted mention of many other useful asymptotic expansions, for example all but a few of those for Bessel functions [226, 228].

Most of the continued fractions mentioned in §4.6 may be found in Abramowitz and Stegun [1]. The classical theory is given in the books by Khinchin [140] and Wall [225]. Continued fractions are used in the manner described in §4.6 in arbitrary-precision packages such as MP [47]. A good recent reference on various aspects of continued fractions for the evaluation of special functions is the *Handbook of Continued Fractions for Special Functions* [83]. In particular, Chapter 7 of this book contains a discussion of error bounds. Our Theorem 4.6.1 is a trivial modification of [83, Theorem 7.5.1]. The asymptotically fast algorithm suggested in Exercise 4.32 was given by Schönhage [196].

A proof of a generalisation of (4.54) is given in [4, §4.9]. Miller's algorithm is due to J. C. P. Miller. It is described, for example, in [1, §9.12, §19.28] and [68, §13.14]. An algorithm is given in [102].

A recurrence based on (4.60) was used to evaluate the scaled Bernoulli numbers $C_k$ in the MP package following a suggestion of Christian Reinsch [48, §12]. Previously, the inferior recurrence (4.59) was widely used, for example in [141] and in early versions of the MP package [47, §6.11]. The idea of using tangent numbers is mentioned in [107, §6.5], where it is attributed to B. F. Logan. Our in-place Algorithms **TangentNumbers** and **SecantNumbers** may be new (see Exercises 4.38–4.40). Kaneko [135] describes an algorithm of Akiyama and Tanigawa for computing Bernoulli numbers in a manner similar to "Pascal's triangle". However, it requires more arithmetic operations than Algorithm **TangentNumbers**. Also, the Akiyama-Tanigawa algorithm is only recommended for exact rational arithmetic, since it is numerically unstable if implemented in floating-point arithmetic. For more on Bernoulli, Tangent and Secant numbers, and a connection with Stirling numbers, see Chen [61] and Sloane [204, A027641, A000182, A000364].

The Von Staudt-Clausen theorem was proved independently by Karl von Staudt and Thomas Clausen in 1840. It can be found in many references. If just a single Bernoulli number of large index is required, then Harvey's modular algorithm [117] can be recommended.

Some references on the Arithmetic-Geometric Mean (AGM) are Brent [43, 46, 51], Salamin [193], the Borweins' book [36], Arndt and Haenel [7]. An early reference, which includes some results that were rediscovered later, is the fascinating report HAKMEM [15]. Bernstein [19] gives a survey of different AGM algorithms for computing the logarithm. Eqn. (4.70) is given in Borwein & Borwein [36, (1.3.10)], and the bound (4.73) is given in [36, p. 11, Exercise 4(c)]. The AGM can be extended to complex starting values provided we take the correct branch of the square root (the one with positive real part): see Borwein & Borwein [36, pp. 15–16]. The use of the complex AGM is discussed in [88]. For theta function identities, see [36, Chapter 2], and for a proof of (4.78), see [36, §2.3].

The use of the exact formula (4.79) to compute $\ln x$ was first suggested by Sasaki and Kanada (see [36, (7.2.5)], but beware the typo). See [46] for Landen transformations, and [43] for more efficient methods; note that the constants given in those papers might be improved using faster square root algorithms (Chapter 3).

The constants in Table 4.1 are justified as follows. We assume we are in the FFT domain, and one Fourier transform costs $M(n)/3$. The $13M(n)/9 \approx 1.444M(n)$ cost for a real reciprocal is from Harvey [116], and assumes $M(n) \sim 3T(2n)$, where $T(n)$ is the time to perform a Fourier transform of size $n$. For the complex reciprocal $1/(v + iw) = (v - iw)/(v^2 + w^2)$, we compute $v^2 + w^2$ using two forward transforms and one backward transform, equivalent in cost to $M(n)$, then one real reciprocal to obtain say $x = 1/(v^2 + w^2)$, then two real multiplications to compute $vx, wx$, but take advantage of the fact that we already know the forward transforms of $v$ and $w$, and the transform of $x$ only needs to be computed once, so these two multiplications cost only $M(n)$. Thus the total cost is $31M(n)/9 \approx 3.444M(n)$. The $1.666M(n)$ cost for real division is from [125, Remark 6], and assumes $M(n) \sim 3T(2n)$ as above for the real reciprocal. For complex division, say $(t + iu)/(v + iw)$, we first compute the complex reciprocal $x + iy = 1/(v + iw)$, then perform a complex multiplication $(t + iu)(x + iy)$, but save the cost of two transforms by observing that the transforms of $x$ and $y$ are known as a byproduct of the complex reciprocal algorithm. Thus the total cost is $(31/9 + 4/3)M(n) \approx 4.777M(n)$. The $4M(n)/3$ cost for the real square root is from Harvey [116], and assumes $M(n) \sim 3T(2n)$ as above. The complex square root uses Friedland's algorithm [97]: $\sqrt{x + iy} = w + iy/(2w)$ where $w = \sqrt{(|x| + (x^2 + y^2)^{1/2})/2}$; as for the complex reciprocal, $x^2 + y^2$ costs $M(n)$, then we compute its square root in $4M(n)/3$, the second square root in $4M(n)/3$, and the division $y/w$ costs $1.666M(n)$, which gives a total of $5.333M(n)$.

The cost of one real AGM iteration is at most the sum of the multiplication cost and of the square root cost, but since we typically perform several iterations it is reasonable to assume that the input and output of the iteration includes the transforms of the operands. The transform of $a+b$ is obtained by linearity from the transforms of $a$ and $b$, so is essentially free. Thus we save one transform or $M(n)/3$ per iteration, giving a cost per iteration of $2M(n)$. (Another way to save $M(n)/3$ is to trade the multiplication for a squaring, as explained in [199, §8.2.5].) The complex AGM is analogous: it costs the same as a complex multiplication $(2M(n))$ and a complex square root $(5.333M(n))$, but we can save two (real) transforms per iteration $(2M(n)/3)$, giving a net cost of $6.666M(n)$. Finally, the logarithm via the AGM costs $2\lg(n) + O(1)$ AGM iterations.

We note that some of the constants in Table 4.1 may not be optimal. For example, it *may* be possible to reduce the cost of reciprocal or square root (Harvey, Sergeev). We leave this as a challenge to the reader (see Exercise 4.46). Note that the constants for operations on power series may differ from the corresponding constants for operations on integers/reals.

There is some disagreement in the literature about "binary splitting" and the "FEE method" of E. A. Karatsuba [137].[12] We choose the name "binary splitting" because it is more descriptive, and let the reader call it the "FEE method" if he/she prefers. Whatever its name, the idea is quite old, since in 1976 Brent [45, Theorem 6.2] gave a binary splitting algorithm to compute $\exp x$ in time $O(M(n)(\log n)^2)$. The CLN library implements several functions with binary splitting [108], and is thus quite efficient for precisions of a million bits or more.

The "bit-burst" algorithm was invented by David and Gregory Chudnovsky [65], and our Theorem 4.9.1 is based on their work. Some references on holonomic func-tions are J. Bernstein [25, 26], van der Hoeven [123] and Zeilberger [234]. See also the Maple GFUN package [194], which allows one, amongst other things, to deduce the recurrence for the Taylor coefficients of $f(x)$ from its differential equation.

---

[12]It is quite common for the same idea to be discovered independently several times. For example, Gauss and Legendre independently discovered the connection between the arithmetic-geometric mean and elliptic integrals; Brent and Salamin independently discovered an application of this to the computation of $\pi$, and related algorithms were known to the authors of [15].

There are several topics that are not covered in this chapter, but might have been if we had more time and space. We mention some references here. A useful resource is the website [144].

The Riemann zeta-function $\zeta(s)$ can be evaluated by the Euler-Maclaurin expansion (4.34)–(4.36), or by Borwein's algorithm [38, 39], but neither of these methods is efficient if $\Im(s)$ is large. On the critical line $\Re(s) = 1/2$, the Riemann-Siegel formula [99] is much faster and in practice sufficiently accurate, although only an asymptotic expansion. If enough terms are taken the error seems to be $O(\exp(-\pi t))$ where $t = \Im(s)$: see Brent's review [82] and Berry's paper [28]. An error analysis is given in [185]. The Riemann-Siegel coefficients may be defined by a recurrence in terms of certain integers $\rho_n$ that can be defined using Euler numbers (see Sloane's sequence A087617 [204]). Sloane calls this the Gabcke sequence but Gabcke credits Lehmer [156] so perhaps it should be called the *Lehmer-Gabcke sequence*. The sequence $(\rho_n)$ occurs naturally in the asymptotic expansion of $\ln(\Gamma(1/4 + it/2))$. The non-obvious fact that the $\rho_n$ are integers was proved by de Reyna [85].

Borwein's algorithm for $\zeta(s)$ can be generalised to cover functions such as the polylogarithm and the Hurwitz zeta-function: see Vepštas [224].

To evaluate the Riemann zeta-function $\zeta(\sigma + it)$ for fixed $\sigma$ and many equally-spaced points $t$, the fastest known algorithm is due to Andrew Odlyzko and Arnold Schönhage [180]. It has been used by Odlyzko to compute blocks of zeros with very large height $t$, see [178, 179]; also (with improvements) by Xavier Gourdon to verify the Riemann Hypothesis for the first $10^{13}$ nontrivial zeros in the upper half-plane, see [105]. The Odlyzko-Schönhage algorithm can be generalised for the computation of other L-functions.

In §4.10 we briefly discussed the numerical approximation of contour integrals, but we omitted any discussion of other forms of numerical quadrature, for example Romberg quadrature, the tanh rule, the tanh-sinh rule, *etc.* Some references are [11, 12, 13, 95, 173, 214], and [37, §7.4.3]. For further discussion of the contour integration method, see [157]. For Romberg quadrature (which depends on Richardson extrapolation), see [58, 189, 192]. For Clenshaw-Curtis and Gaussian quadrature, see [67, 93, 220]. An example of the use of numerical quadrature to evaluate $\Gamma(x)$ is [32, p. 188]. This is an interesting alternative to the method based on Stirling's asymptotic expansion (4.5).

We have not discussed the computation of specific mathematical constants such as $\pi$, $\gamma$ (Euler's constant), $\zeta(3)$, *etc.* $\pi$ can be evaluated using $\pi = 4 \arctan(1)$ and a fast arctan computation (§4.9.2); or by the *Gauss-Legendre* algorithm (also known as the *Brent-Salamin* algorithm), see [43, 46, 193]. This asymptotically fast algorithm is based on the arithmetic-geometric mean and Legendre's relation (4.72). A recent record computation by Bellard [16] used a rapidly-converging series for $1/\pi$ by the Chudnovsky brothers [64], combined with binary splitting. Its complexity is $O(M(n) \log^2 n)$ (theoretically worse than Gauss-Legendre's $O(M(n) \log n)$, but with a small constant factor). There are several popular books on $\pi$: we mention Arndt and Haenel [7]. A more advanced book is the one by the Borwein brothers [36].

For a clever implementation of binary splitting and its application to the fast computation of constants such as $\pi$ and $\zeta(3)$ — and more generally constants defined by hypergeometric series — see Cheng, Hanrot, Thomé, Zima and Zimmermann [62, 63].

The computation of $\gamma$ and its continued fraction is of interest because it is not known whether $\gamma$ is rational (though this is unlikely). The best algorithm for computing $\gamma$ appears to be the "Bessel function" algorithm of Brent and McMillan [54], as modified by Papanikolaou and later Gourdon [106] to incorporate binary splitting. A very useful source of information on the evaluation of constants (including $\pi$, $e$, $\gamma$, $\ln 2$, $\zeta(3)$) and certain functions (including $\Gamma(z)$ and $\zeta(s)$) is Gourdon and Sebah's web site [106].

A nice book on accurate numerical computations for a diverse set of "SIAM 100-Digit Challenge" problems is Bornemann, Laurie, Wagon and Waldvogel [32]. In particular, Appendix B of this book considers how to solve the problems to $10,000$-decimal digit accuracy (and succeeds in all cases but one).

# Chapter 5

# Implementations and Pointers

Here we present a non-exhaustive list of software packages that (in most cases) the authors have tried, together with some other useful pointers. Of course, we can not accept any responsibility for bugs/errors/omissions in any of the software or documentation mentioned here — *caveat emptor!*

Websites change. If any of the websites mentioned here disappear in the future, you may be able to find the new site using a search engine with appropriate keywords.

## 5.1 Software Tools

### 5.1.1 CLN

CLN (Class Library for Numbers, `http://www.ginac.de/CLN/`) is a library for efficient computations with all kinds of numbers in arbitrary precision. It was written by Bruno Haible, and is currently maintained by Richard Kreckel. It is written in C++ and distributed under the GNU General Public License (GPL). CLN provides some elementary and special functions, and fast arithmetic on large numbers, in particular it implements Schönhage-Strassen multiplication, and the binary splitting algorithm [108]. CLN can be configured to use GMP low-level MPN routines, which improves its performance.

## 5.1.2   GNU MP (GMP)

The GNU MP library is the main reference for arbitrary-precision arithmetic. It has been developed since 1991 by Torbjörn Granlund and several other contributors. GNU MP (GMP for short) implements several of the algorithms described in this book. In particular, we recommend reading the "Algorithms" chapter of the GMP reference manual [104]. GMP is written in C, is released under the GNU Lesser General Public License (LGPL), and is available from `http://gmplib.org`.

GMP's MPZ class implements arbitrary-precision integers (corresponding to Chapter 1), while the MPF class implements arbitrary-precision floating-point numbers (corresponding to Chapter 3).[1] The performance of GMP comes mostly from its low-level MPN class, which is well designed and highly optimized in assembly code for many architectures.

As of version 5.0.0, MPZ implements different multiplication algorithms (schoolbook, Karatsuba, Toom-Cook 3-way, 4-way, 6-way, 8-way, and FFT using Schönhage-Strassen's algorithm); its division routine implements Algorithm **RecursiveDivRem** (§1.4.3) in the middle range, and beyond that Newton's method, with complexity $O(M(n))$, and so does its square root, which implements Algorithm **SqrtRem**, since it relies on division. The Newton division first precomputes a reciprocal to precision $n/2$, and then performs two steps of Barrett reduction to precision $n/2$: this is an integer variant of Algorithm **Divide**. It also implements unbalanced multiplication, with Toom-Cook $(3,2)$, $(4,3)$, $(5,3)$, $(4,2)$, or $(6,3)$ [31]. Function `mpn_ni_invertappr`, which is not in the public interface, implements Algorithm **ApproximateReciprocal** (§3.4.1). GMP 5.0.0 does not implement elementary or special functions (Chapter 4), nor does it provide modular arithmetic with an invariant divisor in its public interface (Chapter 2). However, it contains a preliminary interface for Montgomery's **REDC** algorithm.

MPIR is a "fork" of GMP, with a different license, and various other differences that make some functions more efficient with GMP, and some with MPIR; also, the difficulty of compiling under Microsoft operating systems may vary between the forks. Of course, the developers of GMP and MPIR are continually improving their code, so the situation is dynamic. For more on MPIR, see `http://www.mpir.org/`.

---

[1]However, the authors of GMP recommend using MPFR (see §5.1.4) for new projects.

### 5.1.3  MPFQ

MPFQ is a software library developed by Pierrick Gaudry and Emmanuel Thomé for manipulation of finite fields. What makes MPFQ different from other modular arithmetic libraries is that the target finite field is given at *compile time*, thus more specific optimizations can be done. The two main targets of MPFQ are the Galois fields $\mathbb{F}_{2^n}$ and $\mathbb{F}_p$ with $p$ prime. MPFQ is available from `http://www.mpfq.org/`, and is distributed under the GNU Lesser General Public License (LGPL).

### 5.1.4  MPFR

MPFR is a multiple-precision binary floating-point library, written in C, based on the GNU MP library, and distributed under the GNU Lesser General Public License (LGPL). It extends the main ideas of the IEEE 754 standard to arbitrary-precision arithmetic, by providing *correct rounding* and *exceptions*. MPFR implements the algorithms of Chapter 3 and most of those of Chapter 4, including all mathematical functions defined by the ISO C99 standard. These strong semantics are in most cases achieved with no significant slowdown compared to other arbitrary-precision tools. For details of the MPFR library, see `http://www.mpfr.org` and the paper [96].

### 5.1.5  Other Multiple-Precision Packages

Without attempting to be exhaustive, we briefly mention some of MPFR's predecessors, competitors, and extensions.

1. ARPREC is a package for multiple-precision floating-point arithmetic, written by David Bailey *et al.* in C++/Fortran. The distribution includes *The Experimental Mathematician's Toolkit* which is an interactive high-precision arithmetic computing environment. ARPREC is available from `http://crd.lbl.gov/~dhbailey/mpdist/`.

2. MP [47] is a package for multiple-precision floating-point arithmetic and elementary and special function evaluation, written in Fortran77. MP permits any small base $\beta$ (subject to restrictions imposed by the word-size), and implements several rounding modes, though correct rounding-to-nearest is not guaranteed in all cases. MP is now obsolete, and we recommend the use of a more modern package such as MPFR.

However, much of Chapter 4 was inspired by MP, and some of the algorithms implemented in MP are not yet available in later packages, so the source code and documentation may be of interest: see `http://rpbrent.com/pub/pub043.html`.

3. MPC (`http://www.multiprecision.org/`) is a C library for arithmetic using complex numbers with arbitrarily high precision and correct rounding, written by Andreas Enge, Philippe Théveny and Paul Zimmermann [90]. MPC is built on and follows the same principles as MPFR.

4. MPFI is a package for arbitrary-precision floating-point interval arithmetic, based on MPFR. It can be useful to get rigorous error bounds using interval arithmetic. See `http://mpfi.gforge.inria.fr/`, and also §5.3.

5. Several other interesting/useful packages are listed under "Other Related Free Software" at the MPFR website `http://www.mpfr.org/`.

## 5.1.6 Computational Algebra Packages

There are several general-purpose computational algebra packages that incorporate high-precision or arbitrary-precision arithmetic. These include Magma, Mathematica, Maple and Sage. Of these, Sage is free and open-source; the others are either commercial or semi-commercial and not open-source. The authors of this book have often used Magma, Maple and Sage for prototyping and testing algorithms, since it is usually faster to develop an algorithm in a high-level language (at least if one is familiar with it) than in a low-level language like C, where one has to worry about many details. Of course, if speed of execution is a concern, it may be worthwhile to translate the high-level code into a low-level language, but the high-level code will be useful for debugging the low-level code.

1. *Magma* (`http://magma.maths.usyd.edu.au/magma/`) was developed and is supported by John Cannon's group at the University of Sydney. Its predecessor was *Cayley*, a package designed primarily for computational group theory. However, Magma is a general-purpose algebra package with logical syntax and clear semantics. It includes arbitrary-precision arithmetic based on GMP, MPFR and MPC. Although Magma is not open-source, it has excellent online documentation.

2. *Maple* (`http://www.maplesoft.com`) is a commercial package originally developed at the University of Waterloo, now by Waterloo Maple, Inc. It uses GMP for its integer arithmetic (though not necessarily the latest version of GMP, so in some cases calling GMP directly may be significantly faster). Unlike most of the other software mentioned in this chapter, Maple uses radix 10 for its floating-point arithmetic.

3. *Mathematica* is a commercial package produced by Stephen Wolfram's company Wolfram Research, Inc. In the past, public documentation on the algorithms used internally by Mathematica was poor. However, this situation may be improving. Mathematica now appears to use GMP for its basic arithmetic. For information about Mathematica, see `http://www.wolfram.com/products/mathematica/`.

4. *NTL* (`http://www.shoup.net/ntl/`) is a C++ library providing data structures and algorithms for manipulating arbitrary-length integers, as well as vectors, matrices, and polynomials over the integers and over finite fields. For example, it is very efficient for operations on polynomials over the finite field $\mathbb{F}_2$ (that is, GF(2)). NTL was written by and is maintained by Victor Shoup.

5. *PARI/GP* (`http://pari.math.u-bordeaux.fr/`) is a computer algebra system designed for fast computations in number theory, but also able to handle matrices, polynomials, power series, algebraic numbers *etc.* PARI is implemented as a C library, and GP is the scripting language for an interactive shell giving access to the PARI functions. Overall, PARI is a small and efficient package. It was originally developed in 1987 by Christian Batut, Dominique Bernardi, Henri Cohen and Michel Olivier at Université Bordeaux I, and is now maintained by Karim Belabas and many volunteers.

6. *Sage* (`http://www.sagemath.org/`) is a free, open-source mathematical software system. It combines the power of many existing open-source packages with a common Python-based interface. According to the Sage website, its mission is "Creating a viable free open-source alternative to Magma, Maple, Mathematica and Matlab". Sage was started by William Stein and is developed by a large team of volunteers. It uses MPIR, MPFR, MPC, MPFI, PARI, NTL, *etc.* Thus, it is a large system, with many capabilities, but occupying a lot of space and taking a long time to compile.

## 5.2　Mailing Lists

### 5.2.1　The BNIS Mailing List

The BNIS mailing list was created by Dan Bernstein for "Anything of interest to implementors of large-integer arithmetic packages". It has low traffic (a few messages per year only). See `http://cr.yp.to/lists.html` to subscribe. An archive of this list is available at `http://www.nabble.com/cr.yp.to---bnis-f846.html`.

### 5.2.2　The GMP Lists

There are four mailing lists associated with GMP: `gmp-bugs` for bug reports; `gmp-announce` for important announcements about GMP, in particular new releases; `gmp-discuss` for general discussions about GMP; `gmp-devel` for technical discussions between GMP developers. We recommend subscription to `gmp-announce` (very low traffic), to `gmp-discuss` (medium to high traffic), and to `gmp-devel` only if you are interested in the internals of GMP. Information about these lists (including archives and how to subscribe) is available from `http://gmplib.org/mailman/listinfo/`.

### 5.2.3　The MPFR List

There is only one mailing list for the MPFR library. See `http://www.mpfr.org` to subscribe or search through the list archives.

## 5.3　On-line Documents

The *NIST Digital Library of Mathematical Functions* (DLMF) is an ambitious project to completely rewrite Abramowitz and Stegun's classic *Handbook of Mathematical Functions* [1]. It will be published in book form by Cambridge University Press as well as online at `http://dlmf.nist.gov/`. As of February 2010 the project is incomplete, but still very useful. For example, it provides an extensive online bibliography with many hyperlinks at `http://dlmf.nist.gov/bib/`.

The Wolfram Functions Site `http://functions.wolfram.com/` contains a lot of information about mathematical functions (definition, specific values, general characteristics, representations as series, limits, integrals, continued fractions, differential equations, transformations, and so on).

The Encyclopedia of Special Functions (ESF) is another nice web site, whose originality is that all formulæ are automatically generated from very few data that uniquely define the corresponding function in a general class [164]. This encyclopedia is currently being reimplemented in the Dynamic Dictionary of Mathematical Functions (DDMF); both are available from `http://algo.inria.fr/online.html`.

A large amount of information about interval arithmetic (introduction, software, languages, books, courses, applications) can be found on the Interval Computations page `http://www.cs.utep.edu/interval-comp/`.

Mike Cowlishaw maintains an extensive bibliography of conversion to and from decimal arithmetic at `http://speleotrove.com/decimal/`.

Useful if you want to identify an unknown real constant such as $1.414213\cdots$ is the *Inverse Symbolic Calculator* (ISC) by Simon Plouffe (building on earlier work by the Borwein brothers) at `http://oldweb.cecm.sfu.ca/projects/ISC/`.

Finally, an extremely useful site for all kinds of integer/rational sequences is Neil Sloane's *Online Encyclopaedia of Integer Sequences* (OEIS) at `http://www.research.att.com/~njas/sequences/`.

# Index

# Summary of Complexities

| Integer Arithmetic ($n$-bit or $(m, n)$-bit input) | |
|---|---|
| Addition, Subtraction | $O(n)$ |
| Multiplication | $M(n)$ |
| Unbalanced Multiplication ($m \geq n$) | $M(m, n) \leq \lceil \frac{m}{n} \rceil M(n), M(\frac{m+n}{2})$ |
| Division | $O(M(n))$ |
| Unbalanced Division (with remainder) | $D(m + n, n) = O(M(m, n))$ |
| Square Root | $O(M(n))$ |
| $k$-th Root (with remainder) | $O(M(n))$ |
| GCD, extended GCD, Jacobi symbol | $O(M(n) \log n)$ |
| Base Conversion | $O(M(n) \log n)$ |

| Modular Arithmetic ($n$-bit modulus) | |
|---|---|
| Addition, Subtraction | $O(n)$ |
| Multiplication | $M(n)$ |
| Division, Inversion, Conversion to/from RNS | $O(M(n) \log n)$ |
| Exponentiation ($k$-bit exponent) | $O(kM(n))$ |

| Floating-Point Arithmetic ($n$-bit input and output) | |
|---|---|
| Addition, Subtraction | $O(n)$ |
| Multiplication | $M(n)$ |
| Division | $O(M(n))$ |
| Square Root, $k$-th Root | $O(M(n))$ |
| Base Conversion | $O(M(n) \log n)$ |
| Elementary Functions (in a compact set excluding zeros and poles) | $O(M(n) \log n)$ |